%% file: V2_SaYa_Visco-Anelasticity.tex
\documentclass{article}

\usepackage{transparent}
\providecommand{\transparent}[1]{}
\usepackage{authblk}
\usepackage[square]{natbib}
\usepackage{yhmath}
\usepackage{mathtools, amsmath, amsthm, amssymb, amsbsy, amsfonts, amscd}
\usepackage{euscript}
\usepackage{bm}
\usepackage{breqn}
\usepackage{xcolor, soul}
\usepackage{empheq}
\usepackage{rotating}
\usepackage{enumerate}
\usepackage{enumitem}
\usepackage{bbold}

\usepackage{ftnxtra}
\usepackage{fnpos}
\usepackage{appendix}

\usepackage{tikz}
\usepackage{placeins}

\usepackage{graphicx}
\usepackage{pstool}
\usepackage{psfrag}

\usepackage{float}

\usepackage{comment}
\usepackage{transparent}

\numberwithin{equation}{section}

\theoremstyle{plain}

\theoremstyle{definition}	
 \newtheorem{defi}{Definition}[section]
 \newtheorem{remark}{Remark}[section]
 \newtheorem{example}{Example}[section]

\usepackage{subcaption}

\usepackage[all,cmtip]{xy}

\usepackage{hyperref}
\hypersetup{colorlinks=true, linkcolor=blue}
\hypersetup{colorlinks=true,citecolor=blue}

\DeclareMathAlphabet{\mathpzc}{OT1}{pzc}{m}{it}

\usepackage{mathrsfs}

\definecolor{lighter_purple_mathematica}{rgb}{0.6666666666,0.33333333333,0.666666666666}

\makeatletter
\newsavebox{\@brx}
\newcommand{\llangle}[1][]{\savebox{\@brx}{$\m@th{#1\langle}$}%
  \mathopen{\copy\@brx\mkern2mu\kern-0.9\wd\@brx\usebox{\@brx}}}
\newcommand{\rrangle}[1][]{\savebox{\@brx}{$\m@th{#1\rangle}$}%
  \mathclose{\copy\@brx\mkern2mu\kern-0.9\wd\@brx\usebox{\@brx}}}%
\let\oldabs\abs
\def\abs{\@ifstar{\oldabs}{\oldabs*}}
\makeatother

\usepackage{accents}
\newcommand{\Fe}{\accentset{e}{\mathbf{F}}}
\newcommand{\Fa}{\accentset{a}{\mathbf{F}}}
\newcommand{\Fi}{\accentset{i}{\mathbf{F}}}
\newcommand{\Fp}{\accentset{p}{\mathbf{F}}}
\newcommand{\Fv}{\accentset{v}{\mathbf{F}}}
\newcommand{\Fve}{\accentset{ve}{\mathbf{F}}}
\newcommand{\Fth}{\accentset{\scalebox{0.45}{$\vartheta$}}{\mathbf{F}}}
\newcommand{\Fn}{\accentset{n}{\mathbf{F}}}
\newcommand{\cFa}{\accentset{a}{\mathrm F}}
\newcommand{\cFe}{\accentset{e}{\mathrm F}}
\newcommand{\cFv}{\accentset{v}{\mathrm F}}
\newcommand{\cFve}{\accentset{ve}{\mathrm F}}

\newcommand{\Gv}{\accentset{v}{\mathbf{G}}}
\newcommand{\Gi}{\accentset{i}{\mathbf{G}}}
\newcommand{\cGi}{\accentset{i}{\mathrm{G}}}
\newcommand{\Go}{\mathring{\mathbf{G}}}
\newcommand{\cGo}{\mathring{\mathrm{G}}}

\newcommand{\Cve}{\accentset{ve}{\mathbf{C}}}
\newcommand{\cCve}{\accentset{ve}{\mathrm C}}
\newcommand{\cve}{\accentset{ve}{\mathbf{c}}}
\newcommand{\ccve}{\accentset{ve}{\mathrm{c}}}
\newcommand{\bve}{\accentset{ve}{\mathbf{b}}}
\newcommand{\cbve}{\accentset{ve}{\mathrm{b}}}
\newcommand{\Bve}{\accentset{ve}{\mathbf{B}}}
\newcommand{\cBve}{\accentset{ve}{\mathrm{B}}}

\newcommand{\Ce}{\accentset{e}{\mathbf{C}}}
\newcommand{\cCe}{\accentset{e}{\mathrm C}}
\newcommand{\ce}{\accentset{e}{\mathbf{c}}}
\newcommand{\cce}{\accentset{e}{\mathrm{c}}}
\newcommand{\be}{\accentset{e}{\mathbf{b}}}
\newcommand{\cbe}{\accentset{e}{\mathrm{b}}}
\newcommand{\Be}{\accentset{e}{\mathbf{B}}}
\newcommand{\cBe}{\accentset{e}{\mathrm{B}}}

\newcommand{\Cv}{\accentset{v}{\mathbf{C}}}

\newcommand{\Cn}{\accentset{n}{\mathbf{C}}}

\newcommand{\Ive}{\accentset{ve}{I}}
\newcommand{\Ie}{\accentset{e}{I}}

\newcommand{\Bv}{\accentset{v}{\mathbf{B}}}

\newcommand{\Bn}{\accentset{n}{\mathbf{B}}}

\newcommand{\Bth}{\accentset{\scalebox{0.4}{$\vartheta$}}{\mathbf B}}
\newcommand{\Je}{\accentset{e}{J}}
\newcommand{\Jv}{\accentset{v}{J}}
\newcommand{\Ja}{\accentset{a}{J}}
\newcommand{\Jn}{\accentset{n}{J}}
\newcommand{\Jth}{\accentset{\scalebox{0.4}{$\vartheta$}}{J}}
\newcommand{\Lambdav}{\accentset{v}{\boldsymbol{\Lambda}}}
\newcommand{\Lambdai}{\accentset{i}{\boldsymbol{\Lambda}}}
\newcommand{\pv}{\accentset{v}{p}}
\newcommand{\pn}{\accentset{n}{p}}
\newcommand{\Ne}{\accentset{e}{\mathbf{n}}}
\newcommand{\Nve}{\accentset{ve}{\mathbf{n}}}

\newcommand{\sharpo}{\mathring\sharp}
\newcommand{\flato}{\mathring\flat}

\newcommand{\Ds}{\accentset{s}{\mathbf{D}}}
\newcommand{\Da}{\accentset{a}{\mathbf{D}}}

\newcommand{\lvr}{\accentset{v}{\lambda}_R}
\newcommand{\lvt}{\accentset{v}{\lambda}_\Theta}
\newcommand{\lvz}{\accentset{v}{\lambda}_Z}

\usepackage{dsfont}

\DeclareMathOperator\erf{erf}

\usepackage{xcolor}
\definecolor{darkred}{RGB}{180,0,0}

\begin{document}

\title{\textbf{Nonlinear Anisotropic Visco-Anelasticity
}}

\author[1]{Souhayl Sadik\thanks{Corresponding author, e-mail: sosa@mpe.au.dk}}
\author[2,3]{Arash Yavari}
\affil[1]{\small \textit{Department of Mechanical and Production Engineering, Aarhus University, 8000~Aarhus~C, Denmark}}
\affil[2]{\small \textit{School of Civil and Environmental Engineering, Georgia Institute of Technology, Atlanta, GA 30332, USA}}
\affil[3]{\small \textit{The George W. Woodruff School of Mechanical Engineering, Georgia Institute of Technology, Atlanta, GA 30332, USA}}


\maketitle

\vskip 0.0in
{\centering Dedicated to Professor Michael Ortiz on the occasion of his $70$th birthday. \par}
\vskip 0.5in

\begin{abstract}
We formulate a nonlinear geometric theory of visco-anelasticity that unifies viscoelastic and anelastic responses within a single thermodynamic framework. At each material point, the total deformation gradient is multiplicatively decomposed into elastic, viscous, and anelastic distortions, thereby generalizing the Bilby-Kr\"oner-Lee decomposition to visco-anelasticity. The theory explicitly incorporates the material metric, which encodes the evolving natural configuration of the solid, the transformed structural tensors, and provides a consistent formulation of the constitutive equations, the balance laws, the thermodynamic potentials, and the kinetic equations. The first and second laws of thermodynamics are systematically applied to derive the constitutive and evolution equations without invoking observer invariance. Anisotropy is treated in full generality through structural tensors. As illustrative examples, we specialize the general framework to isotropic and transversely isotropic visco-anelastic solids. Two examples within the class of universal deformations, which admit closed-form or partially closed-form solutions, show how the proposed framework can be used to model the coupled viscous and anelastic response of incompressible anisotropic solids with distributed eigenstrains and the associated residual stresses. This geometric framework unifies nonlinear viscoelasticity and anelasticity by coupling time-dependent and eigenstrain-driven effects within a single, fully consistent geometric formulation. In particular, the proposed framework clarifies the geometric structure of the elastic, viscous, and anelastic distortions and resolves ambiguities associated with intermediate configurations in existing formulations of nonlinear viscoelasticity and viscoplasticity.
\end{abstract}

\begin{description}
\item[Keywords:] Nonlinear viscoelasticity; anelasticity; visco-anelastic coupling; multiplicative decomposition; intermediate configuration; anisotropic solids.
\end{description}

\tableofcontents

\section{Introduction}

Anelasticity describes the finite deformation of bodies that, in addition to elastic response, undergo non-elastic distortions or internal microstructural changes driven by physical, chemical, or biological processes such as growth, swelling, plasticity, thermal expansion, diffusion, temperature changes, and line and point defects. These processes induce internal strains referred to as \textit{anelastic strains} or \textit{eigenstrains}, a term derived from the German \textit{Eigenspannungsquellen} meaning ``sources of inherent stresses,” as introduced by \cite{Reissner1931EigenspannungenUE}. Examples include bulk growth, swelling and cavitation, accretion, and various thermal or defect-driven effects \citep{Epstein2000, BenAmarGoriely2005, pence2005swelling, pence2006swelling, pence2007bulk, Yavari2010, Ozakin2010, goriely2010elastic, moulton2011anticavitation, Goriely17, Tomassetti2016, Zurlo2017, Zurlo2018, Truskinovsky2019, Sozio2017, Sozio2019, sadik2016small, Sadik2017Thermoelasticity, YavariGoriely2012a, YavariGoriely2012b, YavariGoriely2013a}. In such models the material undergoes finite deformations under external mechanical forces while also experiencing additional non-elastic distortions or internal microstructural reconfigurations due to these processes, and a defining feature is that stress depends only on the elastic strain, not the total strain. Eigenstrains typically lead to inherent, or residual, stresses.

A key assumption in the modern theory of nonlinear anelasticity is that the total deformation can be locally decomposed multiplicatively as $\mathbf{F} = \Fe\Fa$, where $\Fa$ is a material tensor field encoding anelastic deformations (e.g., due to growth, thermal expansion, or plastic flow), and $\Fe$ denotes the elastic strain relative to the intermediate relaxed configuration. Although the total deformation is compatible, neither $\Fe$ nor $\Fa$ need be. The incompatibility of $\Fe$ and $\Fa$ gives rise to residual stresses in the body. Consequently, unlike elastic materials that admit a stress-free reference configuration in Euclidean space, generic anelastic bodies do not possess such a configuration and are thus referred to as \emph{non-Euclidean solids}. This terminology, originally introduced by Henri Poincar\'e \citep{Poincare1905}, has been used in recent literature to describe bodies with intrinsic geometric incompatibility \citep{Zurlo2017, Zurlo2018, Truskinovsky2019}.
The use of multiplicative decomposition in modeling inelastic deformations has emerged from multiple scientific communities \citep{Sadik2017,YavariSozio2023}. The first systematic theory of nonlinear anelasticity is due to \citet{Eckart1948}, who proposed modifying two foundational assumptions of classical elasticity: what he termed the ``principle of a constant relaxed state” and the ``principle of relaxability-in-the-large.” The first rejects the notion of a fixed, stress-free reference configuration independent of history or loading, while the second challenges the existence of any Euclidean stress-free configuration for anelastic solids.

A significant portion of the modern theory of finite viscoelasticity builds upon the foundational work of \citet{green1946new}, who introduced a thermodynamic theory for rubber-like viscoelastic relaxation. This was extended to finite deformations by \citet{Lubliner1985}, who employed the multiplicative decomposition of the deformation gradient, originally introduced by \citet{bilby1955} and \citet{Kroner1959}. \citet{sidoroff1974} was the first to apply the multiplicative decomposition to viscoelasticity, motivated by earlier geometric formulations of elasto-plasticity \citep{BerdiSedov1967, sidoroff1973geometrical}. In his model, Sidoroff assumed a free energy depending on both the total and elastic parts of the deformation gradient, imposed the second law of thermodynamics via the Clausius--Duhem inequality, and introduced a quadratic dissipation potential following Casimir--Onsager reciprocity to derive the kinetic equations. 
Building on this framework, \citet{Lubliner1985} introduced an internal variable representing the non-elastic part of the deformation gradient and proposed evolution laws governed by linear kinetics. \citet{LeTallec1993}, following assumptions by \citet{Leonov1976}, adopted the multiplicative decomposition for incompressible solids and required both the total and viscous deformation gradients to be volume-preserving. The free energy was additively decomposed into equilibrium and non-equilibrium parts and a dissipation potential was postulated.
\citet{Reese1998} and \citet{ReeseGovindjee1998} extended these ideas by formulating a nonlinear viscoelastic model based on a split of the free energy into equilibrium and non-equilibrium contributions. The equilibrium part depends on the total deformation gradient, while the non-equilibrium part depends solely on the elastic part and vanishes in the relaxed state. The kinetic equations are derived using a positive-definite quadratic dissipation potential ensuring consistency with the second law of thermodynamics.
Other notable contributions to nonlinear viscoelasticity are \citep{holzapfel1996a,Bonet2001,Nguyen2007,Kumar2016}.

A visco-anelastic solid is an anelastic solid which further exhibits a viscoelastic behaviour, i.e.,~time-dependent strain under constant stress and time-dependent stress under constant strain. Examples of such solids include, but are not limited to, viscoplastic solids, thermo-viscoelastic solids, and visco-morphoelastic solids.
It is known for a long time that metallic structures exhibit a combined plastic and viscous response under dynamic loads \citep{Perzyna1971}. The earliest studies of viscoplasticity are due to \citet{Bingham1922} and \citet{Oldroyd1947}, and several reviews of the subject have since appeared \citep{Perzyna1966,Perzyna1971,Chaboche2008}.

Similar to plasticity and viscoelasticity, the popular formulations of nonlinear viscoplasticity have been based on multiplicative decompositions of the deformation gradient.
Similar to viscoelasticity, there have been conceptual mistakes regarding the nature of intermediate configuration(s) in viscoplasticity (see \citet{SaYa2024viscoelasticity} for a detailed critical review of the literature of nonlinear viscoelasticity).
\citet{Nedjar2002} formulated a nonlinear theory of visco-plasticity using two different kinematic assumptions that he called rheological models $1$ and $2$. In the first model two separate multiplicative decompositions are assumed: $\mathbf{F}=\accentset{\infty}{\mathbf{F}}\Fp=\Fe\Fv$, where $\Fp$ is the plastic part of deformation gradient, $\Fv$ is the viscous part of deformation gradient, $\Fe$ is the elastic part of deformation gradient (\citet{Nedjar2002}  called it ``elastic non-equilibrated part"), and $\accentset{\infty}{\mathbf{F}}$ is ``elastic part at equilibrium". If a small piece of material is unloaded independently of the rest of the body, there is an instantaneous elastic unloading, which is due to $\Fe^{-1}$. This is followed by a slow viscous relaxation until one arrives at a local configuration defined by $\Fp$. Thus, $\accentset{\infty}{\mathbf{F}}^{-1}$ is the slow viscoelastic relaxation of the small piece to its stress-free local state defined by $\Fp$. In this sense, the decomposition $\mathbf{F}=\accentset{\infty}{\mathbf{F}}\Fp$ makes sense. However, the decomposition $\mathbf{F}=\Fe\Fv$ does not make sense in the presence of local plastic distortions. In the second model, \citet{Nedjar2002} used the decomposition $\mathbf{F}=\Fe\Fv\Fp$, which is one of the six possible multiplicative decompositions of deformation gradient into elastic, viscous, and plastic distortions.
In formulating nonlinear viscoplaticity and following \citet{Lion2000}, \citet{Shutov2008} used a multiplicative decomposition of the deformation gradient into elastic and inelastic parts: $\mathbf{F}=\Fe\Fi$, and claimed that $\Fi$ corresponds to a stress-free intermediate configuration. However, knowing that $\Fi$ includes both plastic and viscous distortions, this intermediate configuration cannot be stress-free (see a detailed discussion in \citep{SaYa2024viscoelasticity}).
Recently, \citet{Liu2024} formulated a fully nonlinear theory of viscoelastodynamics based on an additive decomposition of the velocity gradient into elastic and viscous parts, drawing motivation from earlier works such as \citep{GreenNaghdi1965,GreenNaghdi1971,Naghdi1990}. They state that ``Moreover, the intermediate configuration can be arbitrarily rotated without affecting the multiplicative decomposition." While this assertion holds for isotropic solids, it is a misconception to assume its general validity. For non-isotropic solids, the symmetry of the intermediate configuration is restricted to the material symmetry group rather than the full $SO(3)$ group. For a detailed discussion, see \citep{YavariSozio2023}.

There have been efforts in the literature on formulating viscoplasticity for anisotropic solids. As an example, \citet{Spencer2001} formulated a theory of viscoplasticity for solids reinforced by two families of inextensible fibers. 
Another example is \citet{Nedjar2007}, who modeled the matrix and each fibre family separately, using a multiplicative decomposition of the deformation gradient and defining evolution equations for the viscous response of both components. Fiber anisotropy and finite extensibility are incorporated through distinct strain energy contributions and internal variables governing the viscoelastic behavior.
\citet{Latorre2016} formulated a finite strain anisotropic visco-hyperelasticity model based on a reverse multiplicative decomposition of the deformation gradient, allowing both the stored energy and viscous contributions to be fully anisotropic.
\citet{Liu2019} formulated a finite strain anisotropic viscoelastic constitutive model for fiber-reinforced soft tissues, based on a multiplicative decomposition of the deformation gradient and distinct anisotropic Helmholtz free energy and dissipation potentials for each fiber family.

In this paper, we will present a general theory of anisotropic visco-anelasticity, which includes visco-anelastic solids reinforced by one or two families of fibers as specific cases.

\paragraph{Contributions of this paper.}
In this paper, we formulate a fully nonlinear geometric theory of visco-anelasticity that unifies viscoelastic and anelastic responses within a single thermodynamic framework. The theory is based on a multiplicative decomposition of the deformation gradient into elastic, viscous, and anelastic distortions, and incorporates the material metric and the transformed structural tensors as geometric objects encoding the natural configuration and anisotropic structure of the body. The main contributions of this work can be summarized as follows:

\begin{itemize}[topsep=0pt,noitemsep, leftmargin=10pt]

\item We formulate a nonlinear geometric theory of visco-anelasticity based on the multiplicative decomposition $\mathbf{F}=\Fe\Fv\Fa$.

\item We clarify the tensorial character of the elastic, viscous, and anelastic distortions and explain the geometric meaning of the corresponding local intermediate configurations.

\item We show that, in the presence of anelasticity, the equilibrium response is governed by the viscoelastic distortion $\Fve=\Fe\Fv$, rather than by the total deformation gradient $\mathbf{F}$.

\item We derive the governing equations from the first and second laws of thermodynamics without invoking observer invariance.

\item We incorporate anisotropy through structural tensors and specialize the theory to isotropic, transversely isotropic, orthotropic, and monoclinic visco-anelastic solids.

\item We derive kinetic equations for the viscous and anelastic distortions in both compressible and incompressible settings.

\item We present examples within the class of universal deformations that illustrate the coupled viscous and anelastic response and the emergence of residual stresses.

\end{itemize}

This paper is organized as follows. In \S\ref{Sec:Kinematics}, the kinematics of visco-anelasticity is formulated, including a discussion of the deformation gradient, the Bilby-Kr\"oner-Lee decomposition, and the introduction of the material metric. \S\ref{Sec:Thermodynamics} presents the thermodynamic formulation, where the balance laws and constitutive equations are derived from the first and second laws of thermodynamics. The implications of material symmetry in visco-anelasticity are discussed in \S\ref{Sec:MaterialSymmetry}, where we consider the effects of structural tensors and classify anisotropic solids based on their symmetry groups. In \S\ref{Sec:Examples}, two examples are solved semi-analytically. Finally, conclusions are given in \S\ref{Sec:Conclusions}.

\section{Kinematics of Visco-Anelasticity} \label{Sec:Kinematics}

In this section, we formulate the kinematics of visco-anelasticity by introducing the necessary geometric and kinematic preliminaries. We begin with a discussion of the deformation gradient and its multiplicative decomposition, which forms the foundation for modeling visco-anelastic solids. The concept of the material metric is then introduced, providing a framework for describing the natural configuration of a body undergoing both viscous and anelastic distortions.

\subsection{Kinematic and geometric preliminaries}
\label{Sec:Kinematics1}

Let us consider a visco-anelastic solid body $B$, and identify it with a $3$-dimensional submanifold $\mathcal{B}$ of the Euclidean ambient space $\mathcal{S}=\mathbb R^3$.
We adopt the standard convention of denoting objects and indices by uppercase characters in the reference configuration $\mathcal{B}$ (e.g.,~$X\in\mathcal{B}$) and by lowercase characters in the Euclidean ambient space $\mathcal{S}$ (e.g.,~$x\in\mathcal{S}$).
Let $\{X^A\}$ and $\{x^a\}$ denote the local coordinate charts on $\mathcal{B}$ and $\mathcal{S}$, respectively. Let $\left\{\partial_A=\frac{\partial}{\partial X^A}\right\}$ and $\left\{\partial_a=\frac{\partial}{\partial x^a}\right\}$ denote the corresponding local coordinate bases, respectively; and let $\left\{\mathrm{d}X^A\right\}$ and $\left\{\mathrm{d}x^a\right\}$ denote the corresponding dual bases, respectively. We also adopt Einstein's repeated index summation convention, e.g.,~$u^i v_i\coloneq \sum_i u^i v_i$.
Let $\operatorname{id}_{\mathcal{B}} : \mathcal{B} \to \mathcal{B}$ ($\operatorname{id}_{\mathcal{S}} : \mathcal{S} \to \mathcal{S}$,~respectively) denote the identity mapping on $\mathcal{B}\,$ ($\mathcal{S}$,~respectively). Its tangent map $\mathbf{I}\coloneq T[\operatorname{id}_{\mathcal{B}}] : T\mathcal{B} \to T\mathcal{B}$ ($\mathbf{i}\coloneq T[\operatorname{id}_{\mathcal{S}}] : T\mathcal{S} \to T\mathcal{S}$,~respectively) acts as the identity tensor on $\mathcal{X}(\mathcal{B})$ ($\mathcal{X}(\mathcal{S})$,~respectively), the set of vector filelds over $\mathcal{B}$ ($\mathcal{S}$,~respectively).

Motion of $\mathcal{B}$ is represented by a time-parameterized family of maps $\varphi_t:\mathcal{B}\to\varphi_t(\mathcal{B})\subset\mathcal{S}$.
The Euclidean ambient space is equipped with the flat metric $\mathbf{g}$, which may be represented as $\mathbf{g}=\mathrm{g}_{ab}\,\mathrm{d}x^a\otimes \mathrm{d}x^b$.
For two vectors $\mathbf{u}\,,\mathbf{w}\in T_x\mathcal{S}$\textemdash the tangent space of $\mathcal S$ at $x$, their dot product is denoted by $\llangle \mathbf{u},\mathbf{w} \rrangle_{\mathbf{g}}=\mathrm{u}^a\,\mathrm{g}_{ab}\,\mathrm{w}^b$. For a vector $\mathbf{u}\in T_x\mathcal{S}$ and a $1$-form $\boldsymbol{\omega}\in T^{*}_x\mathcal{S}$\textemdash the cotangent space of $\mathcal S$ at $x$, their natural pairing is denoted by $\langle \boldsymbol{\omega},\mathbf{u} \rangle=\boldsymbol{\omega}(\mathbf{u})=\omega_a\,\mathrm{u}^a$. The spatial volume form has the coordinate representation ${\mathrm{d}v = \sqrt{\det\mathbf g}\,\mathrm{d}x^1 \wedge \mathrm{d}x^2 \wedge \mathrm{d}x^3}$. We denote by $\bar\nabla$ the Levi-Civita connection of $(\mathcal{S},\mathbf{g})$, and we denote its Christoffel symbols by ${\gamma^a}_{bc}$ in the local coordinate chart $\{x^a\}$.

The background Euclidean metric $\mathbf{g}$ induces a Euclidean metric $\Go=\iota^*\mathbf{g}$ on $\mathcal{B}$ as the pull-back of the ambient space metric by the inclusion map $\iota:\mathcal{B}\hookrightarrow\mathcal{S}$.\footnote{Note however that due to the presence of eigenstrains (anelasticity) in the solid, $\Go$ is not necessarily the material metric measuring the natural (stress-free) distances in the undeformed body $\mathcal B$. We will introduce the material metric in \S\ref{S:Mat_Metric}.} It may be represented as ${\Go=\mathring{\mathrm{G}}_{AB}\,\mathrm{d}X^A\otimes \mathrm{d}X^B}$.\footnote{Note that $\Go=\iota^*\mathbf{g}=(T\iota)^\star\mathbf{g}\,T\iota$ may be thought of as the restriction of $\mathbf g$ on $\mathcal B \subset \mathcal S$.}
If $\{X^A\}$ is a Cartesian coordinate system, the metric is written as ${\mathbf{G}=\delta_{AB}\,\mathrm{d}X^A\otimes \mathrm{d}X^B}$. For example, in cylindrical coordinates $\{R,\vartheta,Z\}$, it reads ${\mathbf{G} = dR \otimes dR + R^2\,d\vartheta \otimes d\vartheta + dZ \otimes dZ}$.
For two vectors $\mathbf{U}\,,\mathbf{W}\in T_X\mathcal{B}$\textemdash the tangent space of $\mathcal B$ at $X$,  their dot product with respect to $\Go$ is denoted by $\llangle \mathbf{U},\mathbf{W} \rrangle_{\Go}=\mathrm{U}^A\,\mathring{\mathrm{G}}_{AB}\,\mathrm{W}^B$. Given a vector $\mathbf{U}\in T_X\mathcal{B}$ and a $1$-form $\boldsymbol{\Omega}\in T^{*}_X\mathcal{B}$\textemdash the cotangent space of $\mathcal B$ at $X$,  their natural pairing is denoted by $\langle \boldsymbol{\Omega},\mathbf{U} \rangle=\boldsymbol{\Omega}(\mathbf{U})=\Omega_A\,\mathrm{U}^A$. The volume form of the manifold $(\mathcal{B},\Go)$ is written as $ \mathrm{d}\mathring{V} = \sqrt{\det\Go} \,\mathrm{d}X^1 \wedge \mathrm{d}X^2 \wedge \mathrm{d}X^3$.
We denote by $\mathring{J}$ the Jacobian of the motion with respect to the Euclidean reference configuration $(\mathcal{B},\Go)$, which is such that $\varphi^*\mathrm{d}v = \mathring{J} \, \mathrm{d}\mathring{V}$. It can be shown that

\begin{equation}
	\mathring{J}=\sqrt{\frac{\det\mathbf g}{\det\Go}}\,\det\mathbf{F} \,.
\end{equation}
Let $\mathring\nabla$ be the Levi-Civita connection of $(\mathcal{B},\Go)$. We denote its Christoffel symbols by ${\mathring{\Gamma}^A}_{BC}$ in the local coordinate chart $\{X^A\}$.

We define the material velocity $\mathbf V$ of the motion as ${\mathbf{V}:\mathcal{B}\times\mathbb{R}^+\to T \mathcal{S}\,, \mathbf{V}(X,t)\coloneq \partial \varphi(X,t)/\partial t}\,$\textemdash it has components $\mathrm{V}^a = \frac{\partial \varphi^a}{\partial t}$. We define the spatial velocity as ${\mathbf{v}:\varphi_t(\mathcal{B})\times\mathbb{R}^+\to T \mathcal{S}}$,  ${\mathbf{v}(x,t)\coloneq \mathbf{V}(\varphi_t^{-1}(x),t)}$.
We define the material acceleration as ${\mathbf{A}:\mathcal{B}\times\mathbb{R}^+\to T \mathcal{S}}$,  ${\mathbf{A}(X,t)\coloneq \operatorname{D}_t^{\mathbf g}\mathbf{V}(X,t)}$,  where $\operatorname{D}_t^{\mathbf g}$ denotes the covariant derivative along $\varphi_X:t\mapsto \varphi(X,t)$. In components, $\mathrm{A}^a=\frac{\partial \mathrm{V}^a}{\partial t}+{\gamma^a}_{bc}\,\mathrm{V}^b\,\mathrm{V}^c$.\footnote{Let $\mathbf{U}(X,t)$ and $\mathbf{W}(X,t)$ be vector fields along the motion $\varphi_t(X)$, i.e.,~
$\mathbf{U}(X,t),\mathbf{W}(X,t)\in T_{\varphi_t(X)}\mathcal{S}$ for each $t$. 
The covariant time derivative $\operatorname{D}_t^{\mathbf g}$ is defined as the unique derivative along the curve 
$\varphi_t(X)$ that is compatible with the Levi-Civita connection of $\mathbf g$ and satisfies the Leibniz rule
\begin{equation}
	\frac{d}{dt}\llangle \mathbf{U},\mathbf{W}\rrangle_{\mathbf g}
	=\llangle \operatorname{D}_t^{\mathbf g}\mathbf{U},\mathbf{W}\rrangle_{\mathbf g}
	+\llangle \mathbf{U},\operatorname{D}_t^{\mathbf g}\mathbf{W}\rrangle_{\mathbf g}
	\,.
\end{equation}
In local coordinates, for $\mathbf{U}=U^a\,\partial/\partial x^a$, one has
\begin{equation}
	(\operatorname{D}_t^{\mathbf g}\mathbf{U})^a
	=\frac{\partial U^a}{\partial t}
	+\gamma^a{}_{bc}\,U^b\,V^c
	\,,
\end{equation}
where $\mathbf{V}=\partial\varphi_t/\partial t$ and $\gamma^a{}_{bc}$ are the Christoffel symbols of $\mathbf g$. 
This can also be calculated by using the local coordinate representation $\mathbf{U}(X,t)=U^a(X,t)\,\mathbf{e}_a(\varphi_t(X))$. Thus,
\begin{equation}\label{eq:DtU}
	\operatorname{D}_t^{\mathbf g}\mathbf{U}
	=\frac{\partial U^a}{\partial t}\,\mathbf{e}_a +U^a\, \nabla_{\mathbf{e}_b} \mathbf{e}_a\,V^b
	=\frac{\partial U^a}{\partial t}\,\mathbf{e}_a +U^a\, \gamma^c{}_{ba} \,\mathbf{e}_c\,V^b
	=\frac{\partial U^a}{\partial t}\,\mathbf{e}_a +U^c\, \gamma^a{}_{bc} \,\mathbf{e}_a\,V^b
	=\!\left(\frac{\partial U^a}{\partial t} + \gamma^a{}_{bc} \,V^b\,U^c\right)\!\mathbf{e}_a
	\,.
\end{equation}
}
We define the spatial acceleration as $\mathbf{a}:\varphi_t(\mathcal{B})\times\mathbb{R}^+\to T \mathcal{S}$,  $\mathbf{a}(x,t)\coloneq \mathbf{A}(\varphi_t^{-1}(x),t)\in T_x\mathcal{S}\,$\textemdash it has components $\mathrm{a}^a=\frac{\partial\mathrm{v}^a}{\partial t}+\frac{\partial\mathrm{v}^a}{\partial x^b}\,\mathrm{v}^b+{\gamma^a}_{bc}\,\mathrm{v}^b\,\mathrm{v}^c$.\footnote{
Let $u(x,t)=U(\varphi_t^{-1}(x),t)$, which amounts to writing $u(\varphi_t(X),t)=U(X,t)$. Hence, one may write following~\eqref{eq:DtU}
\begin{equation}
	\operatorname{D}_t^{\mathbf g}\mathbf{u}
	=\operatorname{D}_t^{\mathbf g}\mathbf{U}
	=\!\left(\frac{\partial U^a}{\partial t} + \gamma^a{}_{bc} \,V^b\,U^c\right)\!\mathbf{e}_a
	=\!\left(\frac{\partial u^a}{\partial t} + \frac{\partial u^a}{\partial x^b} \frac{\partial \varphi^b_t(X)}{\partial t} + \gamma^a{}_{bc} \,v^b\,u^c\right)\!\mathbf{e}_a
	=\!\left(\frac{\partial u^a}{\partial t} + \frac{\partial u^a}{\partial x^b} v^b + \gamma^a{}_{bc} \,v^b\,u^c\right)\!\mathbf{e}_a
	\,.
\end{equation}
}
\subsection{The deformation gradient and the Bilby-Kröner-Lee decomposition}
\label{S:FeFvFa}

\begin{figure}[t!]
\centering
\includegraphics[width=0.80\textwidth]{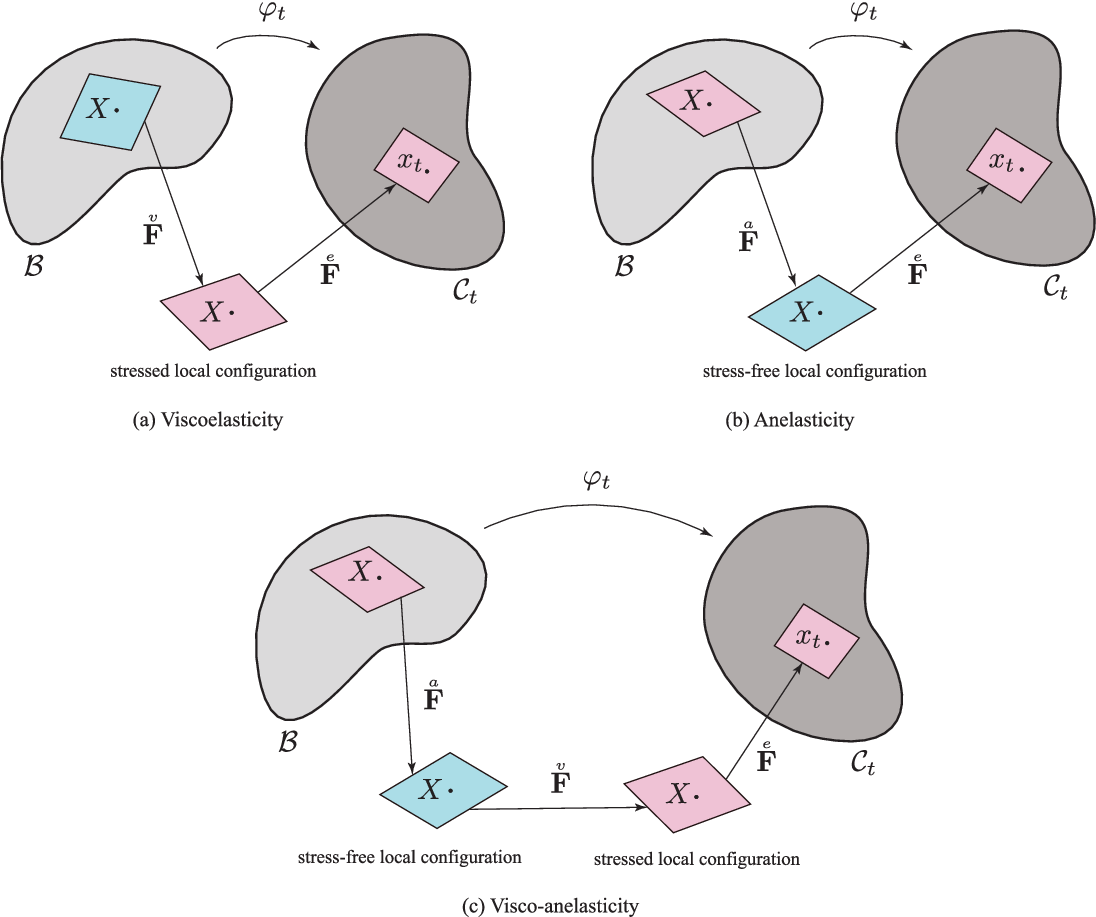}
\vspace*{0.10in}
\caption{The local intermediate configurations in viscoelasticity, anelasticity, and visco-anelasticity. Blue and pink rhombi indicate locally stress-free and locally stressed configurations, respectively.
(a) In viscoelasticity, the local intermediate configuration is stressed, and the material manifold is the Euclidean manifold $(\mathcal{B},\Go)$. (b) In anelasticity, the local intermediate configuration is stress-free, and the material manifold is $(\mathcal{B},\mathbf{G})$, where $\mathbf{G}$ is the non-flat material metric, which is related to the Euclidean metric $\Go$ via pull-back, i.e., $\mathbf{G}=\protect\Fa^\star\Go \protect\Fa$. (c) In visco-anelasticity there are two local intermediate configurations, one stressed, and one stress-free. Similar to anelasticity, the material manifold is $(\mathcal{B},\mathbf{G})$, where $\mathbf{G}$ is the non-flat material metric, which is related to the Euclidean metric $\Go$ via pull-back. } 
\label{Local-Configurations}
\end{figure}

To describe the local strain in an elastic solid, we introduce the deformation gradient as the tangent of the deformation mapping, i.e.,~$\mathbf{F}(X,t) = T\varphi_t(X): T_X\mathcal{B} \to T_{\varphi_t(X)}\varphi_t(\mathcal{B})\,$\textemdash it has components ${\mathrm{F}^a}_A=\partial \varphi^a/\partial X^A$. 
The deformation gradient is a two-point tensor, and its coordinate representation is given by
\begin{equation} \label{Deformation-Gradient-Representation}
	\mathbf{F}(X,t) = F^a{}_A(X,t)\, \frac{\partial}{\partial x^a}\otimes \mathrm{d}X^A\,.
\end{equation}
In this subsection we discuss the Bilby--Kr\"oner--Lee (BKL) decomposition \citep{Bilby1957,Kroner1959,Lee1967,Lee1969,Sadik2017,YavariSozio2023} in the context of nonlinear visco-anelasticity.

Let us consider a visco-anelastic solid in its current (deformed) configuration.
If we proceed to unload the body, we observe an instantaneous elastic partial relaxation followed by a slow viscoelastic generally incomplete relaxation into a residually-stressed state due to anelasticity.
Locally, on a volume element (i.e.,~an infinitesimally small neighborhood) in the current configuration, there are tractions acting on its boundary and keeping it in its deformed state. Independently of the rest of the body, we let this element be unloaded. Subsequently, there is an instantaneous elastic relaxation $\Fe^{-1}$ into a stressed state followed by a slow viscoelastic relaxation $\Fv^{-1}$ eventually leading, at large times, to a local stress-free state, unlike the global unloading experiment described above. Due to the presence of eigenstrains (anelastic strains), this locally unloaded stress-free state is, in general, different from the initial undeformed volume element. It is in fact the realization of anelastic effects in the solid, as a local incompatible distortion $\Fa$ of the initial undeformed volume element to this stress-free state.
The thought experiment just described motivates the multiplicative decomposition $\mathbf{F}=\Fe\Fv\Fa$.
Suppose that this local relaxation is performed independently everywhere in the solid.
The local volume elements, after their instantaneous elastic unloading, cannot, in general, be put together into a single compatible configuration in Euclidean space; their collective configuration generally differs from that of the entire body after its instantaneous global partial unloading.
This is expected, since the instantaneous elastic distortion $\Fe$ is, in general, incompatible.
Further, unlike in viscoelastic bodies, the set of unloaded, relaxed volume elements cannot, in general, be assembled into a compatible configuration in Euclidean space; moreover, this set differs from the set of volume elements obtained by unloading the residually stressed body as a whole.
This is expected as well, since the anelastic distortion $\Fa$ is, in general, incompatible. Note that it follows from the above discussion that the viscous distortion $\Fv$ is also, in general, incompatible.
In anelasticity (and also viscoelasticity) there are two possibilities for the multiplicative decomposition: $\mathbf{F}=\Fe\Fa$ or $\mathbf{F}=\accentset{a}{\boldsymbol{\mathds{F}}}\accentset{e}{\boldsymbol{\mathds{F}}}$ \citep{YavariSozio2023}. In visco-anelasticity, there are $3!=6$ possibilities; in this work, we choose the decomposition
\begin{equation}
\label{eq:BKL}
\mathbf{F}=\Fe\Fv\Fa\,.
\end{equation}
We also introduce a shorthand notation for the viscoelastic distortion $\Fve=\Fe\Fv$.

In our study of viscoelasticity \citep{SaYa2024viscoelasticity}, we showed that when assuming an additive decomposition of the free energy into equilibrium and non-equilibrium parts \citep{Reese1998} and using the direct multiplicative decomposition $\mathbf{F}=\Fe\Fv$, the elastic and viscous distortions have the following tensorial characters: ${\Fv(X,t):T_X\mathcal{B}\to T_X\mathcal{B}}$, and ${\Fe(X,t):T_X\mathcal{B}\to T_x\mathcal{C}}$, where $x=\varphi_t(X)$.
Using similar arguments, one finds in nonlinear visco-anelasticity that when assuming the additive decomposition of the free energy\footnote{The additive decomposition of the free energy is later discussed in \S\ref{Additive_decomp}.} and the multiplicative decomposition $\mathbf{F}=\Fe\Fv\Fa$, the elastic, viscous, and anelastic distortions must have the following tensorial characters (see Fig.~\ref{Local-Configurations}):
\begin{equation}
	\Fa(X,t):T_X\mathcal{B}\to T_X\mathcal{B}\,, \qquad
	\Fv(X,t):T_X\mathcal{B}\to T_X\mathcal{B}\,, \qquad  \Fe(X,t):T_X\mathcal{B}\to T_x\mathcal{C}\,.
\end{equation}
Note, however, that anelasticity in a solid may emerge from different sources, e.g., temperature changes, growth, remodeling, swelling, plastic deformation, etc. Assuming $N$ sources of anelasticity, each with its corresponding distortion $\accentset{j}{\mathbf{F}}$, the anelastic distortion may hence be written as $\Fa=\prod_{j=1}^{N} \accentset{j}{\mathbf{F}}\,=\accentset{1}{\mathbf{F}}\hdots\accentset{N}{\mathbf{F}}$.

\begin{figure}[hbt!]
	\begin{center}
	\vskip 0.0 in
	\includegraphics[scale=0.75,angle=0]{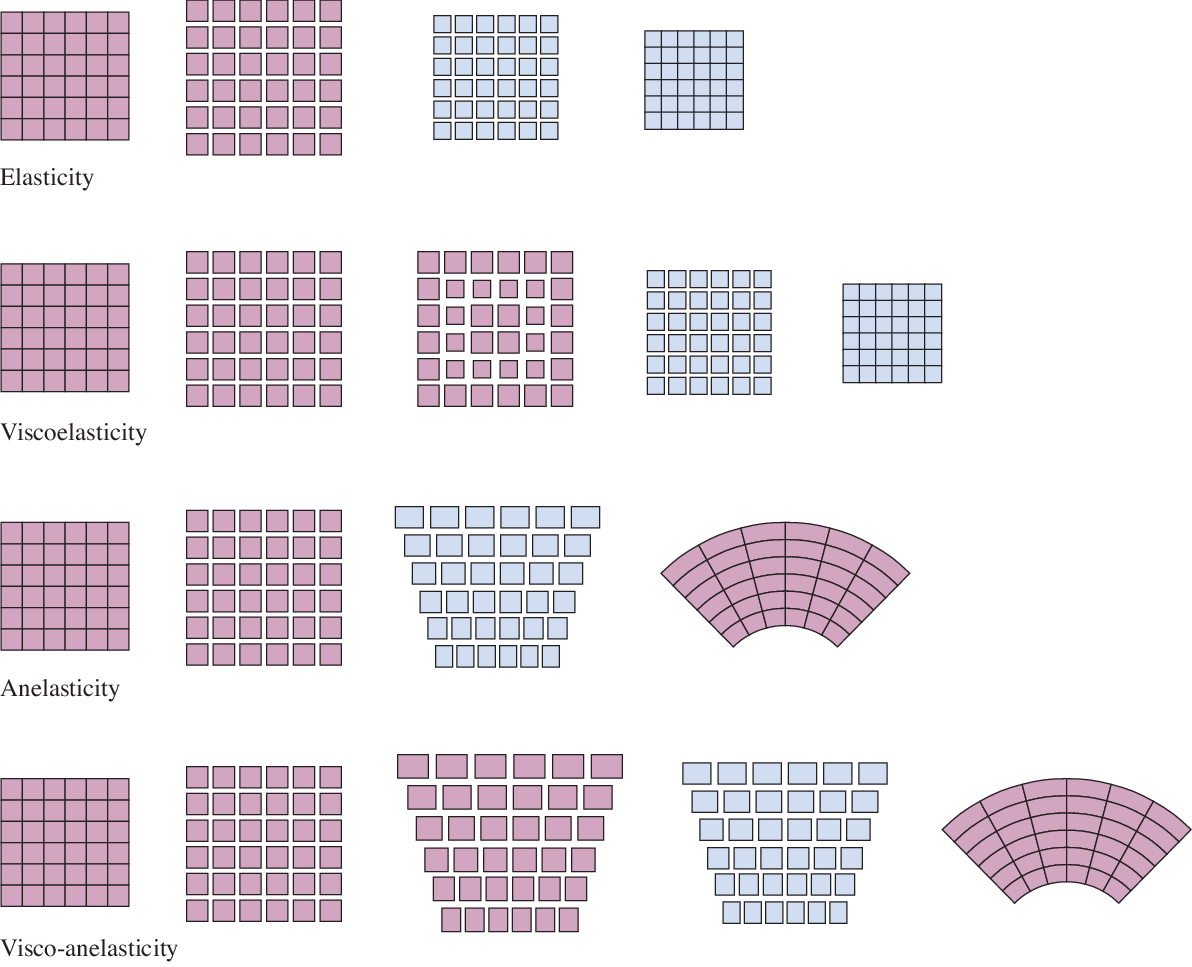}
	\end{center}
	\vskip -0.0 in
	\caption{\footnotesize Schematic illustration of the elastic, viscoelastic, anelastic, and visco-anelastic responses of a sheet, illustrating the relation between local relaxation, compatibility, and the emergence of residual stress. Blue and pink indicate locally stress-free and locally stressed states, respectively.}
\label{Visco-Elastic-Relaxation}
\end{figure}

In order to visualize the differences between elasticity, viscoelasticity, anelasticity, and visco-anelasticity, consider the square sheet shown in Fig.~\ref{Visco-Elastic-Relaxation}. In each case, the leftmost configuration represents a loaded, stressed state (the applied loads are omitted for clarity), and the subsequent stages illustrate different forms of unloading.

\noindent\textbf{Elasticity.} Consider an elastic sheet in a loaded, stressed configuration. We imagine partitioning the sheet into a collection of small square elements. Next, we cut these elements from the sheet while applying the necessary boundary tractions to keep each element in its deformed square shape. We then elastically unload each element by removing these tractions. Since the body is purely elastic, each unloaded element relaxes to its natural stress-free shape. The relaxed elements remain compatible and can be reassembled to recover a single global stress-free sheet.

\noindent\textbf{Viscoelasticity.} Consider a loaded viscoelastic sheet in its current configuration, and imagine partitioning it into a collection of small square elements. Next, we cut these elements from the sheet while applying the necessary boundary tractions to keep each element in its deformed square shape. We then remove these tractions. The instantaneous elastic unloading gives the third state, in which each element is still stressed. In other words, after this instantaneous elastic relaxation, the elements have not yet reached their natural stress-free shapes. As time passes, each element undergoes viscous relaxation, and for sufficiently long times one obtains a completely relaxed configuration. In this final configuration, the elements are stress-free, remain compatible, and can be reassembled to recover a single global relaxed sheet.

\noindent\textbf{Anelasticity.} Consider a loaded anelastic sheet with a distribution of eigenstrains, and imagine partitioning it into a collection of small elements. Next, we cut these elements from the sheet and let each element relax independently. Each element relaxes to its own local stress-free configuration determined by the prescribed eigenstrain. However, these locally relaxed elements are not, in general, compatible. In other words, they cannot be put back together to form a single global stress-free sheet unless residual stresses are induced. In the present sketch, because of the particular distribution of eigenstrains, the configuration adopted by the sheet in the absence of external forces is a bent residually-stressed sheet.

\noindent\textbf{Visco-anelasticity.} Consider a loaded visco-anelastic sheet with a distribution of eigenstrains, and imagine partitioning it into a collection of small elements. Next, we cut these elements from the sheet while applying the necessary boundary tractions to keep each element in its deformed shape, and then remove these tractions. The instantaneous elastic unloading leads to a state in which each element is still stressed. As time passes, each element undergoes viscous relaxation, and after sufficiently long times it reaches its fully relaxed local stress-free configuration. However, these fully relaxed elements are not, in general, compatible. In other words, they cannot be reassembled into a single global stress-free sheet without residual stresses. In the present sketch, because of the particular distribution of eigenstrains, the configuration adopted after sufficiently long time, in the absence of external forces, is a bent residually-stressed sheet.

\begin{remark}[Thermal distortion]\label{rmrk:thermal_distortion}
In particular, if the solid is subject to a non-isothermal process, i.e.,~temperature is time dependent $\vartheta = \vartheta(X, t)$, thermal distortion arises as a distinct form of anelastic distortion.
This effect yields local, typically incompatible, distortions driven by temperature variations \citep{stojanovic1964finite, Ozakin2010, Sadik2017Thermoelasticity}.
Accordingly, the total anelastic distortion $\Fa$ can be decomposed into a thermal distortion component $\Fth$, and an athermal distortion $\Fn$ encapsulating all other coexisting anelastic mechanisms within the material, such that
\begin{equation} \label{eq:FaFnFth}
	\Fa = \Fn \Fth\,.
\end{equation}
It can be shown, under the constitutive assumption of (linear) thermal expansion, that thermal distortion relates to temperature as follows \citep{SaYa2025GenColemanNoll}:
\begin{equation} \label{eq:Th_distort}
	\boldsymbol\alpha = \frac{\partial\Fth}{\partial\vartheta} \,\Fth^{-1} = \Fth^{-1} \frac{\partial\Fth}{\partial\vartheta}\,,
\end{equation}
where $\boldsymbol\alpha$ denotes the (linear) thermal expansion tensor of the material. Note that it can further be shown that \citep{SaYa2025GenColemanNoll}
\begin{equation}
\label{eq:Fth}
	\Fth(X,\vartheta) 
	= \exp\!\left[\boldsymbol{\omega}(X,\vartheta)\right]\!
	= \exp\!\left[\int_{\vartheta_0}^\vartheta \boldsymbol{\alpha}(X,\tilde\vartheta)\, d\tilde\vartheta\right]\!
	\,.
\end{equation}
\end{remark}

\begin{remark}[Thermal coupling]
\label{rmrk:ind_fields}
Note that the deformation gradient $\mathbf{F}$ and the temperature field $\vartheta$ are \emph{a priori} independent fields of the theory, i.e., independent before any constitutive assumption, constraint, or balance law is imposed. The BKL decomposition $\mathbf{F}=\Fe \Fv \Fn \Fth$ (following~\eqref{eq:BKL} and~\eqref{eq:FaFnFth}) is a kinematic ansatz that assigns a distortion to each physical mechanism and enlarges the set of kinematic fields. Since each factor is invertible, any one of the four distortions is determined by $\mathbf{F}$ and the remaining three. Hence, for simple materials in the sense of \citet{noll1958}, the kinematic state is specified by $\mathbf{F}$, $\vartheta$, and any three distortions from $\{\Fe,\Fv,\Fn,\Fth\}$. Note that $\mathbf{F}$ cannot be traded for the four distortions without constraining their product to be compatible. Kinematics places no relation among these fields, and any coupling arises \emph{a posteriori} through the constitutive relations, the kinematic constraints, and the balance laws. In particular, writing $\Fth=\Fth(X,\vartheta)$ as in~\eqref{eq:Fth} is a constitutive choice, and one that should not be made prior to the derivation of the governing equations: $\vartheta$ and $\Fth$ enter the theory as independent arguments.
\end{remark}

\subsection{Natural configuration and material metric}
\label{S:Mat_Metric}

In a visco-anelastic body\textemdash just as in an anelastic body,\footnote{See for example \citep{sidoroff1973geometrical, Simo1988, lubarda2004constitutive, YavariSozio2023}.} due to the presence of eigenstrains, the natural configuration cannot, in general, be isometrically embedded in the Euclidean ambient space, i.e.,~it cannot be realized as a stress-free body in the physical space. In fact, the resulting residually-stressed state is an expression of the non-Euclidean character of the natural (material) configuration of the body. Hence, the manifold $\mathcal B$ ought to be endowed with a material metric $\mathbf G$ that is not necessarily equal to $\Go$, the restriction of the Euclidean metric on $\mathcal B$.
Let us consider a curve in the reference configuration, i.e.,~$\gamma: I \to \mathcal{B}$, where $I$ is an open interval in $\mathbb{R}$.
The arc-length element squared of $\gamma$ at a point $p=\gamma(t)\in\mathcal B$ is written as $\llangle\gamma'(t),\gamma'(t)\rrangle_{\Go}$.
Recall that, due to the presence of eigenstrains, the curve $\gamma$ is generally stressed in the Euclidean reference configuration $(\mathcal{B},\Go)$. However, its push-forward by the anelastic distortion $\Fa_*\gamma$ is locally stress-free (see Fig.~\ref{Local-Configurations}), and its arc-length element\textemdash the natural arc-length of $\gamma$\textemdash squared at $p=\gamma(t)\in\mathcal B$ is given by $\llangle\Fa_*\gamma'(t),\Fa_*\gamma'(t)\rrangle_{\Go}=\llangle\Fa\gamma'(t),\Fa\gamma'(t)\rrangle_{\Go}=\llangle\gamma'(t),\gamma'(t)\rrangle_{\Fa^*\Go}$. Therefore, the material metric measuring the natural distances in the material (stress-free) configuration is given by the pull-back of the Euclidean metric $\Go$ by the anelastic distortion $\Fa$, i.e.,~$\mathbf{G}=\Fa^*\Go\,$ \citep{Yavari2021Eshelby}. 
As discussed in \S\ref{S:FeFvFa}, the natural (material) configuration is the realization of anelasticity in the solid, as a distortion of the initial undeformed configuration---where distances are measured by $\Go$---into a stress-free configuration---where natural distances are measured by $\mathbf{G}=\Fa^*\Go$.
The volume form of the material manifold $(\mathcal{B},\mathbf{G})$ is denoted by $ \mathrm{d}V = \sqrt{\det\mathbf G} \,\mathrm{d}X^1 \wedge \mathrm{d}X^2 \wedge \mathrm{d}X^3$.
We denote by $J$ the Jacobian of the motion with respect to the material (stress-free) configuration $(\mathcal{B},\mathbf{G})$, which is such that $\varphi^*\mathrm{d}v = J \mathrm  \mathrm{d}V$. It can be shown that
\begin{equation}
\label{eq:Jacob}
	J=\sqrt{\frac{\det\mathbf g}{\det\mathbf G}}\det\mathbf{F}=\sqrt{\frac{\det\mathbf g}{\det\Go}}\det\Fe\det\Fv\,.
\end{equation}
Let us denote the Levi-Civita connection of $(\mathcal{B},\mathbf{G})$ by $\nabla$, and let ${\Gamma^A}_{BC}$ be its Christoffel symbols in the local coordinate chart $\{X^A\}$.

\subsection{Dual, adjoint, transpose, and derived measures of strain}

\paragraph{Dual.}
In a Riemannian manifold $(\mathcal M, \mathbf m)$,  one can define the dual of a vector field $\mathbf v$ by the metric $\mathbf m$ as the $1$-form given by $\mathbf v^\flat \coloneq \mathbf m(\mathbf v,.)$, which also defines the operation $(.)^\flat$, the musical isomorphism for lowering indices with respect to the metric $\mathbf m$. 
In components $\mathrm v_i = \mathrm m_{ij} \mathrm v^j$. Conversely, one defines the inverse dual of a $1$-form $\boldsymbol \omega$ as the vector field given by $\boldsymbol \omega^\sharp \coloneq {\mathbf m}^{-1}(\boldsymbol \omega,.)$,\footnote{The inverse metric $\mathbf m^{-1}$ is also denoted by $\mathbf m^\sharp$, since $\mathbf m^\sharp$ is obtained by raising both indices of $\mathbf m$. Note that $(\mathrm m^\sharp)^{ik}\mathrm m_{kj}=\mathrm m^{ia}\mathrm m^{kb}\mathrm m_{ab}\mathrm m_{kj}=\mathrm m^{ia}\delta^b_j\mathrm m_{ab}=\mathrm m^{ia}\mathrm m_{aj}=\delta^i_j$. Therefore, $\mathbf m^\sharp=\mathbf m^{-1}$.} which also defines the operation $(.)^\sharp$, the musical isomorphism for raising indices with respect to the metric $\mathbf m$.
In our setting, we may define two different sets of musical isomorphisms: $\flato$ and $\sharpo$ with respect to the reference Euclidean metric $\Go$ and the ambient space metric $\mathbf g\,$; and $\flat$ and $\sharp$ with respect to the material metric $\mathbf G$ and the ambient space metric $\mathbf g$.
Note that $\flat$ and $\sharp$ are respectively identical to $\flato$ and $\sharpo$ for spatial tensors, i.e.,~tensors in the target configuration, but they differ for referential and two-point tensors.
Also, note that $\mathbf{G}^\sharp=\mathbf{G}^{-1}$, $\Go^\sharpo=\Go^{-1}$, and $\mathbf{g}^\sharp=\mathbf{g}^\sharpo=\mathbf{g}^{-1}$.

\paragraph{Adjoint.}
We define the adjoint of $\mathbf{F}$ as the operator ${\mathbf{F}^\star(X,t):T_{\varphi_t(X)}\mathcal{C}_t \to T_X\mathcal{B}}$, such that ${\langle\boldsymbol\alpha,\mathbf{F} \mathbf U\rangle=\langle\mathbf{F}^\star \boldsymbol\alpha,\mathbf U\rangle}$,  $\forall\, \mathbf U \in T_X\mathcal B$,  $\forall\, \boldsymbol\alpha \in T^{*}_{\varphi(X)}\mathcal S$. It follows that it has components ${\!\left(\mathrm{F}^\star\right)\!_{\!A}}^{\!a}={\left.{\mathrm F}\right.^a}_A$, and has the following coordinate representation
\begin{equation}
	\mathbf{F}^{\star}=\mathrm F^a{}_A \,\mathrm{d}X^A \otimes \frac{\partial}{\partial x^a}\,.
\end{equation}

\paragraph{Transpose.}
Unlike the adjoint, the transpose requires a metric to be defined; hence, there are two different ways of defining the transpose of $\mathbf{F}$\textemdash either using the Euclidean metric $\Go$ or the material metric $\mathbf G$.
Let us first define the transpose of $\mathbf{F}$ with respect to the metrics $\Go$ and $\mathbf{g}\,$: it is given as the operator ${\mathbf{F}^{\mathring{\mathsf T}}(X,t):T_{\varphi_t(X)}\mathcal{C}_t \to T_X\mathcal{B}}$, such that ${\llangle \mathbf{F}\mathbf{U},\mathbf{u} \rrangle_{\mathbf{g}}=\llangle \mathbf{U},\mathbf{F}^{\mathring{\mathsf T}}\mathbf{u} \rrangle_{\Go}}$,  $\forall\, \mathbf U \in T_X\mathcal B$,  $\forall\, \mathbf u \in T_{\varphi(X)}\mathcal S\,$\textemdash it has components ${(\mathrm{F}^{\mathring{\mathsf T}})^A}_a = \mathring{\mathrm{G}}^{AB}\, {\mathrm{F}^b}_B\, \mathrm{g}_{ba}$, and in tensorial form $\mathbf{F}^{\mathring{\mathsf T}}=\Go^\sharp\, \mathbf{F}^\star \mathbf g$.
Similarly, the transpose of $\mathbf{F}$ with respect to the metrics $\mathbf G$ and $\mathbf{g}$ is given as the operator ${\mathbf{F}^{\mathsf T}(X,t):T_{\varphi_t(X)}\mathcal{C}_t \to T_X\mathcal{B}}$, such that ${\llangle \mathbf{F}\mathbf{U},\mathbf{u} \rrangle_{\mathbf{g}}=\llangle \mathbf{U},\mathbf{F}^{\mathring{\mathsf T}}\mathbf{u} \rrangle_{\mathbf G}}$,  $\forall\, \mathbf U \in T_X\mathcal B$,  $\forall\, \mathbf u \in T_{\varphi(X)}\mathcal S\,$\textemdash it reads in components ${(\mathrm{F}^{\mathsf T})^A}_a = \mathrm{G}^{AB}\, {\mathrm{F}^b}_B\, \mathrm{g}_{ba}$, and in tensorial form $\mathbf{F}^{\mathsf T}=\mathbf G^{\sharp} \mathbf{F}^\star \mathbf g$.
The two transposes have the following coordinate representations
\begin{equation}
	\mathbf{F}^{{\mathring{\mathsf T}}}
	= \mathring{\mathrm{G}}^{AB}\, {\mathrm{F}^b}_B\, \mathrm{g}_{ba}
	\frac{\partial}{\partial X^A}\otimes \mathrm{d}x^a\,,\qquad
	\mathbf{F}^{\textsf{T}}
	= \mathrm{G}^{AB}\, {\mathrm{F}^b}_B\, \mathrm{g}_{ba}
	\frac{\partial}{\partial X^A} \otimes \mathrm{d}x^a	\,.
\end{equation}

\paragraph{Derived measures of strain.}
From the total deformation gradient and using the adjoint, dual, and transpose operators, the following measures of strain are defined:
\begin{itemize}[topsep=2pt,noitemsep, leftmargin=12pt]
	
\item {Right Cauchy-Green tensor}: $ \mathbf C = \varphi^* \mathbf g = \mathbf{F}^\star \mathbf g \mathbf{F} $,\footnote{Traditionally, the right Cauchy-Green deformation tensor is defined as the product of the transpose of the deformation gradient with itself, i.e.,~$\mathbf{F}^{\mathsf{T}} \mathbf{F}$ or $\mathbf{F}^{\mathring{\mathsf{T}}} \mathbf{F}$. This definition requires a choice of referential metric, either the material metric $\mathbf{G}$ or the reference Euclidean metric $\Go$. Observing that
$$\!\left(\mathbf{F}^{\mathsf T} \mathbf{F}\right)^\flat = \!\left(\mathbf{F}^{\mathring{\mathsf T}} \mathbf{F}\right)^{\flato}=\mathbf{F}^\star \mathbf g \mathbf{F}\,,$$
we choose to define the Cauchy-Green deformation tensor as the pull-back of the spatial metric $\mathbf{g}$ by the deformation mapping $\varphi$, i.e.,
$$\mathbf{C} = \varphi^* \mathbf{g} = \mathbf{F}^\star \mathbf{g} \mathbf{F}\,.$$
This alternative definition avoids dependence on a specific referential metric and yields a tensorial quantity that is independent of the choice between $\mathbf{G}$ and $\Go$.} with components $ \mathrm C_{AB} = \mathrm F^a{}_A \,\mathrm g_{ab}\, \mathrm F^b{}_B $.
	
\item {Piola deformation tensor}: $ \mathbf{B} = \varphi^* \mathbf g^\sharp = \mathbf{F}^{-1} \mathbf g^\sharp \mathbf{F}^{-\star} $, with components $ \mathrm B^{AB} = (\mathrm F^{-1})^A{}_a \,\mathrm g^{ab}\, (\mathrm F^{-1})^B{}_b $.

\item {Left Cauchy-Green (Finger) tensor}: $ \mathbf{b} = \varphi_* \mathbf G^\sharp = \mathbf{F} \mathbf G^\sharp \mathbf{F}^\star $, with components $ \mathrm b^{ab} = \mathrm F^a{}_A \,\mathrm G^{AB} \,\mathrm F^b{}_B $.

\item {Inverse Finger tensor}: $ \mathbf{c} = \varphi_* \mathbf G = \mathbf{F}^{-\star} \mathbf G \mathbf{F}^{-1} $, with components $ \mathrm c_{ab} = (\mathrm F^{-1})^A{}_a \,\mathrm G_{AB} \,(\mathrm F^{-1})^B{}_b$.
\end{itemize}

\vskip 0.1 in \noindent
We also define their viscoelastic counter-parts as follow:
\begin{itemize} [topsep=2pt,noitemsep, leftmargin=12pt]

\item {Viscoelastic right Cauchy-Green tensor}: $ \Cve = \Fve^* \mathbf g = \Fve^\star \mathbf g \Fve $, with components $ \cCve_{AB} = \cFve^a{}_A \,\mathrm g_{ab} \,\cFve^b{}_B $.

\item {Viscoelastic Piola deformation tensor}: $ \Bve = \Fve^* \mathbf g^\sharp = \Fve^{-1} \mathbf g^\sharp \,\Fve^{-\star} $, with components $ \cBve^{AB} = (\cFve^{-1})^A{}_a \,\mathrm g^{ab} \,(\cFve^{-1})^B{}_b $.

\item {Viscoelastic left Cauchy-Green (Finger) tensor}: $ \bve = \Fve_* \Go^\sharpo = \Fve \Go^\sharpo \,\Fve^\star $, with components $ \cbve^{ab} = \cFve^a{}_A \,\mathring{\mathrm G}^{AB} \,\cFve^b{}_B $.

\item {Viscoelastic inverse Finger tensor}: $ \cve = \Fve_* \Go = \Fve^{-\star} \,\Go \,\Fve^{-1} $, with components $ \ccve_{ab} = (\cFve^{-1})^A{}_a \,\mathring{\mathrm G}_{AB} \,(\cFve^{-1})^B{}_b $.
\end{itemize}
\vskip 0.1 in \noindent
Their (instantaneous) elastic counter-parts are defined as:
\begin{itemize}[topsep=2pt,noitemsep, leftmargin=12pt]

\item {Elastic right Cauchy-Green tensor}: $\Ce = \Fe^* \mathbf g = \Fe^\star \,\mathbf g \,\Fe $, with components $ \cCe_{AB} = \cFe^a{}_A \,\mathrm g_{ab} \,\cFe^b{}_B $.

\item {Elastic Piola deformation tensor}: $\Be = \Fe^* \mathbf g^\sharp = \Fe^{-1} \,\mathbf g^\sharp \,\Fe^{-\star} $, with components $ \cBe^{AB} = (\cFe^{-1})^A{}_a \,\mathrm g^{ab} \,(\cFe^{-1})^B{}_b$.

\item {Elastic left Cauchy-Green (Finger) tensor}: $ \be = \Fe_* \Go^\sharpo = \Fe \,\Go^\sharpo \,\Fe^\star $, with components $ \cbe^{ab} = \cFe^a{}_A \,\mathring{\mathrm G}^{AB} \,\cFe^b{}_B$.

\item {Elastic inverse Finger tensor}: $\ce = \Fe_* \Go = \Fe^{-\star} \,\Go \,\Fe^{-1} $, with components $\cce_{ab} = (\cFe^{-1})^A{}_a \,\mathring{\mathrm G}_{AB} \,(\cFe^{-1})^B{}_b$.
\end{itemize}

\begin{remark}
Note that since $\mathbf{G}=\Fa^*\Go$, it follows that $\cve=\mathbf{c}$ and $\bve=\mathbf{b}$, while $\mathbf{C}=\Fa_*\Cve$ and $\mathbf{B}=\Fa^*\Bve$.
\end{remark}

\section{Thermodynamics of Visco-Anelasticity} \label{Sec:Thermodynamics}

In this section, assuming a hyperelastic constitutive model, we discuss the derivation of the constitutive equations, balance laws, and kinetic equations of visco-anelasticity from the first and second laws of thermodynamics without assuming any (observer) invariance. This section is an extension to hyper-visco-anelasticity of the generalized Coleman-Noll procedure that was recently introduced by the authors in the context of hyperelasticity and hyper-anelasticity \citep{SaYa2025GenColemanNoll}.

\subsection{The first law of thermodynamics}

The first law of thermodynamics postulates the existence of the internal energy as a state function satisfying the following energy balance \citep{truesdell1952mechanical,gurtin1974modern,MarsdenHughes1983,EpsteinMaugin2000,LubardaHoger2002}\footnote{Note that the last term in~\eqref{eq:Thermo_First} is added to account for the change in the internal and kinetic energies of the system due to bulk growth or resorption with a material rate of change of mass $S_m$.}
\begin{equation}\label{eq:Thermo_First}
\begin{split}
	\frac{d}{ dt}\int_{\mathcal{U}} \rho\!\left(\mathscr{E}
	+\frac{1}{2} \Vert\mathbf V\Vert^2_{\mathbf g}\right)\!
	 \mathrm{d}V
	=\int_{\mathcal{U}}\rho \!\left(\llangle\boldsymbol{\mathsf{B}},\mathbf{V}\rrangle_{\mathbf{g}}
	+R\right)\! \mathrm{d}V+\int_{\partial \mathcal{U}}
	\!\left(\llangle\boldsymbol{\mathsf{T}},\mathbf{V}\rrangle_{\mathbf{g}}
	-\llangle \mathbf{Q},\hat{\mathbf N}\rrangle_{\mathbf{G}}\right)\!dA
	+\int_{\mathcal{U}} S_m\!\left(\mathscr{E}+\frac{1}{2} \Vert\mathbf V\Vert^2_{\mathbf g}\right)\!
	 \mathrm{d}V\,,
\end{split}
\end{equation}
for any open set $\mathcal{U}\subset\mathcal{B}$, where $\mathscr{E}$ is the specific internal energy (i.e.,~internal energy per unit mass), $\rho$ is the material mass density, ${S_m=S_m(X,t)}$ is the material rate of change of mass per unit (stress-free) volume\textemdash it is identically equal to zero in the absence of bulk growth or resorption, $\boldsymbol{\mathsf{B}}$ is the specific body force, $\boldsymbol{\mathsf{T}}$ is the boundary traction vector field per unit (stress-free) material area, ${R=R(X,t)}$ is the specific external heat supply, ${\mathbf Q=\mathbf Q(X,\vartheta,d\vartheta,\mathbf C,\mathbf G)}$ is the external heat flux per unit material (stress-free) area, and $\hat{\mathbf N}$ is the $\mathbf{G}$-unit normal to the boundary $\partial\mathcal U$.

We use a version of Cauchy's theorem by \citet[\S2.1(1.9)]{MarsdenHughes1983} which assumes a master balance law that need not be  the balance of linear momentum. Indeed, at this point, we have not even introduced the concept of a stress tensor.
Taking the balance of energy~\eqref{eq:Thermo_First} as the master balance law, it follows that there exists a unique material vector field $ \boldsymbol{\mathsf U} $ satisfying $\llangle \boldsymbol{\mathsf U}, \hat{\mathbf N} \rrangle_{\mathbf{G}} = \llangle \boldsymbol{\mathsf{T}}, \mathbf{V} \rrangle_{\mathbf{g}}$.
By linearity of $ \llangle \boldsymbol{\mathsf{T}}, \mathbf{V} \rrangle_{\mathbf{g}} $ with respect to $ \mathbf{V} $, the vector field $ \boldsymbol{\mathsf U} $ is also linear in $ \mathbf{V} $, implying the existence of a second-order two-point tensor field $ \boldsymbol{\mathsf M} $ such that $ \boldsymbol{\mathsf U} = \boldsymbol{\mathsf M} \mathbf{V} $. The uniqueness of $ \boldsymbol{\mathsf U} $ ensures the uniqueness of $ \boldsymbol{\mathsf M} $.
Consequently, $\llangle \boldsymbol{\mathsf M} \mathbf{V}, \hat{\mathbf N} \rrangle_{\mathbf{G}} = \llangle \boldsymbol{\mathsf{T}}, \mathbf{V} \rrangle_{\mathbf{g}},$
which can be recast as $\llangle \boldsymbol{\mathsf M}^{\mathsf{T}} \hat{\mathbf N}, \mathbf{V} \rrangle_{\mathbf{g}} = \llangle \boldsymbol{\mathsf{T}}, \mathbf{V} \rrangle_{\mathbf{g}}.$
By arbitrariness of $ \mathbf{V} $, we deduce that $\boldsymbol{\mathsf{T}} = \mathbf{P} \hat{\mathbf N}^\flat,$
where $ \mathbf{P} \coloneqq \boldsymbol{\mathsf M}^{\mathsf{T}} \mathbf{g}^\sharp $ is well-defined due to the existence and uniqueness of $ \boldsymbol{\mathsf M} $. This allows us to express the energy balance~\eqref{eq:Thermo_First} in local form as\footnote{
Recall that $\mathbf{G}=\Fa^*\Go=\Fa^\star\Go\Fa$ and $ \mathrm{d}V=\sqrt{\det\mathbf{G}}\, \mathrm{d}X^1\wedge \mathrm{d}X^2\wedge \mathrm{d}X^3$. Hence, we ought to consider the implicit time dependance of $ \mathrm{d}V$ through $\det\mathbf{G}$ on the left-hand-side of~\eqref{eq:Thermo_First} and we use the identity $\frac{ d}{ d t}\!\left[\det\mathbf A \right]\!=\!\left[\dot{\mathbf A}\!:\!\mathbf A^{-1}\right]\!\det\mathbf A$.}
\begin{equation}\label{eq:loc_Thermo_First}
	\rho\,\dot{\mathscr{E}} =
	\mathbf P \!:\!  \mathbf{g} \bar\nabla\mathbf{V}
	+\rho R - \operatorname{Div} \mathbf Q
	+ \llangle \operatorname{Div}\mathbf{P}+ \rho(\boldsymbol{\mathsf{B}} 
	- \mathbf{A}),\mathbf{V} \rrangle_{\mathbf{g}}
	+ \!\left(S_m-\dot{\rho} - \rho\,\dot{\Fa}\!:\!\Fa^{-\star} \right)\!\!\left(\mathscr{E}+\frac{1}{2} \Vert\mathbf V\Vert^2_{\mathbf g}\right)\!\,,
\end{equation}
where a dotted quantity denotes its total time derivative, $\operatorname{Div}$ denotes the divergence operator with respect to the connections $\nabla$ and $\bar\nabla$,\footnote{Note that in components one has
\begin{equation}
	\operatorname{Div} \mathbf{Q} = \frac{\partial Q^A}{\partial X^A} 
	+ \Gamma^A{}_{AB} \, Q^B\,,\qquad
	(\operatorname{Div} \mathbf{P})^a =\frac{\partial P^{aA}}{\partial X^A}
	+ \gamma^a{}_{bc} \, F^b{}_A \, P^{cA}
	+ \Gamma^A{}_{AB} \, P^{aB}\,.
\end{equation}
}
and a colon denotes the double contraction product, i.e.,~$\mathbf P\!:\! \mathbf{g} \bar\nabla \mathbf V = \textrm{P}^{aA}\textrm{V}_{a|A}$.\footnote{The vertical stroke denotes covariant differentiation with respect to $\mathbf{g}$, i.e.,~$\textrm{V}_{a|A} = \textrm{V}_{a,A} - \textrm{V}_b \,\gamma^b_{ac} \,{\textrm{F}^c}_{A}=\textrm{V}_{a,c}\,F^c{}_A - \textrm{V}_b \,\gamma^b_{ac} \,{\textrm{F}^c}_{A}=\textrm{V}_{a|c}\,F^c{}_A$.}
Following the symmetry lemma \citep{Nishikawa2002}, one may write\footnote{In components, we have
\begin{equation}
\begin{aligned}
	{V^a}_{|A} 
	&=V^a{}_{|b}\,F^b{}_A \\
	& =(V^a{}_{,b}+\gamma^{a}{}_{bc}\,V^c)\,F^b{}_A \\
	& =V^a{}_{,b}\,F^b{}_A+\gamma^{a}{}_{bc}\,V^c\,F^b{}_A \\
	& =V^a{}_{,A}+\gamma^{a}{}_{bc}\,V^c\,F^b{}_A \\
	& =\!\left(\frac{\partial \varphi^a}{\partial t}\right)\!_{\!,A}
	+\gamma^{a}{}_{bc}\,V^c\,F^b{}_A \\
	& =\!\left(\frac{\partial \varphi^a}{\partial X^A}\right)\!_{\!,t}
	+\gamma^{a}{}_{bc}\,V^c\,F^b{}_A \\
	& =\frac{\partial}{\partial t}F^a{}_A+\gamma^{a}{}_{bc}\,V^c\,F^b{}_A\\
	&= (\operatorname{D}_t^{\mathbf g}\mathbf{F})^a{}_A\,.
\end{aligned}
\end{equation}
}
\begin{equation}
	{V^a}_{|A}=\!\left(\frac{\partial \varphi^a}{\partial t}\right)\!{\!}_{|A} 
	= \operatorname{D}_t^{\mathbf g}\!\left(\frac{\partial \varphi^a}{\partial X^A}\right)\!\,, 
	\quad\textrm{i.e.,}\quad
	\bar\nabla \mathbf V = \operatorname{D}_t^{\mathbf g}\mathbf{F}\,,
\end{equation}
where $\operatorname{D}_t^{\mathbf g}(.)$ denotes the covariant time derivative in $(\mathcal S,\mathbf g)$.
We denote the symmetric and anti-symmetric parts of $\mathbf{F}^\star \mathbf{g} \, \operatorname{D}_t^{\mathbf g}\mathbf{F}$, as $\Ds$ and $\Da$, respectively:
\begin{equation}
\label{Eq:Def_rates}
	\Ds
	= \frac{1}{2}\!\left[\mathbf{F}^\star \mathbf{g} \, 
	\operatorname{D}_t^{\mathbf g}\mathbf{F}
	+ \operatorname{D}_t^{\mathbf g}\mathbf{F}^\star \mathbf{g} \mathbf{F}\right]\!\,,
	\qquad
	\Da
	= \frac{1}{2}\!\left[\mathbf{F}^\star \mathbf{g} \, 
	\operatorname{D}_t^{\mathbf g}\mathbf{F}
	- \operatorname{D}_t^{\mathbf g}\mathbf{F}^\star \mathbf{g} \mathbf{F}\right]\!
	 \,.
\end{equation}
Hence, we may write
\begin{equation} \label{eq:rate-decomp}
	\mathbf P\!:\! (\mathbf{g} \bar\nabla \mathbf V) 
	= \mathbf P\!:\! (\mathbf{g} \operatorname{D}_t^{\mathbf g}\mathbf{F}) 
	= (\mathbf{F}^{-1} \mathbf P) \!:\! (\Ds+ \Da)\,.
\end{equation}

\subsection{The second law of thermodynamics}

The second law of thermodynamics postulates the existence of the entropy as a state function satisfying the material Clausius-Duhem inequality \citep{truesdell1952mechanical,gurtin1974modern,MarsdenHughes1983,EpsteinMaugin2000,LubardaHoger2002}\footnote{Note that the last term in~\eqref{eq:Thermo_Second} is added to account for the change in the entropy of the system due to bulk growth or resorption with a material rate of change of mass $S_m$.}
\begin{equation}\label{eq:Thermo_Second}
	\frac{d}{ dt} \int_{\mathcal{U}} \rho \mathscr{N} \mathrm{d}V
	\geq\int_{\mathcal{U}}\rho \frac{R}{\vartheta} \mathrm{d}V-\int_{\partial\mathcal{U}}\frac{1}{\vartheta}\llangle \mathbf{Q},\hat{\mathbf N}\rrangle_{\mathbf{G}}dA
	+\int_{\mathcal{U}} S_m\, \mathscr{N} \mathrm{d}V\,,
\end{equation}
for any open set $\mathcal{U}\subset\mathcal{B}$, where $\mathscr{N}$ is the specific entropy (i.e.,~entropy per unit mass).
In localized form, the material Clausius-Duhem inequality~\eqref{eq:Thermo_Second} is written as
\begin{equation} \label{loc_Thermo_Second}
	\dot\eta = \rho\, \dot{\mathscr{N}}\vartheta 
	+ \vartheta\operatorname{Div}\!\left(\frac{\mathbf Q}{\vartheta}\right)\! - \rho R
	- \!\left(S_m-\dot{\rho} - \rho\,\dot{\Fa}\!:\!\Fa^{-\star} \right)\! \vartheta \mathscr N 
	\geq 0\,,
\end{equation}
where $\dot\eta$ is the rate of energy dissipation.

\subsection{The governing equations}
\label{sec:gov_eqs}

\subsubsection{The free energy}
\label{Additive_decomp}

The specific free energy $\Psi$ (i.e.,~free energy per unit mass) is the Legendre transform of the specific internal energy $\mathscr E$ with respect to the conjugate pair: specific entropy $\mathscr N$ and temperature $\vartheta$
\begin{equation}\label{eq:Legendre_trans}
	\Psi = \mathscr E - \vartheta \mathscr N\,.
\end{equation}
We first examine the objectivity of the energy function, as this result will be used later in this subsection. We then discuss the additive decomposition of the free energy function in viscoelasticity and extend it to the case of visco-anelasticity.

\paragraph{Objectivity (material-frame-indifference).}
Let us consider an energy function $\mathsf{W}=\mathsf{W}(X,\boldsymbol{\mathsf{A}},\Go,\mathbf{g})$, where $\boldsymbol{\mathsf{A}}$ is any invertible two-point tensor from $\mathcal B$ to $\mathcal C$, i.e.,~evaluated at $X\in \mathcal B$, and $\boldsymbol{\mathsf{A}}$ is a multilinear mapping from $T_X\mathcal B$ to $T_{\varphi(X)}\mathcal C$.  
We assume that $\mathsf{W}$ is objective,\footnote{Note that in the case of a Euclidean ambient space, material frame indifference (objectivity) and invariance under superposed ambient space rigid body motions are equivalent.} i.e.,~$\mathsf{W}(X,\mathbf q_{*}\boldsymbol{\mathsf{A}},\Go, \mathbf q_{*}\mathbf{g})=\mathsf{W}(X,\boldsymbol{\mathsf{A}},\Go,\mathbf{g})$
for all such $\boldsymbol{\mathsf{A}}$ and any arbitrary $\mathbf g$-orthogonal second-order tensor $\mathbf q:T_x\mathcal S \to T_x\mathcal S$,  i.e.,~$\mathbf{q}^{\mathsf T} \mathbf{q} = \mathrm{id}_{\mathcal{S}}$,  which is equivalent to writing $\mathbf q_{*} \mathbf g = \mathbf g$.
Let us consider two two-point tensors $\boldsymbol{\mathsf{A}}_1$ and $\boldsymbol{\mathsf{A}}_2$ such that 
$\boldsymbol{\mathsf{A}}_1^*\mathbf{g}=\boldsymbol{\mathsf{A}}_2^*\mathbf{g}$. 
Let us define $\boldsymbol{\mathsf{B}}=\boldsymbol{\mathsf{A}}_2\boldsymbol{\mathsf{A}}_1^{-1}:T_x\mathcal{C}\to T_x\mathcal{C}$.
For arbitrary vectors $\mathbf{u}_1, \mathbf{u}_2\in T_x\mathcal{C}$, as $\boldsymbol{\mathsf{A}}_1$ is invertible, one has $\mathbf{u}_1=\boldsymbol{\mathsf{A}}_1\mathbf{U}_1$ and $\mathbf{u}_2=\boldsymbol{\mathsf{A}}_1\mathbf{U}_2$, where $\mathbf{U}_1, \mathbf{U}_2\in T_X\mathcal{B}$.
Thus 
\begin{equation}
\begin{aligned}
	\llangle \boldsymbol{\mathsf{B}}\mathbf{u}_1 , \boldsymbol{\mathsf{B}}\mathbf{u}_2 
	\rrangle_{\mathbf{g}}
	&=\llangle \boldsymbol{\mathsf{B}}\boldsymbol{\mathsf{A}}_1\mathbf{U}_1 , 
	\boldsymbol{\mathsf{B}}\boldsymbol{\mathsf{A}}_1\mathbf{U}_2\rrangle_{\mathbf{g}} 
	=\llangle \boldsymbol{\mathsf{A}}_2\mathbf{U}_1 , 
	\boldsymbol{\mathsf{A}}_2\mathbf{U}_2\rrangle_{\mathbf{g}} 
	=\llangle \mathbf{U}_1 , \mathbf{U}_2\rrangle_{\boldsymbol{\mathsf{A}}_2^*\mathbf{g}}
	=\llangle \mathbf{U}_1 , 
	\mathbf{U}_2\rrangle_{\boldsymbol{\mathsf{A}}_1^*\mathbf{g}} \\
	&=\llangle \boldsymbol{\mathsf{A}}_1\mathbf{U}_1 , 
	\boldsymbol{\mathsf{A}}_1\mathbf{U}_2\rrangle_{\mathbf{g}}=\llangle \mathbf{u}_1 , 
	\mathbf{u}_2 \rrangle_{\mathbf{g}}		
	\,.
\end{aligned}
\end{equation}
This implies that $\boldsymbol{\mathsf{B}}$ is a local isometry, and hence material frame-indifference tells us that $\mathsf{W}(X,\boldsymbol{\mathsf{A}}_1,\Go,\mathbf{g})=\mathsf{W}(X,\boldsymbol{\mathsf{B}}\boldsymbol{\mathsf{A}}_1,\Go,\mathbf{g})=\mathsf{W}(X,\boldsymbol{\mathsf{A}}_2,\Go,\mathbf{g})$.
Thus, $\mathsf{W}=\hat{\mathsf{W}}(X,\boldsymbol{\mathsf{A}}^*\mathbf{g},\Go)$ is well-defined and is equal to the common value of $\mathsf{W}(X,\boldsymbol{\mathsf{A}},\Go,\mathbf{g})$ for any invertible two-point tensor $\boldsymbol{\mathsf{A}}$.
This is the argument given by \cite{MarsdenHughes1983} regarding the dependence of an elastic energy function on the deformation gradient. However, it holds for any objective function depending on any invertible two-point tensor $\boldsymbol{\mathsf{A}}:T_X\mathcal{B}\to T_x\mathcal{C}$.

\paragraph{Free energy function in viscoelasticity.}
In hyper-viscoelasticity, the free energy depends on the elastic distortion $\Fe$, which drives the instantaneous recoverable stresses during viscous flow, and on the total deformation gradient $\mathbf{F}$, which captures the long-term equilibrated behavior after viscous relaxation.
\begin{equation}\label{eq:Psi_ViscoElast}
	\Psi=\Psi(X,\vartheta,\mathbf{F}, \Fe, \Go,\mathbf{g})
\,.
\end{equation}
In particular, it is customary in viscoelasticity literature to additively decompose the free energy into an non-equilibrium part $\Psi_{\textrm{neq}}$ and a equilibrium part $\Psi_{\textrm{eq}}$, reflecting the transient viscous stresses and the long-term relaxed response, respectively: $\Psi = \Psi_{\textrm{neq}} + \Psi_{\textrm{eq}}$ ~\citep{Reese1998, Kumar2016, SaYa2024viscoelasticity}. Here, the non-equilibrium part depends on the elastic distortion $\Fe$ and the equilibrium part depends on the total deformation gradient $\mathbf{F}$. The free energy thus takes the form
\begin{equation}\label{eq:Psi_Fv_decomp}
	\Psi
	=\Psi_{\textrm{eq}}(X,\vartheta,\mathbf{F}, \Go,\mathbf{g})
	+\Psi_{\textrm{neq}}(X,\vartheta,\Fe,\Go,\mathbf{g})\,.
\end{equation}
Since $\mathbf{F}$ and $\Fe$ are invertible two-point tensors, it follows by the above objectivity result, for $\Psi_{\textrm{eq}}$ and $\Psi_{\textrm{neq}}$ separately, that
\begin{equation}\label{eq:Psi_C_decomp}
	\Psi_{\textrm{eq}}(X,\vartheta,\mathbf{F}, \Go,\mathbf{g})
	=\hat\Psi_{\textrm{eq}}(X,\vartheta,\mathbf C, \Go)\,,
	\quad\textrm{and}\qquad
	\Psi_{\textrm{neq}}(X,\vartheta,\Fe,\Go,\mathbf{g})
	=\hat\Psi_{\textrm{neq}}(X,\vartheta,\Ce, \Go)		\,,
\end{equation}
where we recall $\mathbf C=\varphi^*\mathbf{g}=\mathbf{F}^{\star}\mathbf{g}\mathbf{F}$ and introduce $\Ce=\Fe^*\mathbf{g}=\Fe^{\star}\mathbf{g}\Fe$.
Thus
\begin{equation}\label{eq:Psi_C_decomp}
	\Psi=\hat\Psi(X,\vartheta,\mathbf C, \Ce, \Go)
	=\hat\Psi_{\textrm{eq}}(X,\vartheta,\mathbf C, \Go) + \hat\Psi_{\textrm{neq}}(X,\vartheta,\Ce, \Go)
	\,.
\end{equation}

\paragraph{Free energy function in visco-anelasticity.}
In viscoelasticity, the representation~\eqref{eq:Psi_ViscoElast} relies on $\mathbf{F}$ mapping from a globally stress-free reference configuration to the deformed configuration~\citep{SaYa2024viscoelasticity}. However, in visco-anelasticity, anelastic effects introduce non-elastic distortions or microstructural changes: the distortion $\Fa$ first defines a locally stress-free copy of the reference configuration, and the viscoelastic distortion $\Fve=\Fe\Fv$ then maps it to the current deformed state.
Thus, $\Fve$ replaces $\mathbf{F}$ as the primary internal variable driving the equilibrium response and the free energy hence takes the form
\begin{equation}\label{eq:Psi_ViscoAnElast}
	\Psi=\Psi(X,\vartheta,\Fve, \Fe, \Go,\mathbf{g})\,.
\end{equation}
Under the additive free energy split, this specializes to
\begin{equation}\label{eq:Psi_Fve_decomp}
	\Psi
	=\Psi_{\textrm{eq}}(X,\vartheta,\Fve, \Go,\mathbf{g})
	+\Psi_{\textrm{neq}}(X,\vartheta,\Fe, \Go,\mathbf{g})\,.
\end{equation}
Since $\Fve$ and $\Fe$ are invertible two-point tensors, objectivity implies that
\begin{equation}
\begin{gathered}
	\Psi_{\textrm{eq}}(X,\vartheta,\Fve, \Go,\mathbf{g})
	= \hat{\Psi}_{\textrm{eq}}(X,\vartheta,\Fve^*\mathbf{g}, \Go)
	= \hat{\Psi}_{\textrm{eq}}(X,\vartheta,\Cve, \Go)\,,\\
	\Psi_{\textrm{neq}}(X,\vartheta,\Fe, \Go,\mathbf{g})
	=\hat{\Psi}_{\textrm{neq}}(X,\vartheta,\Fe^*\mathbf{g}, \Go)
	=\hat{\Psi}_{\textrm{neq}}(X,\vartheta,\Ce, \Go)\,,
\end{gathered}
\end{equation}
where $\Cve=\Fve^*\mathbf{g}=\Fv^*\Ce$. Therefore
\begin{equation}\label{eq:Psi_Cve_decomp}
	\Psi
	=\hat{\Psi}(X,\vartheta,\Cve, \, \Ce, \Go)
	=\hat{\Psi}_{\textrm{eq}}(X,\vartheta,\Cve, \Go)
	+\hat{\Psi}_{\textrm{neq}}(X,\vartheta,\Ce, \Go)\,.
\end{equation}

Recalling Remark~\ref{rmrk:ind_fields}, the kinematic state for simple materials is specified by the following \emph{a priori} independent fields: the deformation gradient $\mathbf{F}$, any three distortions from $\{\Fe,\Fv,\Fn,\Fth\}$, and the temperature field $\vartheta$. The free energy representation~\eqref{eq:Psi_ViscoAnElast} displays $\vartheta$, $\Fe$, and $\Fve$; two further distortions are needed to complete the description. These cannot trivially be taken to be $\Fn$ and $\Fth$: the resulting set $\{\Fe,\Fve,\Fn,\Fth\}$ should further be constrained to be compatible. Alternatively, we explicitly include the deformation gradient among the primary fields. The same is in fact required by semi-inverse methods---widely used in solid mechanics---which employ \emph{a priori} symmetry assumptions that simplify the kinematics, working directly with a reduced deformation gradient $\mathbf{F}$. This motivates the equivalent representation
\begin{equation}\label{eq:psi_tilde_psi}
	\Psi = \tilde{\Psi}(X, \vartheta, \mathbf{F}, \Fv, \Fn, \Fth, \Go, \mathbf{g}) 
	= \Psi \!\left(X, \vartheta, \mathbf{F}\Fth^{-1}\Fn^{-1}, \mathbf{F}\Fth^{-1}\Fn^{-1}\Fv^{-1}, 
	\Go, \mathbf{g} \right)\!\,,
\end{equation}
whose arguments $\mathbf{F}$, $\Fv$, $\Fn$, and $\Fth$ are an admissible choice of independent distortions, and which is hereby used as the main constitutive representation in subsequent derivations in the remainder of this section. Recall that $\vartheta$ and $\Fth$ enter $\tilde\Psi$ as independent arguments. Substituting $\Fth=\Fth(X,\vartheta)$ at this stage would conflate, in the temperature derivative of the free energy, the entropy (conjugate to $\vartheta$) with the driving force conjugate to $\Fth$.

\begin{figure}[t!]
\centering
\includegraphics[width=0.45\textwidth]{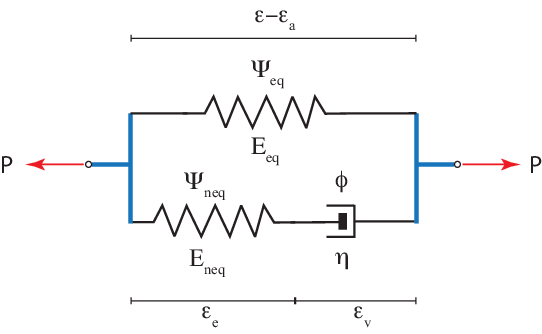}
\vspace*{0.0in}
\caption{Standard anelastic solid: $\Psi_{\text{eq}}\,$, $\Psi_{\text{neq}}\,$, and $\phi\,$ are the nonlinear analogues of $E_{\text{eq}}\,$, $E_{\text{neq}}\,$, and $\eta\,$, respectively. The total strain is decomposed as $\varepsilon = \varepsilon_a + \varepsilon_e + \varepsilon_v\,$, or equivalently $\varepsilon - \varepsilon_a = \varepsilon_e + \varepsilon_v\,$, where $\varepsilon_e$ denotes the linearized elastic distortion,  $\varepsilon_v$ denotes the linearized viscous distortion, and  $\varepsilon_a$ denotes the linearized anelastic distortion. This additive decomposition is the linearization of any of the six possible multiplicative decompositions of the deformation gradient into elastic, anelastic, and viscous distortions, including the decomposition used in this paper, namely $\mathbf{F}=\protect\Fe\protect\Fv\protect\Fa$.}
\label{Standard-Anelastic-Solid}
\end{figure}

\begin{remark}
The standard linear solid (also known as the Maxwell-Kelvin-Voigt model or the Zener solid) is the classical linear viscoelastic model consisting of an equilibrium spring in parallel with a Maxwell element \citep{Zener1948}. Fig.~\ref{Standard-Anelastic-Solid} introduces its anelastic counterpart, which we call the standard anelastic solid: in addition to the elastic and viscous strains, the total strain contains an anelastic strain $\varepsilon_a$, so that $\varepsilon = \varepsilon_a + \varepsilon_e + \varepsilon_v$, or equivalently $\varepsilon-\varepsilon_a=\varepsilon_e+\varepsilon_v$. In this sense, the standard anelastic solid is the natural linear analogue of the nonlinear visco-anelastic model developed here, and the additive strain decomposition in Fig.~\ref{Standard-Anelastic-Solid}  is the linearization of the multiplicative decomposition of the deformation gradient into elastic, viscous, and anelastic distortions.
\end{remark}

\begin{remark}
It should be emphasized that the total free energy densities do not necessarily reduce to the forms~\eqref{eq:Psi_C_decomp} and~\eqref{eq:Psi_Cve_decomp} without the additive split assumption~\eqref{eq:Psi_Fv_decomp}.
\end{remark}

\begin{remark}
Note that the results in the remainder of this section hold for a general free-energy function $\Psi$, even without assuming the additive split~\eqref{eq:Psi_Fve_decomp}. One must, however, assume that the distortions retain their tensorial character, with $\Fe$ a two-point tensor and $\Fv$ and $\Fa$ material tensors. These properties are established following the energy additive split assumption, but the conclusions in the remainder of this section do not depend on the split.
\end{remark}

\subsubsection{Balance laws and constitutive relations for compressible visco-anelastic solids}
\label{S:Bal_Const}

In the remainder of this section, we consider the constitutive functional representation given in~\eqref{eq:psi_tilde_psi} for the free energy, i.e.,
\begin{equation}\label{eq:Psi_F_const}
	\Psi=\tilde\Psi(X,\vartheta,\mathbf{F},\Fv,\Fn,\Fth,\Go,\mathbf{g})\,.
\end{equation}
Expanding $\dot{\Psi}$ yields\footnote{Note that, in~\eqref{eq:Psi_dot}, the term associated with the partial derivative with respect to $\mathbf g$ vanishes since $\operatorname{D}_t^{\mathbf g}\mathbf g = \mathbf 0$. Indeed, by definition, $\operatorname{D}_t^{\mathbf g}\mathbf g = \frac{\partial \mathbf g}{\partial t} + \bar\nabla_{\mathbf v}\mathbf g$. The first term on the right-hand side vanishes since the background metric does not depend explicitly on time\textemdash any time dependence of $\mathbf{g}$ is implicit through a time-dependent embedding of the deformed body. The second term vanishes by metric compatibility of the Levi-Civita connection $\bar\nabla$ with $\mathbf{g}$.}

\begin{equation}\label{eq:Psi_dot}
\dot{\Psi}
	= \frac{\partial \tilde\Psi}{\partial \vartheta} \dot{\vartheta}
	+\frac{\partial \tilde\Psi}{\partial \mathbf F} \!:\! \operatorname{D}_t^{\mathbf g}\mathbf F
	+\frac{\partial \tilde\Psi}{\partial \Fv} \!:\! \dot\Fv
	+\frac{\partial \tilde\Psi}{\partial \Fn} \!:\! \dot\Fn
	+\frac{\partial \tilde\Psi}{\partial \Fth} \!:\! \dot\Fth
	\,.
\end{equation}
Similarly to~\eqref{eq:rate-decomp}, one finds
\begin{equation}
\frac{\partial \tilde\Psi}{\partial \mathbf F} \!:\! \operatorname{D}_t^{\mathbf g}\mathbf F
= \mathbf{F}^{-1} \mathbf g^\sharp \frac{\partial \tilde\Psi}{\partial \mathbf F} \!:\! (\Da +\Ds)\,.
\end{equation}
One may prove that\footnote{This is shown as follows. By the Leibniz rule, one may write
\begin{equation} 
	\frac{\partial \tilde\Psi}{\partial \mathrm{F}^a{}_A}
	=\frac{\partial \tilde\Psi}{\partial (\mathrm{F}^i{}_K \,\mathrm{F}^j{}_L \,\textrm{g}_{ij})}
	\frac{\partial (\mathrm{F}^k{}_K \,\mathrm{F}^l{}_L \,\textrm{g}_{kl})}{\partial \mathrm{F}^a{}_A}
	=\frac{\partial \tilde\Psi}{\partial (\mathrm{F}^i{}_K \,\mathrm{F}^j{}_L \,\textrm{g}_{ij})}
	\!\left(
	\delta^k{}_a \delta^A{}_K \,\mathrm{F}^l{}_L \,\textrm{g}_{kl}
	+ \mathrm{F}^k{}_K \,\delta^l{}_a \delta^A{}_L \,\textrm{g}_{kl}
	\right)\!
	\,.
\end{equation}
By contracting the Kronecker delta symbols, it follows that
\begin{equation} 
	\frac{\partial \tilde\Psi}{\partial \mathrm{F}^a{}_A}
	= 2 \textrm{g}_{al}\, \mathrm{F}^l{}_K \,
	\frac{\partial \tilde\Psi}{\partial \mathrm{C}_{KA}}\,,
	\quad\textrm{and}\qquad
	\frac{\partial \tilde\Psi}{\partial \mathrm{F}^a{}_A}
	=2\frac{\partial \tilde\Psi}{\partial \mathrm{C}_{AL}}
	\,  \mathrm{F}^k{}_L \,\textrm{g}_{ka}\,,
\end{equation}
which one may rewrite as
\begin{equation}
	\frac{\partial \tilde\Psi}{\partial \mathbf{F}}
	= 2 \mathbf{g} \mathbf{F}\frac{\partial \tilde\Psi}{\partial \mathbf{C}}\,,
	\quad\textrm{and}\qquad
	\!\left(\frac{\partial \tilde\Psi}{\partial \mathbf{F}}\right)^\star
	= 2 \frac{\partial \tilde\Psi}{\partial \mathbf{C}} \mathbf{F}^\star \mathbf{g} \,.
\end{equation}
Hence
\begin{equation}
	\mathbf{F}^{-1} \mathbf{g}^\sharp \frac{\partial \Psi}{\partial \mathbf{F}}
	= 2 \frac{\partial \tilde\Psi}{\partial \mathbf{C}}
	= \!\left(\frac{\partial \Psi}{\partial \mathbf{F}}\right)^\star\!\mathbf{g}^\sharp \mathbf{F}^{-\star}\,,
\end{equation}
which then yields~\eqref{eq:F-symmetry}.}
\begin{equation}
\label{eq:F-symmetry}
\!\left[ \mathbf{F}^{-1} \mathbf g^\sharp \frac{\partial \tilde\Psi}{\partial \mathbf F} \right]^\star = \mathbf{F}^{-1} \mathbf g^\sharp \frac{\partial \tilde\Psi}{\partial \mathbf F}\,,
\end{equation}
and it follows that\footnote{$\mathbf{F}^{-1} \mathbf g^\sharp \dfrac{\partial \tilde\Psi}{\partial \mathbf F} \!:\! \Da=0$ since $\mathbf{F}^{-1} \mathbf g^\sharp \dfrac{\partial \tilde\Psi}{\partial \mathbf F}$ is symmetric (cf.~\eqref{eq:F-symmetry}) and $\Da$ is anti-symmetric.}
\begin{equation}
\frac{\partial \tilde\Psi}{\partial \mathbf F} \!:\! \operatorname{D}_t^{\mathbf g}\mathbf F
= \mathbf{F}^{-1} \mathbf g^\sharp \frac{\partial \tilde\Psi}{\partial \mathbf F} \!:\! \Ds\,.
\end{equation}
Thus,~\eqref{eq:Psi_dot} can be rewritten as
\begin{equation}\label{eq:Psi_dot1}
\begin{split}
	\dot{\Psi}
	= \frac{\partial \tilde\Psi}{\partial \vartheta} \dot{\vartheta}
	+\mathbf{F}^{-1} \mathbf g^\sharp \frac{\partial \tilde\Psi}{\partial \mathbf F} \!:\! \Ds
	+\frac{\partial \tilde\Psi}{\partial \Fv} \!:\! \dot\Fv
	+\frac{\partial \tilde\Psi}{\partial \Fn} \!:\! \dot\Fn
	+\frac{\partial \tilde\Psi}{\partial \Fth} \!:\! \dot\Fth
	\,.
\end{split}
\end{equation}

\begin{remark}\label{rmrk:Fdot}
Note that, unlike the two-point tensors $\mathbf{F}$ and $\Fe$, which are sections of the product of the tangent bundle of $\mathcal{B}$ and the cotangent bundle of $\varphi_t(\mathcal{B})$, the referential tensors $\Fv$, $\Fn$, and $\Fth$ are sections of the product of the tangent and cotangent bundles of $\mathcal{B}$. As a result, while we can express the time rates of the $\Fv$, $\Fn$, and $\Fth$ by their total time derivatives $\dot\Fv$, $\dot\Fn$, and $\dot\Fth$ since they are well-defined on the fixed material manifold $\mathcal{B}$, the total time derivatives of $\mathbf{F}$ and $\Fe$ are not well-defined due to the time-dependent nature of the spatial manifold $\varphi_t(\mathcal{B})$, and their material rates must instead be expressed using the spatial covariant derivative $\operatorname{D}_t^{\mathbf g}(.)$ to account for the evolving geometry of the spatial configuration $\varphi_t(\mathcal{B})$.
\end{remark}

\begin{remark}\label{rmrk:partial_temp}
Note that, in view of Remark~\ref{rmrk:ind_fields} and the constitutive representation~\eqref{eq:Psi_F_const} herein considered, the fields $\{\vartheta, \mathbf{F}, \Fv, \Fn, \Fth, \Go, \mathbf{g}\}$ are treated as mutually independent.
Consequently, each partial derivative in~\eqref{eq:Psi_dot} and~\eqref{eq:Psi_dot1} is taken with all remaining independent fields held fixed; for instance, ${\partial\tilde\Psi}/{\partial\vartheta}$ is evaluated at fixed $\{\mathbf{F}, \Fv, \Fn, \Fth, \Go, \mathbf{g}\}$, and ${\partial\tilde\Psi}/{\partial\Fth}$ is evaluated at fixed $(\vartheta, \mathbf{F}, \Fv, \Fn, \Go, \mathbf{g})$, treating the thermal distortion as a kinematically independent quantity of temperature $\vartheta$.
Any coupling between $\vartheta$ and $\Fth$---arising from the constitutive prescription ${\Fth = \Fth(X,\vartheta)}$ of~\eqref{eq:Fth}---enters only \emph{a posteriori}. 
\end{remark}

\begin{remark}
For the equivalent representation for the free energy $\Psi = {\Psi}(X,\vartheta,\Fve, \Fe, \Go,\mathbf{g})$, one finds, following~\eqref{eq:psi_tilde_psi}, that
\begin{subequations}
\label{eq:psi_psi-tilde_EQ}
\begin{align}
\label{eq:psi_psi-tilde_EQ-F}
\frac{\partial \tilde\Psi}{\partial \mathbf{F}} \mathbf{F}^{\star}
	&= \frac{\partial \Psi}{\partial \Fve} \Fve^{\star} + \frac{\partial \Psi}{\partial \Fe} \Fe^{\star}
	\,\\
\label{eq:psi_psi-tilde_EQ-Fv}
\frac{\partial \tilde\Psi}{\partial \Fv} \Fv^{\star}
	&= - \Fe^{\star} \frac{\partial \Psi}{\partial \Fe}
	\,\\
\label{eq:psi_psi-tilde_EQ-Fn}
\frac{\partial \tilde\Psi}{\partial \Fn} \Fn^{\star}
	&= -\Fve^\star\!\left(\frac{\partial \Psi}{\partial \Fve} + \frac{\partial \Psi}{\partial \Fe} \Fv^{-\star} \right)\!
	\,\\
\label{eq:psi_psi-tilde_EQ-Fth}
\frac{\partial \tilde\Psi}{\partial \Fth} \Fth^{\star}
	&= -\Fn^\star\Fve^\star\!\left(\frac{\partial \Psi}{\partial \Fve} + \frac{\partial \Psi}{\partial \Fe} \Fv^{-\star} \right)\!\Fn^{-\star}
	\,.
\end{align}
\end{subequations}
Indeed, one may find, by direct differentiation of $\Psi = {\Psi}(X,\vartheta,\Fve, \Fe, \Go,\mathbf{g})$, that
\begin{equation}
\begin{split}
	\dot{\Psi}
	 & =   \frac{\partial \tilde\Psi}{\partial \vartheta} \dot{\vartheta}
	+\!\left(\frac{\partial \Psi}{\partial \Fve} \Fve^{\star} 
	+ \frac{\partial \Psi}{\partial \Fe} \Fe^{\star}\right)\! \mathbf{F}^{-\star}
	\!:\! \operatorname{D}_t^{\mathbf g}\mathbf{F}
	-\Fe^\star \frac{\partial \Psi}{\partial \Fe} \Fv^{-\star} \!:\! \dot\Fv
	\\
	& \quad -\Fve^\star\!\left(\frac{\partial \Psi}{\partial \Fve} 
	+ \frac{\partial \Psi}{\partial \Fe} \Fv^{-\star} \right)\! \Fn^{-\star}
	\!:\! \dot\Fn
	-\Fn^\star\Fve^\star\!\left(\frac{\partial \Psi}{\partial \Fve} 
	+ \frac{\partial \Psi}{\partial \Fe} \Fv^{-\star} \right)\!\Fn^{-\star} \Fth^{-1} \!:\! \dot\Fth
	\,,
\end{split}
\end{equation}
which mirrors~\eqref{eq:Psi_dot} by means of~\eqref{eq:psi_psi-tilde_EQ}. Further note that~\eqref{eq:psi_psi-tilde_EQ} gives the following relations
\begin{subequations}
\label{eq:Psi_tilde_rel}
\begin{align}
\label{eq:Psi_tilde_rel_a}
\frac{\partial \tilde\Psi}{\partial \Fn}
	&= - \Fve^{\star} \frac{\partial \tilde\Psi}{\partial \mathbf{F}} \Fth^{\star}
	\,,\\
\label{eq:Psi_tilde_rel_b}
\Fth^\star \frac{\partial \tilde\Psi}{\partial \Fth}
	&= - \mathbf{F}^{\star} \frac{\partial \tilde\Psi}{\partial \mathbf{F}}
	\,.
\end{align}
\end{subequations}
\end{remark}

Let us now revisit the entropy inequality~\eqref{loc_Thermo_Second} by substituting into~\eqref{eq:Legendre_trans}
\begin{equation}\label{eq:loc_Thermo_Second_1}
\dot\eta = \rho\dot{\mathscr E} -\rho\dot\Psi - \rho \dot\vartheta \mathscr{N} + \operatorname{Div}\mathbf Q - \frac{1}{\vartheta} \langle d\vartheta, \mathbf Q \rangle -\rho R - \!\left(S_m-\dot{\rho} - \rho\,\dot{\Fa}\!:\!\Fa^{-\star} \right)\! \vartheta \mathscr N \geq 0\,.
\end{equation}
By further substituting in equations~\eqref{eq:loc_Thermo_First},~\eqref{eq:rate-decomp}, and~\eqref{eq:Psi_dot1}, one finds
\begin{equation}\label{eq:loc_Thermo_Second_3}
\begin{aligned}
	\dot\eta & = 
	\mathbf{F}^{-1} \!\left[\mathbf P
	- \rho \mathbf g^\sharp \frac{\partial \tilde\Psi}{\partial \mathbf{F}} \right]\!:\!\Ds + \mathbf{F}^{-1} \mathbf P \!:\! \Da
	- \rho \frac{\partial \tilde\Psi}{\partial \Fv} \!:\! \dot\Fv
	- \rho \frac{\partial \tilde\Psi}{\partial \Fn} \!:\! \dot\Fn
	- \rho \frac{\partial \tilde\Psi}{\partial \Fth} \!:\! \dot\Fth
	- \rho \!\left( \mathscr{N} +\frac{\partial \tilde\Psi}{\partial \vartheta} \right)\!  \dot\vartheta
	\\&\quad
	+\llangle \operatorname{Div}\mathbf{P}+ \rho(\boldsymbol{\mathsf{B}} - \mathbf{A}),\mathbf{V} \rrangle_{\mathbf{g}}
	+ \!\left(S_m-\dot{\rho} - \rho\,\dot{\Fa}\!:\!\Fa^{-\star} \right)\!\!\left(\Psi+\frac{1}{2} \Vert\mathbf V\Vert^2_{\mathbf g}\right)\!
	- \frac{1}{\vartheta} \langle d\vartheta, \mathbf Q \rangle
	\geq 0\,.
\end{aligned}
\end{equation}
The inequality above must hold for all thermoelastic configurations $(\varphi,\vartheta)$ such that $\Ds$, $\Da$, $\mathbf{V}$, $\Vert\mathbf V\Vert^2_{\mathbf g}$ and $\dot\vartheta$ may be chosen independently.\footnote{Note that introducing any additive constant to the free energy has no effect on the thermodynamic equilibrium. This means that the inequality~\eqref{eq:loc_Thermo_Second_3} must remain unchanged under the transformations $\Psi \to \Psi+a$, $\forall~a\in\mathbb{R}$, and hence,~\eqref{eq:Bal_M} must hold.}
It follows that~\eqref{eq:loc_Thermo_Second_3} can hold only if
\begin{subequations}\label{eq:Govern}
\begin{empheq}[left={\empheqlbrace }]{align}
	\label{eq:Bal_M}
	\dot{\rho} + \rho\,\dot{\Fa}\!:\!\Fa^{-\star} &= S_m\,,\\
	\label{eq:Lin_M}
	\operatorname{Div}\mathbf{P}+ \rho\boldsymbol{\mathsf{B}} &= \rho \mathbf{A}\,,\\
	\label{eq:Ang_M}
	\mathbf{F}\mathbf{P}^\star &=\mathbf{P}\mathbf{F}^\star\,,
\end{empheq}
\end{subequations}
and
\begin{subequations}
\begin{align}
\label{eq:Cons_Eq_stress}
	&\mathbf{P} = \rho \mathbf g^\sharp \frac{\partial \tilde\Psi}{\partial \mathbf{F}}
	\,,\\
\label{eq:Legendre_conj}
	&\mathscr N = -\frac{\partial \tilde\Psi}{\partial \vartheta}
	\,.
\end{align}
\end{subequations}
Therefore, we effectively obtain the balance of mass~\eqref{eq:Bal_M}, the balances of linear and angular momenta~\eqref{eq:Lin_M} and~\eqref{eq:Ang_M}, respectively, the Doyle-Ericksen formula~\eqref{eq:Cons_Eq_stress}, which identifies $\mathbf P$ as the first Piola-Kirchhoff stress tensor, and the conjugacy condition~\eqref{eq:Legendre_conj} for the pair $\{\mathscr{N},\vartheta\}$ arising from the Legendre transform~\eqref{eq:Legendre_trans}.\footnote{The transform~\eqref{eq:Legendre_trans} of $\mathscr{E}$ to $\Psi$ with respect to the conjugate pair $\{\mathscr{N},\vartheta\}$ is essentially a change of variable satisfying~\eqref{eq:Legendre_conj}. See \citep{Arnold1989ClassMech, Goldstein2002ClassMech} for further details in the context of Lagrangian mechanics and thermodynamics.}
Note that, by recalling~\eqref{eq:psi_psi-tilde_EQ-F},~\eqref{eq:Cons_Eq_stress} may alternatively be written as
\begin{equation} \label{eq:Pstress_decomp}
	\mathbf{P}= \rho \mathbf g^\sharp\!\left(\frac{\partial \Psi}{\partial \Fve} \Fve^{\star} 
	+ \frac{\partial \Psi}{\partial \Fe} \Fe^{\star}\right)\! \mathbf{F}^{-\star}\,,
\end{equation}
which may be cast as $\mathbf{P} = \mathbf{P}_{\textrm{eq}} + \mathbf{P}_{\textrm{neq}}$
with
\begin{equation}
\label{eq:P_eq_neq}
	\mathbf{P}_{\textrm{eq}} 
	= \rho \mathbf g^\sharp\frac{\partial \Psi}{\partial \Fve} \Fve^{\star} \mathbf{F}^{-\star}\,,
	\quad \textrm{and} \qquad
	\mathbf{P}_{\textrm{neq}} 
	= \rho \mathbf g^\sharp \frac{\partial \Psi}{\partial \Fe} \Fe^{\star} \mathbf{F}^{-\star}\,,
\end{equation}
such that $\mathbf{P}_{\textrm{neq}}$ represents the instantaneous (non-equlibrium) elastic stress response conjugate to the instantaneous elastic distortion $\Fe$ and $\mathbf{P}_{\textrm{eq}}$ represents the long-term (equilibrium) viscoelastic stress conjugate to the viscoelastic distortion $\Fve$.\footnote{
In the particular case where we assume the energy additive split~\eqref{eq:Psi_Fve_decomp}, one finds that
\begin{equation}
	\mathbf{P}_{\textrm{eq}} 
	= \rho \mathbf g^\sharp
	\frac{\partial \Psi_{\textrm{eq}}}{\partial \Fve} \Fve^{\star} \mathbf{F}^{-\star}\,,
	\quad \textrm{and} \qquad
	\mathbf{P}_{\textrm{neq}} 
	= \rho \mathbf g^\sharp 
	\frac{\partial \Psi_{\textrm{neq}}}{\partial \Fe} \Fe^{\star} \mathbf{F}^{-\star}\,.
\end{equation}
}

\begin{remark}
From the decomposition $\Fa=\Fn\Fth$, it follows that $\dot{\Fa}\Fa^{-1}=\dot{\Fn}\Fn^{-1}+\Fn\dot\Fth\Fth^{-1}\Fn^{-1}$. Moreover, applying the chain rule to the constitutively prescribed relation $\Fth=\Fth(X,\vartheta)$ in~\eqref{eq:Fth} gives $\dot\Fth=\boldsymbol\alpha\Fth\,\dot\vartheta$. Finally, using $\dot\rho=\dfrac{\partial\rho}{\partial\vartheta}\dot\vartheta+\dfrac{\partial\rho}{\partial\Fn}\!:\!\dot\Fn$, the mass balance~\eqref{eq:Bal_M} can be rewritten as
\begin{equation} \label{eq:Trans_Bal_M}
	\!\left(\frac{\partial\rho}{\partial\vartheta}
	+\rho\,\operatorname{tr}\,\boldsymbol{\alpha}\right)\!\dot\vartheta
	+\!\left(\frac{\partial\rho}{\partial\Fn}+\rho\,\Fn^{-\star}\right)\!:\!\dot\Fn= S_m\,.
\end{equation}
In many settings one may assume that the rate of added mass $S_m$ does not depend on the temperature rate $\dot\vartheta$ (although it may depend on $\vartheta$), i.e.,~$S_m = S_m(X,\vartheta,\Fn,\dot\Fn)$. Since $\dot\vartheta$ and $\dot\Fn$ may be prescribed independently, it forces the coefficient of $\dot\vartheta$ to vanish and the $\dot\Fn$ term to match $S_m$, i.e.
\begin{subequations}
\begin{empheq}[left={\empheqlbrace\,}]{align}
	\label{eq:Temp_MassBal}
	& \frac{\partial\rho}{\partial\vartheta} + \rho\,\operatorname{tr}\,\boldsymbol{\alpha} = 0\,,\\
	& \!\left(\frac{\partial\rho}{\partial\Fn}+\rho\,\Fn^{-\star}\right)\!:\!\dot\Fn = S_m\,.
\end{empheq}
\end{subequations}
More generally, assuming a linear rate-dependent source
\begin{equation}
	S_m=\boldsymbol{S}_n(\vartheta,\Fn)\!:\!\dot\Fn + S_\vartheta(\vartheta,\Fn)\,\dot\vartheta\,,
\end{equation}
the balance of mass~\eqref{eq:Trans_Bal_M} takes the form
\begin{subequations}
\begin{empheq}[left={\empheqlbrace\,}]{align}
	\frac{\partial\rho}{\partial\vartheta} + \rho\,\operatorname{tr}\,\boldsymbol{\alpha} 
	&= S_\vartheta\,,\\
	\frac{\partial\rho}{\partial\Fn} + \rho\,\Fn^{-\star} &= \boldsymbol{S}_n\,.
\end{empheq}
\end{subequations}
\end{remark}

\begin{remark}[Spatial form of the balance laws and the Doyle-Ericksen formula]
Let $\varrho$ denote the spatial mass density, i.e.,~the mass density in the deformed configuration. Recalling that $J$, the Jacobian of the motion, satisfies $\varphi^{*}\mathrm{d}v = J\,\mathrm{d}V$, it follows that $\varrho J = \rho$. Hence, the material balance of mass~\eqref{eq:Bal_M} in spatial form reads
\begin{equation}
\dot\varrho + \varrho\,\operatorname{div}\mathbf{v} = s_m\,,
\end{equation}
where $s_m=S_m/J$ is the spatial rate of change of mass per unit deformed volume. The Cauchy stress tensor is related to the first Piola--Kirchhoff stress as $\boldsymbol{\sigma}=J^{-1}\mathbf{P}\mathbf{F}^{\star}$. The balance of linear and angular momenta~\eqref{eq:Lin_M} and~\eqref{eq:Ang_M} in spatial form therefore read
\begin{subequations}
\begin{align}
	\label{eq:Lin_M_spt}
	\operatorname{div}\boldsymbol\sigma + \varrho\boldsymbol{\mathsf{b}} &= \varrho\mathbf{a}\,,\\
	\label{eq:Ang_M_spt}
	\boldsymbol\sigma^{\star} &= \boldsymbol\sigma\,,
\end{align}
\end{subequations}
where $\boldsymbol{\mathsf{b}}(x)=\boldsymbol{\mathsf{B}}(\varphi_t^{-1}(x))$ is the spatial body force, and $\mathbf a$ is the spatial acceleration. In terms of the Cauchy stress, the constitutive equation~\eqref{eq:Cons_Eq_stress} takes the form
\begin{equation}
	\label{eq:Cons_Eq_sig}
	\boldsymbol\sigma
	= \varrho \mathbf g^\sharp \frac{\partial \tilde\Psi}{\partial \mathbf{F}} \mathbf{F}^{\star}
	= \varrho \mathbf g^\sharp\!\left(\frac{\partial \Psi}{\partial \Fve} \Fve^{\star} + \frac{\partial \Psi}{\partial \Fe} \Fe^{\star}\right)\! \,,
\end{equation}
which may be written as
$\boldsymbol\sigma = \boldsymbol\sigma_{\textrm{eq}} + \boldsymbol\sigma_{\textrm{neq}}\,,$
where
\begin{equation}\label{eq:sig_eq_neq}
	\boldsymbol\sigma_{\textrm{eq}}
	= \varrho\,\mathbf g^{\sharp}\frac{\partial \Psi}{\partial \Fve}\Fve^{\star}\,,
	\quad \textrm{and} \qquad
	\boldsymbol\sigma_{\textrm{neq}}
	= \varrho\,\mathbf g^{\sharp}\frac{\partial \Psi}{\partial \Fe}\Fe^{\star}\,.
\end{equation}
Recalling the equivalent functional form $\Psi=\hat{\Psi}(X,\vartheta,\Cve,\Ce,\Go)$, one finds
\begin{equation}\label{C-F_transform}
	\mathbf g^{\sharp}\frac{\partial \Psi}{\partial \Fve} = 2\Fve\frac{\partial \hat{\Psi}}{\partial\Cve}
	\,,\qquad
	\mathbf g^{\sharp}\frac{\partial \Psi}{\partial \Fe} = 2\Fe\frac{\partial \hat{\Psi}}{\partial\Ce}\,.
\end{equation}
Therefore, the Cauchy stress representation may be written equivalently as
\begin{equation}\label{eq:sig_eq_neq2}
	\boldsymbol\sigma_{\textrm{eq}}
	=2\varrho\!\left(\Fve\frac{\partial \hat{\Psi}}{\partial\Cve}\Fve^{\star}\right)\!,
	\quad \textrm{and} \qquad
	\boldsymbol\sigma_{\textrm{neq}}
	=2\varrho\!\left(\Fe\frac{\partial \hat{\Psi}}{\partial\Ce}\Fe^{\star}\right)\!,
\end{equation}
and hence, also as
\begin{equation}\label{eq:sig_spat_rep}
	\boldsymbol\sigma
	=2\varrho\,\frac{\partial \Psi}{\partial \mathbf g}\,.
\end{equation}
Finally, the second Piola--Kirchhoff stress, $\mathbf S=\mathbf{F}^{-1}\mathbf P=J \mathbf{F}^{-1}\boldsymbol\sigma\mathbf{F}^{-\star}$, may be represented as $ \mathbf S = \mathbf S_{\textrm{eq}} +\mathbf S_{\textrm{neq}}$, with
\begin{equation}\label{eq:S_eq_neq}
	\mathbf S_{\textrm{eq}}
	=2\rho\,\mathbf{F}^{-1}\!\left(\Fve\frac{\partial \hat{\Psi}}{\partial\Cve}\Fve^{\star}\right)\!\mathbf{F}^{-\star}\,,
	\quad \textrm{and} \qquad
	\mathbf S_{\textrm{neq}}
	=2\rho\,\mathbf{F}^{-1}\!\left(\Fe\frac{\partial \hat{\Psi}}{\partial\Ce}\Fe^{\star}\right)\!\mathbf{F}^{-\star}\,.
\end{equation}
\end{remark}

Let us now revisit the Clausius-Duhem inequality~\eqref{eq:loc_Thermo_Second_3} by substituting in the balance laws~\eqref{eq:Govern} and the constitutive formulae~\eqref{eq:Cons_Eq_stress} and~\eqref{eq:Legendre_conj}:
\begin{equation}\label{eq:loc_Thermo_Second_4}
\begin{aligned}
	\dot\eta & =
	- \rho \frac{\partial \tilde\Psi}{\partial \Fv} \!:\! \dot\Fv
	- \rho \frac{\partial \tilde\Psi}{\partial \Fn} \!:\! \dot\Fn
	- \rho \frac{\partial \tilde\Psi}{\partial \Fth} \!:\! \dot\Fth
	- \frac{1}{\vartheta} \langle d\vartheta, \mathbf Q \rangle
	\geq 0\,.
\end{aligned}
\end{equation}
In a visco-anelastic solid, the rate of energy dissipation may be written as the sum of a viscoelastic, an athermal-anelastic, and a thermal dissipation \citep{LeTallec1993, Maugin2010ConfForces}
\begin{equation}\label{eq:dissipation}
	\dot\eta = -\Bv \!:\! \dot\Fv - \Bn\!:\!\dot{\Fn} - \Bth\!:\!\dot{\Fth}
	- \frac{1}{\vartheta} \langle d\vartheta, \mathbf Q \rangle\,,
\end{equation}
where $\Bv$, $\Bn$, and $\Bth$ are the generalized configurational forces corresponding to the viscoelastic, athermal-anelastic, and thermal dissipation contributions, respectively.
Noting the arbitrariness of the independent variables $\dot{\Fv}$, $\dot{\Fn}$, and  $\dot\Fth$, it follows by identification of~\eqref{eq:loc_Thermo_Second_4} and~\eqref{eq:dissipation} that
\begin{subequations}\label{eq:Config}
\begin{align}
\label{eq:Config_v}
	\Bv & = \rho \frac{\partial \tilde\Psi}{\partial \Fv} \,,\\
\label{eq:Config_n}
	\Bn & = \rho \frac{\partial \tilde\Psi}{\partial \Fn} \,,\\
\label{eq:Config_th}
	\Bth & = \rho \frac{\partial \tilde\Psi}{\partial \Fth}
	\,.
\end{align}
\end{subequations}

\begin{remark}
Recalling~\eqref{eq:psi_psi-tilde_EQ} and~\eqref{C-F_transform}, the configurational forces may equivalently be written as
\begin{subequations}\label{eq:C_Config}
\begin{align}
\label{eq:C_Config_v}
	\Bv 
	&= -\rho \Fe^\star \frac{\partial \Psi}{\partial \Fe} \Fv^{-\star}
	= - 2 \rho \Ce \frac{\partial \hat{\Psi}}{\partial\Ce} \Fv^{-\star}
	\,,\\ 
\label{eq:C_Config_n}
	\Bn
	&= -\rho \Fve^\star\!\left(\frac{\partial \Psi}{\partial \Fve} + \frac{\partial \Psi}{\partial \Fe} \Fv^{-\star} \right)\!\Fn^{-\star}
	= - 2 \rho \!\left( \Cve \frac{\partial \hat{\Psi}}{\partial\Cve} + \Fv^\star \Ce\frac{\partial \hat{\Psi}}{\partial\Ce} \Fv^{-\star} \right)\!\Fn^{-\star}
	\,,\\
\label{eq:C_Config_th}
	\Bth
	&= -\rho \Fn^\star\Fve^\star\!\left(\frac{\partial \Psi}{\partial \Fve} + \frac{\partial \Psi}{\partial \Fe} \Fv^{-\star} \right)\!\Fn^{-\star} \Fth^{-\star}
	= - 2 \rho \Fn^\star\!\left(\Cve\frac{\partial \hat{\Psi}}{\partial\Cve} + \Fv^\star \Ce \frac{\partial \hat{\Psi}}{\partial\Ce} \Fv^{-\star} \right)\!\Fn^{-\star} \Fth^{-\star}
	\,.
\end{align}
\end{subequations}
\end{remark}

\begin{remark}
Note that the athermal configurational forces contributing to the dissipation~\eqref{eq:dissipation}\textemdash i.e.,~viscous $\Bv$ and athermal-anelastic $\Bn\,$\textemdash are not completely determined by the general thermodynamic structure. Equations~\eqref{eq:Config_v} and~\eqref{eq:Config_n} provide their coupling to the stress state of the solid, but they do not by themselves fix the actual evolution law. As a matter of fact, their evolution has yet to be constitutively supplied through additional constitutive assumptions\textemdash typically by introducing a Rayleigh-type dissipation potential, as in \S~\ref{S:Kinetic}, or, more generally, a kinetic flow rule tailored to the particular anelastic mechanism under consideration. By contrast, the thermal part is already fixed since the evolution equation for temperature $\vartheta$ is given by the heat equation\textemdash discussed in the subsequent subsection~\S\ref{S:HeatEq}, and hence that of $\Fth$ and $\Bth$\textemdash readily read through~\eqref{eq:Fth} and~\eqref{eq:Config_th}\textemdash or~\eqref{eq:C_Config_th}, respectively.
\end{remark}

\subsubsection{Heat equation for compressible visco-anelastic solids}
\label{S:HeatEq}

Let us next derive the heat equation for a nonlinear visco-anelastic solid.
We begin by revisiting the localized energy balance~\eqref{eq:loc_Thermo_First} by using~\eqref{eq:Bal_M}-\eqref{eq:Ang_M} and substituting in the Legendre transform~\eqref{eq:Legendre_trans} to find
\begin{equation}\label{eq:to_HtEq0}
	\rho\,\dot{\Psi}+\rho\,\dot{\vartheta}\mathscr{N}+\rho\,\vartheta\dot{\mathscr{N}} 
	= \mathbf{g} \mathbf P \!:\! \operatorname{D}_t^{\mathbf g}\mathbf{F} +\rho R - \operatorname{Div} \mathbf{Q}\,.
\end{equation}
Recalling~\eqref{eq:Psi_dot}, one writes
\begin{equation}
\label{eq:Psi_dot2}
\begin{split}
	\dot{\Psi}
	= \frac{\partial \tilde\Psi}{\partial \vartheta} \dot{\vartheta}
	+\frac{\partial \tilde\Psi}{\partial \mathbf F} \!:\! \operatorname{D}_t^{\mathbf g}\mathbf F
	+\frac{\partial \tilde\Psi}{\partial \Fv} \!:\! \dot\Fv
	+\frac{\partial \tilde\Psi}{\partial \Fn} \!:\! \dot\Fn
	+\frac{\partial \tilde\Psi}{\partial \Fth} \!:\! \dot\Fth
	\,.
\end{split}
\end{equation}
Using~\eqref{eq:Legendre_conj}, and similarly to~\eqref{eq:Psi_dot2}, it follows that
\begin{equation}\label{eq:N_dot}
\begin{split}
\dot{\mathscr{N}}
	= -\frac{d}{dt}{\frac{\partial \tilde\Psi}{\partial \vartheta}}
	&= - \frac{\partial^2 \tilde\Psi}{\partial \vartheta^2} \dot{\vartheta}
	- \frac{\partial^2 \tilde\Psi}{\partial \mathbf F \partial \vartheta} \!:\! \operatorname{D}_t^{\mathbf g}\mathbf F
	- \frac{\partial^2 \tilde\Psi}{\partial \Fv \partial \vartheta} \!:\! \dot\Fv
	- \frac{\partial^2 \tilde\Psi}{\partial \Fn \partial \vartheta} \!:\! \dot\Fn
	- \frac{\partial^2 \tilde\Psi}{\partial \Fth \partial \vartheta} \!:\! \dot\Fth \\
	&= - \frac{\partial^2 \tilde\Psi}{\partial \vartheta^2} \dot{\vartheta}
	- \frac{\partial^2 \tilde\Psi}{\partial \vartheta \partial \mathbf F} \!:\! \operatorname{D}_t^{\mathbf g}\mathbf F
	- \frac{\partial^2 \tilde\Psi}{\partial \vartheta \partial \Fv} \!:\! \dot\Fv
	- \frac{\partial^2 \tilde\Psi}{\partial \vartheta \partial \Fn} \!:\! \dot\Fn
	- \frac{\partial^2 \tilde\Psi}{\partial \vartheta \partial \Fth} \!:\! \dot\Fth
	\,.
\end{split}
\end{equation}
Plugging~\eqref{eq:Psi_dot2} and~\eqref{eq:N_dot} into~\eqref{eq:to_HtEq0}, and using~\eqref{eq:Legendre_conj} and~\eqref{eq:Cons_Eq_stress}, one finds the heat equation for a nonlinear visco-anelastic solid
\begin{equation}\label{eq:to_HtEq1}
\begin{split}
	&- \rho \vartheta \frac{\partial^2 \tilde\Psi}{\partial \vartheta^2} \dot\vartheta
	+ \rho \!\left[ \frac{\partial \tilde\Psi}{\partial \Fth}
	- \vartheta \frac{\partial^2 \tilde\Psi}{\partial \vartheta \partial \Fth} \right]\! \!:\! \dot\Fth
	-\rho\vartheta \frac{\partial^2 \tilde\Psi}{\partial \vartheta \partial \mathbf{F}} \!:\! \operatorname{D}_t^{\mathbf g}\mathbf{F}
	\\&
	+ \rho \!\left[ \frac{\partial \tilde\Psi}{\partial \Fv}
	- \vartheta \frac{\partial^2 \tilde\Psi}{\partial \vartheta \partial \Fv} \right]\! \!:\! \dot\Fv
	+ \rho \!\left[ \frac{\partial \tilde\Psi}{\partial \Fn}
	- \vartheta \frac{\partial^2 \tilde\Psi}{\partial \vartheta \partial \Fn} \right]\! \!:\! \dot\Fn
	= \rho R - \operatorname{Div} \mathbf{Q}
	\,.
\end{split}
\end{equation}
Recalling~\eqref{eq:Psi_tilde_rel} and using
$\!\left(\operatorname{D}_t^{\mathbf g} \mathbf{F} \right)\! \mathbf{F}^{-1} = \!\left(\operatorname{D}_t^{\mathbf g} \Fe \right)\! \Fv \Fn \Fth + \Fe \dot\Fv \Fn \Fth + \Fe \Fv \dot{\Fn} \Fth + \Fe \Fv \Fn \dot{\Fth}$,
one finds
\begin{equation}
\label{eq:tilde_Psi_F}
\begin{split}
\frac{\partial \tilde\Psi}{\partial \mathbf{F}} \!:\! \operatorname{D}_t^{\mathbf g}\mathbf{F}
	&= \frac{\partial \tilde\Psi}{\partial \mathbf{F}} \!:\! \!\left(\operatorname{D}_t^{\mathbf g} \Fe \right)\! \Fv \Fn \Fth
	+ \frac{\partial \tilde\Psi}{\partial \mathbf{F}} \!:\! \Fe \dot\Fv \Fn \Fth
	+ \frac{\partial \tilde\Psi}{\partial \mathbf{F}} \!:\! \Fe \Fv \dot{\Fn}\Fth
	+ \frac{\partial \tilde\Psi}{\partial \mathbf{F}} \!:\! \Fe \Fv \Fn \dot{\Fth}
	\\&
	= \frac{\partial \tilde\Psi}{\partial \mathbf{F}} \!:\! 
	\!\left(\operatorname{D}_t^{\mathbf g} \Fe \right)\! \Fv \Fn \Fth
	+ \frac{\partial \tilde\Psi}{\partial \mathbf{F}} \!:\! \Fe \dot\Fv \Fn \Fth
	+ \Fv^\star \Fe^\star \frac{\partial \tilde\Psi}{\partial \mathbf{F}} \Fth^\star \!:\! \dot{\Fn}
	+ \Fn^\star \Fv^\star \Fe^\star \frac{\partial \tilde\Psi}{\partial \mathbf{F}} \!:\! \dot{\Fth}
	\\&
	= \frac{\partial \tilde\Psi}{\partial \mathbf{F}} \mathbf{F}^\star \Fe^{-\star} \!:\! 
	\operatorname{D}_t^{\mathbf g} \Fe
	+ \Fe^\star \frac{\partial \tilde\Psi}{\partial \mathbf{F}} \mathbf{F}^\star \Fve^{-\star} \!:\! \dot\Fv
	- \frac{\partial \tilde\Psi}{\partial \Fn} \!:\! \dot{\Fn}
	- \frac{\partial \tilde\Psi}{\partial \Fth} \!:\! \dot{\Fth}\,.
\end{split}
\end{equation}
Noting that the partial derivative with respect to temperature $\vartheta$ is evaluated at fixed $\{\mathbf{F}, \Fv, \Fn, \Fth, \Go, \mathbf{g}\}$, one finds from~\eqref{eq:tilde_Psi_F} that
\begin{equation}
\frac{\partial^2 \tilde\Psi}{\partial \vartheta \partial \mathbf{F}} \!:\! \operatorname{D}_t^{\mathbf g}\mathbf{F}
	= \frac{\partial^2 \tilde\Psi}{\partial \vartheta \partial \mathbf{F}} \mathbf{F}^\star \Fe^{-\star} \!:\! \operatorname{D}_t^{\mathbf g} \Fe
	+ \Fe^\star \frac{\partial^2 \tilde\Psi}{\partial \vartheta \partial \mathbf{F}} \mathbf{F}^\star \Fve^{-\star} \!:\! \dot\Fv
	- \frac{\partial^2 \tilde\Psi}{\partial \vartheta \Fn} \!:\! \dot{\Fn}
	- \frac{\partial^2 \tilde\Psi}{\partial \vartheta \Fth} \!:\! \dot{\Fth}\,.
\end{equation}
This readily recasts the heat equation~\eqref{eq:to_HtEq1} as
\begin{equation}\label{eq:to_HtEq1b}
\begin{split}
	&- \rho \vartheta \frac{\partial^2 \tilde\Psi}{\partial \vartheta^2} \dot\vartheta
	+ \rho \frac{\partial \tilde\Psi}{\partial \Fth} \!:\! \dot\Fth
	-\rho\vartheta \frac{\partial^2 \tilde\Psi}{\partial \vartheta \partial \mathbf{F}} \mathbf{F}^\star \Fe^{-\star} \!:\! \operatorname{D}_t^{\mathbf g} \Fe
	\\&
	+ \rho \!\left[ \frac{\partial \tilde\Psi}{\partial \Fv}
	- \vartheta \!\left( \frac{\partial^2 \tilde\Psi}{\partial \vartheta \partial \Fv} + \Fe^\star \frac{\partial^2 \tilde\Psi}{\partial \vartheta \partial \mathbf{F}} \mathbf{F}^\star \Fve^{-\star} \right)\! \right]\! \!:\! \dot\Fv
	+ \rho \frac{\partial \tilde\Psi}{\partial \Fn} \!:\! \dot\Fn
	= \rho R - \operatorname{Div} \mathbf{Q}
	\,.
\end{split}
\end{equation}

\begin{defi}[Specific heat capacity at constant non-thermal distortions]
\label{def:spec_heat}
We define the \emph{specific heat capacity at constant non-thermal distortions}, denoted $c_{\textrm D}$, as the quantity of heat required, in the absence of external heat supply (i.e.,~$R=0$), to produce a unit temperature increase in a unit mass of material through a purely thermal process\textemdash i.e.,~in which all non-thermal distortions $\{\Fe, \Fv, \Fn\}$ are held fixed.
It is is given by
\begin{equation}\label{eq:c_E_def}
    \operatorname{Div}\mathbf{Q} = -\rho\,c_{\textrm D} \dot{\vartheta}\,,
\end{equation}
and it follows by identification of~\eqref{eq:c_E_def} and~\eqref{eq:to_HtEq1b} with $\operatorname{D}_t^{\mathbf g}\Fe = 0 $, $\dot\Fv = \dot\Fn = 0$, and $ R = 0$, that
\begin{equation}\label{eq:c_D_exp}
c_{\textrm D}
	= - \vartheta \frac{\partial^2 \tilde\Psi}{\partial \vartheta^2} + \frac{1}{\dot\vartheta} \frac{\partial \tilde\Psi}{\partial \Fth} \!:\! \dot\Fth
	= - \vartheta \frac{\partial^2 \tilde\Psi}{\partial \vartheta^2} + \frac{\partial \tilde\Psi}{\partial \Fth} \!:\! \frac{\partial\Fth}{\partial\vartheta}
	\,,
\end{equation}
where, by expansion via the chain rule, $\dot\Fth = ({\partial\Fth}/{\partial\vartheta})\,\dot\vartheta$.\footnote{Note that, if one further adopts the thermal-expansion constitutive equation~\eqref{eq:Fth}, it follows that ${\partial\Fth}/{\partial\vartheta} = \boldsymbol\alpha\Fth$, and \eqref{eq:c_D_exp} reduces to
\begin{equation}\label{eq:c_D_exp2}
c_{\textrm D}
	= - \vartheta \frac{\partial^2 \tilde\Psi}{\partial \vartheta^2} + \frac{\partial \tilde\Psi}{\partial \Fth} \!:\! \boldsymbol\alpha\Fth
	\,.
\end{equation}
}
The first term in \eqref{eq:c_D_exp} represents the purely caloric contribution, namely, the specific heat at fixed distortion, whereas the second term represents thermoelastic coupling. As a matter of fact, \eqref{eq:Psi_tilde_rel_b} shows that $\partial\tilde\Psi/\partial\Fth$ is related by conjugation to ${\partial \tilde\Psi}/{\partial \mathbf{F}}$, and hence to the total stress; its contraction with the temperature derivative of $\Fth$ therefore measures the stress power expended against thermal expansion per unit temperature increase and vanishes for a stress-free motion.
\end{defi}

Using the specific heat introduced in Definition~\ref{def:spec_heat}, the heat equation~\eqref{eq:to_HtEq1b} may be rewritten as
\begin{equation}
\label{eq:to_HtEq2}
\begin{split}
	&\rho c_{\text D} \dot\vartheta
	-\rho\vartheta \frac{\partial^2 \tilde\Psi}{\partial \vartheta \partial \mathbf{F}} \mathbf{F}^\star \Fe^{-\star} \!:\! \operatorname{D}_t^{\mathbf g} \Fe
	+ \rho \!\left[ \frac{\partial \tilde\Psi}{\partial \Fv}
	- \vartheta \!\left( \frac{\partial^2 \tilde\Psi}{\partial \vartheta \partial \Fv} + \Fe^\star \frac{\partial^2 \tilde\Psi}{\partial \vartheta \partial \mathbf{F}} \mathbf{F}^\star \Fve^{-\star} \right)\! \right]\! \!:\! \dot\Fv
	+ \rho \frac{\partial \tilde\Psi}{\partial \Fn} \!:\! \dot\Fn
	= \rho R - \operatorname{Div} \mathbf{Q}
	\,.
\end{split}
\end{equation}
Recalling that the partial derivative with respect to temperature $\vartheta$ is evaluated at fixed $\{\mathbf{F}, \Fv, \Fn, \Fth, \Go, \mathbf{g}\}$, following~\eqref{eq:Cons_Eq_sig}  one finds that
\begin{equation}
\label{eq:sig_term_trans}
\frac{\partial^2 \tilde\Psi}{\partial \vartheta \partial \mathbf{F}} \mathbf{F}^\star \Fe^{-\star} \!:\! \operatorname{D}_t^{\mathbf g}\Fe
	= \mathbf g \frac{\partial}{\partial \vartheta}\!\left[ \mathbf{g}^\sharp \frac{\partial \tilde\Psi}{\partial \mathbf{F}} \mathbf{F}^\star \right]\! \!:\! (\operatorname{D}_t^{\mathbf g}\Fe) \Fe^{-\star}
	= \mathbf g \frac{\partial}{\partial \vartheta}\!\left[ \frac{ \boldsymbol\sigma}{\rho/J} \right]\! \!:\! (\operatorname{D}_t^{\mathbf g}\Fe) \Fe^{-\star}
\,,
\end{equation}
and following~\eqref{eq:psi_psi-tilde_EQ-F},~\eqref{eq:psi_psi-tilde_EQ-Fv}, and~\eqref{eq:sig_eq_neq}, one finds
\begin{equation}
\frac{\partial^2 \tilde\Psi}{\partial \vartheta \partial \Fv} + \Fe^\star \frac{\partial^2 \tilde\Psi}{\partial \vartheta \partial \mathbf{F}} \mathbf{F}^\star \Fve^{-\star}
	= \frac{\partial}{\partial \vartheta}\!\left[\frac{\partial \tilde\Psi}{\partial \Fv} + \Fe^\star \frac{\partial \tilde\Psi}{\partial \mathbf{F}} \mathbf{F}^\star \Fve^{-\star}\right]\!
	= \Fe^\star \mathbf{g} \frac{\partial}{\partial \vartheta}\!\left[\mathbf{g}^\sharp \frac{\partial \Psi}{\partial \Fve} \Fve^\star \right]\! \Fve^{-\star}
	= \Fe^\star \mathbf{g} \frac{\partial}{\partial \vartheta}\!\left[ \frac{\boldsymbol\sigma_{\textrm{eq}}}{\rho/J} \right]\! \Fve^{-\star}
	\,.
\end{equation}
Thus, the heat equation~\eqref{eq:to_HtEq2} may be recast as\footnote{
Note that, similarly to~\eqref{eq:rate-decomp}, one may recast the second term in~\eqref{eq:to_HtEq_S} as
\begin{equation}\label{eq:HEq_term_re}
\mathbf{g} \frac{\partial}{\partial \vartheta}\!\left[ \frac{ \boldsymbol\sigma}{\rho/J} \right]\! \!:\! (\operatorname{D}_t^{\mathbf g}\Fe) \Fe^{-1}
	= \frac{\partial}{\partial \vartheta}\!\left[ \frac{ \boldsymbol\sigma}{\rho/J} \right]\! \!:\! \frac{1}{2} \Fe^{-\star} \dot{\Ce}\, \Fe^{-1}\,.
\end{equation}
}
\begin{equation}
\label{eq:to_HtEq_S}
\begin{split}
	&\rho c_{\text D} \dot\vartheta
	-\rho\vartheta \mathbf{g} \frac{\partial}{\partial \vartheta}\!\left[ \frac{ \boldsymbol\sigma}{\rho/J} \right]\! \!:\! (\operatorname{D}_t^{\mathbf g}\Fe) \Fe^{-1}
	- \rho \vartheta \mathbf{g} \frac{\partial}{\partial \vartheta}\!\left[ \frac{\boldsymbol\sigma_{\textrm{eq}}}{\rho/J} \right]\! \!:\! \Fe \dot\Fv \Fv^{-1} \Fe^{-1}
	+ \Bv \!:\! \dot\Fv
	+ \Bn \!:\! \dot\Fn
	= \rho R - \operatorname{Div} \mathbf{Q}
	\,,
\end{split}
\end{equation}
or alternatively as\footnote{
Note that, similarly to~\eqref{eq:HEq_term_re}, one may recast the second and third terms in~\eqref{eq:to_HtEq_Sb} as
\begin{subequations}
\begin{align}
\mathbf{g} \frac{\partial}{\partial \vartheta}\!\left[ \frac{\boldsymbol\sigma_{\textrm{eq}}}{\rho/J} \right]\! \!:\! (\operatorname{D}_t^{\mathbf g}\Fve) \Fve^{-1}
	& = \frac{\partial}{\partial \vartheta}\!\left[ \frac{ \boldsymbol\sigma_{\textrm{eq}}}{\rho/J} \right]\! \!:\! \frac{1}{2} \Fve^{-\star} \dot{\Cve}\, \Fve^{-1}\,,\\
\mathbf{g} \frac{\partial}{\partial \vartheta}\!\left[ \frac{\boldsymbol\sigma_{\textrm{neq}}}{\rho/J} \right]\! \!:\! (\operatorname{D}_t^{\mathbf g}\Fe) \Fe^{-1}
	& = \frac{\partial}{\partial \vartheta}\!\left[ \frac{ \boldsymbol\sigma_{\textrm{neq}}}{\rho/J} \right]\! \!:\! \frac{1}{2} \Fe^{-\star} \dot{\Ce}\, \Fe^{-1}\,.
\end{align}
\end{subequations}
}
\begin{equation}
\label{eq:to_HtEq_Sb}
\begin{split}
	&\rho c_{\text D} \dot\vartheta
	- \rho \vartheta \mathbf{g} \frac{\partial}{\partial \vartheta}\!\left[ \frac{\boldsymbol\sigma_{\textrm{eq}}}{\rho/J} \right]\! \!:\! (\operatorname{D}_t^{\mathbf g}\Fve) \Fve^{-1}
	-\rho\vartheta \mathbf{g} \frac{\partial}{\partial \vartheta}\!\left[ \frac{ \boldsymbol\sigma_{\textrm{neq}}}{\rho/J} \right]\! \!:\! (\operatorname{D}_t^{\mathbf g}\Fe) \Fe^{-1}
	+ \Bv \!:\! \dot\Fv
	+ \Bn \!:\! \dot\Fn
	= \rho R - \operatorname{Div} \mathbf{Q}
	\,.
\end{split}
\end{equation}

\begin{remark}[Corrigendum to \citep{SaYa2025GenColemanNoll}]
In \citep[\S4(d)(i)]{SaYa2025GenColemanNoll}, the expression obtained for the specific heat capacity, Eq.~(4.27) therein, is incorrect. The error has two related sources, both arising from prescribing the thermal distortion as a function of temperature, $\Fth=\Fth(X,\vartheta)$, before deriving the governing equations, as was done in \citep[Remark~2.2 and Eq.~(2.1)]{SaYa2025GenColemanNoll}.
\begin{enumerate}[label=\roman*., leftmargin=*]
\item \emph{Amalgamation of the entropy with the thermal configurational force.} As explained in Remark~\ref{rmrk:ind_fields} and in the discussion preceding the alternative constitutive representation~\eqref{eq:psi_tilde_psi}, $\Fth$ should be treated as an independent field during the derivation of the governing equations. If $\Fth$ is prescribed in advance as a function of temperature, then $\Fe=\mathbf F\Fth^{-1}\Fn^{-1}$ inherits a temperature dependence. Consequently, every temperature derivative of the free energy contains a contribution arising from the dependence of $\Fe$ on $\Fth$. The quantity $-\partial\Psi/\partial\vartheta$ then combines the entropy, which is conjugate to $\vartheta$, with the configurational force conjugate to $\Fth$. This does not alter the balance laws or the dissipation inequality, but it obscures the separate thermal and mechanical origins of the terms appearing in the heat equation~(4.28) of \citep{SaYa2025GenColemanNoll}.
\item \emph{Improper implementation of the constant non-thermal distortion condition.} By definition, the specific heat capacity is the heat absorbed per unit temperature rise at constant non-thermal distortions, i.e., $\operatorname{D}_t^{\mathbf g}\Fe=\mathbf 0$ and $\dot{\Fn}=\mathbf 0$, as stated in Definition~\ref{def:spec_heat}. In \citep{SaYa2025GenColemanNoll}, this condition was instead imposed as $\operatorname{D}_t^{\mathbf g}\mathbf F=\mathbf 0$ and $\dot{\Fn}=\mathbf 0$. Because $\mathbf F=\Fe\Fn\Fth$ contains the thermal distortion and $\Fth$ was assumed to depend on temperature, holding the total deformation fixed during heating does not hold the elastic distortion fixed. Rather,
\begin{equation}
\operatorname{D}_t^{\mathbf g}\Fe
= \operatorname{D}_t^{\mathbf g}\mathbf F\,(\Fn\Fth)^{-1} - \Fe\dot\Fn \Fn^{-1} - \Fe \Fn \dot\Fth \Fth^{-1} \Fn^{-1}
= -\,\Fe\Fn\boldsymbol\alpha\Fn^{-1}\,\dot\vartheta \;\neq\; \mathbf 0 \,.
\end{equation}
The elastic distortion therefore evolves to compensate for the evolution of the thermal distortion. As a result, the stress power associated with this compensating distortion is included in the quantity identified as heat storage. Equation~(4.27) of \citep{SaYa2025GenColemanNoll} consequently represents the heat absorbed at constant total deformation rather than the specific heat capacity defined at constant non-thermal distortions. The amalgamation described in part~(i) causes the associated mechanical contribution to appear within an entropy derivative.
\end{enumerate}
Treating $\Fth$ as an independent field during the derivation, and hence replacing $\boldsymbol\alpha\,\dot\vartheta$ by $\dot{\Fth}\Fth^{-1}$, Eqs.~(4.24a,b) of \citep{SaYa2025GenColemanNoll} become
\begin{align}
\dot\Psi &= -\mathcal N\dot\vartheta
+ \frac{\partial\Psi}{\partial\Fe}(\Fn\Fth)^{-\star}\!:\operatorname{D}_t^{\mathbf g}\mathbf F
- \Fe^{\star}\frac{\partial\Psi}{\partial\Fe}\!:\dot{\Fn}\Fn^{-1}
- \Fn^{\star}\Fe^{\star}\frac{\partial\Psi}{\partial\Fe}\Fn^{-\star}\!:\dot{\Fth}\Fth^{-1},
\\
\dot{\mathcal N} &= -\frac{\partial^2\Psi}{\partial\vartheta^2}\dot\vartheta
- \frac{\partial^2\Psi}{\partial\Fe\partial\vartheta}(\Fn\Fth)^{-\star}\!:\operatorname{D}_t^{\mathbf g}\mathbf F
+ \Fe^{\star}\frac{\partial^2\Psi}{\partial\Fe\partial\vartheta}\!:\dot{\Fn}\Fn^{-1}
+ \Fn^{\star}\Fe^{\star}\frac{\partial^2\Psi}{\partial\Fe\partial\vartheta}\Fn^{-\star}\!:\dot{\Fth}\Fth^{-1},
\end{align}
where $\mathcal N=-\partial\Psi/\partial\vartheta$ is the entropy. Substitution into the energy balance~\citep[Eq.~(4.23)]{SaYa2025GenColemanNoll}, followed by replacing $\operatorname{D}_t^{\mathbf g}\mathbf F$ in favor of $\operatorname{D}_t^{\mathbf g}\Fe$, gives the corrected form of Eq.~(4.25):
\begin{equation}\label{eq:corr_425}
-\rho\,\vartheta\frac{\partial^2\Psi}{\partial\vartheta^2}\,\dot\vartheta
-\rho\,\vartheta\frac{\partial^2\Psi}{\partial\Fe\partial\vartheta}\!:\operatorname{D}_t^{\mathbf g}\Fe
-\rho\,\Fe^{\star}\frac{\partial\Psi}{\partial\Fe}\!:\dot{\Fn}\Fn^{-1}
-\rho\,\Fn^{\star}\Fe^{\star}\frac{\partial\Psi}{\partial\Fe}\Fn^{-\star}\!:\dot{\Fth}\Fth^{-1}
= \rho R - \operatorname{Div}\mathbf Q ,
\end{equation}
in which the thermal and mechanical contributions are separated. Evaluating~\eqref{eq:corr_425} under the constant non-thermal distortion conditions $\operatorname{D}_t^{\mathbf g}\Fe=\mathbf 0$ and $\dot{\Fn}=\mathbf 0$, with $R=0$ and with the thermal distortion allowed to evolve, gives the corrected specific heat capacity:
\begin{equation}
c_E = -\vartheta\,\frac{\partial^2\Psi}{\partial\vartheta^2}
- \Fn^{\star}\Fe^{\star}\frac{\partial\Psi}{\partial\Fe}\Fn^{-\star}\!:\frac{\partial \Fth}{\partial \vartheta}\Fth^{-1}\,,
\end{equation}
which replaces Eq.~(4.27) of \citep{SaYa2025GenColemanNoll}. Accordingly, Eq.~(4.28) becomes
\begin{equation}
\rho\, c_E\,\dot\vartheta = -\operatorname{Div}\mathbf Q
+ \rho\,\vartheta\frac{\partial^2\Psi}{\partial\Fe\partial\vartheta}\!:\operatorname{D}_t^{\mathbf g}\Fe
+ \rho\,\Fe^{\star}\frac{\partial\Psi}{\partial\Fe}\!:\dot{\Fn}\Fn^{-1} + \rho R ,
\end{equation}
and its recast form, Eq.~(4.34), becomes
\begin{equation}
\rho\, c_E\,\dot\vartheta = -\operatorname{Div}\mathbf Q
+ \vartheta\,\mathbf g\,\frac{\partial\mathbf P}{\partial\vartheta}(\Fn\Fth)^{\star}\!:\operatorname{D}_t^{\mathbf g}\Fe
- \Bn\!:\dot{\Fn} + \rho R ,
\end{equation}
where $\Bn$ is the athermal configurational force defined in Eq.~(4.30a) of \citep{SaYa2025GenColemanNoll}, and $\partial\mathbf P/\partial\vartheta$ is the partial derivative at fixed distortions. The correction term $-\mathbf P\boldsymbol\alpha^{-\star}$ appearing in Eq.~(4.33a) of \citep{SaYa2025GenColemanNoll} is therefore absent.
All other results of \citep{SaYa2025GenColemanNoll}, including the balance laws~(4.15), the dissipation inequalities~(4.16) and~(4.18), the configurational forces~(4.30), and the kinetic equations~(4.35)--(4.38), remain valid. The relation $\Fth=\Fth(X,\vartheta)$ may be imposed as a constitutive assumption after the governing equations have been derived with $\Fth$ treated as an independent field. Its premature imposition affects only the interpretation of the temperature derivative and the identification of the specific heat capacity described above.
\end{remark}

\paragraph{Fourier heat conduction.}
It is often assumed that heat conduction is governed by Fourier's law,\footnote{As discussed by \cite{coleman1963Fourier}, Fourier's law should not be regarded as a fundamental law of nature, but rather a limiting constitutive assumption for the heat flux.} whereby the material heat flux has the form
\begin{equation}\label{eq:Fourier}
	\boldsymbol{Q}=-\boldsymbol{K}\,d\vartheta\,,
\end{equation}
where $\boldsymbol{K}$ is a $\binom{2}{0}$-rank conductivity tensor.
Recalling the Clausius-Duhem inequality~\eqref{eq:loc_Thermo_Second_4}, and observing that $\dot{\Fv}$, $\dot{\Fn}$, and $\dot{\vartheta}$ may be prescribed independently of $d\vartheta$, it follows by inserting~\eqref{eq:Fourier} into~\eqref{eq:loc_Thermo_Second_4} that\footnote{The inequality~\eqref{eq:C-D_Q} follows by setting $\dot{\Fv}=\dot{\Fn}=\dot{\vartheta}=0$ in~\eqref{eq:loc_Thermo_Second_4}.}
\begin{equation}\label{eq:C-D_Q}
	\frac{1}{\vartheta}\,\langle d\vartheta,\;\boldsymbol{K}\,d\vartheta\rangle \;\ge 0\,,
\end{equation}
which shows that $\boldsymbol{K}$ is positive semi-definite.
Furthermore, by Onsager’s reciprocity principle---under microscopic reversibility and in the absence of magnetic or Coriolis effects---the conductivity tensor is symmetric, i.e.,~$\mathbf{K}=\mathbf{K}^{\star}$ \citep{onsager1931reciprocal,casimir1945onsager}.
For a Fourier heat conductor (i.e.,~under the assumption~\eqref{eq:Fourier}), the heat conduction term in the heat equation~\eqref{eq:to_HtEq_S} is written as
\begin{equation}\label{eq:Fourier_Q}
	\operatorname{Div} \mathbf{Q} =
	-\boldsymbol{K} \!: \mathbf{Hess}\, \vartheta
	- \langle \operatorname{Div} \boldsymbol{K} , d\vartheta \rangle
	\,,
\end{equation}
where $\mathbf{Hess}\,$ denotes the Hessian operator and is given by
\begin{equation}
\label{eq:Hessian}
\mathbf{Hess}\,\vartheta=\!\left[\frac{\partial^2 \vartheta}{\partial X^A\partial X^B}-\Gamma^C{}_{AB}\frac{\partial \vartheta}{\partial X^C}\right]\!dX^A\otimes dX^B\,.
\end{equation}
If one further assumes an isotropic Fourier heat conductor, i.e.,~$\boldsymbol{K}=K\,\mathbf{G}^{\sharp}$, so that in components
$Q^{A}=-K\,G^{AB}\,\partial_{B}\vartheta$, it follows that
\begin{equation}\label{eq:iso_Fourier_Q}
	\operatorname{Div} \mathbf{Q} =
	- K\Delta \vartheta
	- \llangle dK^\sharp , d\vartheta^\sharp \rrangle_{\mathbf{G}}
	\,,
\end{equation}
where $\Delta$ is the material Laplace-Beltrami operator, given by
\begin{equation}
\label{eq:Laplacian}
\Delta \vartheta
= G^{AB}\!\left(\frac{\partial^2 \vartheta}{\partial X^A\partial X^B}
- \Gamma^{C}{}_{AB}\,\frac{\partial \vartheta}{\partial X^C}\right)\!,
\end{equation}
where $\Gamma^{C}{}_{AB}$ denote the Christoffel symbols of $\mathbf G$.

\subsubsection{Kinetic equations for compressible visco-anelastic solids}
\label{S:Kinetic}

While the evolution of temperature\textemdash and consequently that of the thermal distortion $\Fth$\textemdash is governed by the heat equation~\eqref{eq:to_HtEq_S}, one ought to prescribe a constitutive model for the generalized configurational forces $\Bv$ and $\Bn$ to provide an evolution equation for the evolution of the corresponding distortions\textemdash namely the viscous distortion $\Fv$ and the athermal-anelastic distortion $\Fn$, respectively\textemdash in order to complete the set of governing equations for visco-anelasticity.
Such a constitutive model may be obtained by assuming the existence of a dissipation potential density (a Rayleigh function) $\phi=\phi(X,\vartheta,\Fve,\Fv,\dot\Fv,\Fn,\dot\Fn,\Go,\mathbf{g})$, such that the generalized configurational forces are given by
\begin{equation}\label{Dissipative_B}
	\Bv= -\frac{\partial \phi}{\partial \dot\Fv}\,,\qquad
	\Bn= -\frac{\partial \phi}{\partial \dot\Fn}\,.
\end{equation}
It is assumed that the dissipation potential is convex with respect to $\dot{\Fv}\,$ and $\dot{\Fn}\,$~\citep{ziegler1958attempt, ziegler1987derivation, Germain1983, Goldstein2002ClassMech, Kumar2016}. This is equivalent to
\begin{equation}\label{eq:convex}
	\!\left(\frac{\partial \phi}{\partial \dot{\Fv}_2}-\frac{\partial \phi}{\partial \dot{\Fv}_1}\right)\!
	\!:\!\left(\dot{\Fv}_2-\dot{\Fv}_1\right)\! \geq 0 \,,\qquad
	\!\left(\frac{\partial \phi}{\partial \dot{\Fn}_2}-\frac{\partial \phi}{\partial \dot{\Fn}_1}\right)\!
	\!:\!\left(\dot{\Fn}_2-\dot{\Fn}_1\right)\! \geq 0 \,,
\end{equation}
for any $\dot{\Fv}_1$, $\dot{\Fv}_2$, $\dot{\Fn}_1$, and $\dot{\Fn}_2$.

\begin{remark}
\label{rmrk:Phi_dep}
It should be noted that for a general visco-anelastic solid, similarly to the viscoelastic case as discussed in \citep{SaYa2024viscoelasticity}, while objectivity implies that the dependence of $\phi$ on $\mathbf{F}$ can be reduced to a dependence on the symmetric tensor $\Cve$, i.e.,~$\phi=\hat{\phi}(X,\vartheta,\Cve,\Fv,\dot\Fv,\Fn,\dot\Fn,\Go)$, the dependence of $\phi$ on {$\Fv$}, $\dot{\Fv}$, {$\Fn$}, and $\dot{\Fn}$ cannot generally be reduced to a dependence on the symmetric tensors $\Cv$, $\dot{\Cv}$, $\Cn$, and $\dot{\Cn}$.
\end{remark}

Following the introduction of a dissipation potential density $\phi$, one substitutes~\eqref{Dissipative_B} into~\eqref{eq:Config_v} and~\eqref{eq:Config_n} to find the following set of kinetic equations governing the evolution of the distortions $\Fv$ and $\Fn$:
\begin{subequations}\label{eq:Kinetic}
\begin{align}
\label{eq:Kinetic_v}
	& \frac{\partial \phi}{\partial \dot\Fv}
	+ \rho \frac{\partial \tilde\Psi}{\partial \Fv}
	= \mathbf{0}\,,\\
\label{eq:Kinetic_o}
	& \frac{\partial \phi}{\partial \dot\Fn}
	+ \rho \frac{\partial \tilde\Psi}{\partial \Fn}
	=\mathbf{0}\,.
\end{align}
\end{subequations}

\begin{remark}
Recalling~\eqref{eq:psi_psi-tilde_EQ}, the kinetic equations~\eqref{eq:Kinetic} may equivalently be written as
\begin{subequations}\label{eq:Kinetic_FeFve}
\begin{align}
\label{eq:Kinetic_v}
	& \frac{\partial \phi}{\partial \dot\Fv}
	-\rho \,\Fe^\star \frac{\partial \Psi}{\partial \Fe} \Fv^{-\star}
	= \mathbf{0}\,,\\
\label{eq:Kinetic_o}
	& \frac{\partial \phi}{\partial \dot\Fn}
	-\rho \,\Fve^\star\!\left(\frac{\partial \Psi}{\partial \Fve} 
	+ \frac{\partial \Psi}{\partial \Fe} \Fv^{-\star} \right)\!\Fn^{-\star}
	=\mathbf{0}\,,
\end{align}
\end{subequations}
and by recalling~\eqref{C-F_transform}, the kinetic equations~\eqref{eq:Kinetic} may equivalently be written as
\begin{subequations}\label{eq:Kinetic_CeCve}
\begin{align}
\label{eq:Kinetic_v}
	& \frac{\partial \phi}{\partial \dot\Fv}
	- 2 \rho\, \Ce \frac{\partial \hat{\Psi}}{\partial\Ce} \Fv^{-\star}
	= \mathbf{0}\,,\\
\label{eq:Kinetic_o}
	& \frac{\partial \phi}{\partial \dot\Fn}
	- 2 \rho \!\left( \Cve \frac{\partial \hat{\Psi}}{\partial\Cve} 
	+ \Fv^\star \Ce\frac{\partial \hat{\Psi}}{\partial\Ce} \Fv^{-\star} \right)\!\Fn^{-\star}
	=\mathbf{0}\,.
\end{align}
\end{subequations}
\end{remark}

\subsubsection{The incompressible visco-anelastic solids}
\label{S:inc}

In this subsection, we revisit the governing equations derived in \S\ref{sec:gov_eqs} and record their form under incompressibility conditions. We begin by specifying the relevant notions of incompressibility before specializing the governing equations accordingly.
\paragraph{Incompressibility in visco-anelasticity.}
For a visco-anelastic solid, we can define the following incompressibility constraints:

\begin{itemize} [topsep=2pt,noitemsep, leftmargin=12pt]
\item \emph{Visco-elastic incompressibility} (volume-preserving motion with respect to the stress-free material state) corresponds to
\begin{equation}\label{eq:Inc}
	J = \sqrt{\frac{\det \mathbf{g}}{\det \mathbf{G}}} \det \mathbf{F} 
	=\sqrt{\frac{\det\mathbf g}{\det\Go}}\det\Fe\det\Fv = 1\,.
\end{equation}

\item \emph{Instantaneous elastic incompressibility} (volume-preserving instantaneous elastic response) implies that
\begin{equation}\label{eq:Ela_Inc}
	J_e = \sqrt{\frac{\det \mathbf{g}}{\det \Go}} \det \Fe = 1\,.
\end{equation}

\item \emph{Viscous incompressibility} (volume-preserving viscous flow) is an internal constraint corresponding to 
\begin{equation}\label{eq:V_Inc}
	\Jv = \det \Fv = 1\,.
\end{equation}

\item \emph{Anelastic (athermal) incompressibility} (volume-preserving athermal-anelastic distortion) implies that
\begin{equation}\label{eq:A_Inc}
	\Jn = \det \Fn = 1\,.
\end{equation}
\end{itemize}

\begin{remark}\label{rmrk:Ane-Jac}
Thermal incompressibility ($\Jth = 1$) cannot generally be imposed as an additional independent constraint. Although the thermal distortion and the temperature field are \emph{a priori} independent, the model is closed by prescribing a constitutive thermal-expansion equation such as~\eqref{eq:Th_distort}, which determines $\Fth$ from the temperature field, itself governed by the heat equation~\eqref{eq:to_HtEq_S}. Under such a constitutive equation, the thermal distortion is no longer an independent field, and imposing thermal incompressibility instead restricts the thermal expansion tensor. For~\eqref{eq:Th_distort}, preservation of $\Jth = 1$ during a temperature change requires $\operatorname{tr}\boldsymbol\alpha = 0$, together with an initially isochoric thermal distortion. This constitutes a restriction on the prescribed thermal-expansion data and is generally incompatible with materials exhibiting volumetric thermal expansion. In contrast, the athermal-anelastic distortion is governed by its own kinetic equation and remains an independent field; anelastic incompressibility ($\Jn = 1$) may accordingly be imposed as a constitutive constraint and enforced by the Lagrange multiplier $\pn$. Whether this constraint is appropriate depends on the underlying mechanism. It may be natural for volume-preserving processes such as plastic slip, but is generally inappropriate for swelling and volumetric growth; for defects, its suitability depends on the particular defect type.
\end{remark}

\begin{remark}\label{rmrk:Jacobian}
As discussed before \eqref{eq:Jacob}, the Jacobian of the motion $J$ is measured with respect to the material metric $\mathbf G = \Fa^*\mathring{\mathbf G}$ rather than the Euclidean reference metric $\mathring{\mathbf G}$. The anelastic distortion is therefore already accounted for in its definition. Visco-elastic incompressibility, $J=1$, asserts that the deformation preserves volume relative to the natural state of the material element and places no restriction on the volume change induced by the anelastic distortion. The multiplicative identity $J = \Je\,\Jv$ shows that $J$ constrains only the elastic and viscous distortions. Of the three conditions $J=1$, $\Jv=1$, and $\Je=1$, any two imply the third, and hence only two are independent. These conditions concern the volumetric response of the material: $J=1$ constrains the combined elastic and viscous response, whereas $\Jv=1$ and $\Je=1$ constrain its viscous and elastic parts, respectively. They impose no restriction on the volumetric character of the anelastic distortion. Conversely, anelastic incompressibility, $\Jn=1$, when imposed, places no restriction on the elastic or viscous volume changes.
\end{remark}

The total time derivatives of the Jacobians above read\footnote{In computing $\dot{J}$ in~\eqref{eq:Jac_diff}, we use $\mathbf{F}^{-\star} \!:\! \operatorname{D}_t^{\mathbf g}\mathbf{F} = \!\left(\mathbf{F}^{-1} \mathbf g^\sharp \mathbf{F}^{-\star}\right)\! \!:\! (\Ds+ \Da) = \!\left(\mathbf{F}^{-1} \mathbf g^\sharp \mathbf{F}^{-\star}\right)\! \!:\! \Ds$.}
\begin{equation}\label{eq:Jac_diff}
	\dot J = J \!\left(\mathbf{F}^{-1} \mathbf g^\sharp \mathbf{F}^{-\star}\right)\! \!:\! \Ds
	- J \Fn^{-\star}\!:\!\dot\Fn
	- J \Fth^{-\star}\!:\!\dot\Fth
	\,,\qquad
	\dot\Jv= \Jv \,\Fv^{-\star}\!:\!\dot\Fv\,,\qquad
	\dot\Jn= \Jn \,\Fn^{-\star}\!:\!\dot\Fn\,.
\end{equation}

\paragraph{The free energy.}
Assuming incompressibility, the Legendre transform~\eqref{eq:Legendre_trans} is modified to take into account the corresponding volume preserving constraint(s), such that each constraint introduces its own Lagrange multiplier as follows\footnote{One may for example only assume global and anelastic incompressibility, i.e.,~$J=\Jn=1$, and the modified free energy in this case reads $\rho\Psi_{\text{inc}}=\rho\Psi -p(J-1) -\pn(\Jn -1)$.}
\begin{equation}\label{eq:Legendre_trans_inc}
\rho\Psi -p(J-1) - \pv(\Jv-1) - \pn(\Jn-1)= \rho \mathscr E - \rho \vartheta \mathscr N\,,
\end{equation}
where $p=p(X,t)$, $\pv=\pv(X,t)$, and $\pn=\pn(X,t)$ are the Lagrange multipliers corresponding to the constraints~\eqref{eq:Inc},~\eqref{eq:V_Inc}, and~\eqref{eq:A_Inc}, respectively, and $\Psi$ is the free energy without any incompressibility constraint.

\paragraph{Balance laws and constitutive relations.}
Let us substitute the incompressibility-augmented Legendre transform~\eqref{eq:Legendre_trans_inc} in the Clausius-Duhem inequality~\eqref{loc_Thermo_Second} to find\footnote{The time derivative is taken prior to enforcing the incompressibility constraints. Upon subsequently imposing $J=1$, $\Jv=1$, and $\Jn=1$, the terms involving the time derivatives of the associated Lagrange multipliers vanish, i.e.
$$\dot{p}(J-1)=\dot{\pv}(\Jv-1)=\dot{\pn}(\Jn-1)=0\,.$$}
\begin{equation}\label{eq:loc_Thermo_Second_1_inc}
\dot\eta = \rho\dot{\mathscr E} -\rho\dot\Psi + p \dot{J} + \pv \dot\Jv + \pn \dot\Jn- \rho \dot\vartheta \mathscr N+ \operatorname{Div}\mathbf Q - \frac{1}{\vartheta} \langle d\vartheta, \mathbf Q \rangle -\rho R - \!\left(S_m-\dot{\rho} - \rho\,\dot{\Fa}\!:\!\Fa^{-\star} \right)\! \vartheta \mathscr N \geq 0\,.
\end{equation}
Proceeding similarly to the derivation carried out in \S\ref{S:Bal_Const}, taking~\eqref{eq:loc_Thermo_Second_1_inc} as a starting point instead of~\eqref{eq:loc_Thermo_Second_1} and using~\eqref{eq:Jac_diff}, it follows that for incompressible visco-anelastic solids, the Clausius-Duhem inequality~\eqref{eq:loc_Thermo_Second_3} is rewritten as
\begin{equation}\label{eq:loc_Thermo_Second_inc}
\begin{aligned}
	\dot\eta & = 
	\mathbf{F}^{-1} \!\left[\mathbf P
	- \rho \mathbf g^\sharp \frac{\partial \tilde\Psi}{\partial \mathbf{F}} + p \,\mathbf{g}^\sharp \mathbf{F}^{-\star} \right]\!:\!\Ds + \mathbf{F}^{-1} \mathbf P \!:\! \Da
	- \!\left[ \rho \frac{\partial \tilde\Psi}{\partial \Fv} - \pv\,\Fv^{-\star} \right]\! \!:\! \dot\Fv
	\\&\quad
	- \!\left[ \rho \frac{\partial \tilde\Psi}{\partial \Fn} - (\pn-p) \Fn^{-\star} \right]\! \!:\! \dot\Fn
	- \!\left[ \rho \frac{\partial \tilde\Psi}{\partial \Fth} + p\,\Fth^{-\star} \right]\! \!:\! \dot\Fth
	- \rho \!\left( \mathscr{N} +\frac{\partial \tilde\Psi}{\partial \vartheta} \right)\!  \dot\vartheta
	\\&\quad
	+\llangle \operatorname{Div}\mathbf{P}+ \rho(\boldsymbol{\mathsf{B}} - \mathbf{A}),\mathbf{V} \rrangle_{\mathbf{g}}
	+ \!\left(S_m-\dot{\rho} - \rho\,\dot{\Fa}\!:\!\Fa^{-\star} \right)\!\!\left(\Psi+\frac{1}{2} \Vert\mathbf V\Vert^2_{\mathbf g}\right)\!
	- \frac{1}{\vartheta} \langle d\vartheta, \mathbf Q \rangle
	\geq 0\,.
\end{aligned}
\end{equation}
Recall that by arbitrariness of the thermoelastic configuration, $\Ds$, $\Da$, $\mathbf{V}$, $\Vert\mathbf V\Vert^2_{\mathbf g}$, and $\dot\vartheta$ may all be chosen independently. Hence, it follows from~\eqref{eq:loc_Thermo_Second_inc} that the Doyle-Ericksen formula~\eqref{eq:Cons_Eq_stress} is rewritten as
\begin{equation}\label{eq:P_represent_inc}
	\mathbf{P} 
	= \rho \mathbf g^\sharp\!\left(\frac{\partial \Psi}{\partial \Fve} \Fve^{\star} + \frac{\partial \Psi}{\partial \Fe} \Fe^{\star}\right)\! \mathbf{F}^{-\star} - p \mathbf{g}^\sharp \mathbf{F}^{-\star}\,.
\end{equation}
while the Legendre conjugacy condition~\eqref{eq:Legendre_conj} remains unchanged.
It also follows that the balance of linear and angular momenta~\eqref{eq:Lin_M} and~\eqref{eq:Ang_M} remain unchanged so long as the stress tensor expression includes the Lagrange multiplier term as in~\eqref{eq:P_represent_inc}.
The balance of mass~\eqref{eq:Bal_M} also remains unchanged, but it may be further simplified in the case of anelastic (athermal) incompressibility ($\Jn=1$), by observing that $\dot{\Fa}\!:\!\Fa^{-\star}=\dot{\Fn}\!:\!\Fn^{-\star}+\dot{\Fth}\!:\!\Fth^{-\star} = \operatorname{tr}(\boldsymbol\alpha) \dot\vartheta$,\footnote{If $\Jn=1$, it folows that $0=\dot\Jn= \Jn \,\Fn^{-\star}\!:\!\dot\Fn$ and hence $\Fn^{-\star}\!:\!\dot\Fn = 0$.} to yield
\begin{equation}
	\dot{\rho} + \rho\,\operatorname{tr}(\boldsymbol\alpha)\, \dot\vartheta = S_m\,.
\end{equation}
Finally, the Calusius-Duhem inequality~\eqref{eq:loc_Thermo_Second_inc} for incompressible visco-anelastic solids simplifies to read
\begin{equation}\label{eq:loc_Thermo_Second_4_inc}
\begin{aligned}
	\dot\eta & = 
	- \!\left[ \rho \frac{\partial \tilde\Psi}{\partial \Fv} - \pv\,\Fv^{-\star} \right]\! \!:\! \dot\Fv
	- \!\left[ \rho \frac{\partial \tilde\Psi}{\partial \Fn} - (\pn-p) \Fn^{-\star} \right]\! \!:\! \dot\Fn
	- \!\left[ \rho \frac{\partial \tilde\Psi}{\partial \Fth} + p\,\Fth^{-\star} \right]\! \!:\! \dot\Fth
	- \frac{1}{\vartheta} \langle d\vartheta, \mathbf Q \rangle
	\geq 0\,.
\end{aligned}
\end{equation}
It follows by identification of~\eqref{eq:loc_Thermo_Second_4_inc} with~\eqref{eq:dissipation} that the configurational forces~\eqref{eq:Config} should be rewritten as
\begin{subequations}\label{eq:Config_inc}
\begin{align}
	\label{eq:Config_v_inc}
	\Bv & = \rho \frac{\partial \tilde\Psi}{\partial \Fv} - \pv\,\Fv^{-\star}\,,\\
	\label{eq:Config_n_inc}
	\Bn & = \rho \frac{\partial \tilde\Psi}{\partial \Fn} - (\pn-p) \Fn^{-\star} \,,\\
	\label{eq:Config_t_inc}
	\Bth & = \rho \frac{\partial \tilde\Psi}{\partial \Fth} + p\,\Fth^{-\star}\,.
\end{align}
\end{subequations}

\begin{remark} \label{rmrk:inc_eq_vs_neq}
Note that, similarly to the compressible case, the stress admits an additive decomposition into equilibrium and non-equilibrium parts:
\begin{equation}
	\mathbf{P} = \mathbf{P}_{\textrm{eq}} + \mathbf{P}_{\textrm{neq}}\,,
\end{equation}
where
\begin{equation}
	\mathbf{P}_{\textrm{eq}} = \rho\, \mathbf{g}^{\sharp} 
	\frac{\partial \Psi}{\partial \Fve} \Fve^{\star} \mathbf{F}^{-\star} 
	- p_{\textrm{eq}}\, \mathbf{g}^{\sharp} \mathbf{F}^{-\star}\,, \qquad
	\mathbf{P}_{\textrm{neq}} = \rho \,\mathbf{g}^{\sharp} 
	\frac{\partial \Psi}{\partial \Fe} \Fe^{\star} \mathbf{F}^{-\star} 
	- p_{\textrm{neq}}\, \mathbf{g}^{\sharp} \mathbf{F}^{-\star}\,.
\end{equation}
Here, the Lagrange multiplier is decomposed into equilibrium and non-equilibrium contributions, ${p=p_{\textrm{eq}}+p_{\textrm{neq}}}$. This decomposition follows from the additive structure of the stress~\eqref{eq:P_represent_inc}: the contribution to $p$ associated with ${\partial\Psi}/{\partial\Fve}$ is denoted by $p_{\textrm{eq}}$, while that associated with ${\partial\Psi}/{\partial\Fe}$ is denoted by $p_{\textrm{neq}}$. Furthermore, in terms of the second Piola--Kirchhoff stress, one has
\begin{equation}\label{eq:S_represent_inc}
	\mathbf{S}
	= \mathbf{S}_{\textrm{eq}} + \mathbf{S}_{\textrm{neq}}\,,
\end{equation}
where
\begin{equation}
	\mathbf{S}_{\textrm{eq}}
	=2\rho\mathbf{F}^{-1} \Fve\frac{\partial \hat{\Psi}}{\partial\Cve}\Fve^\star \mathbf{F}^{-\star} 
	- p_{\textrm{eq}} \mathbf C^{-1}\,,
	\qquad
	\mathbf{S}_{\textrm{neq}}
	=2\rho\mathbf{F}^{-1}\Fe\frac{\partial \hat{\Psi}}{\partial \Ce}\,\Fe^{\star} \mathbf{F}^{-\star} 
	- p_{\textrm{neq}} \mathbf C^{-1}\,.
\end{equation}
Similarly, in terms of the Cauchy stress tensor we have
\begin{equation}\label{eq:sig_represent_inc}
	\boldsymbol\sigma
	= \boldsymbol\sigma_{\textrm{eq}} + \boldsymbol\sigma_{\textrm{neq}}
	= 2\varrho\frac{\partial \Psi}{\partial \mathbf g} - p \mathbf g^\sharp
	\,,
\end{equation}
where
\begin{equation}
\label{eq:sig_eq-neq_represent_inc}
\boldsymbol\sigma_{\textrm{eq}}
	= \varrho \mathbf g^\sharp\frac{\partial \Psi}{\partial \Fve} \Fve^{\star} 
	- p_{\textrm{eq}} \, \mathbf g^\sharp
	= 2\varrho\Fve\frac{\partial \hat{\Psi}}{\partial\Cve}\Fve^\star 
	- p_{\textrm{eq}} \, \mathbf g^\sharp\,,\qquad \boldsymbol\sigma_{\textrm{neq}}
	= \varrho \mathbf g^\sharp\frac{\partial \Psi}{\partial \Fe} \Fe^{\star} 
	- p_{\textrm{neq}} \, \mathbf g^\sharp
	= 2\varrho\Fe\frac{\partial \hat{\Psi}}{\partial \Ce}\,\Fe^{\star} - p_{\textrm{neq}} \, \mathbf g^\sharp\,.
\end{equation}
\end{remark}

\paragraph{Heat equation.}
For an incompressible visco-anelastic solid, the energy balance~\eqref{eq:to_HtEq0} is modified to read
\begin{equation}\label{eq:to_HtEq0_inc}
	\rho\,\dot{\Psi}
	- p\, \mathbf{F}^{-\star} \!:\! \operatorname{D}_t^{\mathbf g}\mathbf{F}
	+ p\, \Fth^{-\star}\!:\!\dot\Fth
	-\pv\, \Fv^{-\star}\!:\!\dot\Fv
	-(\pn-p) \Fn^{-\star}\!:\!\dot\Fn
	+\rho\,\dot{\vartheta}\mathscr{N}+\rho\,\vartheta\dot{\mathscr{N}}
	= \mathbf{g} \mathbf P \!:\! \operatorname{D}_t^{\mathbf g}\mathbf{F} +\rho R - \operatorname{Div} \mathbf{Q}\,.
\end{equation}
Following the derivation of the heat equation for compressible visco-anelastic solids in \S\ref{S:HeatEq}, but starting from~\eqref{eq:to_HtEq0_inc} rather than~\eqref{eq:to_HtEq0}, the heat equation for an incompressible visco-anelastic solid takes the form
\begin{equation}
\label{eq:to_HtEq2_inc}
\begin{split}
	\rho c_{\text D} \dot\vartheta
	-\rho\vartheta \frac{\partial^2 \tilde\Psi}{\partial \vartheta \partial \mathbf{F}} 
	\mathbf{F}^\star \Fe^{-\star} \!:\! \operatorname{D}_t^{\mathbf g} \Fe
	-\rho \vartheta \!\left[ \frac{\partial^2 \tilde\Psi}{\partial \vartheta \partial \Fv} 
	+ \Fe^\star \frac{\partial^2 \tilde\Psi}{\partial \vartheta \partial \mathbf{F}} 
	\mathbf{F}^\star \Fve^{-\star} \right]\! \!:\! \dot\Fv
	&\\
	+ \!\left[ \rho \frac{\partial \tilde\Psi}{\partial \Fv} - \pv\, \Fv^{-\star} \right]\! \!:\! \dot\Fv
	+ \!\left[ \rho \frac{\partial \tilde\Psi}{\partial \Fn} - (\pn-p) \Fn^{-\star} \right]\! \!:\! \dot\Fn
	&= \rho R - \operatorname{Div} \mathbf{Q}
	\,,
\end{split}
\end{equation}
where $c_{\text{D}}$, the specific heat capacity at constant distortions, is given by
\begin{equation}
\begin{split}
	c_{\textrm D}
	= - \vartheta \frac{\partial^2 \tilde\Psi}{\partial \vartheta^2}
	+ \frac{1}{\dot\vartheta} \!\left[\frac{\partial \tilde\Psi}{\partial \Fth} - \frac{p}{\rho} \Fth^{-\star}\right]\! \!:\! \dot\Fth
	\,.
	= - \vartheta \frac{\partial^2 \tilde\Psi}{\partial \vartheta^2}
	+ \!\left[\frac{\partial \tilde\Psi}{\partial \Fth} - \frac{p}{\rho} \Fth^{-\star}\right]\! \!:\! \boldsymbol\alpha \Fth
	\,.
\end{split}
\end{equation}
Similiarly to \S\ref{S:HeatEq}, the heat equation~\eqref{eq:to_HtEq2_inc} may be recast as follows
\begin{equation}
\label{eq:to_HtEq_S_inc}
\begin{split}
	\rho c_{\text D} \dot\vartheta
	- \rho \vartheta \mathbf{g} \frac{\partial}{\partial \vartheta}\!\left[ \frac{\boldsymbol\sigma_{\textrm{eq}}-p_{\textrm{eq}} \, \mathbf g^\sharp}{\rho} \right]\! \!:\! (\operatorname{D}_t^{\mathbf g}\Fve) \Fve^{-1}
	-\rho\vartheta \mathbf{g} \frac{\partial}{\partial \vartheta}\!\left[ \frac{ \boldsymbol\sigma_{\textrm{neq}}-p_{\textrm{neq}} \, \mathbf g^\sharp}{\rho} \right]\! \!:\! (\operatorname{D}_t^{\mathbf g}\Fe) \Fe^{-\star}&\\
	+ \Bv \!:\! \dot\Fv
	+ \Bn \!:\! \dot\Fn
	&= \rho R - \operatorname{Div} \mathbf{Q}
	\,.
\end{split}
\end{equation}

\paragraph{Kinetic equations.}
For an incompressible visco-anelastic solid with a dissipation potential density (a Rayleigh function) $\phi=\phi(X,\vartheta,\Fve,\Fv,\dot\Fv,\Fn,\dot\Fn,\Go,\mathbf{g})$, the kinetic equations~\eqref{eq:Kinetic} are rewritten as
\begin{subequations}\label{eq:Kinetic_inc}
\begin{align}
\label{eq:Kinetic_v_inc}
	\frac{\partial \phi}{\partial \dot\Fv} 
	+ \rho \frac{\partial \tilde\Psi}{\partial \Fv}
	 &= \pv \,\Fv^{-\star}\,,\\
\label{eq:Kinetic_n_inc}
	\frac{\partial \phi}{\partial \dot\Fn} 
	+ \rho \frac{\partial \tilde\Psi}{\partial \Fn}
	 &= (\pn-p) \,\Fn^{-\star}\,.
\end{align}
\end{subequations}

\section{Material Anisotropy in Visco-Anelasticity}
\label{Sec:MaterialSymmetry}

In this section, we develop a framework for incorporating anisotropy in visco-anelasticity. We begin by reviewing the notion of structural tensors and their role in defining material symmetry. The formulation of anisotropy in nonlinear elasticity, anelasticity, and viscoelasticity is then discussed, providing the necessary foundation for extending these concepts to visco-anelastic solids.

\subsection{Preamble}
\label{Sec:MaterialSymmetry_Pre}

In this subsection, we recall structural tensors and briefly review material symmetry for elasticity, anelasticity, and viscoelasticity, which prepares the ground for the detailed visco-anelastic formulation in the next subsection.

\subsubsection{Structural tensors}
\label{sec:strucTensors}

Let us consider an objective energy function $\mathsf{W}=\mathsf{W}(X,\vartheta,\boldsymbol{\mathsf{A}},\Go,\mathbf{g})$, where $\boldsymbol{\mathsf{A}}$ is some invertible two-point tensor defined on $\mathcal B$ and $\mathcal C$. The material symmetry group $\mathring{\mathcal G}_X$ of $\mathsf{W}$ at $X\in\mathcal{B}$ with respect to $(\mathcal{B},\Go)$ is the set of those $\mathring{\mathbf{K}}\in\mathrm{Orth}(\Go)=\{\mathbf{Q}: T_X\mathcal{B}\to T_X\mathcal{B}~|~ {\mathbf{Q}}^{\star}\Go{\mathbf{Q}}=\Go \}$ such that
\begin{equation} \label{Elasticity-Sym-Group}
\mathring{\mathbf{K}}_*\mathsf{W}(X,\vartheta,\boldsymbol{\mathsf{A}},\Go,\mathbf{g})=
	\mathsf{W}(X,\vartheta,\mathring{\mathbf{K}}^*\boldsymbol{\mathsf{A}},\mathring{\mathbf{K}}^*\Go,\mathbf{g})
	=\mathsf{W}(X,\vartheta,\boldsymbol{\mathsf{A}},\Go,\mathbf{g})
	\,,
\end{equation}
for all such $\boldsymbol{\mathsf{A}}$, i.e.,~invertible two-point tensor defined on $\mathcal B$ and $\mathcal C$. Note that $\mathring{\mathbf{K}}^*\boldsymbol{\mathsf{A}}=\boldsymbol{\mathsf{A}}\mathring{\mathbf{K}}$ and $\mathring{\mathbf{K}}^*\Go=\mathring{\mathbf{K}}^{\star}\Go\mathring{\mathbf{K}}=\Go$. $\mathring{\mathcal{G}}_X$ is a subgroup of $\mathrm{Orth}(\Go)$, which we denote by $\mathring{\mathcal{G}}_X\leqslant \mathrm{Orth}(\Go)$.

The subgroup $\mathring{\mathcal G}_X \leqslant\mathrm{Orth}(\Go)$ can be fully described by a finite collection of \emph{structural tensors} $\mathring{\boldsymbol{\Lambda}}_i$, $i=1,\dots,N$: a basis for the space of $\mathring{\mathcal G}_X$-invariant tensors that characterizes the material anisotropy of the body in its Euclidean reference configuration $(\mathcal B, \Go)$~\citep{liu1982, boehler1987, zheng1993, zheng1994theory, lu2000covariant, MazzucatoRachele2006}. 
The collection of structural tensors is denoted by $\mathring{\boldsymbol{\Lambda}}=\{\mathring{\boldsymbol\Lambda}_1,\hdots,\mathring{\boldsymbol\Lambda}_N\}$. 
Including $\mathring{\boldsymbol{\Lambda}}$ among the arguments turns the energy function isotropic. 
This is called the \emph{principle of isotropy of space}~\citep{Boehler1979}.  Now, given an isotropic energy function, a set of isotropic invariants\textemdash \emph{integrity basis}\textemdash may be used to simplify its dependence on its arguments \citep{Hilbert1993, Olive2017, Spencer1971}. Objectivity implies that $\mathsf{W}=\hat{\mathsf{W}}(X,\vartheta,\boldsymbol{\mathsf{C}},\Go)$, where $\boldsymbol{\mathsf{C}}=\boldsymbol{\mathsf{A}}^*\mathbf{g}$. Upon adjoining the structural tensors, this representation may be written as $\mathsf{W}=\widetilde{\mathsf{W}}(X,\vartheta,\boldsymbol{\mathsf{C}},\Go,\mathring{\boldsymbol{\Lambda}})$, where $\widetilde{\mathsf{W}}$ is isotropic in its tensorial arguments.
Therefore, it follows that $\mathsf{W}=\overline{\mathsf{W}}(X,\vartheta,I_1,\hdots,I_m)$, where $\left\{I_1,\hdots,I_m\right\}$ is an integrity basis for the set of isotropic invariants of $\{\boldsymbol{\mathsf{C}},\Go,\mathring{\boldsymbol\Lambda}_1,\hdots,\mathring{\boldsymbol{\Lambda}}_N\}$.

\subsubsection{Material symmetry in nonlinear elasticity}
For a hyperelastic solid, it is assumed that there is an energy function of the form $\mathring{W}=\mathring{W}(X,\vartheta,\mathbf{F},\Go,\mathbf{g})$, where $\mathbf{g}$ is the metric of the Euclidean ambient space and $\Go=\iota^*\mathbf{g}$ is the induced metric on the body---the material metric in the absence of eigenstrains.
Recall that, by objectivity, one may write
\begin{equation}
	\mathring{W}=\hat{\mathring{W}}(X,\vartheta,\mathbf{C},\Go)\,.
\end{equation}
At a point $X\in\mathcal{B}$, the material symmetry group $\mathring{\mathcal{G}}_X$ with respect to the Euclidean reference configuration $(\mathcal{B},\Go)$ is defined as the set of $\mathring{\mathbf{K}}\in \mathrm{Orth}(\Go)$ such that
\begin{equation} \label{Elasticity-Sym-Group}
\mathring{\mathbf{K}}_*\mathring{W}(X,\vartheta,\mathbf{F},\Go,\mathbf{g})=
	\mathring{W}(X,\vartheta,\mathring{\mathbf{K}}^*\mathbf{F},\mathring{\mathbf{K}}^*\Go,\mathbf{g})
	=\mathring{W}(X,\vartheta,\mathbf{F},\Go,\mathbf{g})
	\,,
\end{equation}
for all deformation gradients $\mathbf{F}$.
It should be noted that $\mathring{\mathbf{K}}^*\mathbf{F}=\mathbf{F}\mathring{\mathbf{K}}$ (see Fig.~\ref{Symmetry-Transformation}a), $\mathring{\mathbf{K}}^*\Go=\mathring{\mathbf{K}}^{\star}\Go\mathring{\mathbf{K}}=\Go$, and hence
\begin{equation}
	\mathring{\mathbf{K}}_*\mathring{W}(X,\vartheta,\mathbf{F},\Go,\mathbf{g})=\mathring{W}(X,\vartheta,\mathbf{F}\mathring{\mathbf{K}},\Go,\mathbf{g})\,.
\end{equation}

\subsubsection{Material symmetry in anelasticity}
\begin{figure}[t!]
\centering
\includegraphics[width=1.0\textwidth]{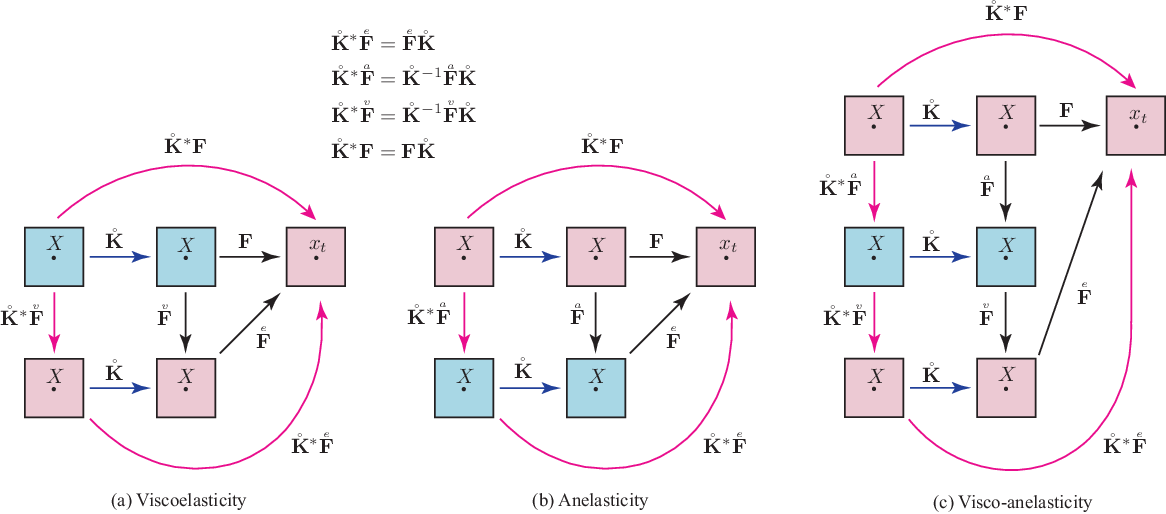}
\vspace*{0.10in}
\caption{The actions of the symmetry group on the total, viscous, elastic, and anelastic deformation gradients. The blue and pink squares indicate locally stress-free and locally stressed configurations, respectively.} 
\label{Symmetry-Transformation}
\end{figure}

In anelasticity, assuming the Bilby-Kröner-Lee direct decomposition of the deformation gradient $\mathbf{F}=\Fe\Fa$,\footnote{See \citep{Sadik2017,YavariSozio2023} for a short historical account on the Bilby-Kröner-Lee decomposition of the deformation gradient.} where $\Fe$ and $\Fa$ are the local elastic and anelastic distortions, respectively, energy explicitly depends on the elastic distortion $\mathring{W}=\mathring{W}(X,\vartheta,\Fe,\Go,\mathbf{g})$, and one can write \citep{YavariSozio2023} 
\begin{equation} 
	W(X,\vartheta,\mathbf{F},\Fa,\Go,\mathbf{g})
	=\mathring{W}(X,\vartheta,\mathbf{F}\Fa^{-1},\Go,\mathbf{g})
	=\mathring{W}(X,\vartheta,\Fe,\Go,\mathbf{g})\,.
\end{equation}
Note that, by objectivity, one may write
\begin{equation} 
	\mathring{W}=\hat{\mathring{W}}(X,\vartheta,\Ce,\Go)\,.
\end{equation}
At a point $X\in\mathcal{B}$, the material symmetry group $\mathring{\mathcal{G}}_X$ with respect to the Euclidean reference configuration $(\mathcal{B},\Go)$ is defined as the set of those $\mathring{\mathbf{K}}\in \mathrm{Orth}(\Go)$ such that
\begin{equation}  \label{Anelasticity-Sym-Group}
\begin{gathered}
	\mathring{\mathbf{K}}_*\mathring{W}(X,\vartheta,\Fe,\Go,\mathbf{g})
	=\mathring{W}(X,\vartheta,\Fe\mathring{\mathbf{K}},\Go,\mathbf{g})
	= \mathring{W}(X,\vartheta,\Fe,\Go,\mathbf{g})\,,\\
	\textrm{or}\\ 
	\mathring{\mathbf{K}}_*W(X,\vartheta,\Fe\Fa,\Fa,\Go,\mathbf{g})
	=W(X,\vartheta,\mathbf{F} \mathring{\mathbf{K}},\mathring{\mathbf{K}}^{-1}\Fa \mathring{\mathbf{K}},
	\Go,\mathbf{g}) 
	= W(X,\vartheta,\Fe\Fa,\Fa,\Go,\mathbf{g}) \,,
\end{gathered}
\end{equation}
for all deformation gradients $\mathbf{F}$ (see Fig.~\ref{Symmetry-Transformation}b).
Introducing a set of structural tensors $\mathring{\boldsymbol\Lambda}$ for $\mathring{\mathcal G}_X$, it follows that $\mathring{W}=\mathring{W}(X,\vartheta,\Fe,\mathring{\boldsymbol\Lambda},\Go,\mathbf{g})$ is materially covariant, i.e.,~for any linear transformation $\boldsymbol{\mathsf{T}}:T_X\mathcal{B}\to T_X\mathcal{B}$, ${\mathring{W}}(X,\vartheta,\Fe,\mathring{\boldsymbol\Lambda},\Go,\mathbf{g})={\mathring{W}}(X,\vartheta,\boldsymbol{\mathsf{T}}^*\Fe,\boldsymbol{\mathsf{T}}^*\mathring{\boldsymbol\Lambda},\boldsymbol{\mathsf{T}}^*\Go,\mathbf{g})$. Let us choose $\boldsymbol{\mathsf{T}}=\Fa$, and hence $\mathring{W}={\mathring{W}}(X,\vartheta,\Fe,\mathring{\boldsymbol\Lambda},\Go,\mathbf{g})={\mathring{W}}(X,\vartheta,\Fa^*\Fe,\Fa^*\mathring{\boldsymbol\Lambda},\Fa^*\Go,\mathbf{g})$. But notice that $\Fa^*\Fe=\Fe\Fa=\mathbf{F}$. Therefore 
\begin{equation}\label{eq:W_transported}
	\mathring{W}={\mathring{W}}(X,\vartheta,\mathbf{F},\boldsymbol\Lambda,\mathbf{G},\mathbf{g})\,,
\end{equation}
where $\mathbf{G}=\Fa^*\Go=\Fa^\star\Go\Fa$ is the material metric, and $\boldsymbol\Lambda=\Fa^*\mathring{\boldsymbol\Lambda}$. Objectivity implies that
\begin{equation} 
	\mathring{W}=\tilde{\mathring{W}}(X,\vartheta,\mathbf{C},\boldsymbol\Lambda,{\mathbf{G}})\,.
\end{equation}

\begin{remark}[On the thermal distortion and the classical description of thermoelasticity]
\label{rmrk:Fth_classical}
The representation \eqref{eq:W_transported} clarifies the status of the thermal distortion introduced in Remark~\ref{rmrk:thermal_distortion}. The energy is expressed in terms of the total deformation gradient $\mathbf F$, with the anelastic content carried by the transported material metric $\mathbf G=\Fa^*\mathring{\mathbf G}$ and structural tensors $\boldsymbol\Lambda=\Fa^*\mathring{\boldsymbol\Lambda}$. For purely thermal anelasticity, $\Fa=\Fth$, this representation reduces to the classical description of nonlinear thermoelasticity, in which the total deformation gradient is used and temperature enters the constitutive function directly. The thermal distortion is not absent from this description; rather, it is encoded in $\mathbf G(\vartheta)=\Fth^*\mathring{\mathbf G}$ and $\boldsymbol\Lambda(\vartheta)=\Fth^*\mathring{\boldsymbol\Lambda}$. \citet{Ozakin2010} showed that, after equipping the body with the temperature-dependent metric $\mathbf G(\vartheta)$ and passing to a suitably chosen $\mathbf G$-orthonormal frame, the components of the total deformation gradient coincide with those of the elastic distortion in the multiplicative description. In this sense, the thermal distortion is represented by the change of frame, and the multiplicative decomposition provides an explicit representation of the same constitutive information. This viewpoint permits a geometric characterization of thermal stress and a systematic determination of stress-free temperature distributions and their associated thermal expansion coefficients \citep{Ozakin2010,Sadik2017Thermoelasticity}. The equivalence requires the transport of both $\mathbf G$ and $\boldsymbol\Lambda$. For an isotropic solid, $\mathring{\boldsymbol{\Lambda}}=\{\mathbf I\}$ and the structural tensors carry no additional information, so the material metric alone suffices; this is the setting commonly considered in classical thermoelasticity. For an anisotropic solid, however, the metric determines the natural lengths in the relaxed state but does not specify how the preferred material directions are transported by the thermal distortion. This additional information must be carried by the structural tensors, because anisotropic thermal expansion generally reorients preferred directions that are not aligned with its principal axes.
\end{remark}

\subsubsection{Material symmetry in viscoelasticity}
Let us consider a viscoelastic solid whose free energy function $\Psi=\Psi(X,\vartheta,\mathbf{F},\Fe,\Go,\mathbf{g})$ is additively decomposed into an equilibrium free energy function $\Psi_{\textrm{eq}}=\Psi_{\textrm{eq}}(X,\vartheta,\mathbf{F},\Go,\mathbf{g})$ and a non-equilibrium free energy $\Psi_{\textrm{neq}}=\Psi_{\textrm{neq}}(X,\vartheta,\Fe,\Go,\mathbf{g})$, and whose dissipation potential is given by $\phi=\phi(X,\vartheta,\Fve,\Fv,\dot\Fv,\Go,\mathbf{g})$.
Since $\mathbf{F}$ and $\Fe$ are invertible two-point tensors, objectivity implies that
\begin{equation}
\begin{split}
\Psi_{\textrm{eq}}(X,\vartheta,\mathbf{F},\Go,\mathbf{g})
	&= \hat{\Psi}_{\textrm{eq}}(X,\vartheta,\mathbf{F}^*\mathbf{g},\Go)
	= \hat{\Psi}_{\textrm{eq}}(X,\vartheta,\mathbf{C} ,\Go)\,,\\
\Psi_{\textrm{neq}}(X,\vartheta,\Fe, \Go,\mathbf{g})
	&=\hat{\Psi}_{\textrm{neq}}(X,\vartheta,\Fe^*\mathbf{g}, \Go)
	=\hat{\Psi}_{\textrm{neq}}(X,\vartheta,\Ce, \Go)\,.
\end{split}
\end{equation}
Therefore
\begin{equation}
	\Psi
	=\hat{\Psi}(X,\vartheta,\mathbf{C}, \, \Ce, \Go)
	=\hat{\Psi}_{\textrm{eq}}(X,\vartheta,\mathbf{C}, \Go)
	+\hat{\Psi}_{\textrm{neq}}(X,\vartheta,\Ce, \Go)\,.
\end{equation}
At a point $X\in\mathcal{B}$, the material symmetry group $\mathring{\mathcal{G}}_X$ with respect to the Euclidean reference configuration $(\mathcal{B},\Go)$ is defined as the set of $\mathring{\mathbf{K}}\in \mathrm{Orth}(\Go)$ such that
\begin{equation} \label{Material-Sym}
\begin{dcases}
	\mathring{\mathbf{K}}_*\Psi_{\textrm{eq}}(X,\vartheta,\mathbf{F},\Go,
	\mathbf{g})
	=\Psi_{\textrm{eq}}(X,\vartheta,\mathbf{F}\mathring{\mathbf{K}},\Go,
	\mathbf{g})
	=\Psi_{\textrm{eq}}(X,\vartheta,\mathbf{F},\Go,\mathbf{g})\,,\\
	\mathring{\mathbf{K}}_*\Psi_{\textrm{neq}}(X,\vartheta,\Fe,\Go,\mathbf{g})
	=\Psi_{\textrm{neq}}(X,\vartheta,\Fe\mathring{\mathbf{K}},\Go,\mathbf{g}) 
	=\Psi_{\textrm{neq}}(X,\vartheta,\Fe,\Go,\mathbf{g})\,,\\
	\mathring{\mathbf{K}}_*\phi(X,\vartheta,\mathbf{F},\Fv,\dot\Fv,\Go,
	\mathbf{g})
	=\phi(X,\vartheta,\mathbf{F}\mathring{\mathbf{K}},\mathring{\mathbf{K}}^*\Fv,
	\mathring{\mathbf{K}}^*\dot\Fv,
	\Go,\mathbf{g})
	=\phi(X,\vartheta,\mathbf{F},\Fv,\dot\Fv,\Go,\mathbf{g})\,,
\end{dcases}
\end{equation}
for all deformation gradients $\mathbf{F}$ and viscous distortions $\Fv$, where $\mathring{\mathbf{K}}^*\Fv=\mathring{\mathbf{K}}^{-1}\Fv \mathring{\mathbf{K}}$ (see Fig.~\ref{Symmetry-Transformation}a).

The non-equilibrium free energy can be written in terms of the total deformation gradient by introducing a set of structural tensors $\mathring{\boldsymbol\Lambda}$ for $\mathring{\mathcal G}_X\,$ leading to material covariance of $\Psi_{\textrm{neq}}=\Psi_{\textrm{neq}}(X,\vartheta,\Fe,\mathring{\boldsymbol\Lambda},\Go,\mathbf{g})\,$\textemdash i.e.,~for any linear transformation $\boldsymbol{\mathsf{T}}:T_X\mathcal{B}\to T_X\mathcal{B}$, $\Psi_{\textrm{neq}}(X,\vartheta,\Fe,\mathring{\boldsymbol\Lambda},\Go,\mathbf{g})=\Psi_{\textrm{neq}}(X,\vartheta,\boldsymbol{\mathsf{T}}^*\Fe,\boldsymbol{\mathsf{T}}^*\mathring{\boldsymbol\Lambda},\boldsymbol{\mathsf{T}}^*\Go,\mathbf{g})$. Choosing $\boldsymbol{\mathsf{T}}=\Fv$, it follows that
\begin{equation} 
	\Psi_{\textrm{neq}}
	=\Psi_{\textrm{neq}}(X,\vartheta,\Fe,\mathring{\boldsymbol\Lambda},\Go,\mathbf{g})
	=\Psi_{\textrm{neq}}(X,\vartheta,\Fv^*\Fe,\Fv^*\mathring{\boldsymbol\Lambda},
	\Fv^*\Go,\mathbf{g})
	=\Psi_{\textrm{neq}}(X,\vartheta,\mathbf{F},\Lambdav,\Gv,\mathbf{g})
	\,,
\end{equation}
where $\Gv=\Fv^*\Go=\Fv^\star\Go\Fv$ is the viscous metric and $\Lambdav=\Fv^*\mathring{\boldsymbol\Lambda}$ are viscous structural tensors \citep{SaYa2024viscoelasticity}.
Objectivity implies that
\begin{equation}
\begin{split}
	\Psi_{\textrm{neq}}(X,\vartheta,\mathbf{F},\Lambdav,\Gv,\mathbf{g}) 
	&=\tilde{\Psi}_{\textrm{neq}}(X,\vartheta,\mathbf{C},\Lambdav,\Gv) \,.
\end{split}
\end{equation}
Therefore
\begin{equation}\label{eq:Psi_C_L}
	\Psi =\tilde{\Psi}(X,\vartheta,\mathbf{C},\mathring{\boldsymbol{\Lambda}},\Lambdav,\Go,\Gv) =\tilde{\Psi}_{\textrm{eq}}(X,\vartheta,\mathbf{C},\mathring{\boldsymbol{\Lambda}},\Go)
	+\tilde{\Psi}_{\textrm{neq}}(X,\vartheta,\mathbf{C},\Lambdav, \Gv) 
	\,.
\end{equation}

\subsection{Material symmetry in visco-anelasticity}
\label{SubS:MatSym}

Let us now consider a visco-anelastic solid whose energy function $\Psi=\Psi(X,\vartheta,\Fve,\Fe,\Go,\mathbf{g})$ is additively decomposed into an equilibrium energy function $\Psi_{\textrm{eq}}=\Psi_{\textrm{eq}}(X,\vartheta,\Fve,\Go,\mathbf{g})$ and a non-equilibrium energy $\Psi_{\textrm{neq}}=\Psi_{\textrm{neq}}(X,\vartheta,\Fe,\Go,\mathbf{g})$, 
with a dissipation potential of the form $\phi=\phi(X,\vartheta,\Fve,\Fv,\dot\Fv,\Fn,\dot\Fn,\mathbf{G},\mathbf{g})$.
At a point $X\in\mathcal{B}$, the material symmetry group $\mathring{\mathcal{G}}_X$ with respect to the Euclidean reference configuration $(\mathcal{B},\Go)$ is defined as the set of $\mathring{\mathbf{K}}\in \mathrm{Orth}(\Go)$ such that
\begin{equation} \label{Material-Sym}
\begin{dcases}
	\mathring{\mathbf{K}}_*\Psi_{\textrm{eq}}(X,\vartheta,\Fe\Fv,\Go,\mathbf{g})
	=\Psi_{\textrm{eq}}(X,\vartheta,(\Fe\mathring{\mathbf{K}})(\mathring{\mathbf{K}}^{-1}\Fv\,	
	\mathring{\mathbf{K}}),\Go,\mathbf{g})
	=\Psi_{\textrm{eq}}(X,\vartheta,\Fe\Fv,\Go,\mathbf{g})\,,\\
	\mathring{\mathbf{K}}_*\Psi_{\textrm{neq}}(X,\vartheta,\Fe,\Go,\mathbf{g})
	=\Psi_{\textrm{neq}}(X,\vartheta,\Fe\mathring{\mathbf{K}},\Go,\mathbf{g})
	=\Psi_{\textrm{neq}}(X,\vartheta,\Fe,\Go,\mathbf{g})\,,\\
	\mathring{\mathbf{K}}_*\phi(X,\vartheta,\Fve,\Fv,\dot\Fv,\Fn,\dot\Fn,
	\Go,\mathbf{g})
	=\phi(X,\vartheta,\Fve\mathring{\mathbf{K}},\mathring{\mathbf{K}}^*\Fv,
	\mathring{\mathbf{K}}^*\dot\Fv,
	\mathring{\mathbf{K}}^*\Fn,\mathring{\mathbf{K}}^*\dot\Fn,\Go,
	\mathbf{g}) 
	=\phi(X,\vartheta,\Fve,\Fv,\dot\Fv,\Fn,\dot\Fn,\Go,\mathbf{g})\,,
\end{dcases}
\end{equation}
for all deformation gradients $\mathbf{F}$, viscous distortions $\Fv$, anelastic distortions $\Fn$, and temperature fields $\vartheta$, where $\mathring{\mathbf{K}}^*\Fv=\mathring{\mathbf{K}}^{-1}\Fv \mathring{\mathbf{K}}$ and $\mathring{\mathbf{K}}^*\Fn=\mathring{\mathbf{K}}^{-1}\Fn \mathring{\mathbf{K}}$ (see Fig.~\ref{Symmetry-Transformation}c).\footnote{Recall that $\Fa=\Fn\Fth$ and the thermal distortion $\Fth$ is such that $\boldsymbol\alpha = ({d\Fth}/{d\vartheta})\,\Fth^{-1}$, where $\boldsymbol\alpha$ is the (linear) thermal expansion coefficient tensor.}

We denote the structural tensors of the material symmetry group $\mathring{\mathcal{G}}_X$ by $\mathring{\boldsymbol{\Lambda}}.$\footnote{In the case of an isotropic solid, one has $\mathring{\boldsymbol{\Lambda}}=\{\mathbf{I}\}$.} Adding this collection of structural tensors to the list of arguments of the free energy, one writes
\begin{equation}
\label{eq:Psi_Fve_L}
	\Psi=\Psi(X,\vartheta,\Fve, \Fe, \mathring{\boldsymbol{\Lambda}},\Go,\mathbf{g})
	=\Psi_{\textrm{eq}}(X,\vartheta,\Fve, \mathring{\boldsymbol{\Lambda}},\Go,\mathbf{g})
	+\Psi_{\textrm{neq}}(X,\vartheta,\Fe, \mathring{\boldsymbol{\Lambda}},\Go,\mathbf{g})\,.
\end{equation}
Since $\Fve$ and $\Fe$ are invertible two-point tensors, objectivity implies that
\begin{equation}
\begin{split}
	\Psi_{\textrm{eq}}(X,\vartheta,\Fve,\mathring{\boldsymbol{\Lambda}},\Go,\mathbf{g})
	&= \hat{\Psi}_{\textrm{eq}}(X,\vartheta,\Fve^*\mathbf{g},\mathring{\boldsymbol{\Lambda}},\Go)
	= \hat{\Psi}_{\textrm{eq}}(X,\vartheta,\Cve ,\mathring{\boldsymbol{\Lambda}},\Go)\,,\\
	\Psi_{\textrm{neq}}(X,\vartheta,\Fe, \mathring{\boldsymbol{\Lambda}},\Go,\mathbf{g})
	&=\hat{\Psi}_{\textrm{neq}}(X,\vartheta,\Fe^*\mathbf{g}, \mathring{\boldsymbol{\Lambda}},\Go)
	=\hat{\Psi}_{\textrm{neq}}(X,\vartheta,\Ce, \mathring{\boldsymbol{\Lambda}},\Go)\,,
\end{split}
\end{equation}
where $\Cve=\Fve^*\mathbf{g}=\Fv^*\Ce$. Therefore
\begin{equation}
\label{eq:Psi_Cve_L}
	\Psi
	=\hat{\Psi}(X,\vartheta,\Cve, \, \Ce, \Go)
	=\hat{\Psi}_{\textrm{eq}}(X,\vartheta,\Cve, \Go)
	+\hat{\Psi}_{\textrm{neq}}(X,\vartheta,\Ce, \Go)\,.
\end{equation}

Recall that, with the structural tensors added, both the equilibrium and non-equilibrium free energies are now materially covariant functions, i.e.,~for any invertible linear transformation $\mathbf{T}:T_X\mathcal{B}\to T_X\mathcal{B}$, one has
\begin{equation}
\begin{aligned}
	& \Psi_{\textrm{eq}}(X,\vartheta,\mathbf{T}^*\Fve,\mathbf{T}^*\mathring{\boldsymbol{\Lambda}},
	\mathbf{T}^*\Go,\mathbf{g})
	=\Psi_{\textrm{eq}}(X,\vartheta,\Fve,\mathring{\boldsymbol{\Lambda}},\Go,
	\mathbf{g})\,, \\
	& \Psi_{\textrm{neq}}(X,\vartheta,\mathbf{T}^*\Fe,
	\mathbf{T}^*\mathring{\boldsymbol{\Lambda}},\mathbf{T}^*\Go,\mathbf{g})
	=\Psi_{\textrm{neq}}(X,\vartheta,\Fe,\mathring{\boldsymbol{\Lambda}},
	\Go,\mathbf{g})\,.
\end{aligned}
\end{equation}
Let us choose $\mathbf{T}=\Fa$ and $\mathbf{T}=\Fv\Fa$, for the equilibrium and non-equilibrium free energies, respectively, and write
\begin{equation}\nonumber
\begin{aligned}
	\Psi &=\Psi_{\textrm{eq}}(X,\vartheta,\Fve,\mathring{\boldsymbol{\Lambda}}	,\Go,\mathbf{g})
	+\Psi_{\textrm{neq}}(X,\vartheta,\Fe,\mathring{\boldsymbol{\Lambda}},\Go,\mathbf{g}) \\
	&=\Psi_{\textrm{eq}}(X,\vartheta,\Fa^*\Fve, \Fa^*\mathring{\boldsymbol{\Lambda}},\Fa^*\Go,\mathbf{g})
	+\Psi_{\textrm{neq}}(X,\vartheta,(\Fv\Fa)^*\Fe, (\Fv\Fa)^*\mathring{\boldsymbol{\Lambda}},(\Fv\Fa)^*\Go,\mathbf{g})\,.
\end{aligned}
\end{equation}
Recalling that $\Fve=\Fa_*\mathbf{F}$ and $\Fe=(\Fv\Fa)_*\mathbf{F}$, it follows that $\Fa^*\Fve=(\Fv\Fa)^*\Fe=\mathbf{F}$. Thus
\begin{equation}
\label{eq:Psi_F_L}
\Psi=\Psi_{\textrm{eq}}(X,\vartheta,\mathbf{F},\boldsymbol{\Lambda},\mathbf{G},\mathbf{g})
+\Psi_{\textrm{neq}}(X,\vartheta,\mathbf{F},\Lambdai,\Gi,\mathbf{g}) 
\,,
\end{equation}
with
\begin{equation}
	\mathbf{G}=\Fa^*\Go\,,\quad
	\boldsymbol{\Lambda}=\Fa^*\mathring{\boldsymbol{\Lambda}}\,,\quad
	\Gi=(\Fv\Fa)^*\Go\,,\quad
	\Lambdai=(\Fv\Fa)^*\mathring{\boldsymbol{\Lambda}}\,.
\end{equation}
The metric $\mathbf{G}$ is used to compute the local natural distances and angles in the body's natural stress-free state; this is the material metric indeed. Objectivity implies that
\begin{equation}
\begin{split}
	\Psi_{\textrm{eq}}(X,\vartheta,\mathbf{F},\boldsymbol{\Lambda},\mathbf{G},\mathbf{g})
	&= \tilde{\Psi}_{\textrm{eq}}(X,\vartheta,\mathbf{C},\boldsymbol{\Lambda},\mathbf{G})\,,\\
	\Psi_{\textrm{neq}}(X,\vartheta,\mathbf{F},\Lambdai, \Gi,\mathbf{g}) 
	&=\tilde{\Psi}_{\textrm{neq}}(X,\vartheta,\mathbf{C},\Lambdai, \Gi) \,.
\end{split}
\end{equation}
Therefore
\begin{equation}
\label{eq:Psi_C_L}
	\Psi =\tilde{\Psi}(X,\vartheta,\mathbf{C},\boldsymbol{\Lambda},\Lambdai,\mathbf{G},\Gi) 
	=\tilde{\Psi}_{\textrm{eq}}(X,\vartheta,\mathbf{C},\boldsymbol{\Lambda},\mathbf{G})
	+\tilde{\Psi}_{\textrm{neq}}(X,\vartheta,\mathbf{C},\Lambdai, \Gi) 
	\,.
\end{equation}

\begin{remark}
\label{rmrk:Equi_Rep}
By the arguments above, the representations~\eqref{eq:Psi_Fve_L}, \eqref{eq:Psi_Cve_L}, \eqref{eq:Psi_F_L} and \eqref{eq:Psi_C_L} are equivalent. Consequently, any integrity basis (minimal set of invariants) constructed from $\{\Cve,\mathring{\boldsymbol\Lambda},\Go\}$ (or $\{\Fve,\mathring{\boldsymbol\Lambda},\Go,\mathbf{g}\}$) coincides, via the push by $\Fa$, with one constructed from $\{\mathbf C,\boldsymbol\Lambda,\mathbf G\}$ (or $\{\mathbf F,\boldsymbol\Lambda,\mathbf G,\mathbf{g}\}$); and the invariants of $\{\Ce,\mathring{\boldsymbol\Lambda},\Go\}$ (or $\{\Fe,\mathring{\boldsymbol\Lambda},\Go,\mathbf{g}\}$) coincide, via the push by $\Fv\Fa$, with those of $\{\mathbf C,\Lambdai,\Gi\}$ (or $\{\mathbf F,\Lambdai,\Gi,\mathbf{g}\}$).
In particular, for an isotropic solid ($\mathring{\boldsymbol\Lambda}=\{\mathbf{I}\}$), one has 
\begin{equation}
	\Ive_i(\Cve,\Go) = I_i(\mathbf C,\mathbf G)\,,\qquad 
	\Ie_i(\Ce,\Go) = I_i(\mathbf C,\Gi)\,,\qquad i=1,2,3
	\,.
\end{equation}
This can be shown directly by recalling that $\mathbf C=\mathbf{F}^*\mathbf g=\mathbf{F}^\star \mathbf g \mathbf{F}$, $\Cve=\Fve^*\mathbf g=\Fve^\star \mathbf g \Fve$, and $\mathbf{G}=\Fa^*\Go=\Fa^\star \Go\Fa$. For the first invariants $\Ive_1$ and $I_1$, one finds
\begin{equation}
\begin{split}
	\Ive_1
	&=\operatorname{tr}_{\Go}\Cve
	=\cCve_{AB}\,\mathring{\mathrm G}^{BA}
	=\cFve^i{}_A \,\mathrm g_{ij} \,\cFve^j{}_B \mathring{\mathrm G}^{BA}
	=\mathrm{F}^i{}_I \,\cFa^I{}_A \,\mathrm g_{ij} \,\mathrm{F}^j{}_J \,\cFa^J{}_B \,\mathring{\mathrm G}^{BA}\\
	&=(\mathrm{F}^i{}_I \,\mathrm g_{ij} \,\mathrm{F}^j{}_J) \,(\cFa^J{}_B \,\mathring{\mathrm G}^{BA} \,\cFa^I{}_A)
	=\mathrm C_{IJ}\,{\mathrm G}^{JI}
	=\operatorname{tr}_{\mathbf G}\mathbf{C}
	=I_1\,.
\end{split}
\end{equation}
For the second invariants, similarly to the above computation, one finds
\begin{equation}
	\operatorname{tr}_{\Go}\Cve^2
	= \mathring{\mathrm G}^{AB}\,\cCve_{BC}\,\mathring{\mathrm G}^{CD}\,\cCve_{DA}
	= \mathrm G^{IJ}\,\mathrm C_{JK}\,\mathrm G^{KL}\,\mathrm C_{LI}
	= \operatorname{tr}_{\mathbf G}\mathbf{C}^2\,,
\end{equation}
and hence
\begin{equation}
	\Ive_2
	=\frac{1}{2}\!\left(\Ive_1^{\,2}-\operatorname{tr}_{\Go}\Cve^2\right)\!
	=\frac{1}{2}\!\left(I_1^{\,2}-\operatorname{tr}_{\mathbf G}\mathbf{C}^2\right)\!
	= I_2\,.
\end{equation}
For the third invariant, one uses
\begin{equation}
	J
	=\sqrt{\frac{\det \mathbf{g}}{\det \mathbf{G}}}\,\det \mathbf{F}
	=\sqrt{\frac{\det\mathbf g}{\det\Go}}\,\det\Fve\,,
\end{equation}
to obtain
\begin{equation}
	\Ive_3
	=\det{\!}_{\Go}\Cve
	=\frac{\det\Cve}{\det\Go}
	= \frac{\det\mathbf g}{\det\Go} \det{\!}^2\Fve
	= \frac{\det \mathbf{g}}{\det \mathbf{G}} \det{\!}^2 \mathbf{F}
	=\frac{\det\mathbf{C}}{\det\mathbf G}
	=\det{\!}_{\mathbf G}\mathbf C
	=I_3\,.
\end{equation}
\end{remark}

\begin{remark}[Physical components]
\label{rmrk:phys_comp}
In curvilinear coordinates, the coordinate components of a tensor are not, in general, dimensionally homogeneous---in cylindrical coordinates $(r,\theta,z)$, the coordinates $r$ and $z$ have the dimension of length, whereas $\theta$ is dimensionless---and the associated coordinate basis vectors are not generally of unit length.
Physical components address both issues by referring to orthonormal frames so that every entry carries the physical dimension of the tensor field itself \citep{Truesdell1953physical, SaYa2026PhysComp}.
On a Riemannian manifold $(\mathcal{M},\mathbf{m})$ with $\mathbf{m}$-orthogonal coordinates, such a frame may be given by the normalized coordinate basis, and the physical components are obtained by scaling every contravariant index by $\sqrt{\textrm{m}_{aa}}$ and every covariant one by $1/\sqrt{\textrm{m}_{aa}}$; for a vector $\mathbf{u}$ and a covector $\boldsymbol\alpha$, they read \citep{Truesdell1953physical}
\begin{equation}
\label{eq:phys_comp_def}
	\hat{\textrm{u}}^a = \sqrt{\textrm{m}_{aa}}\,\textrm{u}^a\,,
	\qquad
	\hat{\alpha}_a = \frac{1}{\sqrt{\textrm{m}_{aa}}}\,\alpha_a\,,
	\quad\text{(no summation on $a$).}
\end{equation}
For a two-point tensor, each index is referred to an orthonormal frame associated with the metric on its corresponding space. See \citep{SaYa2026PhysComp} for the general construction in arbitrary curvilinear coordinates.

Turning to the deformation gradient, the two-point tensor $\mathbf{F}:T_X\mathcal{B}\rightarrow T_x\mathcal{S}$ is metric-independent. The relation $\mathbf{F}\mathbf{U}=\mathbf{u}$ is purely kinematic, as are the coordinate components $\textrm{F}^a{}_A$. The physical components of $\mathbf{F}$, however, depend on the metrics because the normalization in \eqref{eq:phys_comp_def} requires one metric for each leg.
The spatial leg is normalized using the spatial metric $\mathbf{g}$. No metric is canonically prescribed for the reference leg; it is yet to be assigned. Denoting this metric by $\mathbf{M}$, the corresponding physical components of $\mathbf{F}$ are
\begin{equation} \label{eq:F_phys_generic}
	\widehat{\textrm{F}}{}^a{}_A = \sqrt{\frac{\textrm{g}_{aa}}{\textrm{M}_{AA}}}\,\textrm{F}^a{}_A\,,
	\quad\text{(no summation on $a$ or $A$),}
\end{equation}
provided $\{X^A\}$ is $\mathbf{M}$-orthogonal and $\{x^a\}$ is $\mathbf{g}$-orthogonal.
The present setting offers three candidates for the referential metric. The first is the Euclidean metric $\Go=\iota^*\mathbf{g}$ induced by the embedding $\iota:\mathcal{B}\rightarrow\mathcal{S}$. The second is the material metric $\mathbf{G}=\Fa^*\Go$, which encodes the stress-free state associated with the eigenstrains. The third is the instantaneous metric $\Gi=(\Fv\Fa)^*\Go$, relative to which the instantaneous response $\Fe$ is measured. Each choice is admissible and, in general, defines a different set of physical components $\widehat{\textrm{F}}{}^a{}_A$ for the same tensor $\mathbf{F}$.

Geometry alone does not determine which metric should be assigned to a given distortion; the appropriate choice follows from the constitutive structure. The equivalent representations~\eqref{eq:Psi_Fve_L} and \eqref{eq:Psi_F_L} indicate the following: the equilibrium, or long-term, response is described by $\{\mathbf{F},\boldsymbol{\Lambda},\mathbf{G},\mathbf{g}\}$, or equivalently by $\{\Fve,\mathring{\boldsymbol{\Lambda}},\Go,\mathbf{g}\}$; the non-equilibrium, or instantaneous, response is described by $\{\mathbf{F},\Lambdai,\Gi,\mathbf{g}\}$, or equivalently by $\{\Fe,\mathring{\boldsymbol{\Lambda}},\Go,\mathbf{g}\}$.
Indeed, this observation further clarifies the local intermediate configurations discussed in Fig.~\ref{Local-Configurations}(c) and the roles played by the different metrics in visco-anelasticity. The local geometry of visco-anelasticity is represented schematically by the two pointwise commutative diagrams in Fig.~\ref{Fig:ViscoAnelasticMetricMaps}. The dotted arrows represent local isometries, whereas the solid arrows are generally not isometries. For the equilibrium response, $\Fa$ is a local isometry from $(T_X\mathcal{B},\mathbf{G})$ to $(T_X\mathcal{B},\Go)$, and commutativity gives $\mathbf{F}=\Fve\Fa=\Fe\Fv\Fa$. For the non-equilibrium response, the combined distortion $\Fv\Fa$ is a local isometry from $(T_X\mathcal{B},\Gi)$ to $(T_X\mathcal{B},\Go)$, and commutativity gives $\mathbf{F}=\Fe\Fv\Fa$.

\begin{figure}[t!]
\centering
\begin{tikzpicture}[>=stealth]
\begin{scope}[xshift=-3.7cm]
\node (E1) at (-2,1.2)
  {$(T_X\mathcal{B},\mathbf{G})$};
\node (E2) at (2,1.2)
  {$(T_X\mathcal{B},\Go)$};
\node (E3) at (0,-0.5)
  {$(T_x\mathcal{C},\mathbf{g})$};
\draw[->,dash pattern=on 3pt off 2pt,line width=0.5pt]
  (E1) --
  node[midway,above] {$\Fa$}
  (E2);
\draw[->]
  (E1) --
  node[midway,anchor=east,xshift=-3pt] {$\mathbf{F}$}
  (E3);
\draw[->]
  (E2) --
  node[midway,anchor=west,xshift=4pt] {$\Fve$}
  (E3);
\node at (0,-1.35)
  {\footnotesize (a) Equilibrium response};
\end{scope}
\begin{scope}[xshift=3.7cm]
\node (N1) at (-2,1.2)
  {$(T_X\mathcal{B},\Gi)$};
\node (N2) at (2,1.2)
  {$(T_X\mathcal{B},\Go)$};
\node (N3) at (0,-0.5)
  {$(T_x\mathcal{C},\mathbf{g})$};
\draw[->,dash pattern=on 3pt off 2pt,line width=0.5pt]
  (N1) --
  node[midway,above] {$\Fv\Fa$}
  (N2);
\draw[->]
  (N1) --
  node[midway,anchor=east,xshift=-3pt] {$\mathbf{F}$}
  (N3);
\draw[->]
  (N2) --
  node[midway,anchor=west,xshift=4pt] {$\Fe$}
  (N3);
\node at (0,-1.35)
  {\footnotesize (b) Non-equilibrium response};
\end{scope}
\end{tikzpicture}
\caption{
Pointwise commutative diagrams between the relevant tangent
spaces for the equilibrium and non-equilibrium descriptions of the
visco-anelastic response. Dashed arrows denote local isometries,
whereas solid arrows are generally not isometries.}
\label{Fig:ViscoAnelasticMetricMaps}
\end{figure}

In the notation of \eqref{eq:F_phys_generic}, the constitutively motivated kinematics of Fig.~\ref{Fig:ViscoAnelasticMetricMaps} determine the metrics assigned for normalization. For the equilibrium response, the physical components of $\mathbf{F}$ are defined using $\mathbf{G}$, whereas those of $\Fve$ are defined using $\Go$. For the non-equilibrium response, the physical components of $\mathbf{F}$ are defined using $\Gi$, whereas those of $\Fe$ are defined using $\Go$. We denote the equilibrium and non-equilibrium physical components by a hat, $\widehat{(\cdot)}$, and a tilde, $\widetilde{(\cdot)}$, respectively.

Assume that the coordinate bases associated with $\{X^A\}$ and $\{x^a\}$ are orthogonal with respect to $\Go$ and $\mathbf{g}$, respectively, and that $\Fv$ and $\Fa$ are diagonal in the material coordinate basis.\footnote{See \citep{SaYa2026PhysComp} for the general construction in arbitrary curvilinear coordinates.} Under these assumptions, the two representations of each constitutive branch are equivalent.
More explicitly, the following relations hold $\text{(no summation on $a$ or $A$)}$:
\begin{equation}
	\textrm{F}^a{}_A=\cFve^a{}_A\cFa^A{}_A=\cFe^a{}_A\cFv^A{}_A\cFa^A{}_A\,.
\end{equation}
Moreover,
\begin{equation}
	\textrm{G}_{AA}=(\cFa^A{}_A)^2\cGo_{AA}\,,\qquad
	\cGi_{AA}=(\cFa^A{}_A)^2(\cFv^A{}_A)^2\cGo_{AA}\,.
\end{equation}
It follows that
\begin{subequations}\label{eq:F_phys_coincide}
\begin{align}
	\widehat{\textrm{F}}{}^a{}_A
	&=\sqrt{\frac{\textrm{g}_{aa}}{\textrm{G}_{AA}}}\,\textrm{F}^a{}_A
	=\sqrt{\frac{\textrm{g}_{aa}}{\cGo_{AA}}}\,\cFve^a{}_A
	=\widehat{\cFve}{}^a{}_A\,,\\
	\widetilde{\textrm{F}}{}^a{}_A
	&=\sqrt{\frac{\textrm{g}_{aa}}{\cGi_{AA}}}\,\textrm{F}^a{}_A
	=\sqrt{\frac{\textrm{g}_{aa}}{\cGo_{AA}}}\,\cFe^a{}_A
	=\widetilde{\cFe}{}^a{}_A\,.
\end{align}
\end{subequations}
Thus, for each constitutive response, the physical components of the total deformation gradient coincide with those of the corresponding distortion.
The referential distortions do not appear explicitly in these physical components because they act as isometries: $\Fa:(\mathcal{B},\mathbf{G})\rightarrow(\mathcal{B},\Go)$ and $\Fv\Fa:(\mathcal{B},\Gi)\rightarrow(\mathcal{B},\Go)$, and their physical components form the identity matrix; however, their geometric effects are encoded in the induced metrics $\mathbf{G}$ and $\Gi$ \citep{SaYa2026PhysComp}. Accordingly, the equilibrium and non-equilibrium descriptions differ only in the referential metric used for normalization,
\begin{equation}
\label{eq:F_phys_FeFv}
\widehat{\textrm{F}}{}^a{}_A
	= \widetilde{\textrm{F}}{}^a{}_A\,\cFv^A{}_A
	= \widetilde{\cFe}{}^a{}_A\,\cFv^A{}_A\,.
\end{equation}
\end{remark}

\begin{remark}
\label{rmrk:longtime_distort}
Consider a visco-anelastic body undergoing a motion on the time interval $[0,\infty)$. At some time $t=T$, the applied loads and/or prescribed boundary displacements are held fixed. Thus, for all $t > T$, the total deformation gradient remains constant, i.e.,~$\mathbf{F}(X,t)=\mathbf{F}(X,T) \eqqcolon \mathbf{F}_{\infty}(X)$. The elastic, viscous, and anelastic distortions, however, continue to evolve in time:
\begin{equation}
	\mathbf{F}_{\infty}(X)=\Fe(X,t)\,\Fv(X,t)\,\Fa(X,t)\,,\qquad t > T
	\,.
\end{equation}
In physical components, assuming orthonormal bases (and consequently diagonal matrix representations) and recalling \eqref{eq:F_phys_FeFv}, one may write
\begin{equation}
	\widehat{\mathbf F}_{\infty}(X)=\widetilde\Fe(X,t)\, \Fv(X,t)\,,\qquad t > T\,.
\end{equation}
Recalling~\eqref{eq:Pstress_decomp}, the stress admits the additive decomposition
\begin{equation}
	\mathbf{P}=\mathbf{P}_{\textrm{eq}}+\mathbf{P}_{\textrm{neq}}\,,\qquad
	\mathbf{P}_{\textrm{eq}}
	= \rho \mathbf g^\sharp \frac{\partial \Psi}{\partial \Fve}
	\,\Fve^{\star} \,\mathbf{F}^{-\star}\,,\qquad
	\mathbf{P}_{\textrm{neq}}
	= \rho \mathbf g^\sharp \frac{\partial \Psi}{\partial \Fe}
	\,\Fe^{\star} \,\mathbf{F}^{-\star}	\,.
\end{equation}
As $t\to\infty$, the stress relaxes to its equilibrium value, i.e.,~$\mathbf{P}=\mathbf{P}_{\textrm{eq}}$ and therefore $\mathbf{P}_{\textrm{neq}}=\mathbf{0}$.
Hence
\begin{equation}
	\lim_{t\to\infty}\frac{\partial \Psi_{\textrm{neq}}}{\partial \Fe}=\mathbf{0}\,.
\end{equation}
Thus, if $\Fe(X,t)$ admits a limit as $t\to\infty\,$, then one may write in physical components that
\begin{equation}
	\lim_{t\to\infty}\widetilde{\Fe}(X,t)=\mathbf{I}
	\,.
\end{equation}
Therefore,
\begin{equation}
	\lim_{t\to\infty} \Fv(X,t) = \widehat{\mathbf F}_{\infty}(X)
	\,.
\end{equation}
\end{remark}

In what follows, we explicitly write the constitutive relations and kinetic equations for isotropic, transversely isotropic, orthotropic, and monoclinic visco-anelastic solids, by using the respective integrity basis corresponding to the symmetry group as discussed in \S\ref{sec:strucTensors}. 

\subsubsection{Isotropic solids}

An isotropic solid does not have any material preferred directions, i.e.,~$\mathring{\mathcal{G}}_X=\mathrm{Orth}(\Go)\,,\forall X\in \mathcal B$.

\paragraph{Constitutive equations.}
For isotropic solids, $\Psi_{\textrm{eq}}$ and $\Psi_{\textrm{neq}}$ depend only on the principal invariants of $\{\Cve,\Go\}$ (or equivalently $\{\mathbf C,\mathbf G\}$) and $\{\Ce,\Go\}$ (or equivalently $\{\mathbf C,\Gi\}$), respectively, i.e.,
\begin{equation}
\label{Psi_const_iso}
	\Psi_{\textrm{eq}} = \overline{\Psi}(X,\vartheta,I_1,I_2,I_3)   \,,\qquad
	\Psi_{\textrm{neq}}  = \widetilde{\Psi}(X,\vartheta,\Ie_1,\Ie_2,\Ie_3)  \,,
\end{equation}
where
\begin{equation}
\label{eq:iso_inv}
\begin{aligned}
	& I_1=\operatorname{tr}_{\Go}\Cve=\cCve_{AB}\,\mathring{\mathrm G}^{BA}, &&\!
	I_2=\frac{1}{2}\!\left(I_1^{\,2}-\operatorname{tr}_{\Go}\Cve^2\right)\!
	=\frac{1}{2}\!\left(I_1^{\,2}-\mathring{\mathrm G}^{AB}\,\cCve_{BC}\,\mathring{\mathrm G}^{CD}\cCve_{DA}\right)\!,
	&&\!  I_3 =J^2=\frac{\det\Cve}{\det\Go},\\
	& \Ie_1=\operatorname{tr}_{\Go}\Ce=\cCe_{AB}\,\mathring{\mathrm G}^{BA},&&\!
	\Ie_2=\frac{1}{2}\!\left(\Ie_1^2-\operatorname{tr}_{\Go}\Ce^2\right)\!
	=\frac{1}{2}\!\left(\Ie_1^2-\mathring{\mathrm G}^{AB}\,\cCe_{BC}\,\mathring{\mathrm G}^{CD}\,\cCe_{DA}\right)\!,
	&&\!  \Ie_3 =\Je^2=\frac{\det\Ce}{\det\Go}.
\end{aligned}
\end{equation}
Following~\eqref{Psi_const_iso} and~\eqref{eq:iso_inv},\footnote{The derivatives of the principal invariants of $\Cve$ and $\Ce$ are computed in Appendix \ref{appendix:derivatives}.} one may write
\begin{subequations}\label{eq:iso_part_repr}
\begin{align}
	\frac{\partial \hat\Psi}{\partial \Ce}
	=\frac{\partial \hat\Psi_{\textrm{neq}}}{\partial \Ce}
	= \sum_{j=1}^{3} \widetilde\Psi_j\frac{\partial \Ie_j}{\partial \Ce} 
	&= \widetilde\Psi_1\Go^{-1} 
	+\!\left(\widetilde\Psi_2 \Ie_2 + \widetilde\Psi_3 \Ie_3 \right)\!\Ce^{-\sharpo}
	-\widetilde\Psi_2\Ie_3\Ce^{-2\sharpo} 
	\,,\\
	\frac{\partial \hat\Psi}{\partial \Cve}
	=\frac{\partial \hat\Psi_{\textrm{eq}}}{\partial \Cve}
	= \sum_{j=1}^{3} \overline\Psi_j\frac{\partial I_j}{\partial \Cve} 
	&= \overline\Psi_1\Go^{-1} 
	+ \!\left(\overline\Psi_2 I_2+ \overline\Psi_3 I_3\right)\!\Cve^{-\sharpo}
	- \overline\Psi_2 I_3\Cve^{-2\sharpo}
	\,,
\end{align}
\end{subequations}
where
\begin{equation}
\label{eq:Psi_i}
	\overline{\Psi}_j=\overline{\Psi}_j(X,\vartheta,I_1,I_2,I_3)\coloneqq\frac{\partial \overline{\Psi}}{\partial I_j}\,,\qquad
	\widetilde{\Psi}_j=\widetilde{\Psi}_j(X,\vartheta,\Ie_1,\Ie_2,\Ie_3)\coloneqq\frac{\partial \widetilde{\Psi}}{\partial \Ie_j}\,, 		\qquad j=1,2,3\,.
\end{equation}
Therefore, substituting~\eqref{eq:iso_part_repr} into~\eqref{eq:S_eq_neq} leads to
\begin{equation}\label{eq:S_iso_represent1}
\begin{aligned}
	\mathbf{S}
	&=2\rho\mathbf{F}^{-1}\left\{\Fve\!\left[\overline\Psi_1\Go^{-1} 
	+ \!\left(\overline\Psi_2 I_2+ \overline\Psi_3 I_3\right)\!\Cve^{-\sharpo}
	- \overline\Psi_2 I_3\Cve^{-2\sharpo}\right]\!\Fve^\star\right.
	\\&\quad
	\left.
	+\Fe\!\left[\widetilde\Psi_1\Go^{-1} 
	+\!\left(\widetilde\Psi_2 \Ie_2 + \widetilde\Psi_3 \Ie_3 \right)\!\Ce^{-\sharpo}
	-\widetilde\Psi_2\Ie_3\Ce^{-2\sharpo}\right]\!\,\Fe^{\star}\right\}\mathbf{F}^{-\star}\,.
\end{aligned}
\end{equation}
Note that
\begin{equation}
\begin{aligned}
	\Fa^{-1}\Cve^{-1} \Fa^{-\star}
	&=\Fa^{-1}\Fv^{-1}\Ce^{-1}\Fv^{-\star}\Fa^{-\star}=\mathbf{C}^{-1}\,,\\
	\Fa^{-1}\Cve^{-2} \Fa^{-\star}
	&=\Fa^{-1}\Cve^{-1} \Go \Cve^{-1} \Fa^{-\star}
	=(\Fa^{-1}\Cve^{-1} \Fa^{-\star})(\Fa^{\star} \Go\Fa)(\Fa^{-1} 
	\Cve^{-1} \Fa^{-\star})
	=\mathbf{C}^{-1}\mathbf{G}\mathbf{C}^{-1}\,,\\
	\Fa^{-1}\Fv^{-1}\Ce^{-2} \Fv^{-\star}\Fa^{-\star}
	&=\Fa^{-1}\Fv^{-1}\Ce^{-1}  \Go \Ce^{-1} \Fv^{-\star} \Fa^{-\star} \\ 
	&=(\Fa^{-1}\Fv^{-1}\Ce^{-1} \Fv^{-\star} \Fa^{-\star})(\Fa^{\star}\Fv^{\star} \Go	
	\Fv\Fa)(\Fa^{-1} \Fv^{-1}\Ce^{-1} \Fv^{-\star} \Fa^{-\star})
	=\mathbf{C}^{-1}\Gi\mathbf{C}^{-1}\,,
\end{aligned}
\end{equation}
and recall that $\mathbf{G}=\Fa^*\Go=\Fa^\star\Go\Fa$ and $\Gi=(\Fv\Fa)^*\Go=\Fa^*\Fv^*\Go=\Fa^\star\Fv^\star\Go\Fv\Fa$.
Thus,~\eqref{eq:S_iso_represent1} is rewritten as
\begin{equation}\label{eq:S_iso_represent2}
	\mathbf{S} = 2\rho\left\{
	\!\left(\overline{\Psi}_2I_2+\overline{\Psi}_3I_3
	+\widetilde{\Psi}_2\Ie_2+\widetilde{\Psi}_3\Ie_3\right)\!\mathbf{C}^{-1}
	+\overline{\Psi}_1\mathbf{G}^{-1}
	+\widetilde{\Psi}_1\Gi^{-1}
	-\mathbf{C}^{-1}\!\left(\overline{\Psi}_2I_3\mathbf{G}
	+\widetilde{\Psi}_2\Ie_3\Gi\right)\!\mathbf{C}^{-1}
	\right\}
	\,.
\end{equation}
Recalling that the Cauchy stress is written in terms of the second Piola-Kirchhoff stress as {$\boldsymbol{\sigma}=J^{-1} \mathbf{F}\mathbf{S}\mathbf{F}^{\star}$,} and noting that
\begin{equation}
\begin{aligned}
\mathbf{F} \mathbf{C}^{-1} \mathbf{F}^\star
&=\mathbf{F} \mathbf{F}^{-1} \mathbf{g}^\sharp \mathbf{F}^{-\star} \mathbf{F}^\star
=\mathbf{g}^\sharp
\,,\quad
\mathbf{F} \mathbf{G}^{-1} \mathbf{F}^\star
=\mathbf{b}
\,,\\
\mathbf{F} \Gi^{-1} \mathbf{F}^\star
&=\mathbf{F} \Fa^{-1} \Fv^{-1} \Go^{-1} \Fv^{-\star} \Fa^{-\star} \mathbf{F}^\star
= \Fe \Go^{-1} \Fe^\star
= \be
\,,\\
\mathbf{F} \mathbf{C}^{-1}\mathbf{G}\mathbf{C}^{-1} \mathbf{F}^\star
&= \mathbf{F} (\mathbf{F}^{-1} \mathbf{g}^\sharp \mathbf{F}^{-\star}) \mathbf{G} (\mathbf{F}^{-1} \mathbf{g}^\sharp \mathbf{F}^{-\star}) \mathbf{F}^\star
= \mathbf{g}^\sharp \mathbf{F}^{-\star} \mathbf{G} \mathbf{F}^{-1} \mathbf{g}^\sharp
= \mathbf c^\sharp
\,,\\
\mathbf{F} \mathbf{C}^{-1}\Gi\mathbf{C}^{-1} \mathbf{F}^\star
&=\mathbf{F} (\mathbf{F}^{-1} \mathbf{g}^\sharp \mathbf{F}^{-\star}) (\Fa^\star \Fv^\star \Go \Fv \Fa) (\mathbf{F}^{-1} \mathbf{g}^\sharp \mathbf{F}^{-\star}) \mathbf{F}^\star
= \mathbf{g}^\sharp \Fe^{-\star} \Go \Fe^{-1} \mathbf{g}^\sharp 
= \ce^\sharp
\,,
\end{aligned}
\end{equation}
it follows from~\eqref{eq:S_iso_represent2} that
\begin{equation}\label{eq:sig_iso_represent}
\begin{aligned}
	\boldsymbol{\sigma} &= 2\varrho \left\{
	\!\left(\overline{\Psi}_2I_2+\overline{\Psi}_3I_3+\widetilde{\Psi}_2\Ie_2+\widetilde{\Psi}_3\Ie_3\right)\!\mathbf{g}^{\sharpo}
	+\overline{\Psi}_1\mathbf{b}
	+\widetilde{\Psi}_1\be
	-\overline{\Psi}_2I_3\,\mathbf{c}^\sharp
	-\widetilde{\Psi}_2\Ie_3\ce^\sharp
	\right\}
	\,.
\end{aligned}
\end{equation}
Assuming viscoelastic incompressibility, i.e.,~$J=\Jv=1$, the Cauchy stress has the following representation
\begin{equation}
\label{eq:sigma_iso}
	\boldsymbol{\sigma} = -p\,\mathbf{g}^\sharp 
	+2\rho\overline{\Psi}_1\mathbf{b}
	+2\rho\widetilde{\Psi}_1\be
	-2\rho\overline{\Psi}_2\,\mathbf{c}^\sharp
	-2\rho\widetilde{\Psi}_2\ce^\sharp
 \,.
\end{equation}

\paragraph{Dissipation potential.}
For an isotropic visco-anelastic solid, the dissipation potential must be invariant under the orthogonal group, i.e.,
\begin{equation}\label{eq:phi_sym}
	\phi(X,\vartheta,\Fve\mathring{\mathbf{K}},\mathring{\mathbf{K}}^*\Fv,
	\mathring{\mathbf{K}}^*\dot\Fv,\mathring{\mathbf{K}}^*\Fn,\mathring{\mathbf{K}}^*\dot\Fn,
	\Go,\mathbf{g})=
	\phi(X,\vartheta,\Fve,\Fv,\dot\Fv,\Fn,\dot\Fn,\Go,\mathbf{g})\,,
	\qquad \forall\,\,\mathring{\mathbf K}\in \mathrm{Orth}(\Go)\,,
\end{equation}
for all viscoelastic distortions $\Fve$, viscous distortions $\Fv$, and anelastic distortions $\Fn$.
Notice that, as previously observed for the general case in Remark~\ref{rmrk:Phi_dep}, even for an isotropic viscoelastic solid, the dependence of $\phi$ on {$\Fv$}, $\dot{\Fv}$, {$\Fn$}, and $\dot{\Fn}$ cannot generally be reduced to a dependence on the symmetric tensors $\Cv$, $\dot{\Cv}$, $\Cn$, and $\dot{\Cn}$.
As a real symmetric tensor, $\Cve^\sharpo$ may be decomposed as $\Cve^{\sharpo}=\lambda_1^2\,\mathbf{W}_1\otimes\mathbf{W}_1+\lambda_2^2\,\mathbf{W}_2\otimes\mathbf{W}_2+\lambda_3^2\,\mathbf{W}_3\otimes\mathbf{W}_3\,$ where $\lambda_i^2$  and $\mathbf{W}_i$ ($i=1, 2, 3$) are its eigenvalues and corresponding eigenvectors, respectively.
Following \cite{Shariff2023}, the dissipation potential may be written as a function of $I_1, I_2, I_3$, and the following $36$ spectral invariants:
\begin{equation}
\begin{aligned}
	V_{ij}&=\llangle\mathbf{W}_i,\Fv \mathbf{W}_j\rrangle_{\Go}\,,\qquad
	\widetilde{V}_{ij}&&=\llangle\mathbf{W}_i,\dot\Fv \mathbf{W}_j\rrangle_{\Go}\,,
	\qquad i,j=1,2,3\,,\\
	O_{ij}&=\llangle\mathbf{W}_i,\Fn \mathbf{W}_j\rrangle_{\Go}\,,\qquad
	\widetilde{O}_{ij}&&=\llangle\mathbf{W}_i,\dot\Fn \mathbf{W}_j\rrangle_{\Go}\,,
	\qquad i,j=1,2,3\,,
\end{aligned}
\end{equation}
i.e.,~$\hat{\phi}=\bar{\phi}\big(I_1,I_2,I_3,V_{11},V_{12},\cdots,V_{33},\widetilde{V}_{11},\widetilde{V}_{12},\cdots,\widetilde{V}_{33},O_{11},O_{12},\cdots,O_{33},\widetilde{O}_{11},\widetilde{O}_{12},\cdots,\widetilde{O}_{33}\big)$.\footnote{This functional form essentially means that, even in the isotropic case, while the dissipation potential depends only on the three principal invariants of the right Cauchy-Green deformation tensor instead of its $6$ components, there is no reduction in its dependence on the non-symmetric tensors $\Fv$, $\dot\Fv$, $\Fn$, and $\dot\Fn\,$; it still depends on all their $36$ components. Note, however, that when these components are written with respect to the eigenbasis $\{\mathbf \overline{\Psi}_1,\mathbf \overline{\Psi}_2,\mathbf W_3\}$, they are invariant under the orthogonal group.}

\begin{remark}
If one assumes that the dissipation potential has the form $\phi(X,\Fv,\dot\Fv,\Fn,\dot\Fn,\Go)$, then for an isotropic solid,  $\phi$ is an isotropic function of four non-symmetric material tensors $\Fv$, $\dot\Fv$, $\Fn$, and $\dot\Fn$. Let us consider the spectral representation of the real symmetric tensor $\Cv^{\sharpo}$: $\Cv^{\sharpo}=\mu_1\mathbf{U}_1\otimes\mathbf{U}_1+\mu_2\mathbf{U}_2\otimes\mathbf{U}_2+\mu_3\mathbf{U}_3\otimes\mathbf{U}_3$, where $\mu_i$ and $\mathbf{U}_i$ ($i=1, 2, 3$) are its eigenvalues and corresponding eigenvectors, respectively.
The dissipation potential may hence be written as a function of the following $36$ spectral invariants \citep{Shariff2023}:
\begin{equation}
\begin{aligned}
	V_{ij}&=\llangle\mathbf{U}_i,\Fv \mathbf{U}_j\rrangle_{\Go}\,,\qquad
	\widetilde{V}_{ij}&&=\llangle\mathbf{U}_i,\dot\Fv \mathbf{U}_j\rrangle_{\Go}\,,
	\qquad i,j=1,2,3\,,\\
	O_{ij}&=\llangle\mathbf{U}_i,\Fn \mathbf{U}_j\rrangle_{\Go}\,,\qquad
	\widetilde{O}_{ij}&&=\llangle\mathbf{U}_i,\dot\Fn \mathbf{U}_j\rrangle_{\Go}\,,
	\qquad i,j=1,2,3\,,
\end{aligned}
\end{equation}
i.e.,~$\phi=\tilde{\phi}\big(V_{11},V_{12},\cdots,V_{33},\widetilde{V}_{11},\widetilde{V}_{12},\cdots,\widetilde{V}_{33},O_{12},\cdots,O_{33},\widetilde{O}_{11},\widetilde{O}_{12},\cdots,\widetilde{O}_{33}\big)$.
\end{remark}

\paragraph{Kinetic equation.}
Using~\eqref{eq:iso_part_repr}, the kinetic equations~\eqref{eq:Kinetic_CeCve} for an isotropic solid read
\begin{subequations}
\begin{align}
	& \frac{\partial \phi}{\partial \dot\Fv} -2\rho \!\left[
	\widetilde\Psi_1\Ce\Go^\sharpo 
	+\!\left(\widetilde\Psi_2 \Ie_2 + \widetilde\Psi_3 \Ie_3 \right)\!\mathbf{I}
	-\widetilde\Psi_2 \Ie_3\Go\Ce^{-1}
	\right]\! \Fv^{-\star} = \mathbf{0}\,,\\
	&\begin{aligned}
	\frac{\partial \phi}{\partial \dot\Fn} &-2\rho \Big\{
	\overline\Psi_1\Cve\Go^\sharpo 
	+ \!\left(\overline\Psi_2 I_2 + \overline\Psi_3 I_3 \right)\!\mathbf{I}
	-\overline\Psi_2I_3\Go\Cve^{-1}
	\\[-5pt]&
	+\Fv^\star \!\left[
	\widetilde\Psi_1\Ce\Go^\sharpo
	+\!\left(\widetilde\Psi_2 \Ie_2 + \widetilde\Psi_3 \Ie_3 \right)\!\mathbf{I}
	-\widetilde\Psi_2\Ie_3\Go\Ce^{-1}
	\right]\! \Fv^{-\star} \Big\}\Fn^{-\star} = \mathbf{0}\,.
	\end{aligned}
\end{align}
\end{subequations}
Assuming visco-anelastic incompressibility, i.e.,~$J=\Jv=\Jn=1$, the kinetic equations above are rewritten as
\begin{subequations}
\begin{align}
	& \frac{\partial \phi}{\partial \dot\Fv} -2\rho \!\left[
	\widetilde\Psi_1\Ce\Go^\sharpo 
	+\widetilde\Psi_2 \!\left(\Ie_2 \mathbf{I} -\Go\Ce^{-1}\right)\!
	\right]\! \Fv^{-\star} = \pv \,\Fv^{-\star}\,,\\
	&\begin{aligned}
	\frac{\partial \phi}{\partial \dot\Fn} &-2\rho \Big\{
	\overline\Psi_1\Cve\Go^\sharpo 
	+ \overline\Psi_2 \!\left(I_2 \mathbf{I} - \Go\Cve^{-1}\right)\!
	\\[-5pt]&
	+\Fv^\star \!\left[
	\widetilde\Psi_1\Ce\Go^\sharpo
	+\widetilde\Psi_2 \!\left(\Ie_2 \mathbf{I} - \Go\Ce^{-1} \right)\!
	\right]\! \Fv^{-\star} \Big\}\Fn^{-\star} =  (\pn-p) \,\Fn^{-\star}\,.
	\end{aligned}
\end{align}
\end{subequations}

\subsubsection{Transversely isotropic solids}
\label{sec:trans}

A transversely isotropic solid is characterized by a material preferred direction normal to a plane of isotropy. At each material point $X\in\mathcal{B}$, this material preferred direction is specified by a unit vector $\mathring{\mathbf N}(X)$.

\paragraph{Constitutive equations.}
For a transversely isotropic solid, the set of structural tensors may be given by the singleton $\mathring{\boldsymbol\Lambda}=\{\mathring{\mathbf N}\otimes\mathring{\mathbf N}\}$. When $\mathring{\boldsymbol\Lambda}$ is added to their list of their arguments, the equilibrium and non-equilibrium free energies become isotropic functions \citep{Doyle1956, spencer1982formulation, lu2000covariant}. In this case, the integrity bases consist each of five arguments as follow
\begin{equation}
\label{Psi_const_trans-iso}
	\Psi_{\textrm{eq}} = \overline{\Psi}(X,\vartheta,I_1,I_2,I_3,I_4,I_5)   \,,\qquad
	\Psi_{\textrm{neq}}  = \widetilde{\Psi}(X,\vartheta,\Ie_1,\Ie_2,\Ie_3,\Ie_4,\Ie_5)  \,,
\end{equation}
where the additional invariants, in addition to the isotropic invariants~\eqref{eq:iso_inv}, are
\begin{equation}\label{Trans_inv}
\begin{aligned}
	&I_4=\mathring{\mathbf N}\cdot\Cve\cdot\mathring{\mathbf N}
	=\mathrm N^A \mathrm N^B\,\cCve_{AB}\,,&&
	I_5=\mathring{\mathbf N}\cdot\Cve^{2}\cdot\mathring{\mathbf N}
	=\mathrm N^A\,\cCve_{AM}\,\mathring{\mathrm G}^{MN}\,\cCve_{NB} \mathrm N^B\,,\\
	&\Ie_4=\mathring{\mathbf N}\cdot\Ce\cdot\mathring{\mathbf N}
	=\mathrm N^A \mathrm N^B\,\cCe_{AB}\,,&&
	\Ie_5=\mathring{\mathbf N}\cdot\Ce^{2}\cdot\mathring{\mathbf N}
	=\mathrm N^A\,\cCe_{AM}\,\mathring{\mathrm G}^{MN}\,\cCe_{NB} \mathrm N^B\,.
\end{aligned}
\end{equation}
Note that
\begin{equation}
	\frac{\partial I_4}{\partial\Cve}=\mathring{\mathbf N}\otimes\mathring{\mathbf N}\,,\qquad
	\frac{\partial I_5}{\partial\Cve}
	=\mathring{\mathbf N}\otimes(\Cve\mathring{\mathbf N})^\sharpo
	+(\Cve\mathring{\mathbf N})^\sharpo\otimes\mathring{\mathbf N}\,,
\end{equation}
and
\begin{equation}\label{eq:diff_Ie4-5}
	\frac{\partial \Ie_4}{\partial\Ce}=\mathring{\mathbf N}\otimes\mathring{\mathbf N}\,,\qquad
	\frac{\partial \Ie_5}{\partial\Ce}
	=\mathring{\mathbf N}\otimes(\Ce\mathring{\mathbf N})^\sharpo
	+(\Ce\mathring{\mathbf N})^\sharpo\otimes\mathring{\mathbf N}\,.
\end{equation}
Using the invariants~\eqref{Trans_inv}, one finds for a transversely isotropic solid that
\begin{subequations}\label{eq:trans_part_repr}
\begin{align}
	\begin{aligned}
	\frac{\partial \hat\Psi}{\partial \Ce}
	= \frac{\partial \hat\Psi_{\textrm{neq}}}{\partial \Ce}
	= \sum_{j=1}^{5} \widetilde\Psi_j\frac{\partial \Ie_j}{\partial \Ce}
	&= \widetilde\Psi_1\Go^{-1} 
	+\!\left(\widetilde\Psi_2 \Ie_2 + \widetilde\Psi_3 \Ie_3 \right)\!\Ce^{-1}
	-\widetilde\Psi_2\Ie_3\Ce^{-2}
	\\[-7.5pt]
	& \quad + \widetilde\Psi_4 \mathring{\mathbf N}\otimes\mathring{\mathbf N}
	+ \widetilde\Psi_5 \!\left[\mathring{\mathbf N}\otimes(\Ce\mathring{\mathbf N})^{\sharpo}
	+(\Ce\mathring{\mathbf N})^{\sharpo}\otimes\mathring{\mathbf N}\right]\!
	\,,
	\end{aligned}
	\\
	\begin{aligned}
	\frac{\partial \hat\Psi}{\partial \Cve}
	= \frac{\partial \hat\Psi_{\textrm{eq}}}{\partial \Cve}
	= \sum_{j=1}^{5} \overline\Psi_j\frac{\partial I_j}{\partial \Cve}
	&= \overline\Psi_1\Go^{-1}
	+ \!\left(\overline\Psi_2 I_2+ \overline\Psi_3 I_3\right)\!\Cve^{-1}
	- \overline\Psi_2 I_3\Cve^{-2}
	\\[-7.5pt]
	& \quad + \overline\Psi_4 \mathring{\mathbf N}\otimes\mathring{\mathbf N}
	+ \overline\Psi_5 \!\left[\mathring{\mathbf N}\otimes(\Cve\mathring{\mathbf N})^{\sharpo}
	+(\Cve\mathring{\mathbf N})^{\sharpo}\otimes\mathring{\mathbf N} \right]\!\,,
	\end{aligned}
\end{align}
\end{subequations}
where
\begin{equation}
	\overline{\Psi}_j=\overline{\Psi}_j(X,\vartheta,I_1,I_2,I_3,I_4,I_5)\coloneqq\frac{\partial \overline{\Psi}}{\partial I_j}\,,
	\qquad
	\widetilde{\Psi}_j=\widetilde{\Psi}_j(X,\vartheta,\Ie_1,\Ie_2,\Ie_3,\Ie_4,\Ie_5)
	\coloneqq\frac{\partial \widetilde{\Psi}}{\partial \Ie_j}\,, \qquad j=1,\cdots,5\,.
\end{equation}
Similarly to~\eqref{eq:S_iso_represent2}, it follows by~\eqref{eq:trans_part_repr} that
\begin{equation}\label{eq:S_trans_represent}
\begin{aligned}
	\mathbf{S} &= 2\rho\left\{
	\!\left(\overline{\Psi}_2I_2+\overline{\Psi}_3I_3
	+\widetilde{\Psi}_2\Ie_2+\widetilde{\Psi}_3\Ie_3\right)\!\mathbf{C}^{-1}
	+\overline{\Psi}_1\mathbf{G}^{-1}
	+\widetilde{\Psi}_1\Gi^{-1}
	-\mathbf{C}^{-1}\!\left(\overline{\Psi}_2I_3\mathbf{G}
	+\widetilde{\Psi}_2\Ie_3\Gi\right)\!\mathbf{C}^{-1}
	\right.
	\\&\quad
	\left.
	+ \overline\Psi_4 \!\left(\mathbf{F}^{-1}\Fve\mathring{\mathbf N}\right)\! \otimes \!\left(\mathbf{F}^{-1}\Fve\mathring{\mathbf N}\right)\!
	+ \widetilde{\Psi}_4 \!\left(\mathbf{F}^{-1}\Fe\mathring{\mathbf N}\right)\! \otimes \!\left(\mathbf{F}^{-1}\Fe\mathring{\mathbf N}\right)\!
	\right.
	\\&\quad
	\left.
	+ \overline\Psi_5 \!\left[\!\left(\mathbf{F}^{-1}\Fve\mathring{\mathbf N}\right)\!\otimes(\Fa^{-1}\Cve\mathring{\mathbf N})
	+(\Fa^{-1}\Cve\mathring{\mathbf N})\otimes\!\left(\mathbf{F}^{-1}\Fve\mathring{\mathbf N}\right)\!\right]\!
	\right.
	\\&\quad
	\left.
	+ \widetilde\Psi_5 \!\left[\!\left(\mathbf{F}^{-1}\Fe\mathring{\mathbf N}\right)\!\otimes\!\left(\mathbf{F}^{-1}\Fe^{\sharpo}\Ce\mathring{\mathbf N}\right)\!
	+\!\left(\mathbf{F}^{-1}\Fe^{\sharpo}\Ce\mathring{\mathbf N}\right)\!\otimes\!\left(\mathbf{F}^{-1}\Fe\mathring{\mathbf N}\right)\! \right]\!
	\right\}
	\,.
\end{aligned}
\end{equation}
Similarly to~\eqref{eq:sig_iso_represent}, it follows from~\eqref{eq:S_trans_represent} that the Cauchy stress has the following representation
\begin{equation}\label{eq:sig_trans_respresent}
\begin{aligned}
	\boldsymbol{\sigma} &= 2\varrho\left\{
	\!\left(\overline{\Psi}_2I_2+\overline{\Psi}_3I_3+\widetilde{\Psi}_2\Ie_2+\widetilde{\Psi}_3\Ie_3\right)\!\mathbf{g}^{\sharpo}
	+\overline{\Psi}_1\mathbf{b}
	+\widetilde{\Psi}_1\be
	-\overline{\Psi}_2I_3\,\mathbf{c}^\sharp
	-\widetilde{\Psi}_2\Ie_3\ce^\sharp
	\right.
	\\&\quad
	\left.
	+ \overline\Psi_4 \Nve \otimes \Nve
	+ \widetilde{\Psi}_4 \Ne \otimes \Ne
	+ \overline\Psi_5 \!\left[\Nve\otimes(\mathbf{b}\,\Nve^{\,\flat})
	+(\mathbf{b}\,\Nve^{\,\flat})\otimes\Nve\right]\!
	+ \widetilde\Psi_5 \!\left[\Ne\otimes(\be^{}\,\Ne^\flat)
	+(\be^{}\,\Ne^\flat)\otimes \Ne \right]\!
	\right\}
	\,,
\end{aligned}
\end{equation}
where $\Ne=\Fe\mathring{\mathbf N}\,$ and $\Nve=\Fve\mathring{\mathbf N}$. 
Note that $\Fe^{\sharpo}\,\Ce\mathring{\mathbf N}=\be^{}\,\Ne^\flat$ and $\Fve^{\sharpo}\,\Cve\,\mathring{\mathbf N}=\mathbf{b}\,\Nve^{\,\flat}$.
Assuming viscoelastic incompressibility, i.e.,~$J=\Jv=1$, the Cauchy stress is written as
\begin{equation}\label{eq:Trans_Cauchy}
\begin{aligned}
	\boldsymbol{\sigma} &=
	-p\,\mathbf{g}^{\sharpo}
	+2\rho\overline{\Psi}_1\mathbf{b}
	+2\rho\widetilde{\Psi}_1\be
	-2\rho\overline{\Psi}_2\mathbf{c}^\sharp
	-2\rho\widetilde{\Psi}_2\ce^\sharp
	+ 2\rho\overline\Psi_4 \Nve \otimes \Nve
	+ 2\rho\widetilde{\Psi}_4 \Ne \otimes \Ne
	\\&\quad
	+ 2\rho\overline\Psi_5 \!\left[\Nve\otimes(\mathbf{b}\,\Nve^{\,\flat})
	+(\mathbf{b}\,\Nve^{\,\flat})\otimes\Nve\right]\!
	+ 2\rho\widetilde\Psi_5 \!\left[\Ne\otimes(\be^{}\,\Ne^\flat)
	+(\be^{}\,\Ne^\flat)\otimes \Ne \right]\!
	\,.
\end{aligned}
\end{equation}

\paragraph{Dissipation potential.}
For a transversely isotropic visco-anelastic solid, including the structural tensor $ \mathring{\boldsymbol\Lambda} = \{\mathring{\mathbf N}\otimes\mathring{\mathbf N}\} $ among the arguments of the dissipation potential $ \phi $ allows it to be expressed as an isotropic function of its arguments, $ \phi = \phi(X, \Fve, \Fv, \dot{\Fv}, \Fn, \dot{\Fn}, \mathring{\boldsymbol\Lambda}, \Go, \mathbf{g}) $. Although the standard representation theorem used for free energy functions does not directly apply, the dissipation potential can still be formulated in terms of the standard invariants and a set of spectral invariants, similar to the approach for isotropic visco-anelastic solids.

\paragraph{Kinetic equations.}
Using~\eqref{eq:trans_part_repr}, the kinetic equations~\eqref{eq:Kinetic_CeCve} for a transversely isotropic solid read
\begin{subequations}
\begin{align}
&\begin{aligned}\label{eq:Trans_v_Kinetic}
	\frac{\partial \phi}{\partial \dot\Fv} &-2\rho \Big\{
	\widetilde\Psi_1\Ce\Go^\sharpo 
	+\!\left(\widetilde\Psi_2 \Ie_2 + \widetilde\Psi_3 \Ie_3 \right)\!\mathbf{I}
	-\widetilde\Psi_2 \Ie_3\Go\Ce^{-1}
	\\[-5pt]&+\widetilde\Psi_4 (\Ce \mathring{\mathbf N})\otimes\mathring{\mathbf N}
	+ \widetilde\Psi_5 \!\left[(\Ce \mathring{\mathbf N}) \otimes (\Ce \mathring{\mathbf N})^\sharpo
	+(\Ce^2\mathring{\mathbf N}) \otimes\mathring{\mathbf N}\right]\!
	\Big\} \Fv^{-\star} = \mathbf{0}\,,
\end{aligned}
\\
&\begin{aligned}\label{eq:Trans_n_Kinetic}
	\frac{\partial \phi}{\partial \dot\Fn} & -2\rho \Big\{
	\overline\Psi_1\Cve\Go^\sharpo 
	+ \!\left(\overline\Psi_2 I_2 + \overline\Psi_3 I_3 \right)\!\mathbf{I}
	-\overline\Psi_2I_3\Go\Cve^{-1}
	\\[-5pt]
	& + \overline\Psi_4 (\Cve \mathring{\mathbf N})\otimes\mathring{\mathbf N}
	+\overline\Psi_5 \!\left[(\Cve\mathring{\mathbf N}) \otimes (\Cve\mathring{\mathbf N})^\sharpo
	+(\Cve^2\mathring{\mathbf N}) \otimes \mathring{\mathbf N} \right]\! \\
	& + \Fv^\star \Big[
	\widetilde\Psi_1\Go\Ce^\sharpo
	+\!\left(\widetilde\Psi_2 \Ie_2 + \widetilde\Psi_3 \Ie_3 \right)\!\mathbf{I}
	- \widetilde\Psi_2\Ie_3\Go\Ce^{-\sharpo} \\
	& + \widetilde\Psi_4 (\Ce \mathring{\mathbf N}) \otimes\mathring{\mathbf N}
	+ \widetilde\Psi_5 \!\left((\Ce\mathring{\mathbf N})\otimes (\Ce\mathring{\mathbf N})^\sharpo
	+ (\Ce^2\mathring{\mathbf N}) \otimes \mathring{\mathbf N}\right)\!
	\Big] \Fv^{-\star} \Big\} \Fn^{-\star} = \mathbf{0}\,.
\end{aligned}
\end{align}
\end{subequations}
Assuming visco-anelastic incompressibility, i.e.,~$J=\Jv=\Jn=1$, the kinetic equations above are rewritten as
\begin{subequations}
\begin{flalign}
&\begin{aligned}\label{eq:Trans_v_Kinetic_inc}
	\frac{\partial \phi}{\partial \dot\Fv} &-2\rho \Big\{
	\widetilde\Psi_1\Ce\Go^\sharpo 
	+\widetilde\Psi_2 \!\left(\Ie_2 \mathbf{I} -\Go\Ce^{-1}\right)\!
	+\widetilde\Psi_4 (\Ce \mathring{\mathbf N})\otimes\mathring{\mathbf N}
	+ \widetilde\Psi_5 \!\left[(\Ce \mathring{\mathbf N}) \otimes (\Ce \mathring{\mathbf N})^\sharpo
	+(\Ce^2\mathring{\mathbf N}) \otimes\mathring{\mathbf N}\right]\!
	\Big\} \Fv^{-\star}
	\\[-5pt]&
	= \pv \,\Fv^{-\star}\,,
\end{aligned}
\\
&\begin{aligned}\label{eq:Trans_n_Kinetic_inc}
	\frac{\partial \phi}{\partial \dot\Fn} &-2\rho \Big\{
	\overline\Psi_1\Cve\Go^\sharpo
	+ \overline\Psi_2 \!\left(I_2 \mathbf{I} - \Go\Cve^{-1}\right)\!
	+ \overline\Psi_4 (\Cve \mathring{\mathbf N})\otimes\mathring{\mathbf N}
	+\overline\Psi_5 \!\left[(\Cve\mathring{\mathbf N}) \otimes (\Cve\mathring{\mathbf N})^\sharpo
	+(\Cve^2\mathring{\mathbf N}) \otimes \mathring{\mathbf N} \right]\!
	\\[-5pt]&
	+ \Fv^\star \Big[
	\widetilde\Psi_1\Go\Ce^\sharpo
	+\widetilde\Psi_2 \!\left(\Ie_2 \mathbf{I} - \Go\Ce^{-1} \right)\!
	+ \widetilde\Psi_4 (\Ce \mathring{\mathbf N}) \otimes\mathring{\mathbf N}
	+ \widetilde\Psi_5 \!\left[(\Ce\mathring{\mathbf N})\otimes (\Ce\mathring{\mathbf N})^\sharpo
	+ (\Ce^2\mathring{\mathbf N}) \otimes \mathring{\mathbf N}\right]\!
	\Big] \Fv^{-\star} \Big\} \Fn^{-\star}
	\\[-2.5pt]&
	= (\pn-p) \,\Fn^{-\star}\,.
\end{aligned}
\end{flalign}
\end{subequations}

\subsubsection{Orthotropic solids}

An orthotropic solid at a material point $X\in\mathcal{B}$ has reflection symmetry with respect to three mutually perpendicular planes with $\Go$-orthonormal vectors $\mathring{\mathbf N}_1(X)$, $\mathring{\mathbf N}_2(X)$, and $\mathring{\mathbf N}_3(X)$, i.e.,~$\llangle\mathring{\mathbf N}_i(X),\mathring{\mathbf N}_j(X)\rrangle_{\Go}=\delta_{ij}$.
A choice for structural tensors is the set
\begin{equation}
	\mathring{\boldsymbol\Lambda}
	=\{\mathring{\mathbf A}_1, \mathring{\mathbf A}_2, \mathring{\mathbf A}_3\}
	=\{\mathring{\mathbf N}_1\otimes\mathring{\mathbf N}_1, 
	\mathring{\mathbf N}_2\otimes\mathring{\mathbf N}_2, 
	\mathring{\mathbf N}_3\otimes\mathring{\mathbf N}_3\}
	\,.
\end{equation}
However, $\mathring{\mathbf A}_1+\mathring{\mathbf A}_2+\mathring{\mathbf A}_3=\mathbf{I}$, and hence only two of them are independent.

\paragraph{Constitutive equations.}
One can take $\mathring{\mathbf A}_1$ and $\mathring{\mathbf A}_2$ to be the independent structural tensors of the set $\mathring{\boldsymbol\Lambda}$. When these two tensors are added to the list of the arguments of the equilibrium and non-equilibrium free energies, they become isotropic functions of their arguments.\footnote{The functions $\Psi_{\textrm{eq}}(X,\vartheta,\Fve, \mathring{\mathbf A}_1, \mathring{\mathbf A}_2 , \Go)$, $\Psi_{\textrm{neq}}(X,\vartheta,\Fe, \mathring{\mathbf A}_1, \mathring{\mathbf A}_2, \Go)$, and $\Psi(X,\vartheta,\Fve, \Fe, \mathring{\mathbf A}_1, \mathring{\mathbf A}_2, \Go, \mathbf g)$ are isotropic.} This is equivalent to the free energy functions each depending on seven invariants:
\begin{equation}
	\Psi_{\textrm{eq}} = \overline{\Psi}(X,\vartheta,I_1,I_2,I_3,I_4,I_5,I_6,I_7)   \,,\qquad
	\Psi_{\textrm{neq}}  = \widetilde{\Psi}(X,\vartheta,\Ie_1,\Ie_2,\Ie_3,\Ie_4,\Ie_5,\Ie_6,\Ie_7)  \,,
\end{equation}
where the additional invariants, in addition to the isotropic invariants~\eqref{eq:iso_inv}, are given by
\begin{subequations} \label{Orthotropic-Invariants}
\begin{align}
	I_4=\mathring{\mathbf N}_1\cdot\Cve\cdot\mathring{\mathbf N}_1\,,
	\quad I_5=\mathring{\mathbf N}_1\cdot\Cve^2\cdot\mathring{\mathbf N}_1\,,
	\quad I_6=\mathring{\mathbf N}_2\cdot\Cve\cdot\mathring{\mathbf N}_2\,,
	\quad I_7=\mathring{\mathbf N}_2\cdot\Cve^2\cdot\mathring{\mathbf N}_2\,,
	\\
	\Ie_4=\mathring{\mathbf N}_1\cdot\Ce\cdot\mathring{\mathbf N}_1\,,
	\quad \Ie_5=\mathring{\mathbf N}_1\cdot\Ce^2\cdot\mathring{\mathbf N}_1\,,
	\quad \Ie_6=\mathring{\mathbf N}_2\cdot\Ce\cdot\mathring{\mathbf N}_2\,,
	\quad \Ie_7=\mathring{\mathbf N}_2\cdot\Ce^2\cdot\mathring{\mathbf N}_2\,.
\end{align}
\end{subequations}
Similarly to~\eqref{eq:sig_trans_respresent}, the Cauchy stress has the following representation
\begin{equation}
\begin{aligned}
	\boldsymbol{\sigma} &= 2\varrho\left\{
	\!\left(\overline{\Psi}_2I_2+\overline{\Psi}_3I_3+\widetilde{\Psi}_2\Ie_2+\widetilde{\Psi}_3\Ie_3\right)\!\mathbf{g}^{\sharpo}
	+\overline{\Psi}_1\mathbf{b}
	+\widetilde{\Psi}_1\be
	-\overline{\Psi}_2I_3\,\mathbf{c}^\sharp
	-\widetilde{\Psi}_2\Ie_3\ce^\sharp
	\right.
	\\&\quad
	\left.
	+ \overline\Psi_4 \Nve_1 \otimes \Nve_1
	+ \widetilde{\Psi}_4 \Ne_1 \otimes \Ne_1
	+ \overline\Psi_5 \!\left[\Nve_1\otimes(\mathbf{b}\,\Nve_1^{\,\flato})
	+(\mathbf{b}\,\Nve_1^{\,\flato})\otimes\Nve_1\right]\!
	+ \widetilde\Psi_5 \!\left[\Ne_1\otimes(\be^{}\,\Ne_1^\flato)
	+(\be^{}\,\Ne_1^\flato)\otimes \Ne_1 \right]\!
	\right.
	\\&\quad
	\left.
	+ \overline\Psi_6 \Nve_2 \otimes \Nve_2
	+ \widetilde{\Psi}_6 \Ne_2 \otimes \Ne_2
	+ \overline\Psi_7 \!\left[\Nve_2\otimes(\mathbf{b}\,\Nve_2^{\,\flato})
	+(\mathbf{b}\,\Nve_2^{\,\flato})\otimes\Nve_2\right]\!
	+ \widetilde\Psi_7 \!\left[\Ne_2\otimes(\be^{}\,\Ne_2^\flato)
	+(\be^{}\,\Ne_2^\flato)\otimes \Ne_2 \right]\!
	\right\}
	\,,
\end{aligned}
\end{equation}
where
$\Nve_i=\Fve\mathring{\mathbf N}_i$ and
$\Ne_,=\Fe\mathring{\mathbf N}_i$ for $i=1,2$, and
\begin{equation}
	\overline{\Psi}_j=\overline{\Psi}_j(X,\vartheta,I_1,\cdots,I_7)\coloneqq\frac{\partial \overline{\Psi}}{\partial I_j}\,,\qquad
	\widetilde{\Psi}_j=\widetilde{\Psi}_j(X,\vartheta,\Ie_1,\cdots,\Ie_7)
	\coloneqq\frac{\partial \widetilde{\Psi}}{\partial \Ie_j}\,, \qquad j=1,\cdots,7\,.
\end{equation}
Assuming viscoelastic incompressibility, i.e.,~$J=\Jv=1$, the Cauchy stress tensor is written as
\begin{equation}
\begin{aligned}
	\boldsymbol{\sigma} &=
	-p \mathbf{g}^{\sharpo}
	+2\rho\overline{\Psi}_1\mathbf{b}
	+2\rho\widetilde{\Psi}_1\be
	-2\rho\overline{\Psi}_2I_3\,\mathbf{c}^\sharp
	-2\rho\widetilde{\Psi}_2\Ie_3\ce^\sharp
	+ 2\rho\overline\Psi_4 \Nve_1 \otimes \Nve_1
	+ 2\rho\widetilde{\Psi}_4 \Ne_1 \otimes \Ne_1
	\\&\quad
	+ 2\rho\overline\Psi_5 \!\left[\Nve_1\otimes(\mathbf{b}\,\Nve_1^{\,\flato})
	+(\mathbf{b}\,\Nve_1^{\,\flato})\otimes\Nve_1\right]\!
	+ 2\rho\widetilde\Psi_5 \!\left[\Ne_1\otimes(\be^{}\,\Ne_1^\flato)
	+(\be^{}\,\Ne_1^\flato)\otimes \Ne_1 \right]\!
	\\&\quad
	+ 2\rho\overline\Psi_6 \Nve_2 \otimes \Nve_2
	+ 2\rho\widetilde{\Psi}_6 \Ne_2 \otimes \Ne_2
	\\&\quad
	+ 2\rho\overline\Psi_7 \!\left[\Nve_2\otimes(\mathbf{b}\,\Nve_2^{\,\flato})
	+(\mathbf{b}\,\Nve_2^{\,\flato})\otimes\Nve_2\right]\!
	+ 2\rho\widetilde\Psi_7 \!\left[\Ne_2\otimes(\be^{}\,\Ne_2^\flato)
	+(\be^{}\,\Ne_2^\flato)\otimes \Ne_2 \right]\!
	\,.
\end{aligned}
\end{equation}

\paragraph{Dissipation potential.}
For an orthotropic visco-anelastic solid, when two elements of the set of structural tensors $\mathring{\boldsymbol\Lambda}$ are added to the list of the arguments of the dissipation potential $\phi$, it becomes an isotropic function of its arguments, e.g., $\phi=\phi(X,\vartheta,\Fve,\Fv,\dot\Fv,\Fn,\dot\Fn,\mathring{\mathbf A}_1, \mathring{\mathbf A}_2, \Go,\mathbf{g})$. Although one may not use the standard representation theorem as for the free energy functions, the dissipation potential will be a function of some standard invariants and a set of spectral invariants, similarly to the dissipation potential of the isotropic visco-anelastic solids.

\paragraph{Kinetic equation.}
Similarly to~\eqref{eq:trans_part_repr}, one may write
\begin{subequations}
\begin{align}
	\begin{aligned}
	\frac{\partial \hat\Psi_{\textrm{neq}}}{\partial \Ce}
	= \sum_{j=1}^{7} \widetilde\Psi_j\frac{\partial \Ie_j}{\partial \Ce}
	&= \widetilde\Psi_1\Go^{-1} 
	+\!\left(\widetilde\Psi_2 \Ie_2 + \widetilde\Psi_3 \Ie_3 \right)\!\Ce^{-1}
	-\widetilde\Psi_2\Ie_3\Ce^{-2}
	\\[-10pt]&\quad
	+ \widetilde\Psi_4 \mathring{\mathbf N}_1\otimes\mathring{\mathbf N}_1
	+ \widetilde\Psi_5 \!\left[\mathring{\mathbf N}_1\otimes(\Ce\mathring{\mathbf N}_1)^{\sharpo}
	+(\Ce\mathring{\mathbf N}_1)^{\sharpo}\otimes\mathring{\mathbf N}_1\right]\!
	\\&\quad
	+ \widetilde\Psi_6 \mathring{\mathbf N}_2\otimes\mathring{\mathbf N}_2
	+ \widetilde\Psi_7 \!\left[\mathring{\mathbf N}_2\otimes(\Ce\mathring{\mathbf N}_2)^{\sharpo}
	+(\Ce\mathring{\mathbf N}_2)^{\sharpo}\otimes\mathring{\mathbf N}_2\right]\!
	\,.
	\end{aligned}
	\\
	\begin{aligned}
	\frac{\partial \hat\Psi_{\textrm{eq}}}{\partial \Cve}
	= \sum_{j=1}^{7} \overline\Psi_j\frac{\partial I_j}{\partial \Cve}
	&= \overline\Psi_1\Go^{-1} 
	+\!\left(\overline\Psi_2 I_2 + \overline\Psi_3 I_3 \right)\!\Cve^{-1}
	-\overline\Psi_2I_3\Cve^{-2}
	\\[-10pt]&\quad
	+ \overline\Psi_4 \mathring{\mathbf N}_1\otimes\mathring{\mathbf N}_1
	+ \overline\Psi_5 \!\left[\mathring{\mathbf N}_1\otimes(\Cve\mathring{\mathbf N}_1)^{\sharpo}
	+(\Cve\mathring{\mathbf N}_1)^{\sharpo}\otimes\mathring{\mathbf N}_1\right]\!
	\\&\quad
	+ \overline\Psi_6 \mathring{\mathbf N}_2\otimes\mathring{\mathbf N}_2
	+ \overline\Psi_7 \!\left[\mathring{\mathbf N}_2\otimes(\Cve\mathring{\mathbf N}_2)^{\sharpo}
	+(\Cve\mathring{\mathbf N}_2)^{\sharpo}\otimes\mathring{\mathbf N}_2\right]\!
	\,.
	\end{aligned}
\end{align}
\end{subequations}
Hence, it follows from~\eqref{eq:Kinetic_CeCve} that the kinetic equations for orthotropic visco-anelastic solids read
\begin{subequations}
\begin{align}
&\begin{aligned}
	\frac{\partial \phi}{\partial \dot\Fv} &-2\rho \Big\{
	\widetilde\Psi_1\Ce\Go^\sharpo 
	+\!\left(\widetilde\Psi_2 \Ie_2 + \widetilde\Psi_3 \Ie_3 \right)\!\mathbf{I}
	-\widetilde\Psi_2 \Ie_3\Go\Ce^{-1}
	\\[-5pt]&+\widetilde\Psi_4 (\Ce \mathring{\mathbf N}_1)\otimes\mathring{\mathbf N}_1
	+ \widetilde\Psi_5 \!\left[(\Ce \mathring{\mathbf N}_1) \otimes (\Ce \mathring{\mathbf N}_1)^\sharpo
	+(\Ce^2\mathring{\mathbf N}_1) \otimes\mathring{\mathbf N}_1\right]\!
	\\&+\widetilde\Psi_6 (\Ce \mathring{\mathbf N}_2)\otimes\mathring{\mathbf N}_2
	+ \widetilde\Psi_7 \!\left[(\Ce \mathring{\mathbf N}_2) \otimes (\Ce \mathring{\mathbf N}_2)^\sharpo
	+(\Ce^2\mathring{\mathbf N}_2) \otimes\mathring{\mathbf N}_2\right]\!
	\Big\} \Fv^{-\star} = \mathbf{0}\,,
\end{aligned}
\\
&\begin{aligned}
	\frac{\partial \phi}{\partial \dot\Fn} & -2\rho \Big\{
	\overline\Psi_1\Cve\Go^\sharpo 
	+ \!\left(\overline\Psi_2 I_2 + \overline\Psi_3 I_3 \right)\!\mathbf{I}
	-\overline\Psi_2I_3\Go\Cve^{-1}
	\\[-5pt]
	& + \overline\Psi_4 (\Cve \mathring{\mathbf N}_1)\otimes\mathring{\mathbf N}_1
	+\overline\Psi_5 \!\left[(\Cve\mathring{\mathbf N}_1) \otimes (\Cve\mathring{\mathbf N}_1)^\sharpo
	+(\Cve^2\mathring{\mathbf N}_1) \otimes \mathring{\mathbf N}_1 \right]\! \\
	& + \overline\Psi_6 (\Cve \mathring{\mathbf N}_2)\otimes\mathring{\mathbf N}_2
	+\overline\Psi_7 \!\left[(\Cve\mathring{\mathbf N}_2) \otimes (\Cve\mathring{\mathbf N}_2)^\sharpo
	+(\Cve^2\mathring{\mathbf N}_2) \otimes \mathring{\mathbf N}_2 \right]\! \\
	& + \Fv^\star \Big[
	\widetilde\Psi_1\Go\Ce^\sharpo
	+\!\left(\widetilde\Psi_2 \Ie_2 + \widetilde\Psi_3 \Ie_3 \right)\!\mathbf{I}
	- \widetilde\Psi_2\Ie_3\Go\Ce^{-\sharpo} \\
	& + \widetilde\Psi_4 (\Ce \mathring{\mathbf N}_1) \otimes\mathring{\mathbf N}_1
	+ \widetilde\Psi_5 \!\left((\Ce\mathring{\mathbf N}_1)\otimes (\Ce\mathring{\mathbf N}_1)^\sharpo
	+ (\Ce^2\mathring{\mathbf N}_1) \otimes \mathring{\mathbf N}_1\right)\! \\
	& + \widetilde\Psi_6 (\Ce \mathring{\mathbf N}_2) \otimes\mathring{\mathbf N}_2
	+ \widetilde\Psi_7 \!\left((\Ce\mathring{\mathbf N}_2)\otimes (\Ce\mathring{\mathbf N}_2)^\sharpo
	+ (\Ce^2\mathring{\mathbf N}_2) \otimes \mathring{\mathbf N}_2\right)\!
	\Big] \Fv^{-\star} \Big\} \Fn^{-\star} = \mathbf{0}\,.
\end{aligned}
\end{align}
\end{subequations}
Assuming visco-anelastic incompressibility, i.e.,~$J=\Jv=\Jn=1$, the kinetic equations above are rewritten as
\begin{subequations}
\begin{align}
&\begin{aligned}
	\frac{\partial \phi}{\partial \dot\Fv} &-2\rho \Big\{
	\widetilde\Psi_1\Ce\Go^\sharpo 
	+\widetilde\Psi_2 \!\left(\Ie_2 \mathbf{I} -\Go\Ce^{-1}\right)\!
	+\widetilde\Psi_4 (\Ce \mathring{\mathbf N}_1)\otimes\mathring{\mathbf N}_1
	+ \widetilde\Psi_5 \!\left[(\Ce \mathring{\mathbf N}_1) \otimes (\Ce \mathring{\mathbf N}_1)^\sharpo
	+(\Ce^2\mathring{\mathbf N}_1) \otimes\mathring{\mathbf N}_1\right]\!
	\\[-5pt]&
	+\widetilde\Psi_6 (\Ce \mathring{\mathbf N}_2)\otimes\mathring{\mathbf N}_2
	+ \widetilde\Psi_7 \!\left[(\Ce \mathring{\mathbf N}_2) \otimes (\Ce \mathring{\mathbf N}_2)^\sharpo
	+(\Ce^2\mathring{\mathbf N}_2) \otimes\mathring{\mathbf N}_2\right]\!
	\Big\} \Fv^{-\star}
	= \pv \,\Fv^{-\star}\,,
\end{aligned}
\\
&\begin{aligned}
	\frac{\partial \phi}{\partial \dot\Fn} &-2\rho \Big\{
	\overline\Psi_1\Cve\Go^\sharpo
	+ \overline\Psi_2 \!\left(I_2 \mathbf{I} - \Go\Cve^{-1}\right)\!
	+\overline\Psi_4 (\Cve \mathring{\mathbf N}_1)\otimes\mathring{\mathbf N}_1
	+\overline\Psi_5 \!\left[(\Cve\mathring{\mathbf N}_1) \otimes (\Cve\mathring{\mathbf N}_1)^\sharpo
	+(\Cve^2\mathring{\mathbf N}_1) \otimes \mathring{\mathbf N}_1 \right]\!
	\\[-5pt]&
	+\overline\Psi_6 (\Cve \mathring{\mathbf N}_2)\otimes\mathring{\mathbf N}_2
	+\overline\Psi_7 \!\left[(\Cve\mathring{\mathbf N}_2) \otimes (\Cve\mathring{\mathbf N}_2)^\sharpo
	+(\Cve^2\mathring{\mathbf N}_2) \otimes \mathring{\mathbf N}_2 \right]\!
	\\&
	+ \Fv^\star \Big[
	\widetilde\Psi_1\Go\Ce^\sharpo
	+\widetilde\Psi_2 \!\left(\Ie_2 \mathbf{I} -\Go\Ce^{-1}\right)\!
	+ \widetilde\Psi_4 (\Ce \mathring{\mathbf N}_1) \otimes\mathring{\mathbf N}_1
	+ \widetilde\Psi_5 \!\left[(\Ce\mathring{\mathbf N}_1)\otimes (\Ce\mathring{\mathbf N}_1)^\sharpo
	+ (\Ce^2\mathring{\mathbf N}_1) \otimes \mathring{\mathbf N}_1\right]\!
	\\&
	+ \widetilde\Psi_6 (\Ce \mathring{\mathbf N}_2) \otimes\mathring{\mathbf N}_2
	+ \widetilde\Psi_7 \!\left[(\Ce\mathring{\mathbf N}_2)\otimes (\Ce\mathring{\mathbf N}_2)^\sharpo
	+ (\Ce^2\mathring{\mathbf N}_2) \otimes \mathring{\mathbf N}_2\right]\!
	\Big] \Fv^{-\star} \Big\} \Fn^{-\star}
	= (\pn-p) \,\Fn^{-\star}\,.
\end{aligned}
\end{align}
\end{subequations}

\subsubsection{Monoclinic solids}

A monoclinic solid at a material point $X\in\mathcal{B}$ has three material preferred directions $\{\mathring{\mathbf N}_1(X),\mathring{\mathbf N}_2(X),\mathring{\mathbf N}_3(X)\}$ such that $\llangle\mathring{\mathbf N}_1, \mathring{\mathbf N}_2\rrangle_{\Go}\neq 0$ and $\mathring{\mathbf N}_3$ is normal to the plane of $\mathring{\mathbf N}_1$ and $\mathring{\mathbf N}_2$, i.e.,~${\llangle\mathring{\mathbf N}_1, \mathring{\mathbf N}_3\rrangle_{\Go}=\llangle\mathring{\mathbf N}_2, \mathring{\mathbf N}_3\rrangle_{\Go}=0}$~\citep{merodio2020finite}.

\paragraph{Constitutive equations.}
The equilibrium and non-equilibrium free energies of a monoclinic solid depend on nine invariants \citep{Spencer1986}:
\begin{equation}
	\Psi_{\textrm{eq}} = \overline{\Psi}(X,\vartheta,I_1,I_2,I_3,I_4,I_5,I_6,I_7,I_8,I_9)   \,,\qquad
	\Psi_{\textrm{neq}}  
	= \widetilde{\Psi}(X,\vartheta,\Ie_1,\Ie_2,\Ie_3,\Ie_4,\Ie_5,\Ie_6,\Ie_7,\Ie_8,\Ie_9)  \,.
\end{equation}
The first seven invariants for each free energy above are identical to those of orthotropic solids ~\eqref{eq:iso_inv} and~\eqref{Orthotropic-Invariants}. The three extra invariants are
\begin{equation}
	I_8=\mathcal{I}\,\mathring{\mathbf N}_1 \Cve \,\mathring{\mathbf N}_2\,,\qquad
	\Ie_8=\mathcal{I}\,\mathring{\mathbf N}_1 \Ce \,\mathring{\mathbf N}_2\,,\qquad
	I_9=\Ie_9=\mathcal{I}^2\,,
\end{equation}  
where $\mathcal{I}=\llangle\mathring{\mathbf N}_1, \mathring{\mathbf N}_2\rrangle_{\Go}$. Note that
\begin{equation}\label{eq:mono_inv}
	\frac{\partial I_8}{\partial\Cve}
	=\frac{\partial \Ie_8}{\partial\Ce}
	=\frac{1}{2}\mathcal{I}\,(\mathring{\mathbf N}_1\otimes\mathring{\mathbf N}_2+\mathring{\mathbf N}_2\otimes\mathring{\mathbf N}_1)\,,
	\qquad
	\frac{\partial I_9}{\partial\Cve}=\frac{\partial \Ie_9}{\partial\Ce}=\mathbf{0}\,.
\end{equation}
The Cauchy stress has the following representation
\begin{equation}
\begin{aligned}
	\boldsymbol{\sigma} &= 2\varrho\left\{
	\!\left(\overline{\Psi}_2I_2+\overline{\Psi}_3I_3+\widetilde{\Psi}_2\Ie_2+\widetilde{\Psi}_3\Ie_3\right)\!\mathbf{g}^{\sharpo}
	+\overline{\Psi}_1\mathbf{b}
	+\widetilde{\Psi}_1\be
	-\overline{\Psi}_2I_3\,\mathbf{c}^\sharp
	-\widetilde{\Psi}_2\Ie_3\ce^\sharp
	\right.
	\\&\quad
	\left.
	+ \overline\Psi_4 \Nve_1 \otimes \Nve_1
	+ \widetilde{\Psi}_4 \Ne_1 \otimes \Ne_1
	+ \overline\Psi_5 \!\left[\Nve_1\otimes(\mathbf{b}\,\Nve_1^{\,\flato})
	+(\mathbf{b}\,\Nve_1^{\,\flato})\otimes\Nve_1\right]\!
	+ \widetilde\Psi_5 \!\left[\Ne_1\otimes(\be^{}\,\Ne_1^\flato)
	+(\be^{}\,\Ne_1^\flato)\otimes \Ne_1 \right]\!
	\right.
	\\&\quad
	\left.
	+ \overline\Psi_6 \Nve_2 \otimes \Nve_2
	+ \widetilde{\Psi}_6 \Ne_2 \otimes \Ne_2
	+ \overline\Psi_7 \!\left[\Nve_2\otimes(\mathbf{b}\,\Nve_2^{\,\flato})
	+(\mathbf{b}\,\Nve_2^{\,\flato})\otimes\Nve_2\right]\!
	+ \widetilde\Psi_7 \!\left[\Ne_2\otimes(\be^{}\,\Ne_2^\flato)
	+(\be^{}\,\Ne_2^\flato)\otimes \Ne_2 \right]\!
	\right.
	\\&\quad
	\left.
	+ \overline\Psi_8\,\mathcal{I}\,(\Nve_1\otimes\Nve_2+\Nve_2\otimes\Nve_1)/2
	+ \widetilde\Psi_8\,\mathcal{I}\,(\Ne_1\otimes\Ne_2+\Ne_2\otimes\Ne_1)/2
	\right\}
	\,,
\end{aligned}
\end{equation}
where
$\Nve_i=\Fve\mathring{\mathbf N}_i$ and
$\Ne_,=\Fe\mathring{\mathbf N}_i$ for $i=1,3$, and
\begin{equation}
	\overline{\Psi}_j=\overline{\Psi}_j(X,\vartheta,I_1,\cdots,I_9)\coloneqq\frac{\partial \overline{\Psi}}{\partial I_j}\,,\qquad
	\widetilde{\Psi}_j=\widetilde{\Psi}_j(X,\vartheta,\Ie_1,\cdots,\Ie_9)
	\coloneqq\frac{\partial \widetilde{\Psi}}{\partial \Ie_j}\,, \qquad j=1,\cdots,8\,.
\end{equation}
Assuming viscoelastic incompressibility, i.e.,~$J=\Jv=1$, the Cauchy stress tensor is rewritten as
\begin{equation}
\begin{aligned}
	\boldsymbol{\sigma} &=
	-p \mathbf{g}^{\sharpo}
	+2\rho\overline{\Psi}_1\mathbf{b}
	+2\rho\widetilde{\Psi}_1\be
	-2\rho\overline{\Psi}_2I_3\,\mathbf{c}^\sharp
	-2\rho\widetilde{\Psi}_2\Ie_3\ce^\sharp
	\\&\quad
	+ 2\rho\overline\Psi_4 \Nve_1 \otimes \Nve_1
	+ 2\rho\widetilde{\Psi}_4 \Ne_1 \otimes \Ne_1
	\\&\quad
	+ 2\rho\overline\Psi_5 \!\left[\Nve_1\otimes(\mathbf{b}\,\Nve_1^{\,\flato})
	+(\mathbf{b}\,\Nve_1^{\,\flato})\otimes\Nve_1\right]\!
	+ 2\rho\widetilde\Psi_5 \!\left[\Ne_1\otimes(\be^{}\,\Ne_1^\flato)
	+(\be^{}\,\Ne_1^\flato)\otimes \Ne_1 \right]\!
	\\&\quad
	+ 2\rho\overline\Psi_6 \Nve_2 \otimes \Nve_2
	+ 2\rho\widetilde{\Psi}_6 \Ne_2 \otimes \Ne_2
	\\&\quad
	+ 2\rho\overline\Psi_7 \!\left[\Nve_2\otimes(\mathbf{b}\,\Nve_2^{\,\flato})
	+(\mathbf{b}\,\Nve_2^{\,\flato})\otimes\Nve_2\right]\!
	+ 2\rho\widetilde\Psi_7 \!\left[\Ne_2\otimes(\be^{}\,\Ne_2^\flato)
	+(\be^{}\,\Ne_2^\flato)\otimes \Ne_2 \right]\!
	\\&\quad
	+ \rho \overline\Psi_8\,\mathcal{I}\,(\Nve_1\otimes\Nve_2+\Nve_2\otimes\Nve_1)
	+ \rho \widetilde\Psi_8\,\mathcal{I}\,(\Ne_1\otimes\Ne_2+\Ne_2\otimes\Ne_1)
	\,.
\end{aligned}
\end{equation}

\paragraph{Dissipation potential.}
For a monoclinic visco-anelastic solid, when the full set of structural tensors
\begin{equation}
	\boldsymbol\Lambda
	=\{\mathring{\mathbf A}_1, \mathring{\mathbf A}_2, \mathring{\mathbf A}_3\}
	=\{\mathring{\mathbf N}_1\otimes\mathring{\mathbf N}_1,\,
	\mathring{\mathbf N}_2\otimes\mathring{\mathbf N}_2,\,
	\mathring{\mathbf N}_3\otimes\mathring{\mathbf N}_3\}\,.
\end{equation}
is added to the list of the arguments of the dissipation potential $\phi$, it becomes an isotropic function of its arguments, i.e.,~$\phi=\phi(X,\vartheta,\Fve,\Fv,\dot\Fv,\Fn,\dot\Fn,\mathring{\mathbf A}_1, \mathring{\mathbf A}_2, \mathring{\mathbf A}_3, \Go,\mathbf{g})$. Although one may not use the standard representation theorem as for the free energy functions, the dissipation potential will be a function of some standard invariants and a set of spectral invariants, similarly to the dissipation potential of isotropic visco-anelastic solids.

\paragraph{Kinetic equation.}
Using~\eqref{eq:mono_inv} and similarly to~\eqref{eq:trans_part_repr}, one may write
\begin{subequations}
\begin{align}
	\begin{aligned}
	\frac{\partial \hat\Psi_{\textrm{neq}}}{\partial \Ce}
	= \sum_{j=1}^{9} \widetilde\Psi_j\frac{\partial \Ie_j}{\partial \Ce}
	&= \widetilde\Psi_1\Go^{-1} 
	+\!\left(\widetilde\Psi_2 \Ie_2 + \widetilde\Psi_3 \Ie_3 \right)\!\Ce^{-1}
	-\widetilde\Psi_2\Ie_3\Ce^{-2}
	\\[-10pt]&\quad
	+ \widetilde\Psi_4 \mathring{\mathbf N}_1\otimes\mathring{\mathbf N}_1
	+ \widetilde\Psi_5 \!\left[\mathring{\mathbf N}_1\otimes(\Ce\mathring{\mathbf N}_1)^{\sharpo}
	+(\Ce\mathring{\mathbf N}_1)^{\sharpo}\otimes\mathring{\mathbf N}_1\right]\!
	\\&\quad
	+ \widetilde\Psi_6 \mathring{\mathbf N}_2\otimes\mathring{\mathbf N}_2
	+ \widetilde\Psi_7 \!\left[\mathring{\mathbf N}_2\otimes(\Ce\mathring{\mathbf N}_2)^{\sharpo}
	+(\Ce\mathring{\mathbf N}_2)^{\sharpo}\otimes\mathring{\mathbf N}_2\right]\!
	\\&\quad
	+ \frac{1}{2} \widetilde\Psi_8\,\mathcal{I}\,
	(\mathring{\mathbf N}_1\otimes\mathring{\mathbf N}_2+\mathring{\mathbf N}_2\otimes\mathring{\mathbf N}_1)
	\,,
	\end{aligned}
	\\
	\begin{aligned}
	\frac{\partial \hat\Psi_{\textrm{eq}}}{\partial \Cve}
	= \sum_{j=1}^{9} \overline\Psi_j\frac{\partial I_j}{\partial \Cve}
	&= \overline\Psi_1\Go^{-1} 
	+\!\left(\overline\Psi_2 I_2 + \overline\Psi_3 I_3 \right)\!\Cve^{-1}
	-\overline\Psi_2I_3\Cve^{-2}
	\\[-10pt]&\quad
	+ \overline\Psi_4 \mathring{\mathbf N}_1\otimes\mathring{\mathbf N}_1
	+ \overline\Psi_5 \!\left[\mathring{\mathbf N}_1\otimes(\Cve\mathring{\mathbf N}_1)^{\sharpo}
	+(\Cve\mathring{\mathbf N}_1)^{\sharpo}\otimes\mathring{\mathbf N}_1\right]\!
	\\&\quad
	+ \overline\Psi_6 \mathring{\mathbf N}_2\otimes\mathring{\mathbf N}_2
	+ \overline\Psi_7 \!\left[\mathring{\mathbf N}_2\otimes(\Cve\mathring{\mathbf N}_2)^{\sharpo}
	+(\Cve\mathring{\mathbf N}_2)^{\sharpo}\otimes\mathring{\mathbf N}_2\right]\!
	\\&\quad
	+ \frac{1}{2} \overline\Psi_8\,\mathcal{I}\,
	(\mathring{\mathbf N}_1\otimes\mathring{\mathbf N}_2+\mathring{\mathbf N}_2\otimes\mathring{\mathbf N}_1)
	\,.
	\end{aligned}
\end{align}
\end{subequations}
Hence, it follows from~\eqref{eq:Kinetic_CeCve} that the kinetic equations for monoclinic visco-anelastic solids read
\begin{subequations}
\begin{align}
&\begin{aligned}
	\frac{\partial \phi}{\partial \dot\Fv} &-2\rho \Big\{
	\widetilde\Psi_1\Ce\Go^\sharpo 
	+\!\left(\widetilde\Psi_2 \Ie_2 + \widetilde\Psi_3 \Ie_3 \right)\!\mathbf{I}
	-\widetilde\Psi_2 \Ie_3\Go\Ce^{-1}
	\\[-5pt]&+\widetilde\Psi_4 (\Ce \mathring{\mathbf N}_1)\otimes\mathring{\mathbf N}_1
	+ \widetilde\Psi_5 \!\left[(\Ce \mathring{\mathbf N}_1) \otimes (\Ce \mathring{\mathbf N}_1)^\sharpo
	+(\Ce^2\mathring{\mathbf N}_1) \otimes\mathring{\mathbf N}_1\right]\!
	\\&+\widetilde\Psi_6 (\Ce \mathring{\mathbf N}_2)\otimes\mathring{\mathbf N}_2
	+ \widetilde\Psi_7 \!\left[(\Ce \mathring{\mathbf N}_2) \otimes (\Ce \mathring{\mathbf N}_2)^\sharpo
	+(\Ce^2\mathring{\mathbf N}_2) \otimes\mathring{\mathbf N}_2\right]\!
	\\&
	+\frac{1}{2} \widetilde\Psi_8\,\mathcal{I}\,(\Ce\mathring{\mathbf N}_1\otimes\mathring{\mathbf N}_2
	+\Ce\mathring{\mathbf N}_2\otimes\mathring{\mathbf N}_1)
	\Big\} \Fv^{-\star} = \mathbf{0}\,,
\end{aligned}
\\
&\begin{aligned}
	\frac{\partial \phi}{\partial \dot\Fn} & -2\rho \Big\{
	\overline\Psi_1\Cve\Go^\sharpo 
	+ \!\left(\overline\Psi_2 I_2 + \overline\Psi_3 I_3 \right)\!\mathbf{I}
	-\overline\Psi_2I_3\Go\Cve^{-1}
	\\[-5pt]& + \overline\Psi_4 (\Cve \mathring{\mathbf N}_1)\otimes\mathring{\mathbf N}_1
	+\overline\Psi_5 \!\left[(\Cve\mathring{\mathbf N}_1) \otimes (\Cve\mathring{\mathbf N}_1)^\sharpo
	+(\Cve^2\mathring{\mathbf N}_1) \otimes \mathring{\mathbf N}_1 \right]\! \\
	& + \overline\Psi_6 (\Cve \mathring{\mathbf N}_2)\otimes\mathring{\mathbf N}_2
	+\overline\Psi_7 \!\left[(\Cve\mathring{\mathbf N}_2) \otimes (\Cve\mathring{\mathbf N}_2)^\sharpo
	+(\Cve^2\mathring{\mathbf N}_2) \otimes \mathring{\mathbf N}_2 \right]\!
	\\&
	+  \frac{1}{2}\overline\Psi_8\,\mathcal{I}\,(\Cve\mathring{\mathbf N}_1\otimes\mathring{\mathbf N}_2
	+\Cve\mathring{\mathbf N}_2\otimes\mathring{\mathbf N}_1)
	\\&
	+ \Fv^\star \Big[
	\widetilde\Psi_1\Go\Ce^\sharpo
	+\!\left(\widetilde\Psi_2 \Ie_2 + \widetilde\Psi_3 \Ie_3 \right)\!\mathbf{I}
	- \widetilde\Psi_2\Ie_3\Go\Ce^{-\sharpo} \\
	& + \widetilde\Psi_4 (\Ce \mathring{\mathbf N}_1) \otimes\mathring{\mathbf N}_1
	+ \widetilde\Psi_5 \!\left((\Ce\mathring{\mathbf N}_1)\otimes (\Ce\mathring{\mathbf N}_1)^\sharpo
	+ (\Ce^2\mathring{\mathbf N}_1) \otimes \mathring{\mathbf N}_1\right)\! \\
	& + \widetilde\Psi_6 (\Ce \mathring{\mathbf N}_2) \otimes\mathring{\mathbf N}_2
	+ \widetilde\Psi_7 \!\left((\Ce\mathring{\mathbf N}_2)\otimes (\Ce\mathring{\mathbf N}_2)^\sharpo
	+ (\Ce^2\mathring{\mathbf N}_2) \otimes \mathring{\mathbf N}_2\right)\!
	\\&
	+ \frac{1}{2} \widetilde\Psi_8\,\mathcal{I}\,(\Ce\mathring{\mathbf N}_1\otimes\mathring{\mathbf N}_2
	+\Ce\mathring{\mathbf N}_2\otimes\mathring{\mathbf N}_1)
	\Big] \Fv^{-\star} \Big\} \Fn^{-\star} = \mathbf{0}\,.
\end{aligned}
\end{align}
\end{subequations}
Assuming visco-anelastic incompressibility, i.e.,~$J=\Jv=\Jn=1$, the kinetic equations above are rewritten as
\begin{subequations}
\begin{align}
&\begin{aligned}
	\frac{\partial \phi}{\partial \dot\Fv} &-2\rho \Big\{
	\widetilde\Psi_1\Ce\Go^\sharpo 
	+\widetilde\Psi_2 \!\left(\Ie_2 \mathbf{I} -\Go\Ce^{-1}\right)\!
	+\widetilde\Psi_4 (\Ce \mathring{\mathbf N}_1)\otimes\mathring{\mathbf N}_1
	+ \widetilde\Psi_5 \!\left[(\Ce \mathring{\mathbf N}_1) \otimes (\Ce \mathring{\mathbf N}_1)^\sharpo
	+(\Ce^2\mathring{\mathbf N}_1) \otimes\mathring{\mathbf N}_1\right]\!
	\\[-5pt]&+\widetilde\Psi_6 (\Ce \mathring{\mathbf N}_2)\otimes\mathring{\mathbf N}_2
	+ \widetilde\Psi_7 \!\left[(\Ce \mathring{\mathbf N}_2) \otimes (\Ce \mathring{\mathbf N}_2)^\sharpo
	+(\Ce^2\mathring{\mathbf N}_2) \otimes\mathring{\mathbf N}_2\right]\!
	\\&
	+ \frac{1}{2}\widetilde\Psi_8\,\mathcal{I}\,(\Ce\mathring{\mathbf N}_1\otimes\mathring{\mathbf N}_2
	+\Ce\mathring{\mathbf N}_2\otimes\mathring{\mathbf N}_1)
	\Big\} \Fv^{-\star} = \pv \,\Fv^{-\star}\,,
\end{aligned}
\\
&\begin{aligned}
	\frac{\partial \phi}{\partial \dot\Fn} & -2\rho \Big\{
	\overline\Psi_1\Cve\Go^\sharpo 
	+ \overline\Psi_2 \!\left(I_2 \mathbf{I} - \Go\Cve^{-1}\right)\!
	+ \overline\Psi_4 (\Cve \mathring{\mathbf N}_1)\otimes\mathring{\mathbf N}_1
	+\overline\Psi_5 \!\left[(\Cve\mathring{\mathbf N}_1) \otimes (\Cve\mathring{\mathbf N}_1)^\sharpo
	+(\Cve^2\mathring{\mathbf N}_1) \otimes \mathring{\mathbf N}_1 \right]\!
	\\[-7.5pt]& + \overline\Psi_6 (\Cve \mathring{\mathbf N}_2)\otimes\mathring{\mathbf N}_2
	+\overline\Psi_7 \!\left[(\Cve\mathring{\mathbf N}_2) \otimes (\Cve\mathring{\mathbf N}_2)^\sharpo
	+(\Cve^2\mathring{\mathbf N}_2) \otimes \mathring{\mathbf N}_2 \right]\!
	\\&
	+ \frac{1}{2}\overline\Psi_8\,\mathcal{I}\,(\Cve\mathring{\mathbf N}_1\otimes\mathring{\mathbf N}_2
	+\Cve\mathring{\mathbf N}_2\otimes\mathring{\mathbf N}_1)
	\\&
	+ \Fv^\star \Big[
	\widetilde\Psi_1\Go\Ce^\sharpo
	+\widetilde\Psi_2 \!\left(\Ie_2 \mathbf{I} -\Go\Ce^{-1}\right)\!
	+ \widetilde\Psi_4 (\Ce \mathring{\mathbf N}_1) \otimes\mathring{\mathbf N}_1
	+ \widetilde\Psi_5 \!\left((\Ce\mathring{\mathbf N}_1)\otimes (\Ce\mathring{\mathbf N}_1)^\sharpo
	+ (\Ce^2\mathring{\mathbf N}_1) \otimes \mathring{\mathbf N}_1\right)\! \\
	& + \widetilde\Psi_6 (\Ce \mathring{\mathbf N}_2) \otimes\mathring{\mathbf N}_2
	+ \widetilde\Psi_7 \!\left((\Ce\mathring{\mathbf N}_2)\otimes (\Ce\mathring{\mathbf N}_2)^\sharpo
	+ (\Ce^2\mathring{\mathbf N}_2) \otimes \mathring{\mathbf N}_2\right)\!
	\\&
	+ \frac{1}{2}\widetilde\Psi_8\,\mathcal{I}\,(\Ce\mathring{\mathbf N}_1\otimes\mathring{\mathbf N}_2
	+\Ce\mathring{\mathbf N}_2\otimes\mathring{\mathbf N}_1)
	\Big] \Fv^{-\star} \Big\} \Fn^{-\star} = (\pn-p) \,\Fn^{-\star}\,.
\end{aligned}
\end{align}
\end{subequations}

\vskip 0.1in
\noindent 
Table \ref{Table-Summary} summarizes some of the important fields, constitutive equations, and governing equations of nonlinear anisotropic viscoelasticity.

\begin{table}
\begin{center}
\caption{Summary of the main fields, constitutive equations, and governing equations of nonlinear visco-anelasticity.}
\label{Table-Summary}
\hrule
\begin{tabular}{ll}
\multicolumn{2}{c}{\bf Nonlinear Visco-Anelasticity}
\\[7.5pt]
{\bf Kinematics} & $\mathbf{F}:T_X\mathcal{B}\to T_x\mathcal{C}$ (total deformation gradient) \\
${\mathbf{F}=\Fe\Fv\Fa} $ & $\Fa:T_X\mathcal{B}\to T_X\mathcal{B}$ (anelastic distortion) \\
${\Fa=\Fn\Fth} $ &  $\Fv:T_X\mathcal{B}\to T_X\mathcal{B}$ (viscous distortion) \\
$\vartheta$: temperature & $\Fe:T_X\mathcal{B}\to T_x\mathcal{C}$ (elastic distortion) \\
$\boldsymbol\alpha$: coefficient of (linear) thermal expansion & $\Fn:T_X\mathcal{B}\to T_X\mathcal{B}$ (athermal-anelastic distortion) \\
$\boldsymbol\alpha = ({\partial\Fth}/{\partial\vartheta}) \,\Fth^{-1} = \Fth^{-1} ({\partial\Fth}/{\partial\vartheta})$ & $\Fth:T_X\mathcal{B}\to T_X\mathcal{B}$ (thermal distortion)
\\[7.5pt]

{\bf Free energy} $\Psi$ & $\Psi =\Psi(X,\vartheta,\Fve, \Fe, \Go,\mathbf{g}) = \tilde{\Psi}(X, \vartheta, \mathbf{F}, \Fv, \Fn, \Fth, \Go, \mathbf{g})$
\\
Additive decomposition: & $\Psi=\Psi_{\textrm{eq}}(X,\vartheta,\Fe\Fv,\Go,\mathbf{g})+\Psi_{\textrm{neq}}(X,\vartheta,\Fe,\Go,\mathbf{g})$
\\[5pt]

$\Go$ (Euclidean metric) & {\bf Isotropic solids} \\
$\mathbf{G}=\Fa^*\Go$ (material metric) &
$\Psi_{\textrm{eq}}
=\Psi_{\textrm{eq}}(X,\vartheta,\mathbf{F},\mathbf{G},\mathbf{g})
=\hat{\Psi}_{\textrm{eq}}(X,\vartheta,\mathbf{C},\mathbf{G})$ \\
$\Gi=(\Fv\Fa)^*\Go$ (inelastic metric) & $\Psi_{\textrm{neq}}
=\Psi_{\textrm{neq}}(X,\vartheta,\mathbf{F},\Gi,\mathbf{g})
=\hat{\Psi}_{\textrm{neq}}(X,\vartheta,\mathbf{C},\Gi)$
\\[5pt]

$\mathring{\boldsymbol\Lambda}$ (Euclidean structural tensors) & {\bf Anisotropic solids} \\
$\boldsymbol{\Lambda}=\Fa^*\mathring{\boldsymbol{\Lambda}}$ (material structural tensors) & $\Psi_{\textrm{eq}}=\Psi_{\textrm{eq}}(X,\vartheta,\mathbf{F},\boldsymbol{\Lambda},\mathbf{G},\mathbf{g})
=\hat{\Psi}_{\textrm{eq}}(X,\vartheta,\mathbf{C},\boldsymbol{\Lambda},\mathbf{G})$ \\
 $\Lambdai=(\Fv\Fa)^*\mathring{\boldsymbol{\Lambda}}$ (inelastic structural tensors) & $\Psi_{\textrm{neq}}
 =\Psi_{\textrm{neq}}(X,\vartheta,\mathbf{F},\Lambdai,\Gi,\mathbf{g})
 =\hat{\Psi}_{\textrm{neq}}(X,\vartheta,\mathbf{C},\Lambdai,\Gi)$
 \\[7.5pt]

{\bf Dissipation potential} $\phi$ & $\phi=\phi(X,\vartheta,\Fve,\Fv,\dot\Fv,\Fn,\dot\Fn,\Go,\mathbf{g})=\hat{\phi}(X,\vartheta,\mathbf{C},\Fv,\dot\Fv,\Fn,\dot\Fn,\Go)$
\\[7.5pt]

{\bf Material symmetry group} $\mathring{\mathcal G}_X$ at $X\in\mathcal B$ & \\[5pt]
\multicolumn{2}{l}{%
$
\begin{dcases}
    \Psi_{\textrm{eq}}(X,\vartheta,\Fe\Fv\,\mathring{\mathbf{K}},\Go,\mathbf{g})
    =
    \Psi_{\textrm{eq}}(X,\vartheta,\Fe\Fv,\Go,\mathbf{g})\\
    \Psi_{\textrm{neq}}(X,\vartheta,\Fe\mathring{\mathbf{K}},\Go,\mathbf{g})
    =
    \Psi_{\textrm{neq}}(X,\vartheta,\Fe,\Go,\mathbf{g})\\
    \phi(X,\vartheta,\Fve\mathring{\mathbf{K}},\mathring{\mathbf{K}}^*\Fv,\mathring{\mathbf{K}}^*\dot\Fv,\mathring{\mathbf{K}}^*\Fn,\mathring{\mathbf{K}}^*\dot\Fn,\Go,\mathbf{g})
    =\phi(X,\vartheta,\Fve,\Fv,\dot\Fv,\Fn,\dot\Fn,\Go,\mathbf{g})
\end{dcases}
\,,\quad \forall\,\,\mathring{\mathbf{K}} \in \mathring{\mathcal G}_X\leqslant \mathrm{Orth}(\Go)$
}
\\[25pt]

{\bf Balance laws} & {\bf Doyle-Ericksen formula} \\
$\begin{dcases}
	\dot{\rho} + \rho\,\dot{\Fa}\!:\!\Fa^{-\star} = S_m\\[7.5pt]
	\operatorname{Div}\mathbf{P}+ \rho\boldsymbol{\mathsf{B}} = \rho \mathbf{A}\quad\&\quad
	\mathbf{F}\mathbf{P}^\star =\mathbf{P}\mathbf{F}^\star
\end{dcases}$
&
$\displaystyle
\begin{aligned}
\mathbf{P}
&= \rho \mathbf g^\sharp \frac{\partial \tilde\Psi}{\partial \mathbf{F}} - p \,\mathbf{g}^\sharp \mathbf{F}^{-\star}
\\&
= \rho \mathbf g^\sharp\!\left(\frac{\partial \Psi}{\partial \Fve} \Fve^{\star} + \frac{\partial \Psi}{\partial \Fe} \Fe^{\star}\right)\! \mathbf{F}^{-\star}  - p \,\mathbf{g}^\sharp \mathbf{F}^{-\star}
\end{aligned}$
\\[20pt]

{\bf The Clausius-Duhem inequality} & \\[5pt]
$\begin{aligned}
&\dot\eta =
	- \frac{1}{\vartheta} \langle d\vartheta, \mathbf Q \rangle
	- \!\left[ \rho \frac{\partial \tilde\Psi}{\partial \Fn} - (\pn-p) \Fn^{-\star} \right]\! \!:\! \dot\Fn
	\\&
	- \!\left[ \rho \frac{\partial \tilde\Psi}{\partial \Fv} - \pv\,\Fv^{-\star} \right]\! \!:\! \dot\Fv
	- \!\left[ \rho \frac{\partial \tilde\Psi}{\partial \Fth} + p\,\Fth^{-\star} \right]\! \!:\! \dot\Fth
	\geq 0
\end{aligned}
$
&
$\begin{aligned}
p:\,\textrm{Lagrange multiplier associated with } J=1
\\
\pv:\,\textrm{Lagrange multiplier associated with } \Jv=1
\\
\pn:\,\textrm{Lagrange multiplier associated with } \Jn=1
\end{aligned}$
\\[30pt]

{\bf Heat equation} & {\bf Generalized forces}\\
$\begin{aligned}
	\rho c_{\text D} \dot\vartheta
	- \rho \vartheta \mathbf{g} \frac{\partial}{\partial \vartheta}\!\left[ \frac{\boldsymbol\sigma_{\textrm{eq}}-p_{\textrm{eq}} \, \mathbf g^\sharp}{\rho} \right]\! \!:\! (\operatorname{D}_t^{\mathbf g}\Fve) \Fve^{-1}
	\\
	-\rho\vartheta \mathbf{g} \frac{\partial}{\partial \vartheta}\!\left[ \frac{ \boldsymbol\sigma_{\textrm{neq}}-p_{\textrm{neq}} \, \mathbf g^\sharp}{\rho} \right]\! \!:\! (\operatorname{D}_t^{\mathbf g}\Fe) \Fe^{-\star}
	\\
	+ \Bv \!:\! \dot\Fv
	+ \Bn \!:\! \dot\Fn
	= \rho R - \operatorname{Div} \mathbf{Q}
\end{aligned}$
&
$\begin{aligned}
\Bv &= \rho \frac{\partial \tilde\Psi}{\partial \Fv} + (p-\pv) \,\Fv^{-\star}
= -\rho \Fe^\star \frac{\partial \Psi}{\partial \Fe} \Fv^{-\star} + (p-\pv) \,\Fv^{-\star} \\
\Bn &= \rho \frac{\partial \tilde\Psi}{\partial \Fn} + (p-\pn) \,\Fn^{-\star}
\\&= -\rho \Fve^\star\!\left(\frac{\partial \Psi}{\partial \Fve} + \frac{\partial \Psi}{\partial \Fe} \Fv^{-\star} \right)\!\Fn^{-\star} + (p-\pn) \,\Fn^{-\star}
\end{aligned}$
\\[35pt]

{\bf Kinetic equations} \\ [5pt]
$\displaystyle \frac{\partial \phi}{\partial \dot\Fv}  + \rho \frac{\partial \tilde\Psi}{\partial \Fv} = \pv \,\Fv^{-\star}$
&
$\displaystyle \frac{\partial \phi}{\partial \dot\Fn} + \rho \frac{\partial \tilde\Psi}{\partial \Fn} = (\pn-p) \,\Fn^{-\star}$\\
\\[-15pt]
\end{tabular} 
\end{center}
\hrule
\end{table}

\section{Examples} \label{Sec:Examples}

In this section we present two examples to illustrate the capabilities of the visco-anelastic framework developed in the previous sections. First, we analyze an axisymmetric thermal inclusion in an isotropic, infinitely long circular cylinder and study its viscoelastic evolution. Second, we examine a single wedge disclination in a transversely isotropic circular cylinder, comparing the responses under displacement-controlled and force-controlled end loadings. For both examples we assume the existence of a dissipation potential governing the viscous distortion. For both examples, we assume the dissipation potential $\phi$ to be isotropic and quadratic in $\dot\Fv\,$; the most general representation of which may be given by \citep{SaYa2024viscoelasticity}:
\begin{equation}\label{eq:Quad-Rayleigh}
\begin{aligned}
	\phi
    &= \frac{1}{2}\eta_1 \!\left(\operatorname{tr}\dot\Fv\right)^2 + \frac{1}{2}\eta_2 \operatorname{tr}\!\left(\dot\Fv^2\right)\! + \frac{1}{2}\eta_3 \operatorname{tr}\!\left(\dot\Fv \dot\Fv^{\mathring{\mathsf T}}\right)\!\\
    &=\frac{1}{2}\eta_1\!\left(\dot\cFv^A{}_A\right)^2+\frac{1}{2}\eta_2\,\dot\cFv^A{}_B\,\dot\cFv^B{}_A +\frac{1}{2}\eta_3\, \dot\cFv^A{}_B \mathring{\mathrm G}^{BC} \dot\cFv^D{}_C \mathring{\mathrm G}_{DA}\,,
\end{aligned}
\end{equation}
where $\eta_i$ must satisfy\footnote{Note that the conditions~\eqref{eq:eta_cond} ensure that the Rayleigh dissipation potential $\phi$ is convex and thermodynamically admissible, i.e.,~that it does not violate the Clausius-Duhem inequality\textemdash see \citep{SaYa2024viscoelasticity}.}
\begin{equation} \label{eq:eta_cond}
	3 \eta_1 + \eta_2 + \eta_3 > 0\,,\qquad \eta_2 + \eta_3 > 0\,,\qquad
	-\eta_2 + \eta_3 > 0\,.
\end{equation}

\subsection{An incompressible isotropic infinitely-long circular bar with a radial thermal inclusion}
\label{S:eg1}

\paragraph{Kinematics.}
Let us consider an infinitely-long homogeneous solid circular cylinder of radius $R_0$ in its natural stress-free state. We equip the ambient space with a cylindrical coordinate system $(r, \theta, z)$, where the Euclidean metric is hence represented as
\begin{equation}
	\mathbf g=\begin{bmatrix}
	1 & 0 & 0\\
	0 & r^2 & 0\\
	0 & 0 & 1
	\end{bmatrix}\,.
\end{equation}
The Euclidean reference configuration inherits the Euclidean metric as $\Go=\iota^*\mathbf g$,\footnote{Recall that the Euclidean reference metric is obtained as the pull-back of the ambient-space metric by the inclusion map $\iota:\mathcal{B}\hookrightarrow\mathcal{S}$.} 
which has the following representation:
\begin{equation}
	\Go=\begin{bmatrix}
	1 & 0 & 0\\
	0 & R^2 & 0\\
	0 & 0 & 1
	\end{bmatrix}\,,
\end{equation}
with respect to $(R, \Theta, Z)$, a referential cylindrical coordinate system such that the axis of the solid cylinder in its reference state is aligned with $Z$.

Let $\vartheta=\vartheta(R,t)$ denote an axisymmetric temperature distribution.
The cylindrical bar is assumed to have transversely isotropic thermal expansion about the $Z$-axis. The thermal expansion is isotropic in the plane $\{R,\Theta\}$, with coefficient of linear thermal expansion $\alpha=\alpha(\vartheta)$, and vanishes along the $Z$-direction. This is a constitutive assumption on the thermal expansion tensor $\boldsymbol\alpha$, not a plane-strain restriction on the kinematics. In particular, it imposes no corresponding restriction on the viscous distortion.
Following~\eqref{eq:Fth}, the thermal distortion then reads
\begin{equation}
\label{eq:eg1_Fth}
  \Fth =\begin{bmatrix}
	e^{\omega(\vartheta(R,t))} & 0 & 0\\
	0 & e^{\omega(\vartheta(R,t))} & 0\\
	0 & 0 & 1
	\end{bmatrix}\,,
\end{equation}
with
\begin{equation}
	\omega(\vartheta) \coloneqq \int_{\vartheta_0}^{\vartheta} \alpha(\tilde\vartheta)\,d\tilde\vartheta\,,
\end{equation}
where $\vartheta_0$ is a uniform reference temperature at which no thermal distortion is present.
In this problem, the only source of anelasticity is the temperature, the material metric is thus given by $\mathbf{G}=\Fth^*\Go$ and has the following representation
\begin{equation}
\label{eq:G_eg1}
	\mathbf G 
	= \begin{bmatrix}
	e^{2\omega(\vartheta(R,t))} & 0 & 0\\
	0 & R^2 e^{2\omega(\vartheta(R,t))} & 0\\
	0 & 0 & 1
	\end{bmatrix}\,.
\end{equation}

We assume a time-dependent, radially-symmetric embedding of the material manifold into the ambient space, given by
\begin{equation}
	r = r(R,t)\,,\qquad
	\theta = \Theta\,,\qquad
	z = Z\,.
\end{equation}
The deformation gradient $\mathbf{F}$ associated with this embedding then takes the form
\begin{equation}
	\mathbf{F} = 
	\begin{bmatrix}
	r'(R,t) & 0 & 0\\
	0 & 1 & 0\\
	0 & 0 & 1
	\end{bmatrix}\,,
\end{equation}
where $(\cdot)'=(\cdot)_{,R}$ denotes the partial derivative with respect to $R$.
The left Cauchy-Green deformation tensor is given by
\begin{equation}
	\mathbf{C}
	= \begin{bmatrix}
		{r'}^2(R,t) & 0 & 0\\
		0 & r^2(R,t) & 0\\
		0 & 0 & 1
	\end{bmatrix}\,.
\end{equation}
Assuming visco-elastic incompressibility, i.e.,~$J=1$, and recalling that $J=\sqrt{\det \mathbf g / \det \mathbf G}\,\det\mathbf{F}$, one finds
\begin{equation}
r(R,t)\,r'(R,t) = R\,e^{2\omega(\vartheta(R,t))}\,,
\end{equation}
which, after fixing the embedding at the origin, i.e.,~$r(0,t)=0$, integrates to
\begin{equation}\label{eq:eg1_r_solution}
  r(R,t) = \sqrt{\int_{0}^{R} 2\,\xi\,e^{2\omega[\vartheta(\xi,t)]}\,d\xi}\,.
\end{equation}

\begin{remark}\label{rmrk:eg1_incomp}
Equation \eqref{eq:eg1_r_solution} illustrates the point made in Remark~\ref{rmrk:Jacobian}. The bar expands, with its volume element increasing by the factor $e^{2\omega}$, even though the deformation is incompressible. There is no contradiction because incompressibility is imposed relative to the material metric \eqref{eq:G_eg1}, which is determined by the thermal distortion. The natural volume element relative to which the deformation is measured therefore evolves with the temperature field. Equation \eqref{eq:eg1_r_solution} states that, at each instant, the deformed volume element coincides with the thermally distorted natural volume element. Thus, the material accommodates the thermal volume change without an elastic volume change. The thermal distortion is volume-changing, $\Jth = e^{2\omega}$, and is not constrained. The elastic and viscous distortions, by contrast, are volume-preserving: $\Jv = 1$ is imposed constitutively, and $\Je = 1$ follows from $J = 1$ and $\Jv = 1$.
\end{remark}

We assume that the viscous distortion is diagonal and radially symmetric. Note that, unlike the deformation gradient $\mathbf{F}$ and the thermal distortion $\Fth$, the viscous distortion $\Fv$ need not be in-plane. We may hence write
\begin{equation}
	\Fv =
	\begin{bmatrix}
	\lvr(R,t) & 0 & 0\\
	0 & \lvt(R,t) & 0\\
	0 & 0 & \lvz(R,t)
	\end{bmatrix}\,.
\end{equation}
Invoking viscous incompressibility, i.e.,~$\Jv=\det \Fv = 1$, one finds
\begin{equation}
	\lvt(R,t) = \frac{1}{\lvr(R,t)\,\lvz(R,t)}\,.
\end{equation}
Recalling that $\mathbf{F} = \Fe \Fv \Fth$, the elastic distortion reads
\begin{equation}
	\Fe
	= \mathbf{F}\,(\Fv\,\Fth)^{-1}
	= \begin{bmatrix}
		\dfrac{R e^{\omega(\vartheta(R,t))}}{\lvr(R,t) r(R,t)} & 0 & 0\\
		0 & \dfrac{1}{\lvt(R,t)\,e^{\omega(\vartheta(R,t))}} & 0\\
		0 & 0 & \dfrac{1}{\lvz(R,t)}
	\end{bmatrix}\,.
\end{equation}
Therefore, the elastic left Cauchy-Green distortion tensor is given by
\begin{equation}
	\Ce
	= \begin{bmatrix}
		\dfrac{1}{\lvr^2(R,t)} \dfrac{R^2 e^{2\omega(\vartheta(R,t))}}{r^2(R,t)} & 0 & 0\\
		0 & \dfrac{1}{\lvt^2(R,t)} \dfrac{r^2(R,t)}{e^{2\omega(\vartheta(R,t))}} & 0\\
		0 & 0 & \dfrac{1}{\lvz^2(R,t)}
	\end{bmatrix}\,,
\end{equation}
and the viscoelastic left Cauchy-Green distortion tensor is given by
\begin{equation}
	\Cve
	= \begin{bmatrix}
		\dfrac{R^2 e^{2\omega(\vartheta(R,t))}}{r^2(R,t)} & 0 & 0\\
		0 & \dfrac{r^2(R,t)}{e^{2\omega(\vartheta(R,t))}} & 0\\
		0 & 0 & 1
	\end{bmatrix}\,.
\end{equation}
The principal invariants associated with $\{\mathbf C,\mathbf G\}$, or equivalently with $\{\Cve,\Go\}$, are
\begin{equation} \label{eq:eg1_Inv}
	I_1 = I_2 = 1 + \frac{R^2 e^{2\omega(\vartheta(R,t))}}{r^2(R,t)} 
	+ \frac{r^2(R,t)}{R^2e^{2\omega(\vartheta(R,t))}}\,,
\end{equation}
and the principal invariants associated with $\{\mathbf C,\Gi\}$, or equivalently with $\{\Ce,\Go\}$, are
\begin{equation}
\label{eq:eg1_ElasInv}
\begin{split}
\Ie_1 &=
	\frac{R^2 e^{2\omega(\vartheta(R,t))}}{r^2(R,t)\,\lvr^2(R,t)}
	+ \frac{r^2(R,t)}{R^2\,\lvt^2(R,t) e^{2\omega(\vartheta(R,t))}}
	+ \frac{1}{\lvz^2(R,t)}
	\,,\\
\Ie_2 &=
	\frac{r^2(R,t)\,\lvr^2(R,t)}{R^2 e^{2\omega(\vartheta(R,t))}}
	+ \frac{R^2\,\lvt^2(R,t) e^{2\omega(\vartheta(R,t))}}{r^2(R,t)}
	+ \lvz^2(R,t)
	\,.
\end{split}
\end{equation}

\paragraph{Mass balance.}
In the presence of no other source of anelasticity but temperature, the balance of mass reduces to~\eqref{eq:Temp_MassBal}, which reads
\begin{equation}\label{eq:eg1_MassBal}
\frac{\partial\rho}{\partial\vartheta} + 2\rho\,\alpha = 0\,.
\end{equation}
Hence
\begin{equation}\label{eq:eg1_rho}
\rho
= \rho_0 \exp\!\left[-\int_{\vartheta_0}^{\vartheta(R,t)} 2\,\alpha(\tilde\vartheta)\,d\tilde\vartheta\right]\!
= \rho_0 e^{-2\omega(\vartheta(R,t))}
\,,
\end{equation}
where $\rho_0=\rho\big|_{\vartheta=\vartheta_0}$ is the material mass density at the uniform stress-free temperature $\vartheta_0$.

\paragraph{Viscous kinetic equations.}
We assume an isotropic quadratic dissipation potential as given by~\eqref{eq:Quad-Rayleigh}. The system of viscous kinetic equations~\eqref{eq:Trans_v_Kinetic_inc} yields three independent equations in terms of $\lvr(R,t)$, $\lvz(R,t)$, and $\pv(R,t)\,$\textemdash the Lagrange multiplier for viscous incompressibility. We proceed to eliminate $\pv$, and find the following system for the time evolution of $\lvr$ and $\lvz\,$:
\begin{subequations}
\label{eq:eg1_kinetic}
\begin{align}
&\begin{dcases}
&\!\left[
	\eta_1 \!\left( \lvr^2\,\lvz -1 \right)^2
	+ \!\left(\eta_2 + \eta_3\right)\!\!\left(\lvr^4\,\lvz^2 +1 \right)\!
	\right]\!
	\frac{ \dot{\accentset{v}{\lambda}}_R}{\lvr}
\\
&\qquad
+ \!\left[
	\eta_1 \!\left( \lvr^2\,\lvz -1 \right)\!
	\!\left( \lvr\,\lvz^2 -1 \right)\!
	+ \eta_2 + \eta_3
\right]\!
\frac{ \dot{\accentset{v}{\lambda}}_Z}{\lvz}
\\
&\qquad\qquad=
 2 \rho
 \!\left(
   \frac{ R^2 e^{2\omega(\vartheta)} }{ r^2 }
   - \frac{ r^2 }{ R^2 e^{2\omega(\vartheta)} } \lvr^4\,\lvz^2
 \right)\!
 \!\left( \lvz^2\widetilde{\Psi}_1
      + \widetilde{\Psi}_2 \right)\!
\,,
\end{dcases}
\\
&\begin{dcases}
&\!\left[
	\eta_1 \!\left( \lvr - \lvz \right)\!\!\left( \lvr^2\,\lvz -1 \right)\!
	+ (\eta_2 + \eta_3)\,\lvr^3\,\lvz
 \right]\!
 \frac{ \dot{\accentset{v}{\lambda}}_R}{\lvr}
\\
&\qquad
+ \!\left[
	\eta_1 \!\left( \lvr - \lvz \right)\!\!\left( \lvr\,\lvz^2 -1 \right)\!
	- (\eta_2 + \eta_3)\,\lvr\,\lvz^3
	\right]\!
\frac{ \dot{\accentset{v}{\lambda}}_Z}{\lvz}
\\
&\qquad\qquad =
2 \rho \!\left(
	\frac{R^2 e^{2\omega(\vartheta)}}{r^2}\frac{\lvz}{\lvr}
	- \frac{\lvr}{\lvz}
	\right)\!
\!\left(
	\widetilde{\Psi}_1
	+ \frac{ r^2}{ R^2 e^{2\omega(\vartheta)} } \lvr^2\,\lvz^2 \widetilde{\Psi}_2
	\right)\!
\,.
\end{dcases}
\end{align}
\end{subequations}

\paragraph{Stress and equilibrium.}
The cylinder is made of an isotropic material. The stress may hence be computed following~\eqref{eq:sigma_iso}, and one finds the following non-zero components of the Cauchy stress tensor
\begin{subequations}
\label{eq:eg1_stress1}
\begin{align}
\label{eq:eg1_srr1}
\sigma^{rr} &= -p
+ \frac{2 e^{2\omega(\vartheta)} R^2}{r^2} \,\rho \overline{\Psi}_1 
- \frac{2 r^2}{R^2 e^{2\omega(\vartheta)}} \,\rho \overline{\Psi}_2
+ \frac{2 e^{2\omega(\vartheta)} R^2}{r^2\,\lvr^2} \,\rho \widetilde{\Psi}_1
- \frac{2 r^2\,\lvr^2}{R^2 e^{2\omega(\vartheta)}} \,\rho \widetilde{\Psi}_2
\,,
\\
\label{eq:eg1_stt1}
\sigma^{\theta\theta} &= \frac{1}{r^2}\!\left[-p
+ \frac{2 r^2}{R^2 e^{2\omega(\vartheta)}} \,\rho \overline{\Psi}_1
- \frac{2 e^{2\omega(\vartheta)} R^2}{r^2} \,\rho \overline{\Psi}_2
+ \frac{2 r^2}{R^2\,\lvt^2 e^{2\omega(\vartheta)}} \,\rho \widetilde{\Psi}_1
- \frac{2 R^2\,\lvt^2 e^{2\omega(\vartheta)}}{r^2} \,\rho \widetilde{\Psi}_2\right]\!
\,,
\\
\label{eq:eg1_szz1}
\sigma^{zz} &= -p
+ 2 \rho \overline{\Psi}_1 - 2 \rho \overline{\Psi}_2
+ \frac{2}{\lvz^2} \,\rho \widetilde{\Psi}_1
- 2 \lvz^2 \,\rho \widetilde{\Psi}_2
\,,
\end{align}
\end{subequations}
where $p=p(R,t)$ is the Lagrange multiplier enforcing the visco-elastic incompressibility condition ($J=1$).
In terms of the physical components of the Cauchy stress tensor,\footnote{Recall—following \cite{Truesdell1953physical}—that the physical components of the stress are given by $\hat{\sigma}^{rr}=\sigma^{rr}$, $\hat{\sigma}^{\theta\theta}=r^2\sigma^{\theta\theta}$, and $\hat{\sigma}^{zz}=\sigma^{zz}$.} the only non-trivially satisfied equilibrium equation is 
$\hat\sigma^{rr}{}_{,r} + \!\left(\hat\sigma^{rr}-\hat\sigma^{\theta\theta}\right)\!/{r} = 0\,$; it may be written as
\begin{equation}\label{eq:eg1_Equilibrium}
\frac{\partial \hat\sigma^{rr}}{\partial R} 
= \mathcal{I}_\sigma
\,,
\end{equation}
where $\mathcal{I}_\sigma$ is given by
\begin{equation}\label{eq:Is_eg1}
\begin{split}
\mathcal{I}_\sigma &\coloneq 
\frac{2}{R} \!\left( \frac{R^4 e^{4\omega}}{r^4} -1 \right)\!
\rho \!\left( \overline{\Psi}_1 +\overline{\Psi}_2 \right)\!
+ \frac{2}{R} \!\left(
    \frac{R^4 e^{4\omega}}{r^4\,\lvr^2}
    - \frac{1}{\lvt^2}
  \right)\!\rho \widetilde{\Psi}_1
+ \frac{2}{R}\!\left(
    \frac{R^4\,\lvt^2 e^{4\omega}}{r^4}
    - \lvr^2
  \right)\!\rho \widetilde{\Psi}_2
\,.
\end{split}
\end{equation}
Assuming that the lateral boundary of the cylinder is traction-free, i.e.,~$\hat\sigma^{rr}(R_0,t)=0$, it follows from~\eqref{eq:eg1_srr1} and~\eqref{eq:eg1_Equilibrium} that
\begin{equation}
\label{eq:eg1_p}
	p = \int_R^{R_0} \mathcal{I}_\sigma(\xi,t)d\xi + \frac{2 e^{2\omega(\vartheta)} R^2}{r^2} \,\rho \overline{\Psi}_1 
- \frac{2 r^2}{R^2 e^{2\omega(\vartheta)}} \,\rho \overline{\Psi}_2
+ \frac{2 e^{2\omega(\vartheta)} R^2}{r^2\,\lvr^2} \,\rho \widetilde{\Psi}_1
- \frac{2 r^2\,\lvr^2}{R^2 e^{2\omega(\vartheta)}} \,\rho \widetilde{\Psi}_2
	\,.
\end{equation}

\begin{remark}
\label{rmrk:eg1-n-eq-separ}
Recalling~\eqref{eq:Psi_i}, the partial derivatives of the free energy carry the functional dependencies
\begin{equation}
  \overline{\Psi}_j=\overline{\Psi}_j(\vartheta,I_1,I_2) = \frac{\partial \Psi}{\partial I_j}\,,
  \qquad
  \widetilde{\Psi}_j=\widetilde{\Psi}_j(\vartheta,\Ie_1,\Ie_2) = \frac{\partial \Psi}{\partial \Ie_j}\,,
  \qquad j=1,2,3\,.
\end{equation}
Consequently, in~\eqref{eq:eg1_stress1},~\eqref{eq:Is_eg1}, and~\eqref{eq:eg1_p}, the terms involving $\overline{\Psi}_j$ form $\boldsymbol\sigma_{\text{eq}}$, including $p_{\text{eq}}$, while those involving $\widetilde{\Psi}_j$ form $\boldsymbol\sigma_{\text{neq}}$, including $p_{\text{neq}}$, thereby recovering the additive decomposition $\boldsymbol\sigma=\boldsymbol\sigma_{\text{eq}}+\boldsymbol\sigma_{\text{neq}}$ established in~\eqref{eq:sig_eq-neq_represent_inc} for incompressible solids.
\end{remark}

\paragraph{Heat equation.}
Let us assume isotropic Fourier conduction with heat conductivity $K$ (i.e.,~$\mathbf K = K \mathbf{G}^\sharp$), and no external heat supply ($R=0$). 
Therefore, the heat equation~\eqref{eq:to_HtEq_S_inc} specializes to
\begin{equation}
\label{eq:eg1_Heat}
	\rho c_{\text D} \dot\vartheta
	- \vartheta \frac{\partial \boldsymbol\sigma_{\textrm{eq}}}{\partial \vartheta} \!:\! \frac{1}{2} \Fve^{-\star} \dot{\Cve}\, \Fve^{-1}
	- \vartheta \frac{\partial \boldsymbol\sigma_{\textrm{neq}}}{\partial \vartheta} \!:\! \frac{1}{2} \Fe^{-\star} \dot{\Ce}\, \Fe^{-1}
	+ \Bv \!:\! \dot\Fv
	= K\Delta \vartheta + \llangle dK^\sharp , d\vartheta^\sharp \rrangle_{\mathbf{G}}\,,
\end{equation}
where $\Delta$ denotes the Laplace-Beltrami operator defined in~\eqref{eq:Laplacian}, and we recall that that partial derivative with respect to temperature $\vartheta$ is evaluated at fixed $\{\mathbf{F}, \Fv, \Fth, \Go, \mathbf{g}\}$, which in this example translates to fixing $\{R,r,r',\lvr,\lvt,\lvz,\omega\}$.

\paragraph{Boundary-value problem.}
The three unknowns of the problem are $\lvr(R,t)$, $\lvz(R,t)$, and $\vartheta(R,t)$. Once these fields are known, all remaining quantities follow explicitly: $r(R,t)$ from~\eqref{eq:eg1_r_solution}, $\rho$ from~\eqref{eq:eg1_rho}, $p(R,t)$ from~\eqref{eq:eg1_p}, and the Cauchy stress from~\eqref{eq:eg1_stress1}. The three unknowns are simultaneously determined by the viscous kinetic equations~\eqref{eq:eg1_kinetic} and the heat equation~\eqref{eq:eg1_Heat}, subject to the initial conditions
\begin{equation}\label{eq:eg1_Init_Cds}
	\lvr(R,0) = \lvz(R,0) = 1\,,\qquad
	\vartheta(R,0) = \vartheta_{\text{init}}(R)\,,
\end{equation}
where $\vartheta_{\text{init}}=\vartheta_{\text{init}}(R)$ is a given initial temperature distribution. We consider a convection boundary condition on the boundary of the cylinder, i.e.
\begin{equation}
\label{eq:eg1_BdCd}
\!\left[K\frac{\partial \vartheta}{\partial R}\right]\!_{(R_0,t)}=H_0[\vartheta_0-\vartheta(R_0,t)]\,,
\end{equation}
where $H_0$ is the surface heat transfer (constant) coefficient at the boundary of the cylinder.
To close the constitutive description, a specific form of the free
energy must be prescribed; we henceforth specialize to an
incompressible thermoviscoelastic neo-Hookean material.

\begin{example}[Neo-Hookean thermo-viscoelasticity with temperature-independent viscosity]
We consider an isotropic thermo-viscoelastic neo-Hookean solid, with a temperature-dependent equilibrium energy and a temperature-independent non-equilibrium energy, given by
\begin{equation}\label{eq:neo-Hook}
\Psi_{\mathrm{EQ}}=\overline{\Psi}(\vartheta,I_1,I_2)
=\frac{\mu(\vartheta)}{2\rho}(I_1-3)\,,
\qquad
\Psi_{\mathrm{NEQ}}=\widetilde{\Psi}(\Ie_1,\Ie_2)
=\frac{\mu_i}{2\rho}\,(\Ie_1-3)\,,
\end{equation}
where $\mu(\vartheta)=\mu_0\,\vartheta/\vartheta_0$ and $\mu_i$ are the shear moduli associated with the long-term (equilibrium) and instantaneous (non-equilibrium) responses, respectively, such that $\mu_0$ and $\mu_i$ are constants.
For~\eqref{eq:neo-Hook}, the relevant derivatives reduce to
\begin{equation}
\overline{\Psi}_1=\frac{\mu_0}{2\rho}\frac{\vartheta}{\vartheta_0}\,,\qquad
\widetilde{\Psi}_1=\frac{\mu_i}{2\rho}\,,\qquad
\overline{\Psi}_2=\widetilde{\Psi}_2=0\,.
\end{equation}
By collecting and isolating the equilibrium and non-equilibrium parts of the Cauchy stress~\eqref{eq:eg1_stress1} as discussed in Remark~\ref{rmrk:eg1-n-eq-separ}, one finds that
\begin{subequations}
\label{eq:eg1nH_stress}
\begin{align}
&\begin{dcases}
\hat\sigma^{rr}_{\text{eq}} &= - \int_R^{R_0} \mathcal{I}_{eq}(\xi,t)d\xi\,,
\end{dcases}
&&
\begin{dcases}
\hat\sigma^{rr}_{\text{neq}} &= - \int_R^{R_0} \mathcal{I}_{neq}(\xi,t)d\xi
\,,
\end{dcases}
\\
&
\begin{dcases}
\hat\sigma^{\theta\theta}_{\text{eq}} &= \mu_0\frac{\vartheta}{\vartheta_0} \frac{r^2}{R^2 e^{2\omega(\vartheta)}}-\mu_0 \frac{\vartheta}{\vartheta_0} \frac{e^{2\omega(\vartheta(R,t))} R^2}{r^2(R,t)}
\\&- \int_R^{R_0} \mathcal{I}_{eq}(\xi,t)d\xi\,,
\end{dcases}
&&
\begin{dcases}
\hat\sigma^{\theta\theta}_{\text{neq}} &= \mu_i \frac{r^2}{R^2\,\lvt^2 e^{2\omega(\vartheta)}} - \mu_i\frac{e^{2\omega(\vartheta(R,t))} R^2}{r^2(R,t)\,\lvr^2(R,t)}
\\&- \int_R^{R_0} \mathcal{I}_{neq}(\xi,t)d\xi\,,
\end{dcases}
\\
&
\begin{dcases}
\hat\sigma^{zz}_{\text{eq}} &= \mu_0\frac{\vartheta}{\vartheta_0}-\mu_0 \frac{\vartheta}{\vartheta_0} \frac{e^{2\omega(\vartheta(R,t))} R^2}{r^2(R,t)}
\\&- \int_R^{R_0} \mathcal{I}_{eq}(\xi,t)d\xi\,,
\end{dcases}
&&
\begin{dcases}
\hat\sigma^{zz}_{\text{neq}} &= \mu_i \frac{1}{\lvz^2} - \mu_i\frac{e^{2\omega(\vartheta(R,t))} R^2}{r^2(R,t)\,\lvr^2(R,t)}
\\&- \int_R^{R_0} \mathcal{I}_{neq}(\xi,t)d\xi\,,
\end{dcases}
\end{align}
\end{subequations}
with $\mathcal{I}_{\text{eq}}$ and $\mathcal{I}_{\text{neq}}$ given by
\begin{equation}
\mathcal{I}_{\text{eq}}(R,t) \coloneq 
\mu_0 \frac{\vartheta}{\vartheta_0} \frac{1}{R} \!\left( \frac{R^4 e^{4\omega}}{r^4} -1 \right)\!
\,,\qquad
\mathcal{I}_{neq}(R,t) \coloneq 
\mu_i \frac{1}{R} \!\left(
    \frac{R^4 e^{4\omega}}{r^4\,\lvr^2}
    - \frac{1}{\lvt^2}
  \right)\!
\,.
\end{equation}

Following the classical formulations of neo-Hookean thermoelastic solids \citep{ogden1972largecomp, ogden1972largeincomp, chadwick1974thermo, ogden1992thermoelastic, holzapfel1996entropy, Sadik2017Thermoelasticity}, the coefficient of (linear) thermal expansion is assumed to be given by\footnote{Note that \eqref{eq:alpha_theta} is singular at $\vartheta_*=\alpha_0\vartheta_0^2/(1+\alpha_0\vartheta_0)$, where the thermal stretch $e^{\omega}$ vanishes. For the parameter values of Table~3 one finds $\vartheta_*\approx 46$~K, well below the glass transition of a vulcanized rubber and far outside the temperature range considered here; the model is accordingly used well within its domain of validity.}
\begin{equation}\label{eq:alpha_theta}
\alpha(\vartheta)
=\frac{\alpha_0\vartheta_0^2}{(1+\alpha_0\vartheta_0)\vartheta^2-\alpha_0\vartheta_0^2\,\vartheta}\,,
\end{equation}
where $\alpha_0 = \alpha(\vartheta_0)$ is the coefficient of thermal expansion at the reference temperature $\vartheta_0$. Recall that $\vartheta_0$ is a uniform reference temperature at which no thermal distortion is present. Hence, $\omega$ is given by
\begin{equation}
	\omega(\vartheta) = \ln\!\left[1+\alpha_0\vartheta_0-\frac{\alpha_0\vartheta_0^2}{\vartheta}\right]\!\,.
\end{equation}

We assume that the dissipation potential~\eqref{eq:Quad-Rayleigh} is such that
\begin{equation}\label{eg1_Dissip}
\eta_1=\eta_2=0\,, \quad \eta_3 \eqcolon \eta > 0\,,
\end{equation}
which satisfies the convexity conditions~\eqref{eq:eta_cond}.
Thus, the viscous kinetic equations~\eqref{eq:eg1_kinetic} reduce to:
\begin{subequations} \label{eq:eg1_nH_kinetic}
\begin{align}
	&\left(\lvr^4\,\lvz^2 +1 \right)\! \frac{ \dot{\accentset{v}{\lambda}}_R}{\lvr}
	+ \frac{ \dot{\accentset{v}{\lambda}}_Z}{\lvz}=
	\frac{\mu_i}{\eta} \!\left( \frac{ R^2 e^{2\omega(\vartheta)} }{ r^2 }
	- \frac{ r^2 }{ R^2 e^{2\omega(\vartheta)} } \lvr^4\,\lvz^2 \right)\! \lvz^2\,,\\
	& \lvr\,\dot{\accentset{v}{\lambda}}_R- \lvz\,\dot{\accentset{v}{\lambda}}_Z
	=\frac{\mu_i}{\eta} \!\left(\frac{R^2 e^{2\omega(\vartheta)}}{r^2}\frac{1}{\lvr^2}
	- \frac{1}{\lvz^2}	\right)\!\,.
\end{align}
\end{subequations}
By inspection of the kinetic system~\eqref{eq:eg1_nH_kinetic}, the prefactor $\mu_i/\eta$ on the right-hand sides identifies a characteristic time scale for viscous relaxation,
\begin{equation}\label{eq:tau_v}
	\tau_v \coloneq \frac{\eta}{\mu_i}\,,
\end{equation}
which is the time over which the viscous stretches $\bar{\accentset{v}{\lambda}}_R$ and $\bar{\accentset{v}{\lambda}}_Z$ evolve appreciably in response to a mechanical imbalance, and such that for $t\gg\tau_v$ the non-equilibrium stress $\boldsymbol\sigma_{\mathrm{neq}}$ fully relaxes and tends towards $0$.

Let us now look at the heat equation~\eqref{eq:eg1_Heat}. Considering the terms containing the partial derivative with respect to temperature, which we recall is taken with fixed $\{R,r,r',\lvr,\lvt,\lvz,\omega\}$, it follows from~\eqref{eq:eg1nH_stress}\textemdash recalling Remark~\ref{rmrk:eg1-n-eq-separ}\textemdash that
\begin{subequations}
\begin{alignat}{2}
\frac{\partial \sigma^{rr}_{\text{eq}}}{\partial \vartheta} 
=\frac{\partial \!\left(\sigma^{rr}_{\text{eq}} -p_{\text{eq}}\right)\!}{\partial \vartheta} 
&= \mu_0\frac{1}{\vartheta_0} \frac{e^{2\omega(\vartheta)} R^2}{r^2}\,,
&\qquad
\frac{\partial \sigma^{rr}_{\text{neq}}}{\partial \vartheta} 
=\frac{\partial \!\left(\sigma^{rr}_{\text{neq}} -p_{\text{eq}}\right)\!}{\partial \vartheta} 
&= 0 \,,
\\
\frac{\partial \sigma^{\theta\theta}_{\text{eq}}}{\partial \vartheta} 
=\frac{\partial \!\left(\sigma^{\theta\theta}_{\text{eq}} -p_{\text{eq}}\right)\!}{\partial \vartheta} 
&= \mu_0\frac{1}{\vartheta_0} \frac{1}{R^2 e^{2\omega(\vartheta)}}\,,
&\qquad
\frac{\partial \sigma^{\theta\theta}_{\text{neq}}}{\partial \vartheta} 
=\frac{\partial \!\left(\sigma^{\theta\theta}_{\text{neq}} -p_{\text{eq}}\right)\!}{\partial \vartheta} 
&= 0 \,,
\\
\frac{\partial \sigma^{zz}_{\text{eq}}}{\partial \vartheta} 
=\frac{\partial \!\left(\sigma^{zz}_{\text{eq}} -p_{\text{eq}}\right)\!}{\partial \vartheta} 
&= \mu_0\frac{1}{\vartheta_0}\,,
&\qquad
\frac{\partial \sigma^{zz}_{\text{neq}}}{\partial \vartheta} 
=\frac{\partial \!\left(\sigma^{zz}_{\text{neq}} -p_{\text{eq}}\right)\!}{\partial \vartheta} 
&= 0 \,.
\end{alignat}
\end{subequations}
Hence the second term of~\eqref{eq:eg1_Heat} reads
\begin{equation}
\vartheta \frac{\partial \boldsymbol\sigma_{\textrm{eq}}}{\partial \vartheta} \!:\! \frac{1}{2} \Fve^{-\star} \dot{\Cve}\, \Fve^{-1}
	= \mu_0 \frac{\vartheta(R,t)}{\vartheta_0}
	\Bigg(\frac{e^{2\omega(\vartheta(R,t))} R^2}{r^2(R,t)} - \frac{r^2(R,t)}{R^2 e^{2\omega(\vartheta(R,t))}}\Bigg)\!\Bigg(\alpha[\vartheta(R,t)]\dot\vartheta(R,t) - \frac{\dot r(R,t)}{r(R,t)}\Bigg)
\,,
\end{equation}
while the third term of~\eqref{eq:eg1_Heat} in terms of $\boldsymbol\sigma_{\textrm{neq}}$ vanishes. Following the dissipation model~\eqref{eq:Quad-Rayleigh}, the fourth term yields
\begin{equation}
\begin{split}
\Bv \!:\! \dot\Fv
	= - \eta \!\left[ \dot{\accentset{v}{\lambda}}_R^2 + \dot{\accentset{v}{\lambda}}_Z^2 + \frac{\!\left(\lvz\,\dot{\accentset{v}{\lambda}}_R + \lvr\,\dot{\accentset{v}{\lambda}}_Z\right)^2}{\lvr^4\,\lvz^4} \right]\!
	\,.
\end{split}
\end{equation}
We further assume that the heat conduction coefficient is homogeneous and follows the empirical model for elastomer vulcanizates \citep{sircar1982thermal}
\begin{equation}
\label{eq:eg1_conduction_coeff}
K(\vartheta(R,t))=K_0\!\left[1-S(\vartheta(R,t)-\vartheta_0)\right]\!\,,
\end{equation}
where $K_0=K(\vartheta_0)>0$ and $S>0$ is a softening parameter. Therefore, the heat equation~\eqref{eq:eg1_Heat} is simplified to read
\begin{equation}
\begin{split}
\label{eq:eg1nH_Heat}
	&\rho_0 c_{\text D} \dot\vartheta
	- \mu_0 \frac{\vartheta}{\vartheta_0} \bigg(\frac{R^2 e^{2\omega(\vartheta)}}{r^2} - \frac{r^2}{R^2 e^{2\omega(\vartheta)}} \bigg)\!\bigg(\alpha(\vartheta)\dot\vartheta - \frac{\dot r}{r}\bigg) e^{2\omega(\vartheta)}
	\\
	&- \eta \!\left[ \dot{\accentset{v}{\lambda}}_R^2 + \dot{\accentset{v}{\lambda}}_Z^2 + \frac{\!\left(\lvz\,\dot{\accentset{v}{\lambda}}_R + \lvr\,\dot{\accentset{v}{\lambda}}_Z\right)^2}{\lvr^4\,\lvz^4} \right]\!  e^{2\omega(\vartheta)}
	= \!\left[1-S(\vartheta-\vartheta_0)\right]\!\frac{K_0}{R}\frac{\partial}{\partial R}\!\left(R\frac{\partial \vartheta}{\partial R}\right)\!\,.
\end{split}
\end{equation}
Inspection of the heat equation~\eqref{eq:eg1nH_Heat} reveals a characteristic time scale for thermal diffusion across the cylinder,
\begin{equation}\label{eq:tau_D}
  \tau_D \coloneq \frac{\rho_0\,c_{\mathrm D}\,R_0^2}{K_0}\,,
\end{equation}
which is the time over which temperature gradients of wavelength $R_0$ are smoothed out by conduction; for $t\gg\tau_D$ the temperature field equilibrates toward the boundary value $\vartheta_0$. The ratio $\tau_v/\tau_D$ of the two intrinsic time scales~\eqref{eq:tau_v} and~\eqref{eq:tau_D} governs the degree of coupling between the mechanical and thermal evolutions, and emerges naturally as a dimensionless parameter upon non-dimensionalization of the governing system.

Finally, we consider a thermal inclusion of radius $R_i<R_0$ as initial temperature distribution:
\begin{equation}
\label{Cy_Init-Temp}
\vartheta_{\text{init}}(R)=
\begin{dcases}
\vartheta_i & R\leq R_i\,,\\
\vartheta_0 & R>R_i\,,
\end{dcases}
\end{equation}
whereby the inner core $R\leq R_i$ is initialized at the elevated temperature $\vartheta_i>\vartheta_0$, while the outer annulus $R>R_i$ is at the stress-free reference temperature $\vartheta_0$. 
The initial temperature contrast between the inclusion and the surrounding cylinder then decays as heat diffuses through the cylinder and is transferred across the convective boundary, driving the system toward thermal equilibrium.
Our interest lies in the transient response: specifically, how the sudden thermal inclusion at $R=R_i$ generates a non-uniform eigenstrain field that sets the cylinder into a state of residual stress, how the non-equilibrium stress $\boldsymbol\sigma_{\mathrm{neq}}$ builds up and subsequently relaxes on the viscoelastic time scale $\tau_v$, and how these two processes---thermal diffusion on $\tau_D$ and viscous relaxation on $\tau_v$---interact throughout the transient as governed by the ratio $\tau_v/\tau_D$.

\paragraph{Dimensionless formulation.}
We introduce the following dimensionless rescalings
\begin{equation}\label{eq:eg1_dimless_vars}
\begin{gathered}
  \bar{R} = \frac{R}{R_0}\,,\quad
  \bar{r} = \frac{r}{R_0}\,,\quad
  \bar{t} = \frac{t}{\tau_v}\,,\quad
  \bar{\vartheta} = \frac{\vartheta}{\vartheta_0}\,,\quad
  \bar{\alpha} = \alpha\,\vartheta_0\,,\quad
  \bar{S} = S\,\vartheta_0\,,\quad
  \bar{\accentset{v}{\lambda}}_R(\bar{R},\bar{t}) = \lvr(R_0\bar{R},\,\tau_v\bar{t})\,,\\
  \bar{\accentset{v}{\lambda}}_Z(\bar{R},\bar{t}) = \lvz(R_0\bar{R},\,\tau_v\bar{t})\,,
\end{gathered}
\end{equation}
where $\tau_v$ is the viscoelastic relaxation time scale defined in~\eqref{eq:tau_v}.
It follows that the governing system for $\bar{\accentset{v}{\lambda}}_R(\bar{R},\bar{t})$, $\bar{\accentset{v}{\lambda}}_Z(\bar{R},\bar{t})$, and $\bar{\vartheta}(\bar{R},\bar{t})$, i.e.,~\eqref{eq:eg1_nH_kinetic} and~\eqref{eq:eg1nH_Heat}, reads in  dimensionless form as
\begin{subequations}\label{eq:eg1_dimless_system}
\begin{align}
\label{eq:eg1_k1_dimless}
&  \!\left(\bar{\accentset{v}{\lambda}}_R^4\bar{\accentset{v}{\lambda}}_Z^2+1\right)\!
  \frac{\dot{\bar{\accentset{v}{\lambda}}}_R}{\bar{\accentset{v}{\lambda}}_R}
  +\frac{\dot{\bar{\accentset{v}{\lambda}}}_Z}{\bar{\accentset{v}{\lambda}}_Z}
  = \!\left(
      \frac{\bar{R}^2e^{2\omega(\bar{\vartheta})}}{\bar{r}^2}
      -\frac{\bar{r}^2}{\bar{R}^2e^{2\omega(\bar{\vartheta})}}
      \bar{\accentset{v}{\lambda}}_R^4\bar{\accentset{v}{\lambda}}_Z^2
    \right)\!\bar{\accentset{v}{\lambda}}_Z^2\,,
  \\[0.5em]
\label{eq:eg1_k2_dimless}
&  \bar{\accentset{v}{\lambda}}_R\dot{\bar{\accentset{v}{\lambda}}}_R
  -\bar{\accentset{v}{\lambda}}_Z\dot{\bar{\accentset{v}{\lambda}}}_Z
  = \frac{\bar{R}^2e^{2\omega(\bar{\vartheta})}}{\bar{r}^2\bar{\accentset{v}{\lambda}}_R^2}
     -\frac{1}{\bar{\accentset{v}{\lambda}}_Z^2}\,,
  \\[0.5em]
\label{eq:eg1_heat_dimless}
  \begin{split}
&  \dot{\bar{\vartheta}}
  -\beta_0\bar{\vartheta}
	\!\left[\frac{\bar{R}^2 e^{2\omega(\bar{\vartheta})}}{\bar{r}^2}-\frac{\bar{r}^2}{\bar{R}^2 e^{2\omega(\bar{\vartheta})}}\right]\!
  \!\left[\bar{\alpha}(\bar{\vartheta})\dot{\bar{\vartheta}}-\frac{\dot{\bar{r}}}{\bar{r}}\right]\! e^{2\omega(\bar{\vartheta})}
  \\[0.3em]
& \quad -\upsilon_0\!\left[
    \dot{\bar{\accentset{v}{\lambda}}}_R^2
    +\dot{\bar{\accentset{v}{\lambda}}}_Z^2
    +\frac{\Bigl(\bar{\accentset{v}{\lambda}}_Z\dot{\bar{\accentset{v}{\lambda}}}_R+\bar{\accentset{v}{\lambda}}_R\dot{\bar{\accentset{v}{\lambda}}}_Z\Bigr)^2}{\bar{\accentset{v}{\lambda}}_R^4\bar{\accentset{v}{\lambda}}_Z^4}
  \right]\!e^{2\omega(\bar{\vartheta})}
  = \kappa_0\!\left[1-\bar{S}(\bar{\vartheta}-1)\right]\!\frac{1}{\bar{R}}\frac{\partial}{\partial\bar{R}}\!\left(\!\bar{R}\frac{\partial\bar{\vartheta}}{\partial\bar{R}}\right)\!\,,
  \end{split}
\end{align}
\end{subequations}
where $(\dot{\phantom{a}})=\partial/\partial\bar{t}$, and introducing the following dimensional groups:
\begin{equation}
\label{eq:dim_groups}
	\beta_0 = \frac{\mu_0}{\rho_0 c_D \vartheta_0}
	\,, \quad
	\upsilon_0 = \frac{\mu_i}{\rho_0 c_D \vartheta_0}
	\,, \quad
	\kappa_0 = \frac{\tau_v}{\tau_D}= \frac{K_0\eta}{\rho_0 c_D \mu_i R_0^2}\,,
\end{equation}
with $\tau_D$ being the thermal diffusion time scale defined in~~\eqref{eq:tau_D}.
The initial conditions~\eqref{eq:eg1_Init_Cds} read\footnote{Recall that the initial distribution follows from~\eqref{Cy_Init-Temp}.}
\begin{equation}\label{eq:eg1_dimless_ics}
  \bar{\vartheta}(\bar{R},0)=\begin{dcases} \bar{\vartheta}_i & \bar{R}\leq\bar{R}_i\,,\\ 1 & \bar{R}>\bar{R}_i\,,\end{dcases}\qquad
  \bar{\accentset{v}{\lambda}}_R(\bar{R},0) = \bar{\accentset{v}{\lambda}}_Z(\bar{R},0) = 1\,,
\end{equation}
where $\bar{\vartheta}_i=\vartheta_i/\vartheta_0$ and $\bar{R}_i=R_i/R_0$, and the boundary conditions follow from~\eqref{eq:eg1_BdCd} and the axisymmetry of the problem as\footnote{The condition $\left.\partial\bar{\vartheta}/\partial\bar{R}\right|_{\bar{R}=0}=0$ is not an independent physical prescription but follows from the axisymmetry of the problem: regularity of the temperature field at the cylinder axis requires the radial heat flux to vanish there.}
\begin{equation}\label{eq:eg1_dimless_bcs}
  \bigl[1-\bar{S}\,(\bar{\vartheta}(1,\bar{t})-1)\bigr]\left.\frac{\partial\bar{\vartheta}}{\partial\bar{R}}\right|_{\bar{R}=1} = B_i\,\bigl[1-\bar{\vartheta}(1,\bar{t})\bigr]\,,\qquad
  \left.\frac{\partial\bar{\vartheta}}{\partial\bar{R}}\right|_{\bar{R}=0}=0\,,
\end{equation}
where we introduce the Biot number
\begin{equation}
\label{eq:Biot_num}
B_i = \frac{H_0 R_0}{K_0}\,.
\end{equation}

The constitutive functions in terms of the dimensionless temperature $\bar{\vartheta}$ are
\begin{equation}\label{eq:eg1_dimless_constitutive}
  \omega(\bar{\vartheta}) = \ln\!\left[1+\bar{\alpha}_0\!\left(1-\frac{1}{\bar{\vartheta}}\right)\!\right]\!,\qquad
  \bar{\alpha}(\bar{\vartheta}) = \frac{\bar{\alpha}_0}{(1+\bar{\alpha}_0)\,\bar{\vartheta}^2-\bar{\alpha}_0\,\bar{\vartheta}}\,,
\end{equation}
where $\bar{\alpha}_0\coloneq\alpha_0\,\vartheta_0$. The deformed radius and its material time derivative are
\begin{equation}\label{eq:eg1_r_rdot_dimless}
  \bar{r}(\bar{R},\bar{t}) = \sqrt{\int_0^{\bar{R}} 2\,\bar{\xi}\,e^{2\omega[\bar{\vartheta}(\bar{\xi},\bar{t})]}\,d\bar{\xi}}\,,\qquad
  \dot{\bar{r}}(\bar{R},\bar{t}) = \frac{\displaystyle\int_0^{\bar{R}} 2\,\bar{\xi}\;\bar{\alpha}[\bar{\vartheta}(\bar{\xi},\bar{t})]\,\dot{\bar{\vartheta}}(\bar{\xi},\bar{t})\,e^{2\omega[\bar{\vartheta}(\bar{\xi},\bar{t})]}\,d\bar{\xi}}{\displaystyle\bar{r}(\bar{R},\bar{t})}\,.
\end{equation}

The dimensionless problem~\eqref{eq:eg1_dimless_vars}--\eqref{eq:Biot_num} is fully characterized by the eight independent parameters
\begin{equation}\label{eq:eg1_dimless_params}
	\left\{\bar{\alpha}_0,\;\beta_0,\;\upsilon_0,\;\kappa_0,\;\bar{S},\;B_i,\;\bar{\vartheta}_i,\;\bar{R}_i
	\,\right\}\,,
\end{equation}
whose definitions, physical interpretations, and adopted numerical values are collected in Table~\ref{tab:dimless_groups} following the dimensional parameters' values supplied in Table~\ref{tab:dim_params}.

\begin{table}
\centering
\caption{Dimensionless parameters of the thermoviscoelastic inclusion problem.}
\label{tab:dimless_groups}
\renewcommand{\arraystretch}{1.45}
\begin{tabular}{llll}
\hline
Parameter & Definition & Physical meaning & Value \\
\hline
$\bar{\alpha}_0$ & $\alpha_0\,\vartheta_0$ & Dimensionless thermal expansion at $\vartheta_0$ & $0.18$ \\
$\beta_0$ & $\mu_0/(\rho_0\,c_{\mathrm D}\,\vartheta_0)$ & Equilibrium elastic energy vs.\ heat capacity & $1$ \\
$\upsilon_0$ & $\mu_i/(\rho_0\,c_{\mathrm D}\,\vartheta_0)$ & Non-equilibrium elastic energy vs.\ heat capacity & $1.11$ \\
$\kappa_0$ & $\tau_v/\tau_D= K_0\eta/(\rho_0 c_D \mu_i R_0^2)$ & Viscoelastic relaxation vs.\ thermal diffusion & $0.02$ \\
$\bar{S}$ & $S\,\vartheta_0$ & Rate of conductivity decrease with temperature & $0.33$ \\
$B_i$ & $H_0\,R_0/K_0$ & Convective vs.\ conductive thermal resistance at $\bar{R}=1$ & $10$ \\
$\bar{\vartheta}_i$ & $\vartheta_i/\vartheta_0$ & Dimensionless initial inclusion temperature & $1.5$ \\
$\bar{R}_i$ & $R_i/R_0$ & Inclusion-to-cylinder radius ratio & $0.25$ \\
\hline
\end{tabular}
\end{table}

\vspace{1em}
\begin{table}
\centering
\caption{Dimensional material and geometric parameters from which the values in Table~\ref{tab:dimless_groups} are derived.}
\label{tab:dim_params}
\renewcommand{\arraystretch}{1.3}
\begin{tabular}{llll}
\hline
Parameter & Symbol & Value & Unit \\
\hline
Reference temperature        & $\vartheta_0$   & $300$                & K \\
Inclusion temperature        & $\vartheta_i$   & $450$                & K \\
Outer radius                 & $R_0$           & $0.15$                  & m \\
Inclusion radius             & $R_i$           & $0.25\,R_0$          & m \\
Volumetric heat capacity       & $\rho_0\,c_{\mathrm D}$        & $1.8\times10^{6}$                  & J\,m$^{-3}$\,K$^{-1}$ \\
Long-term shear modulus at $\vartheta_0$      & $\mu_0$    & $\approx 0.54$       & GPa \\
Instantaneous shear modulus  & $\mu_i$         & $\approx 0.60$       & GPa \\
Viscosity                    & $\eta$          & $3.24\times10^{12}$                  & Pa\,s \\
Thermal expansion at $\vartheta_0$ & $\alpha_0$ & $6\times10^{-4}$ & K$^{-1}$ \\
Heat conductivity  at $\vartheta_0$  & $K_0$           & $0.15$               & W\,m$^{-1}$\,K$^{-1}$ \\
Conductivity softening       & $S$ & $1.1\times10^{-3}$  & K$^{-1}$ \\
Surface heat-transfer coeff. & $H_0$           & $10$                 & W\,m$^{-2}$\,K$^{-1}$ \\
\hline
\end{tabular}
\end{table}

\vspace{1em}

\begin{remark}
It is worth pointing out some features of the chosen parameter regime.

\smallskip\noindent\emph{Strong thermomechanical coupling.}
The values $\beta_0=1$ and $\upsilon_0=1.11$ indicate that both the equilibrium and non-equilibrium elastic energy densities per unit reference volume are comparable to the thermal energy capacity $\rho_0\,c_{\mathrm D}\,\vartheta_0$. Accordingly, neither the thermoelastic-coupling term nor the viscous-dissipation term in~\eqref{eq:eg1_heat_dimless} is negligible \emph{a priori}, placing this example firmly in the strongly coupled regime. The ratio $\beta_0/\upsilon_0=\mu_0/\mu_i\approx0.90$ indicates that the material retains $90\%$ of its instantaneous stiffness at long times.

\smallskip\noindent\emph{Time-scale separation.}
The parameter $\kappa_0=\tau_v/\tau_D=0.02$ implies that the viscoelastic relaxation time~\eqref{eq:tau_v} is approximately fifty times shorter than the thermal diffusion time~\eqref{eq:tau_D}. In the dimensionless time $\bar{t}=t/\tau_v$, the simulation horizon therefore spans $O(\kappa_0^{-1})=O(50)$ thermal diffusion times, over which the non-equilibrium stress fully relaxes many times before any appreciable redistribution of the temperature field occurs.

\smallskip\noindent\emph{Conductivity softening and admissible temperatures.}
The softening parameter $\bar{S}=0.33$ implies that the effective conductivity $\bar{K}=K_0\bigl[1-\bar{S}\,(\bar{\vartheta}-1)\bigr]$ vanishes at $\bar{\vartheta}^*=1+\bar{S}^{-1} \approx 4$, which defines the upper admissibility bound for this conductivity model. The thermal expansion constitutive equation~\eqref{eq:alpha_theta} defines a lower admissibility bound: it is singular at $\bar{\vartheta}_*=\bar\alpha_0/(1+\bar\alpha_0)\approx0.15$, corresponding to approximately $46$~K, where the thermal stretch $e^{\omega}$ vanishes. Since $\bar{\vartheta}_i=1.5 < \bar{\vartheta}^*$, the conductivity remains strictly positive throughout the transient, although it is reduced by approximately $17\%$ at the center of the inclusion at $\bar{t}=0$.

\smallskip\noindent\emph{Convection-dominated boundary.}
The large Biot number $B_i=10$ means that the outer boundary $\bar{R}=1$ is effectively held close to $\bar{\vartheta}=1$: convective heat removal is ten times faster than conduction across the cylinder, so the rim cools rapidly once the thermal pulse begins to diffuse outward.

\smallskip\noindent\emph{Integro-PDE structure.}
The term $\dot{\bar{r}}/\bar{r}$ in~\eqref{eq:eg1_heat_dimless}, given by the nonlocal spatial integral in~\eqref{eq:eg1_r_rdot_dimless}, renders the heat equation nonlocally coupled in $\bar{R}$, reflecting the fact that the deformed radius at any material point depends on the entire temperature history of the core it encloses.

\end{remark}

\begin{figure}[t!]
    \centering
    \includegraphics[width=0.6\textwidth]{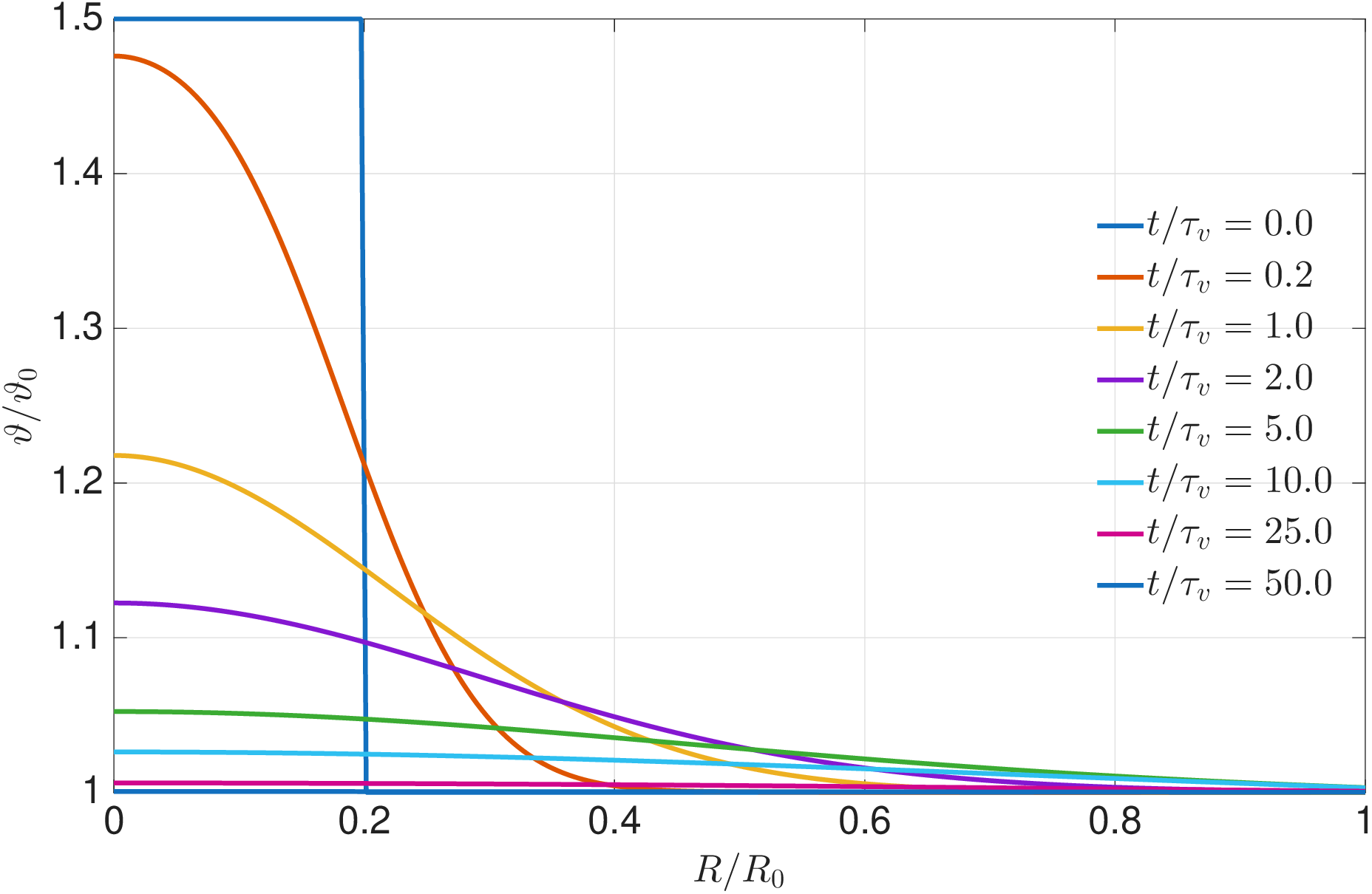}
    \caption{Temperature field $\vartheta(R,t)$ at selected times.}
    \label{fig:temperature}
\end{figure}


\begin{figure}
    \centering
    \begin{subfigure}[b]{0.32\textwidth}
        \centering
        \includegraphics[width=\textwidth]{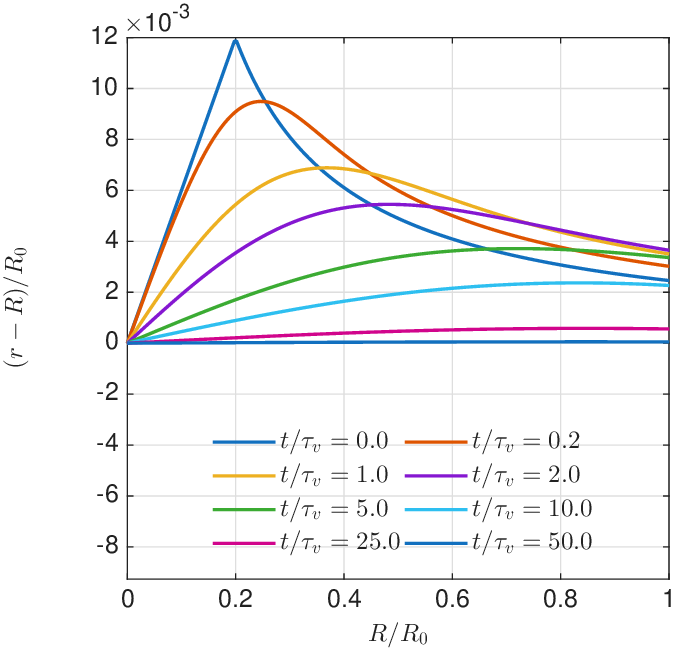}
        \caption{Radial displacement profiles}
        \label{fig:displacement_spatial}
    \end{subfigure}
    \hspace{0.05\textwidth}
    \begin{subfigure}[b]{0.32\textwidth}
        \centering
        \includegraphics[width=\textwidth]{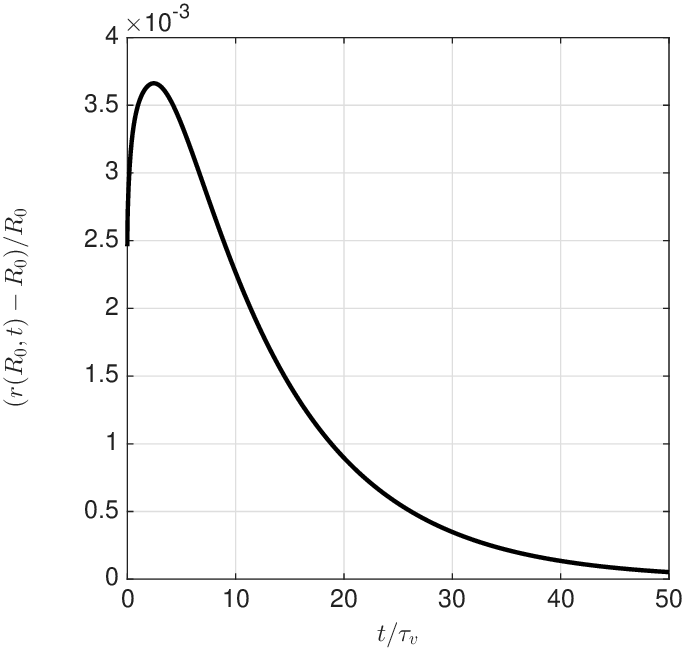}
        \caption{Outer wall radial displacement history}
        \label{fig:displacement_time}
    \end{subfigure}
    \caption{Deformed geometry of the cylinder during thermal relaxation.}
    \label{fig:deformed_geometry}
\end{figure}
\begin{figure}
    \centering
    \begin{subfigure}[b]{0.32\textwidth}
        \includegraphics[width=\textwidth]{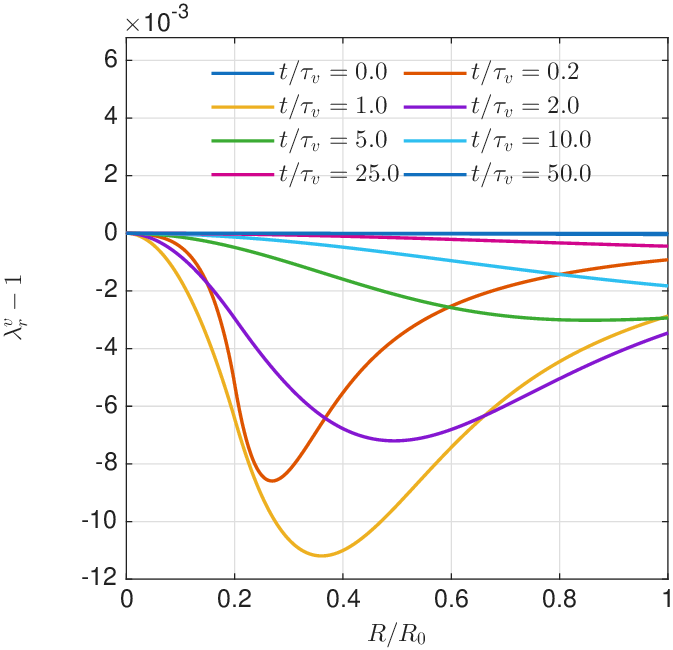}
        \caption{Radial viscous distortion}
        \label{fig:kinematics-a}
    \end{subfigure}
    \hfill
    \begin{subfigure}[b]{0.32\textwidth}
        \includegraphics[width=\textwidth]{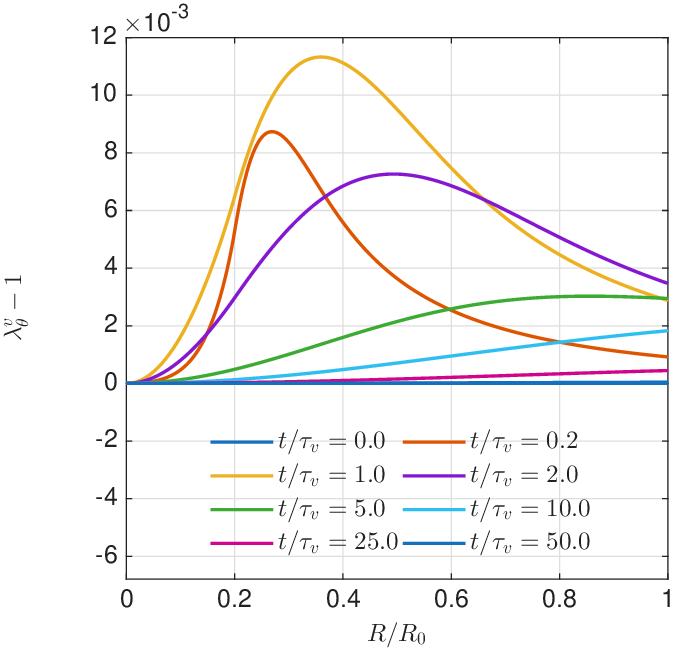}
        \caption{Azimuthal viscous distortion}
        \label{fig:kinematics-b}
    \end{subfigure}
    \hfill
    \begin{subfigure}[b]{0.32\textwidth}
        \includegraphics[width=\textwidth]{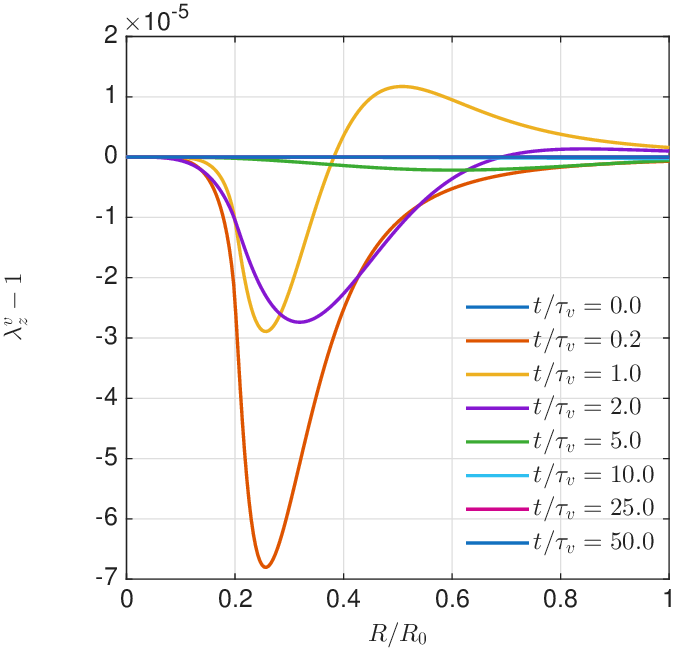}
        \caption{Axial viscous distortion}
        \label{fig:kinematics-c}
    \end{subfigure}
    \\[1em]
    \begin{subfigure}[b]{0.32\textwidth}
        \includegraphics[width=\textwidth]{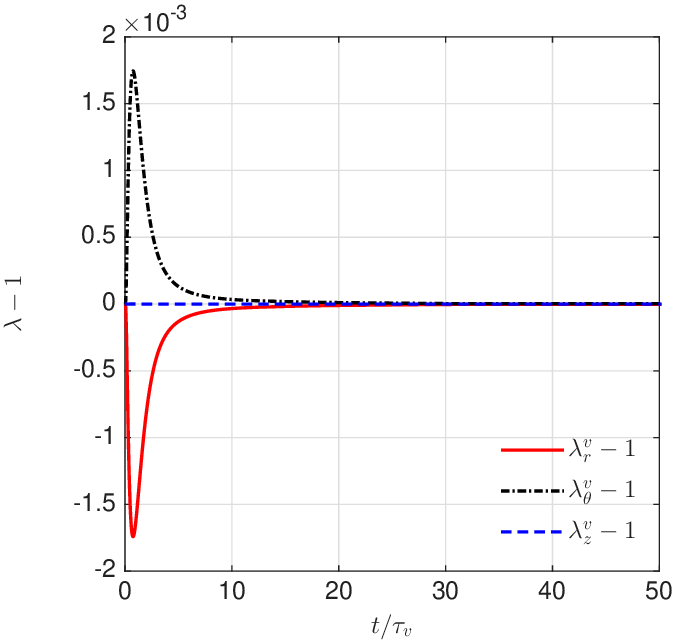}
        \caption{Stretches at $R=0.10\,R_0$}
        \label{fig:kinematics-d}
    \end{subfigure}
    \hfill
    \begin{subfigure}[b]{0.32\textwidth}
        \includegraphics[width=\textwidth]{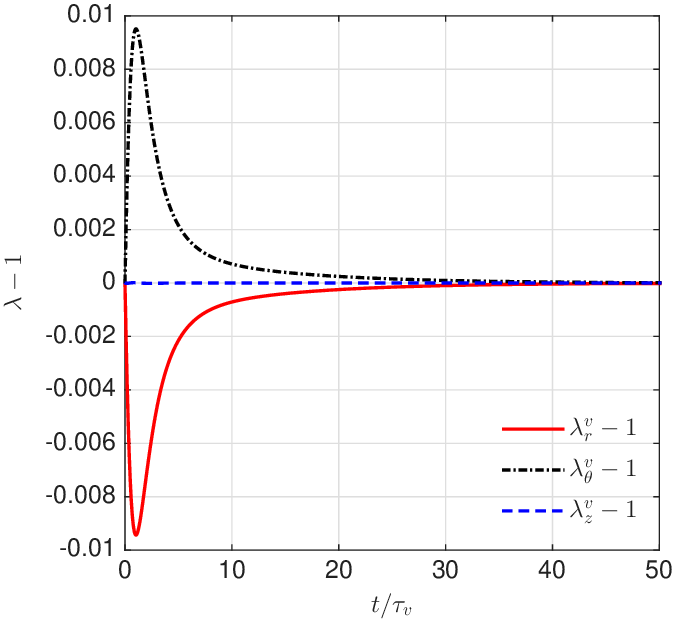}
        \caption{Stretches at $R=0.50\,R_0$}
        \label{fig:kinematics-e}
    \end{subfigure}
    \hfill
    \begin{subfigure}[b]{0.32\textwidth}
        \includegraphics[width=\textwidth]{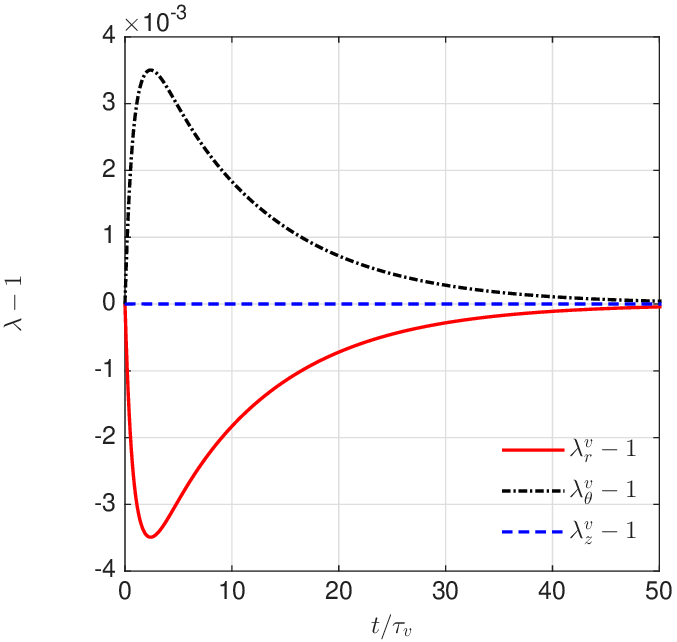}
        \caption{Stretches at $R=1.00\,R_0$}
        \label{fig:kinematics-f}
    \end{subfigure}
    \caption{Viscous stretches.
             Top row: spatial profiles at selected times.
             Bottom row: time evolution at fixed material radii.}
    \label{fig:kinematics}
\end{figure}

\begin{figure}
    \centering
    \begin{subfigure}[b]{0.32\textwidth}
        \includegraphics[width=\textwidth]{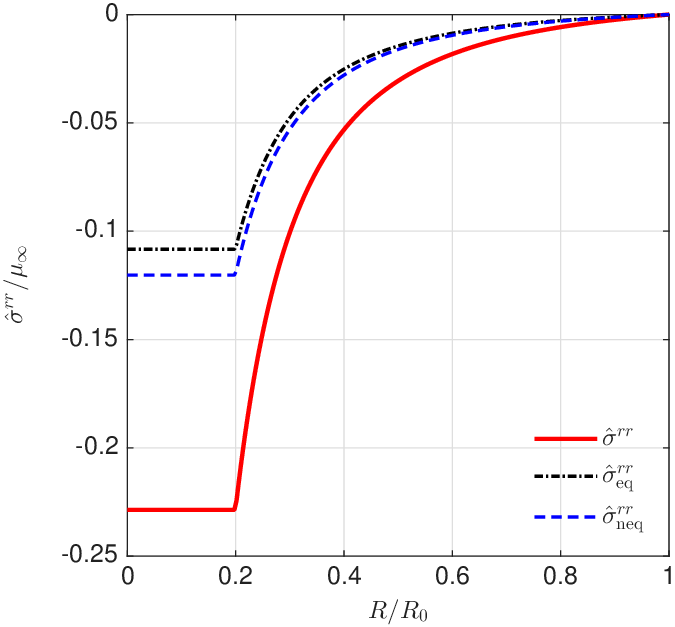}
        \caption{Radial stress}
        \label{fig:snapshot-a}
    \end{subfigure}
    \hfill
    \begin{subfigure}[b]{0.32\textwidth}
        \includegraphics[width=\textwidth]{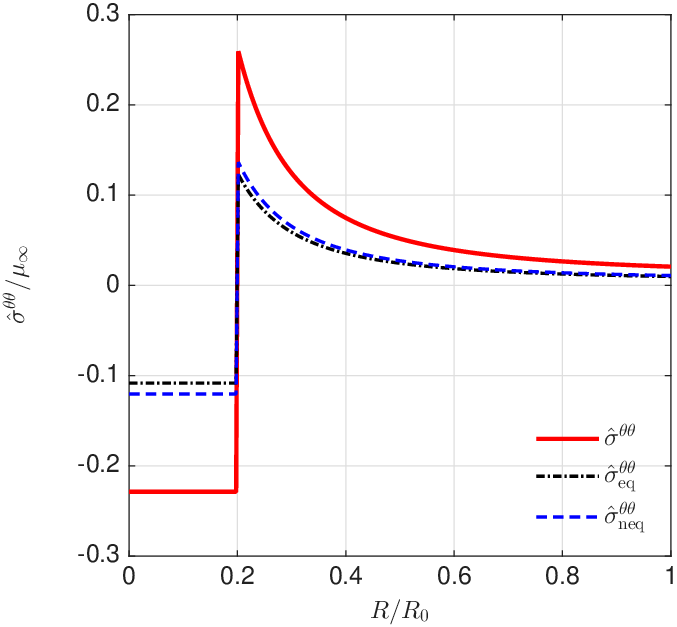}
        \caption{Azimuthal stress}
        \label{fig:snapshot-b}
    \end{subfigure}
    \hfill
    \begin{subfigure}[b]{0.32\textwidth}
        \includegraphics[width=\textwidth]{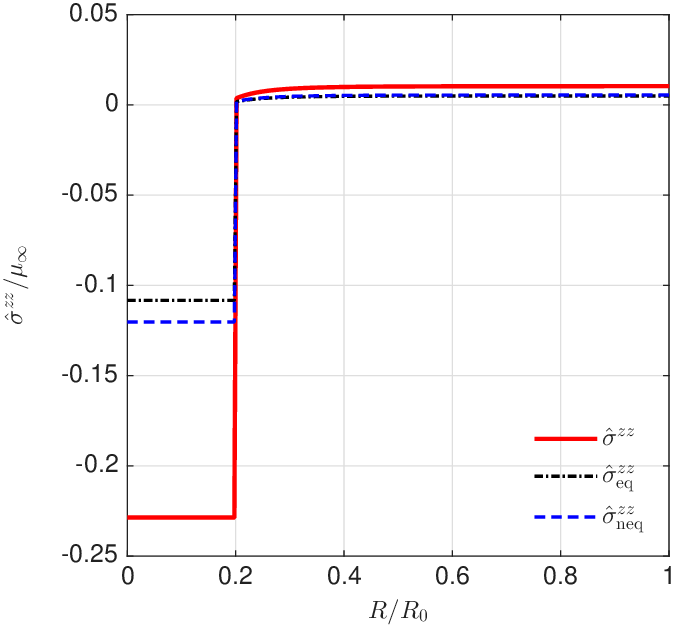}
        \caption{Axial stress}
        \label{fig:snapshot-c}
    \end{subfigure}
    \caption{Snapshot of stresses at $t/\tau_v = 0$.}
    \label{fig:snapshot}
\end{figure}

\vspace{1em}
\begin{figure}
    \centering
    \begin{subfigure}[b]{0.32\textwidth}
        \includegraphics[width=\textwidth]{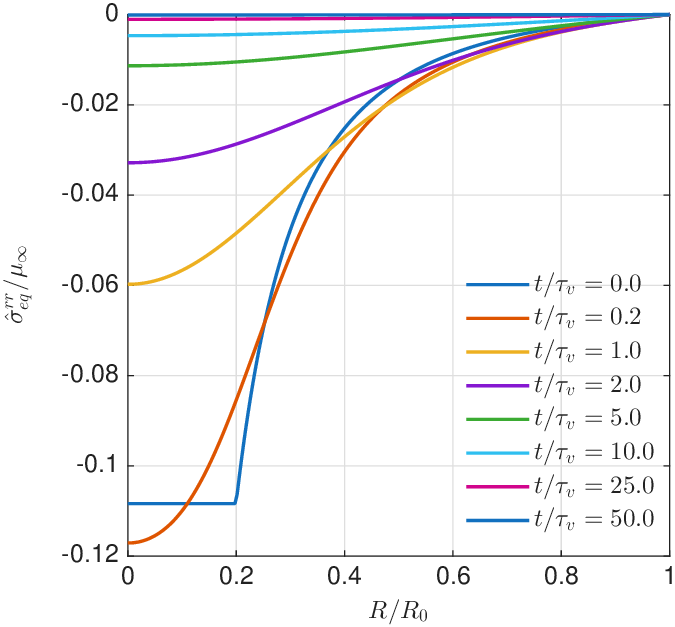}
        \caption{Equilibrium radial stress}
        \label{fig:stress_rr-a}
    \end{subfigure}
    \hfill
    \begin{subfigure}[b]{0.32\textwidth}
        \includegraphics[width=\textwidth]{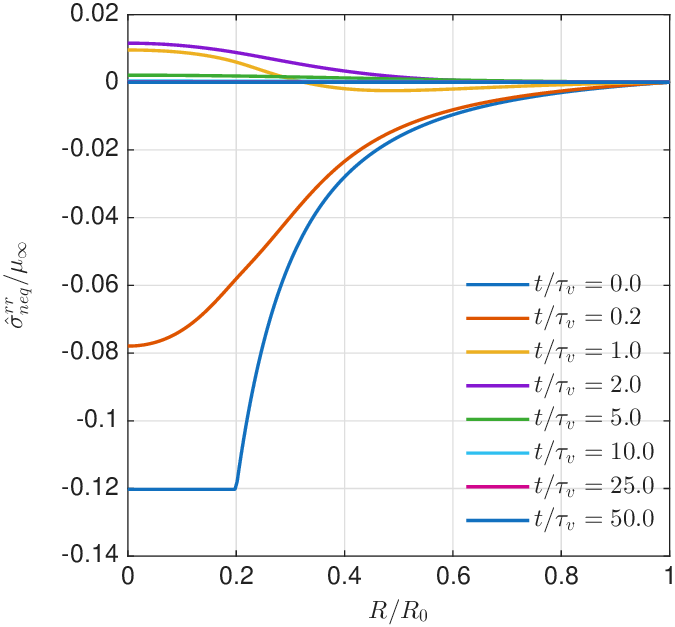}
        \caption{Non-equilibrium radial stress}
        \label{fig:stress_rr-b}
    \end{subfigure}
    \hfill
    \begin{subfigure}[b]{0.32\textwidth}
        \includegraphics[width=\textwidth]{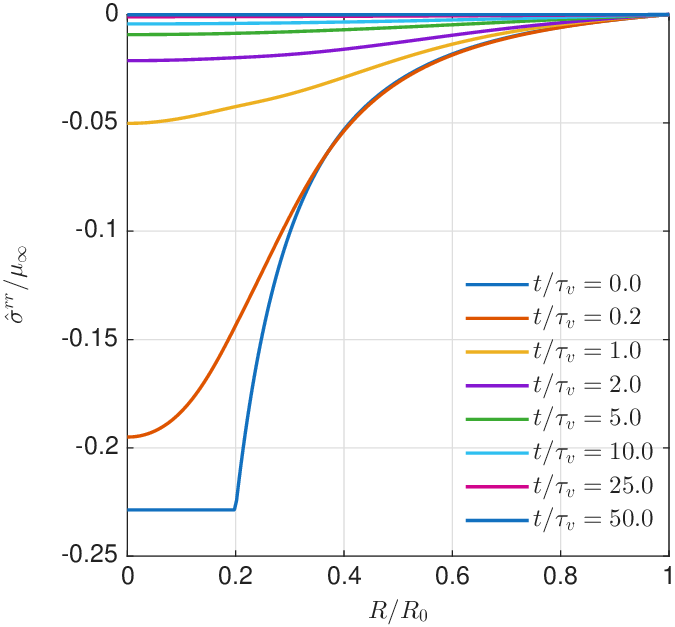}
        \caption{Total radial stress}
        \label{fig:stress_rr-c}
    \end{subfigure}
    \\[1em]
    \begin{subfigure}[b]{0.32\textwidth}
        \includegraphics[width=\textwidth]{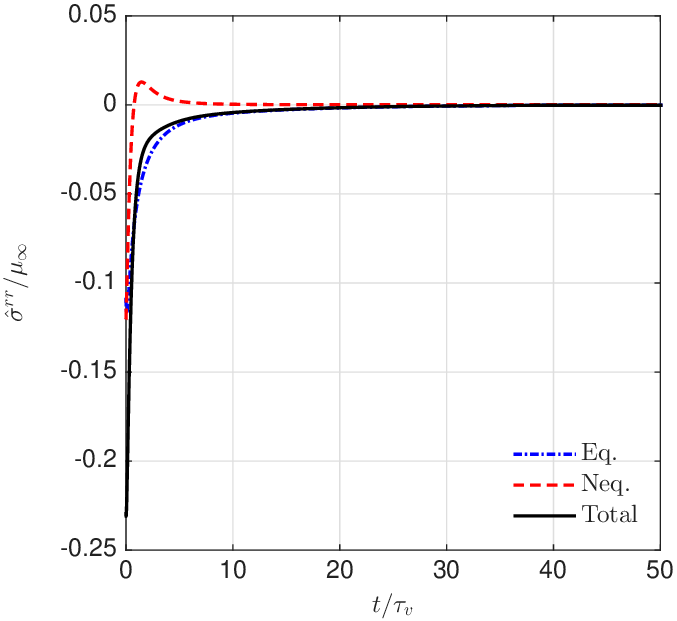}
        \caption{Radial stress at $R=0.10\,R_0$}
        \label{fig:stress_rr-d}
    \end{subfigure}
    \hfill
    \begin{subfigure}[b]{0.32\textwidth}
        \includegraphics[width=\textwidth]{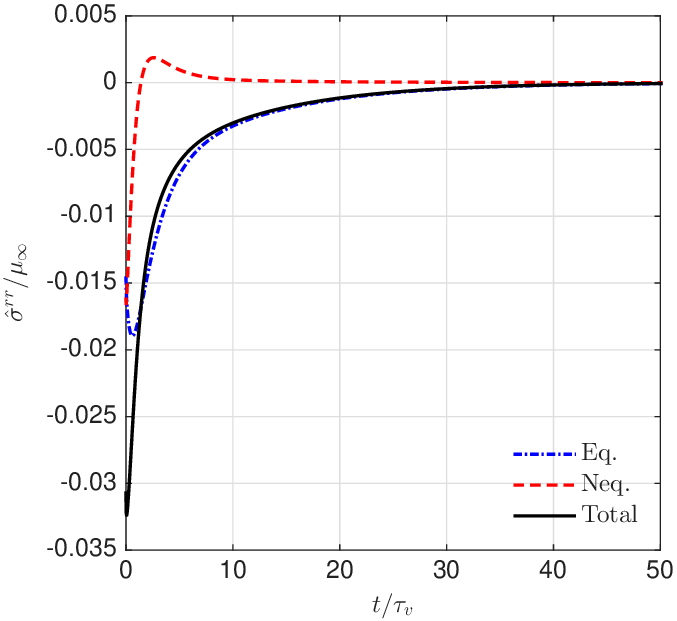}
        \caption{Radial stress at $R=0.50\,R_0$}
        \label{fig:stress_rr-e}
    \end{subfigure}
    \hfill
    \begin{subfigure}[b]{0.32\textwidth}
        \includegraphics[width=\textwidth]{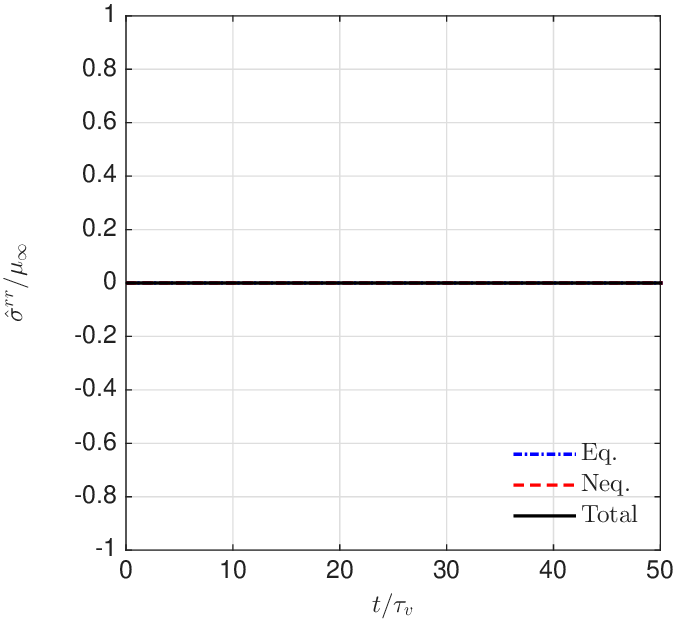}
        \caption{Radial stress at $R=1.00\,R_0$}
        \label{fig:stress_rr-f}
    \end{subfigure}
    \caption{Radial stress.
             Top row: spatial profiles at selected times.
             Bottom row: time evolution at fixed material radii.}
    \label{fig:stress_rr}
\end{figure}

\begin{figure}
    \centering
    \begin{subfigure}[b]{0.32\textwidth}
        \includegraphics[width=\textwidth]{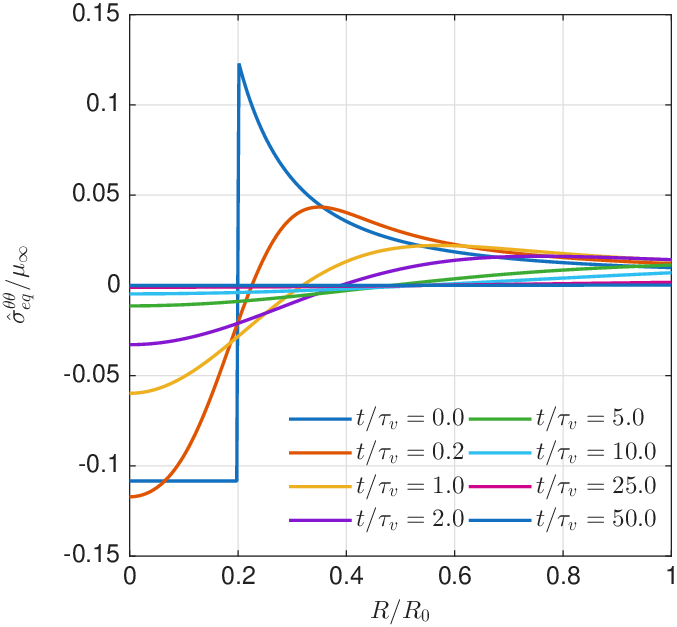}
        \caption{Equilibrium azimuthal stress}
        \label{fig:stress_tt-a}
    \end{subfigure}
    \hfill
    \begin{subfigure}[b]{0.32\textwidth}
        \includegraphics[width=\textwidth]{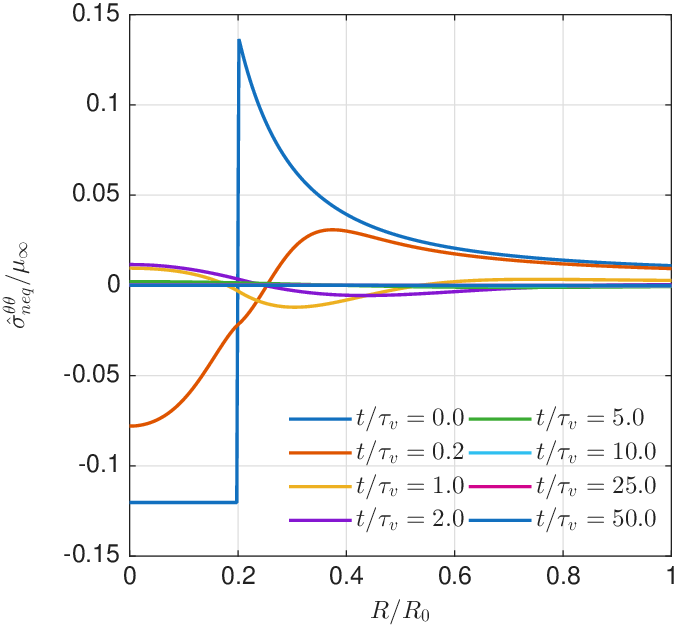}
        \caption{Non-equilibrium azimuthal stress}
        \label{fig:stress_tt-b}
    \end{subfigure}
    \hfill
    \begin{subfigure}[b]{0.32\textwidth}
        \includegraphics[width=\textwidth]{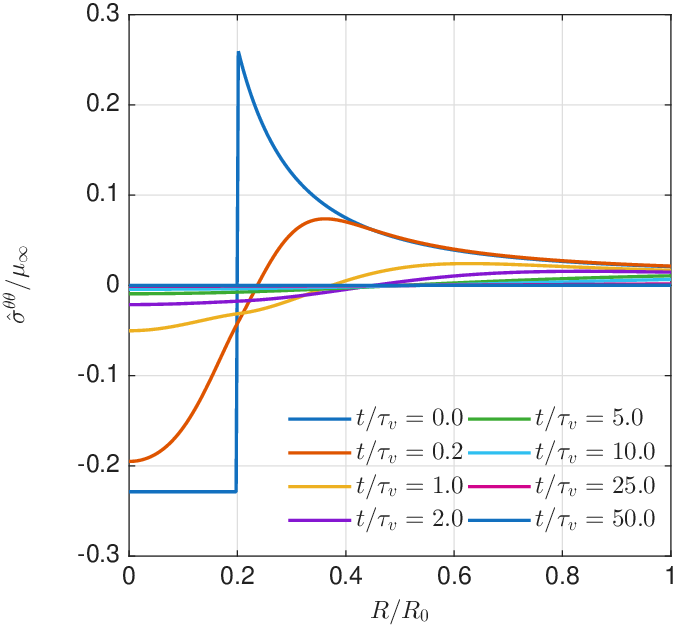}
        \caption{Total azimuthal stress}
        \label{fig:stress_tt-c}
    \end{subfigure}
    \\[1em]
    \begin{subfigure}[b]{0.32\textwidth}
        \includegraphics[width=\textwidth]{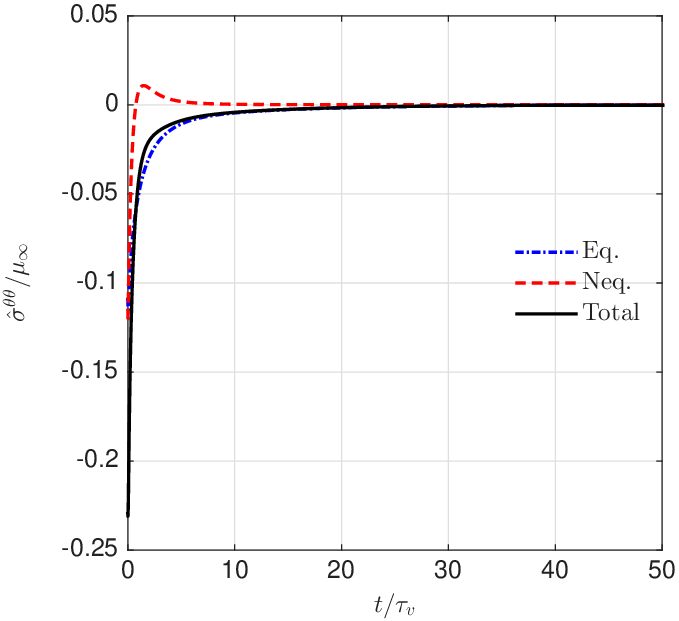}
        \caption{Azimuthal stress at $R=0.10\,R_0$}
        \label{fig:stress_tt-d}
    \end{subfigure}
    \hfill
    \begin{subfigure}[b]{0.32\textwidth}
        \includegraphics[width=\textwidth]{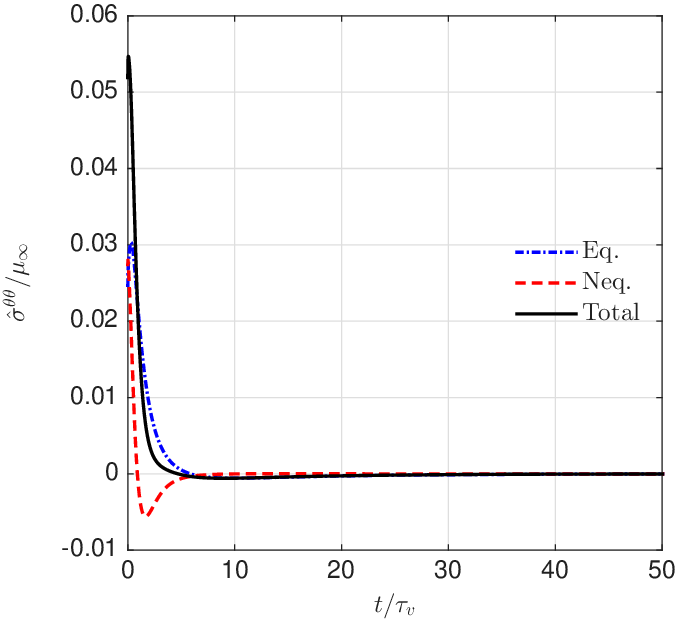}
        \caption{Azimuthal stress at $R=0.50\,R_0$}
        \label{fig:stress_tt-e}
    \end{subfigure}
    \hfill
    \begin{subfigure}[b]{0.32\textwidth}
        \includegraphics[width=\textwidth]{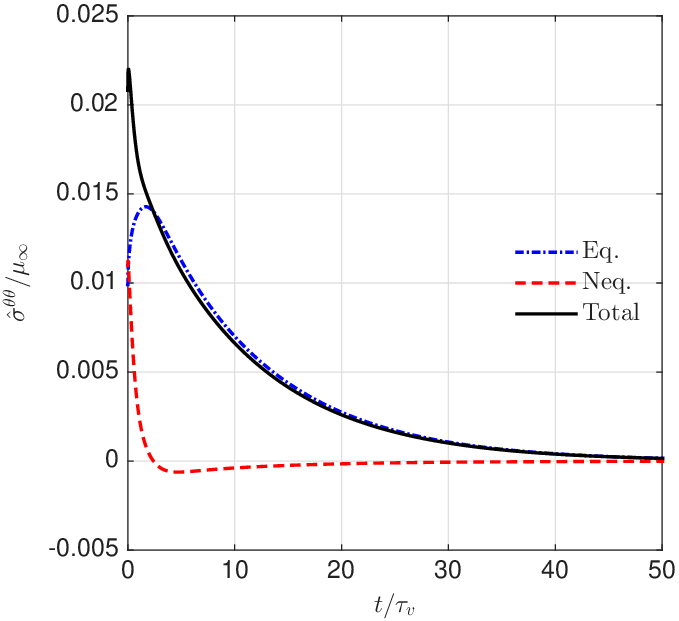}
        \caption{Azimuthal stress at $R=1.00\,R_0$}
        \label{fig:stress_tt-f}
    \end{subfigure}
    \caption{Azimuthal stress.
             Top row: spatial profiles at selected times.
             Bottom row: time evolution at fixed material radii.}
    \label{fig:stress_tt}
\end{figure}

\begin{figure}
    \centering
    \begin{subfigure}[b]{0.32\textwidth}
        \includegraphics[width=\textwidth]{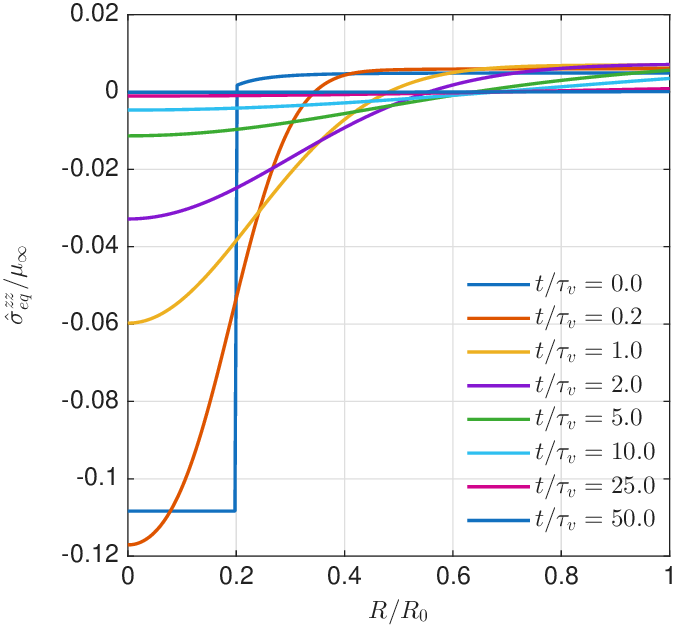}
        \caption{Equilibrium axial stress}
        \label{fig:stress_zz-a}
    \end{subfigure}
    \hfill
    \begin{subfigure}[b]{0.32\textwidth}
        \includegraphics[width=\textwidth]{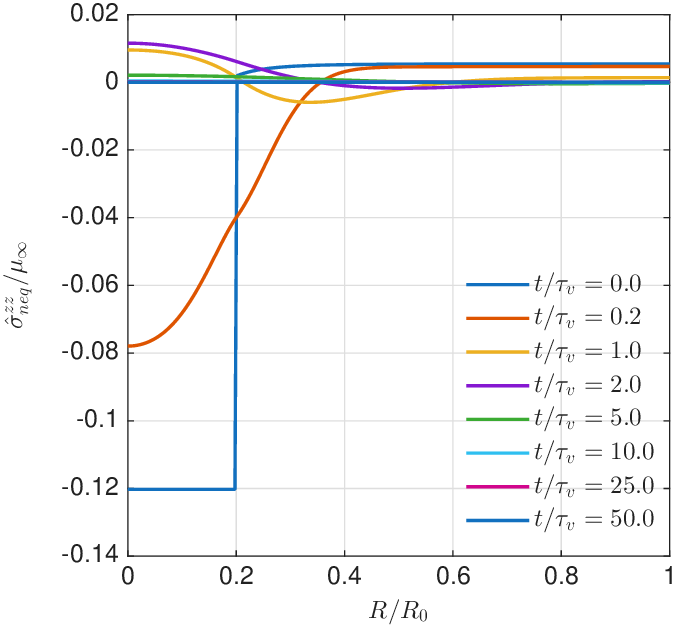}
        \caption{Non-equilibrium axial stress}
        \label{fig:stress_zz-b}
    \end{subfigure}
    \hfill
    \begin{subfigure}[b]{0.32\textwidth}
        \includegraphics[width=\textwidth]{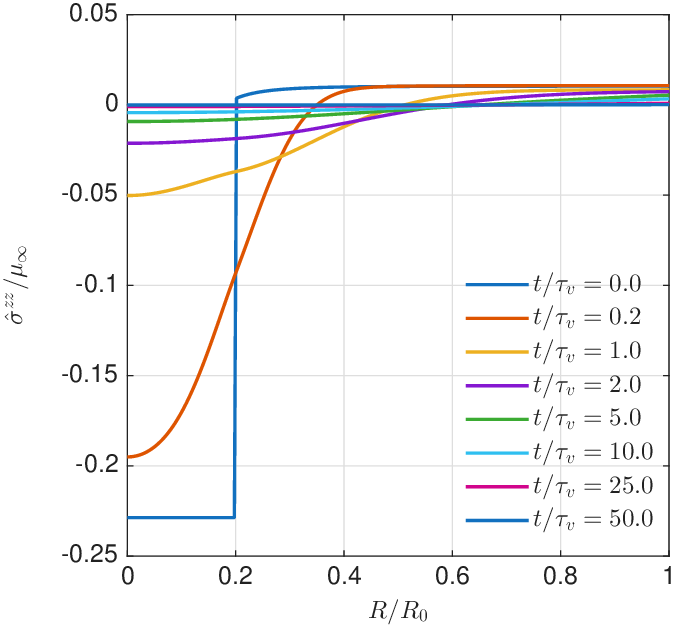}
        \caption{Total axial stress}
        \label{fig:stress_zz-c}
    \end{subfigure}
    \\[1em]
    \begin{subfigure}[b]{0.32\textwidth}
        \includegraphics[width=\textwidth]{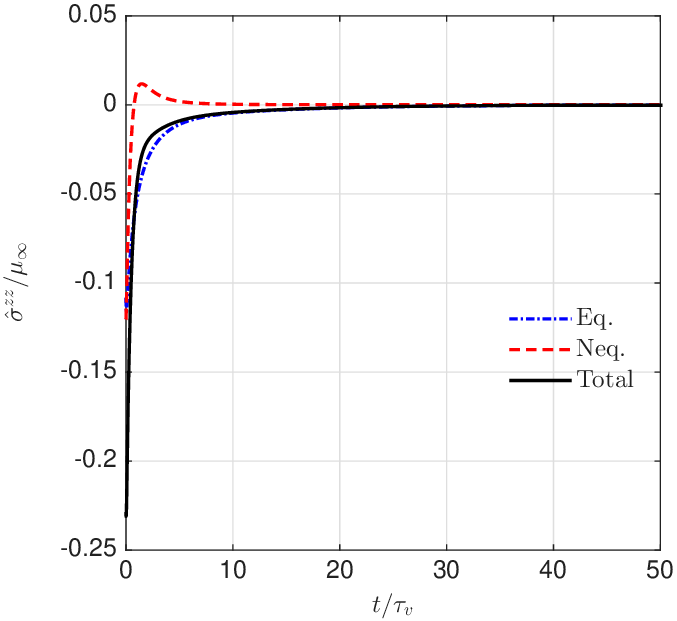}
        \caption{Axial stress at $R=0.10\,R_0$}
        \label{fig:stress_zz-d}
    \end{subfigure}
    \hfill
    \begin{subfigure}[b]{0.32\textwidth}
        \includegraphics[width=\textwidth]{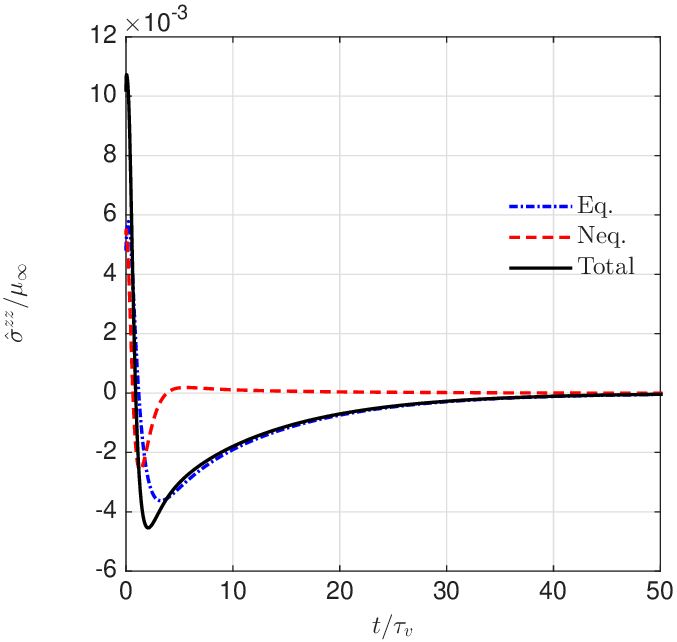}
        \caption{Axial stress at $R=0.50\,R_0$}
        \label{fig:stress_zz-e}
    \end{subfigure}
    \hfill
    \begin{subfigure}[b]{0.32\textwidth}
        \includegraphics[width=\textwidth]{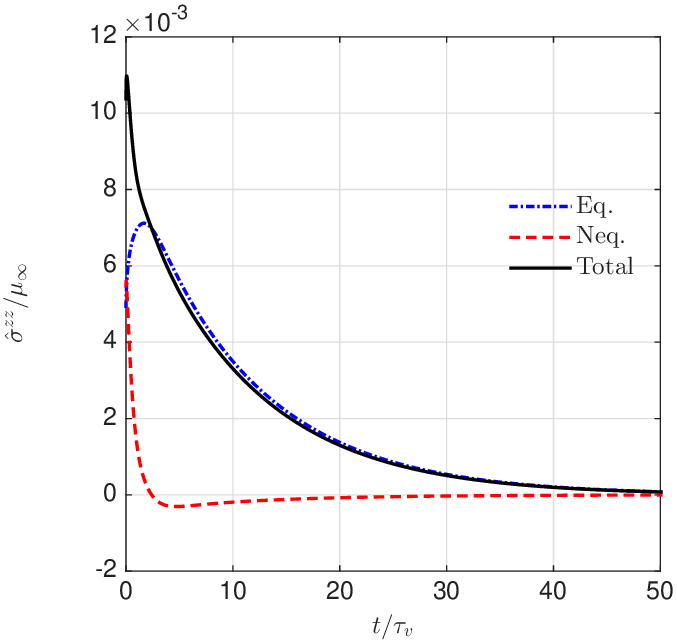}
        \caption{Axial stress at $R=1.00\,R_0$}
        \label{fig:stress_zz-f}
    \end{subfigure}
    \caption{Axial stress.
             Top row: spatial profiles at selected times.
             Bottom row: time evolution at fixed material radii.}
    \label{fig:stress_zz}
\end{figure}

\paragraph{Discussion of the numerical results.}
We solve the dimensionless governing system~\eqref{eq:eg1_dimless_system} numerically by discretizing the spatial domain $\bar R\in[0,1]$ with a uniform finite-difference grid and advancing in time with an implicit backward-Euler scheme; at each time step the nonlinear algebraic system is solved by Newton iteration, with the integrals in~\eqref{eq:eg1_r_rdot_dimless} evaluated by the trapezoidal rule.

\smallskip\noindent\textbf{Temperature.}
The temperature field evolves from the sharp inclusion profile~\eqref{Cy_Init-Temp} toward the spatially uniform reference value $\vartheta=\vartheta_0$ imposed by the convective boundary\textemdash see\ Fig.~\ref{fig:temperature}. The gradient is initially concentrated near the inclusion interface at $R_i=0.25R_0$; the temperature-dependent conductivity~\eqref{eq:eg1_conduction_coeff} with $S\vartheta_0=0.33$ reduces the effective diffusivity inside the hot core, slightly delaying the outward spread of the pulse. The relatively large Biot number $B_i=10$ emerging from the convection boundary condition~\eqref{eq:eg1_BdCd} keeps the outer rim close to $\vartheta=\vartheta_0$ throughout. With $\tau_v/\tau_D=0.02$, the thermal diffusion time is fifty times longer than the viscoelastic relaxation time; so, although the temperature profile is nearly uniform at $t=25\tau_v$, complete thermal equilibration occurs only at $t\sim\tau_D=50\tau_v$, consistent with the slow decay of the outer curves visible in Fig.~\ref{fig:temperature}.

\smallskip\noindent\textbf{Deformed geometry.}
The current radial coordinate $r(R,t)$ encodes the net mechanical response of the cylinder to the transient thermal field.
Figure~\ref{fig:displacement_spatial} shows the normalized radial displacement $(r-R)/R_0$ as a function of the normalized reference coordinate $R/R_0$ at the same time instants as the temperature profiles.
At $t=0$ the displacement profile peaks sharply near the inclusion interface $R_i=0.25R_0$, reflecting the sudden volumetric misfit imposed by the initial temperature jump~\eqref{Cy_Init-Temp}: material points inside the hot core are pushed radially outward, while those in the cooler outer annulus are displaced inward by the radial compression, consistent with the non-positive $\hat\sigma^{rr}$ discussed below.
As heat diffuses outward the peak broadens and migrates toward larger $R$, tracking the retreating temperature gradient, and the displacement amplitude decays monotonically until the cylinder returns to its natural undistorted state at $t\sim\tau_D$.

The outer-wall normalized displacement $r(R_0,t)/R_0 - 1$, shown in Fig.~\ref{fig:displacement_time}, characterizes the global swelling of the cylinder and provides a single scalar observable accessible to experiment.
It rises sharply from zero at $t=0$\textemdash the outer boundary is initially undeformed because the thermal inclusion is entirely interior\textemdash reaches a maximum near $t\approx 3\tau_v$, and subsequently decays on the thermal diffusion time scale $\tau_D = 50\tau_v$.
The early rise reflects the outward propagation of displacement driven by the non-equilibrium stress. The peak occurs approximately when viscous relaxation and thermal diffusion compete most strongly: $\hat{\sigma}^{rr}_{\mathrm{neq}}$ is already decaying, while the equilibrium thermal driving has not yet diminished appreciably.
The long algebraic tail confirms that, once the non-equilibrium part has fully relaxed ($t\gtrsim 5\tau_v$), the outer wall displacement is slaved entirely to the slow thermal decay, providing a direct macroscopic signature of the separation of time scales $\tau_v/\tau_D = 0.02$.

\smallskip\noindent\textbf{Viscous kinematics.}
The viscous stretches $\lambda^v_R$, $\lambda^v_\theta$, and $\lambda^v_z$ start from trivial undeformed state at unity and grow as the initial thermal inclusion~\eqref{Cy_Init-Temp} drives the kinetic equations~\eqref{eq:eg1_nH_kinetic}. Viscous flow is initially confined near the inclusion interface but spreads radially as the temperature gradient diffuses, as seen in Figs.~\ref{fig:kinematics-a}--\ref{fig:kinematics-c}. Radial and azimuthal components have opposite signs, consistent with viscous incompressibility, while the axial component is two orders of magnitude smaller (note the $\times10^{-5}$ scale in Fig.~\ref{fig:kinematics-c}), mirroring the macroscopic plane-strain constraint of the infinite bar. The evolution plots at fixed material points (Figs.~\ref{fig:kinematics-d}--\ref{fig:kinematics-f}) show that the stretches peak near $t\approx\tau_v$ and decay toward unity on the viscoelastic time scale, confirming that non-equilibrium kinematic distortion is confined to the early window of order $\mathcal{O}(\tau_v)$, significantly before any appreciable thermal redistribution has taken place\textemdash inline indeed with $\tau_v/\tau_D=0.02$.

\smallskip\noindent\textbf{Stresses.}
\smallskip\noindent\emph{Initial snapshot.}
At $t=0$ the viscous stretches are identically unity and the non-equilibrium stress response is equivalent to the purely elastic response to the thermal inclusion~\eqref{Cy_Init-Temp}\textemdash cf. \citep{Sadik2017Thermoelasticity}. The snapshots in Figs.~\ref{fig:snapshot-a}--\ref{fig:snapshot-c} reveal two features common to all three stress components: (i)~a hydrostatic state of stress inside the inclusion core\textemdash as previously observed in analogous non-viscous problems \citep{yavari2013nonlinear, Sadik2017Thermoelasticity, golgoon2018nonlinear}, and (ii)~the sharp inclusion interface at $R=R_i$ produces a kink in the stress gradient, arising from the discontinuity of the initial temperature field and the consequent jump in the thermal expansion factor $e^{2\omega}$.

\smallskip\noindent\emph{Radial stress.}
The total radial stress $\hat\sigma^{rr}$ is everywhere non-positive for all $t$: the thermally expanded core is radially compressed against the cooler outer annulus. 
As seen in~\eqref{eq:eg1nH_stress}, the equilibrium contribution $\hat{\sigma}^{rr}_{\mathrm{eq}}$ tracks the instantaneous temperature through the factor $\vartheta/\vartheta_0$ in the temperature-dependent neo-Hookean modulus. It decays monotonically as heat diffuses outward (Fig.~\ref{fig:stress_rr-a}). By contrast, the non-equilibrium contribution $\hat{\sigma}^{rr}_{\mathrm{neq}}$ is driven by the departure of $\bar{\lambda}^v_R$ from unity. It spikes at early times and relaxes near $t\approx\tau_v$. Thus, after a few multiples of $\tau_v$, the radial stress is dominated entirely by its equilibrium part (Fig.~\ref{fig:stress_rr-b}).
Because $\tau_v/\tau_D=0.02$, this mechanical relaxation is completed long before any appreciable thermal redistribution occurs, and the total radial stress (Fig.~\ref{fig:stress_rr-c}) subsequently decays only as fast as the temperature field itself.
At the outer boundary $R=R_0$, the radial stress is identically zero by construction, and the time history there (Fig.~\ref{fig:stress_rr-f}) confirms numerical accuracy of the spatial integration. The time histories at the inclusion center and mid-domain (Figs.~\ref{fig:stress_rr-d}--\ref{fig:stress_rr-e}) clearly display the fast non-equilibrium spike followed by the slow thermal decay.

\smallskip\noindent\emph{Azimuthal stress.}
The azimuthal stress $\hat\sigma^{\theta\theta}$ exhibits a sign change across the inclusion interface (Figs.~\ref{fig:stress_tt-a}--\ref{fig:stress_tt-c}): the hot core is in tension ($\hat\sigma^{\theta\theta}>0$) while the cooler outer annulus is in compression ($\hat\sigma^{\theta\theta}<0$), a pattern typical of a misfitting inclusion. Both $\hat\sigma^{\theta\theta}_{\mathrm{eq}}$ and $\hat\sigma^{\theta\theta}_{\mathrm{neq}}$ are of comparable magnitude at $t=0$, consistent with the ratio $\mu_0/\mu_i\approx0.90$ (Figs.~\ref{fig:stress_tt-a}--\ref{fig:stress_tt-b}). The non-equilibrium part relaxes near $t\approx\tau_v$ and briefly features a spatial oscillatory pattern about zero near the inclusion interface before settling, while the equilibrium part diminishes slowly as the temperature profile flattens (Fig.~\ref{fig:stress_tt-c}). The time histories (Figs.~\ref{fig:stress_tt-d}--\ref{fig:stress_tt-f}) clearly separate these two time scales: a sharp transient within $[0,5\tau_v]$ due to viscous relaxation, followed by a slow algebraic tail driven purely by thermal diffusion.

\smallskip\noindent\emph{Axial stress.}
The equilibrium contribution $\hat\sigma^{zz}_{\mathrm{eq}}$ dominates throughout (Fig.~\ref{fig:stress_zz-a}): the non-equilibrium axial part (Fig.~\ref{fig:stress_zz-b}) is an order of magnitude smaller, consistent with the smallness of $\bar\lambda^v_z-1$ noted in Fig.~\ref{fig:kinematics-c}. As the temperature homogenizes, $\hat\sigma^{zz}_{\mathrm{eq}}$ collapses from its near-piecewise-constant initial shape toward zero, with the discontinuity at $R_i$ smoothing progressively, as seen by comparing successive curves in Fig.~\ref{fig:stress_zz-a}. The time histories (Figs.~\ref{fig:stress_zz-d}--\ref{fig:stress_zz-f}) confirm that the total axial stress at the inclusion center (Fig.~\ref{fig:stress_zz-d}) decays on the thermal time scale, while the mid-domain and outer-wall responses (Figs.~\ref{fig:stress_zz-e}--\ref{fig:stress_zz-f}) are an order of magnitude smaller and decay correspondingly slower.

\smallskip\noindent\textbf{Long-time limit.}
For $ t\gg \tau_v$ the non-equilibrium stress fully relaxes and $\hat{\boldsymbol\sigma}_{\mathrm{neq}} \to 0$; for $ t>\tau_D=50\tau$ the temperature returns to $\vartheta=\vartheta_0$ everywhere (cf. Fig.~\ref{fig:temperature}) and the equilibrium stress vanishes as well, recovering the natural undistorted state (cf.\ Figs.~\ref{fig:stress_rr}--\ref{fig:stress_zz}).

\end{example}

\subsection{An incompressible transversely isotropic finite circular bar with a single wedge disclination}
\label{S:eg2}

\paragraph{Kinematics.}
In this section, we consider a solid circular cylinder of radius $R_0$ and length $L$ in its natural undisturbed state. Similarly to \S\ref{S:eg1}, the Euclidean metric is represented as
\begin{equation}
	\mathbf g=\begin{bmatrix}
	1&0&0\\
	0&r^2&0\\
	0&0&1
	\end{bmatrix}\,,
\end{equation}
with respect to a spatial cylindrical coordinate system $(r, \theta, z)$. The Euclidean reference configuration inherits the following Euclidean metric
\begin{equation}
	\Go=\begin{bmatrix}
	1&0&0\\
	0&R^2&0\\
	0&0&1
	\end{bmatrix}\,,
\end{equation}
with respect to a cylindrical coordinate system $(R,\Theta,Z)$ such that the axis of the cylinder in its reference configuration is aligned with the $Z$-axis. In its reference configuration, the body is assumed to be a transversely isotropic solid whose axis of isotropy coincides with the $Z$-axis and is represented by the uniform unit vector $\mathring{\mathbf N}$.

\begin{figure}
  \centering
  \def\svgwidth{0.5\linewidth}
  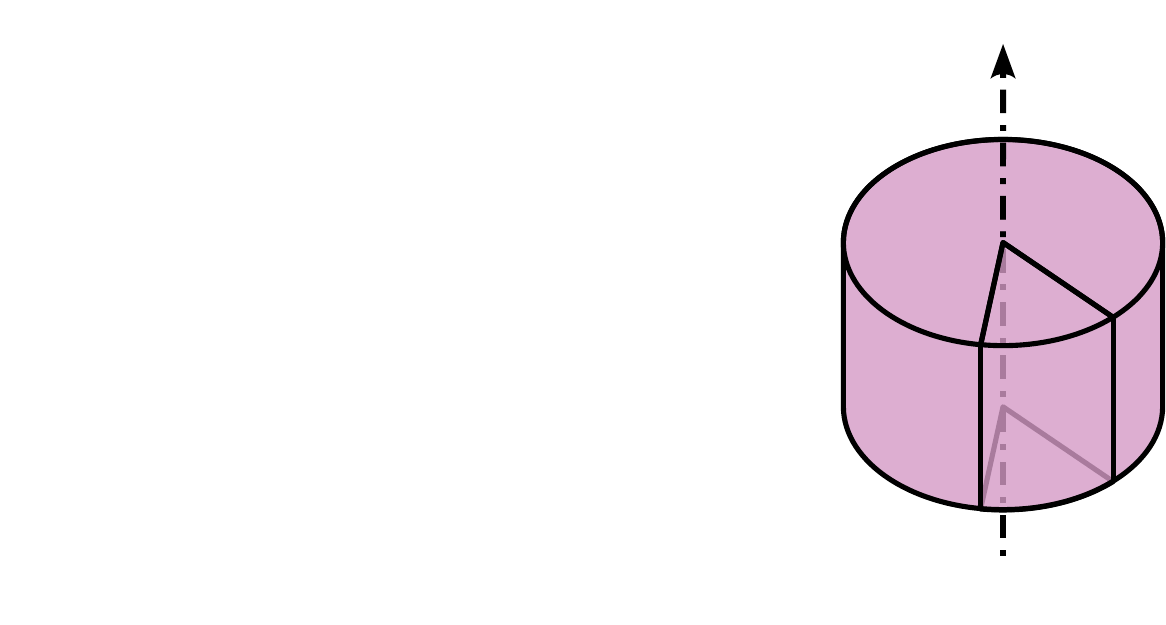
  \vskip 0.1in
  \caption{(a) A negative wedge disclination of magnitude $\Theta_0$ is constructed by making an axial cut in the stress-free cylinder $\mathcal B_0$ and inserting into the cut a wedge of the same material with opening angle $\Theta_0$. (b) The resulting body $\mathcal B$ containing a single wedge disclination in its stressed configuration.}
  \label{fig:disclination}
\end{figure}

The solid cylinder is assumed to contain a wedge disclination of magnitude $\Theta_0$. In the sense of Volterra's construction, this defect is realized by inserting a wedge of angle $\Theta_0$ (negative disclination)\textemdash or removing a wedge of angle $\Theta_0$ (positive disclination)\textemdash made of the same material, into\textemdash or out of, respectively\textemdash an axial cut in the cylinder; see Fig.~\ref{fig:disclination}. For a nonlinear solid containing disclinations, the material manifold is known to be a Riemannian manifold, i.e.,~a manifold with vanishing torsion and non-metricity and a non-flat connection. Following Volterra's construction and using Cartan's moving frames~\citep{YavariGoriely2013a}, one can show that the anelastic distortion $\Fa$ associated with a single disclination of magnitude $\Theta_0$ may be written as
\begin{equation} \label{Distortion-Dislocation}
	\Fa=\begin{bmatrix}
	1&0&0\\
	0&\dfrac{1}{B}&0\\
	0&0&1
	\end{bmatrix}\,,
\end{equation}
where
\begin{equation}
B=
\begin{dcases}
\frac{2\pi}{2\pi+\Theta_0}\,,\quad\textrm{for a negative disclination,}\\
\frac{2\pi}{2\pi-\Theta_0}\,,\quad\textrm{for a positive disclination.}
\end{dcases}
\end{equation}
Recalling that $\mathbf{G}=\Fa^*\Go=\Fa^\star \Go \Fa$, it follows that the material metric has the following representation
\begin{equation} \label{Material-Metric-Disclination}
	\mathbf{G}=\begin{bmatrix}
	1&0&0\\
	0&\dfrac{R^2}{B^2}&0\\
	0&0&1
	\end{bmatrix}\,.
\end{equation}

We assume a time-dependent embedding of the material manifold into the ambient Euclidean space given by 
\begin{equation}
	r = r(R,t)\,,\qquad
	\theta = \Theta\,,\qquad
	z = \lambda(t) Z\,,
\end{equation}
where $\lambda(t)>0$ is the longitudinal stretch. The deformation gradient takes the form 
\begin{equation}
	\mathbf{F}=\begin{bmatrix}
	r'(R,t)&0&0\\
	0&1&0\\
	0&0&\lambda(t)
	\end{bmatrix}\,.
\end{equation}
Following Remark~\ref{rmrk:phys_comp}, the long-term (equilibrium) physical components of $\mathbf{F}$ are given with respect to $\mathbf{G}$ and $\mathbf{g}$ as
\begin{equation}
	\widehat{\mathbf F}=\begin{bmatrix}
	\frac{1}{\sqrt{B\lambda(t)}}&0&0\\
	0& \frac{\sqrt{B}}{\sqrt{\lambda(t)}} &0\\
	0&0&\lambda(t)
	\end{bmatrix}\,.
\end{equation}
Assuming visco-elastic incompressibility, i.e.,~$J=B \lambda r r' /R=1$, and fixing the embedding at the origin, i.e.,~$r(0)=0$, one finds that
\begin{equation}
r(R,t)=\frac{R}{\sqrt{B\lambda(t)}}\,.
\end{equation}
Note that the anelastic distortion \eqref{Distortion-Dislocation} is volume-changing, $\Ja = 1/B \neq 1$, because a wedge of material is inserted into the body. This volume change is fixed by Volterra's construction and is not an independent modeling choice; accordingly, anelastic incompressibility is not imposed. As in the thermal-inclusion problem of \S\ref{S:eg1}, incompressibility is understood relative to the material metric \eqref{Material-Metric-Disclination}; see Remark~\ref{rmrk:eg1_incomp} and Remark~\ref{rmrk:Jacobian}.

We assume that the viscous distortion is radially symmetric. It consequently has a diagonal representation in polar coordinates and we set it to be
\begin{equation}
	\Fv=\begin{bmatrix}
	\lvr(R,t)&0&0\\
	0&\lvt(R,t)&0\\
	0&0&\lvz(R,t)
	\end{bmatrix}\,.
\end{equation}
Using viscous incompressibility, i.e.,~$\Jv=1$, one may write $\lvt(R,t)=(\lvr(R,t) \lvz(R,t))^{-1}$.
Recalling that $\mathbf{F}=\Fe\Fv\Fa$, it follows that the elastic distortion has the following representation
\begin{equation}
	\Fe = \begin{bmatrix}
	\dfrac{1}{\sqrt{B\lambda}\,\lvr} & 0 & 0 \\[1.5ex]
	0 & B \lvr\,\lvz & 0 \\[1.5ex]
	0 & 0 & \dfrac{\lambda}{\lvz}
	\end{bmatrix} \,.
\end{equation}
Following Remark~\ref{rmrk:phys_comp}, the instantaneous (non-equilibrium) physical components of $\Fe$ are given with respect to $\Go$ and $\mathbf{g}$:
\begin{equation}
	\widetilde\Fe = \begin{bmatrix}
	\dfrac{1}{\sqrt{B\lambda}\,\lvr} & 0 & 0 \\[1.5ex]
	0 & \dfrac{B \lvr\,\lvz}{\sqrt{B\lambda(t)}} & 0 \\[1.5ex]
	0 & 0 & \dfrac{\lambda}{\lvz}
	\end{bmatrix} \,.
\end{equation}

The principal viscoelastic invariants are\footnote{These are the invariants of the viscoelastic distortion $\Fve=\Fe\Fv$ with respect to the Euclidian metric $\Go$, and equivalently those of the deformation gradient $\mathbf{F}$ with respect to the material metric $\mathbf{G}$.}
\begin{equation}
I_1 = \frac{1 + B^2 + B \lambda^3}{B\lambda} \,, \qquad
I_2 = \frac{1}{\lambda^2} + \!\left( \frac{1}{B} + B \right)\! \lambda \,, \qquad
I_4 = \lambda^2 \,, \qquad
I_5 = \lambda^4 \,,
\end{equation}
and the elastic principal invariants as\footnote{These are the invariants of the viscoelastic distortion $\Fe$ with respect to the metric $\Go$.}
\begin{equation}
	\Ie_1 = \frac{1}{B\lambda \lvr^2}
	+ \frac{B\lvr^2 \lvz^2}{\lambda}
	+ \frac{\lambda^2}{\lvz^2} \,, \quad
	\Ie_2 = B \lambda \lvr^2 + \frac{\lambda}{B\lvr^2 \lvz^2}
	+ \frac{\lvz^2}{\lambda^2} \,, \quad
	\Ie_4 = \frac{\lambda^2}{\lvz^2} \,, \quad
	\Ie_5 = \frac{\lambda^4}{\lvz^4} \,.
\end{equation}
Recall that, as outlined in \S\ref{sec:trans}, there are five viscoelastic and five elastic invariants in the case of a transversely isotropic solid.

\paragraph{Viscous kinetic equations.}
We assume an isotropic quadratic dissipation potential as given by~\eqref{eq:Quad-Rayleigh}. The system of viscous kinetic equations~\eqref{eq:Trans_v_Kinetic_inc} yields three independent equations in terms of $\lvr(R,t)$, $\lvz(R,t)$, and $\pv(R,t)\,$\textemdash the Lagrange multiplier for viscous incompressibility. We proceed to eliminate $\pv$, and find the following system for the time evolution of $\lvr(R,t)$ and $\lvz(R,t)\,$:
\begin{subequations}
\label{eq:eg2_kinetic}
\begin{align}
&\begin{dcases}
& \!\left[ (\eta_1 + \eta_2 + \eta_3) \!\left(1 + \lvr^4 \lvz^2\right)\! - 2\eta_1 \lvr^2 \lvz \right]\!
\frac{ \dot{\accentset{v}{\lambda}}_R }{\lvr}
+ \!\left[ \eta_1 + \eta_2 + \eta_3 - \eta_1 \lvr \lvz \!\left( \lvr + \lvz - \lvr^2 \lvz^2 \right)\! \right]\!
\frac{ \dot{\accentset{v}{\lambda}}_Z }{\lvz} \\
&\qquad\qquad=
\frac{2\!\left(1 - B^2 \lvr^4 \lvz^2\right)\!}{B\lambda}
\!\left( \lvz^2 \rho\widetilde{\Psi}_1 + \lambda^2 \rho\widetilde{\Psi}_2 \right)\!
\,,
\end{dcases}\\
&\begin{dcases}
& \!\left[ - (\eta_1 + \eta_2 + \eta_3) \lvr^3 \lvz + \eta_1 \!\left( \lvr - \lvz + \lvr^2 \lvz^2 \right)\!
\right]\! \frac{ \dot{\accentset{v}{\lambda}}_R }{\lvr}
\\ & \qquad
+ \!\left[ (\eta_1 + \eta_2 + \eta_3) \lvr \lvz^3 + \eta_1 \!\left( \lvr - \lvz - \lvr^2 \lvz^2 \right)\!
 \right]\!\frac{ \dot{\accentset{v}{\lambda}}_Z }{\lvz}\\
&\qquad=
2\!\left( \frac{\lambda^2 \lvr}{\lvz} - \frac{\lvz}{B\lambda \lvr} \right)\! \rho\widetilde{\Psi}_1
+ 2\lvr\lvz \!\left( B\lambda \lvr^2 - \frac{\lvz^2}{\lambda^2} \right)\! \rho\widetilde{\Psi}_2
+ \frac{2\lambda^2 \lvr}{\lvz} \rho\widetilde{\Psi}_4
+ \frac{4\lambda^4 \lvr}{\lvz^3} \rho\widetilde{\Psi}_5
\,.
\end{dcases}
\end{align}
\end{subequations}

\begin{remark}
For a disclination-free ($B=1$), isotropic viscoelastic solid ($\widetilde{\Psi}_4=\widetilde{\Psi}_5=\overline{\Psi}_4=\overline{\Psi}_5=0$), one finds $\lvr=\lvt=\lvz^{-\frac{1}{2}}$, and the system of kinetic equations~\eqref{eq:eg2_kinetic} reduces to a single kinetic equation. In terms of $\lvz$, this equation coincides with the one obtained for the isotropic viscoelastic case discussed in \citep[Eq.~(5.23)]{SaYa2024viscoelasticity}.
\end{remark}

\paragraph{Stress and equilibrium.}
The solid cylinder is made of a transversely isotropic material with the material preferred direction $\mathring{\mathbf N}$ along the axial direction $Z$. The stress may hence be computed following~\eqref{eq:Trans_Cauchy}, and one finds the following non-zero physical components of the Cauchy stress tensor\footnote{Recall that\textemdash following \cite{Truesdell1953physical}\textemdash the physical components of the stress are given by $\hat{\sigma}^{rr}=\sigma^{rr}$, $\hat{\sigma}^{\theta\theta}=r^2\sigma^{\theta\theta}$, and $\hat{\sigma}^{zz}=\sigma^{zz}$.}
\begin{subequations}
\label{eq:eg2_stress}
\begin{align}
\label{eq:eg2_srr}
\hat\sigma^{rr} &= -p 
+ \frac{2}{B \lambda} \rho\overline{\Psi}_1
- 2 B \lambda \rho\overline{\Psi}_2 
+ \frac{2}{B \lambda \lvr^2} \rho\widetilde{\Psi}_1
- 2 B \lambda \lvr^2 \rho\widetilde{\Psi}_2 \,, \\
\hat\sigma^{\theta\theta} &= -p 
+ \frac{2 B}{\lambda} \rho\overline{\Psi}_1
- \frac{2 \lambda}{B} \rho\overline{\Psi}_2
+ \frac{2 B \lvr^2 \lvz^2}{\lambda} \rho\widetilde{\Psi}_1
- \frac{2 \lambda}{B \lvr^2 \lvz^2} \rho\widetilde{\Psi}_2 \,, \\
\hat\sigma^{zz} &= -p 
+ 2 \lambda^2 \rho\overline{\Psi}_1 
- \frac{2}{\lambda^2} \rho\overline{\Psi}_2
+ 2 \lambda^2 \rho\overline{\Psi}_4 
+ 4 \lambda^4 \rho\overline{\Psi}_5
+ \frac{2 \lambda^2}{\lvz^2} \rho\widetilde{\Psi}_1
- \frac{2 \lvz^2}{\lambda^2} \rho\widetilde{\Psi}_2
+ \frac{2 \lambda^2}{\lvz^2} \rho\widetilde{\Psi}_4
+ \frac{4 \lambda^4}{\lvz^4} \rho\widetilde{\Psi}_5 \,,
\end{align}
\end{subequations}
where we recall $p=p(R,t)$ is the Lagrange multiplier enforcing the visco-elastic incompressibility condition ($J=1$).
The only non-trivially satisfied equilibrium equation is $\hat\sigma^{rr}{}_{,r}+\!\left(\hat\sigma^{rr}-\hat\sigma^{\theta\theta}\right)\! /r = 0\,$, which may be written as
\begin{equation}\label{eq:eg2_Equilibrium}
\frac{\partial \hat\sigma^{rr}}{\partial R} = \mathcal{I}_\sigma(R,t)
\,,
\end{equation}
where 
\begin{equation}\label{eq:Is_eg2}
\mathcal{I}_\sigma =
	\frac{2(1 - B^2)}{B R \lambda} \rho\overline{\Psi}_1
	- \frac{2(1 - B^2) \lambda}{B R} \rho\overline{\Psi}_2
	+ \frac{2\!\left(1 - B^2 \lvr^4 \lvz^2\right)\!}{B R \lambda \lvr^2} \rho\widetilde{\Psi}_1
	+ \frac{2 \lambda \!\left(1 - B^2 \lvr^4 \lvz^2\right)\!}{B R \lvr^2 \lvz^2} \rho\widetilde{\Psi}_2
\,.
\end{equation}
Assuming that the lateral boundary of the cylinder is traction-free, i.e.,~$\sigma^{rr}(R_0,t)=0$, it follows from~\eqref{eq:eg2_srr} and~\eqref{eq:eg2_Equilibrium} that
\begin{equation}
\label{eq:eg2_pe}
\begin{split}
	p(R,t) = \int_R^{R_0} \mathcal{I}_\sigma(\xi,t)d\xi
	+ \frac{2}{B \lambda}  \rho\overline{\Psi}_1
	- 2 B \lambda \rho\overline{\Psi}_2 
	+ \frac{2}{B \lambda \lvr^2}  \rho\widetilde{\Psi}_1
	- 2 B \lambda \lvr^2 \rho\widetilde{\Psi}_2
	\,.
\end{split}
\end{equation}
As in Remark~\ref{rmrk:eg1-n-eq-separ}, the terms involving $\overline{\Psi}_j$ and $\widetilde{\Psi}_j$ constitute the equilibrium and non-equilibrium contributions, respectively, in accordance with the additive decomposition $\boldsymbol\sigma=\boldsymbol\sigma_{\text{eq}}+\boldsymbol\sigma_{\text{neq}}$ discussed in~\eqref{eq:sig_eq-neq_represent_inc}, with the Lagrange multiplier similarly decomposed as $p=p_{\text{eq}}+p_{\text{neq}}$.

Note that the force at the two ends of the bar is written as
\begin{equation}\label{eq:eg2_AxialForce}
	F(t)=2\pi \int_{0}^{r(R_0)}\sigma^{zz}(R,t)r\,dr 
	= 2\pi \int_{0}^{R_0}\frac{\hat\sigma^{zz}(R,t)}{B \lambda(t)}R\,dR\,.
\end{equation}

The formulation is now complete and the governing equations are fully specified. The minimal set of independent fields comprises the viscous stretches $\lvr(R,t)$ and $\lvz(R,t)$, together with the global stretch $\lambda(t)$ when not prescribed. All remaining quantities follow directly: the azimuthal stretch $\lvt(R,t)$ from the viscous incompressibility constraint, and the pressure field $p(R,t)$ from the radial equilibrium relation~\eqref{eq:eg2_pe}. In what follows, we consider an example of a particular nonlinear transversely isotropic model (the reinforced neo-Hookean model). The solution procedure is illustrated for two loading conditions: a displacement-controlled case, in which $\lambda(t)$ is prescribed and the kinetic equations~\eqref{eq:eg2_kinetic} are solved for $\lvr(R,t)$ and $\lvz(R,t)$; and a force-controlled case, in which the axial force $F(t)$ is prescribed and the corresponding stretch $\lambda(t)$ and the viscous stretches $\lvr(R,t)$, $\lvz(R,t)$ are obtained by simultaneously solving the global equilibrium condition~\eqref{eq:eg2_AxialForce} and the kinetic equations~\eqref{eq:eg2_kinetic}.

\begin{example}
We consider an incompressible reinforced neo-Hookean model with free energy
\begin{equation} \label{eg2_neo-Hookean-reinforced}
    \Psi_{\textrm{eq}} = \frac{\mu}{2\rho}(I_1-3)
    +\frac{\kappa}{2\rho}\!\left(I_4-1\right)^2 \,,\qquad
    \Psi_{\textrm{neq}} = \frac{\mu_{i}}{2\rho}(\Ie_1-3)
    +\frac{\kappa_i}{2\rho}\!\left(\Ie_4-1\right)^2 \,,
\end{equation}
where $\mu>0$ and $\mu_i>0$ are the equilibrium and non-equilibrium shear moduli, respectively,
and $\kappa>0$ and $\kappa_i>0$ are the corresponding fiber-reinforcement stiffnesses along the transverse anisotropy direction $Z$, respectively. The relevant partial derivatives read
\begin{equation} \label{eg2_Energy_Inv}
    \overline{\Psi}_1=\frac{\mu}{2\rho}\,, \quad
    \overline{\Psi}_4=\frac{\kappa}{\rho}(I_4-1)\,, \quad
    \overline{\Psi}_2=\overline{\Psi}_5=0\,,\qquad
    \widetilde{\Psi}_1=\frac{\mu_i}{2\rho}\,, \quad
    \widetilde{\Psi}_4=\frac{\kappa_i}{\rho}(\Ie_4-1)\,, \quad
    \widetilde{\Psi}_2=\widetilde{\Psi}_5=0\,,
\end{equation}
recalling that $I_4 = \lambda^2$ and $\Ie_4 = \lambda^2/\lvz^2$.
It follows that the equilibrium and non-equilibrium parts of the Cauchy stress~\eqref{eq:eg2_stress}\textemdash as discussed in Remark~\ref{rmrk:eg1-n-eq-separ} and by using~\eqref{eq:Is_eg2} and~\eqref{eq:eg2_pe}\textemdash reduce to
\begin{subequations}
\label{eq:eg2nH_stress}
\begin{alignat}{2}
&\begin{dcases}
\hat\sigma^{rr}_{\text{eq}} \!\!\!\!&=
- \frac{\mu(1 - B^2)}{B\lambda(t)} \ln\frac{R_0}{R} \,,
\end{dcases}
\quad
&&\begin{dcases}
\hat\sigma^{rr}_{\text{neq}} \!\!\!&=
- \frac{\mu_i}{B\lambda(t)}\!\left(\frac{1}{\lvr^2(t)} - B^2\lvr^2(t)\lvz^2(t)\right)\! \ln\frac{R_0}{R}\,,
\end{dcases}
\\
&\begin{dcases}
\hat\sigma^{\theta\theta}_{\text{eq}} \!\!\!&=
- \frac{\mu(1 - B^2)}{B\lambda(t)} \!\left(1 + \ln\frac{R_0}{R} \right)\!\,,
\end{dcases}
\quad
&&\begin{dcases}
\hat\sigma^{\theta\theta}_{\text{neq}} \!\!\!&=
- \frac{\mu_i}{B\lambda(t)}\!\left(\frac{1}{\lvr^2(t)} - B^2\lvr^2(t)\lvz^2(t)\right)\! \!\left(1 + \ln\frac{R_0}{R} \right)\!\,,
\end{dcases}
\\
&\begin{dcases}
\hat\sigma^{zz}_{\text{eq}} \!\!\!&=
\mu\!\left(\lambda^2(t) - \frac{1}{B\lambda(t)}\right)\!
\\&
- \frac{\mu(1 - B^2)}{B\lambda(t)} \ln\frac{R_0}{R}
\\&
+ 2\kappa\!\left(\lambda^4(t) - \lambda^2(t)\right)\!\,,
\end{dcases}
\quad
&&\begin{dcases}
\hat\sigma^{zz}_{\text{neq}} \!\!\!&=
\mu_i\!\left(\frac{\lambda^2(t)}{\lvz^2(t)} - \frac{1}{B\lambda(t)\lvr^2(t)}\right)\!
\\&
- \frac{\mu_i}{B\lambda(t)}\!\left(\frac{1}{\lvr^2(t)} - B^2\lvr^2(t)\lvz^2(t)\right)\! \ln\frac{R_0}{R}
\\&
+ 2\kappa_i\!\left(\frac{\lambda^4(t)}{\lvz^4(t)} - \frac{\lambda^2(t)}{\lvz^2(t)}\right)\!\,.
\end{dcases}
\end{alignat}
\end{subequations}

We assume a Rayleigh dissipation potential~\eqref{eq:Quad-Rayleigh} with
\begin{equation} \label{eg2_Dissip}
    \eta_1=\eta_2=0\,,\qquad \eta_3 \eqcolon \eta > 0\,,
\end{equation}
which, we recall, satisfies the convexity conditions~\eqref{eq:eta_cond}.
It follows that the kinetic equation~\eqref{eq:eg2_kinetic} simplifies to
\begin{subequations}
\label{eq:eg2_nH_kinetic}
\begin{align}
& \!\left(1 + \lvr^4 \lvz^2\right)\!\frac{ \dot{\accentset{v}{\lambda}}_R }{\lvr}
+ \frac{ \dot{\accentset{v}{\lambda}}_Z }{\lvz}
= \frac{\mu_i}{\eta} \lvz^2 \frac{\!\left(1 - B^2 \lvr^4 \lvz^2\right)\!}{B\lambda}
\,,
\\
& - \lvr^2 \lvz\dot{\accentset{v}{\lambda}}_R
+ \lvr \lvz^2 \dot{\accentset{v}{\lambda}}_Z
= \frac{\mu_i}{\eta} \!\left( \frac{\lambda^2 \lvr}{\lvz} - \frac{\lvz}{B\lambda \lvr} \right)\!
+ 2\frac{\kappa_i}{\eta} \frac{\lambda^2 \lvr}{\lvz} \!\left(\frac{\lambda^2}{\lvz^2}-1\right)\!
\,.
\end{align}
\end{subequations}
Since the kinetic equation does not depend on $R$, and the initial conditions $\lvr(R,0) = \lvz(R,0) = 1$ are likewise $R$-independent, it follows that $\lvr$, $\lvz$, and consequently the viscous distortion $\Fv$, are functions of $t$ alone.
We assume that the cylinder has a single negative disclination along its axis, i.e. $B=2\pi/(2\pi+\Theta_0)>1$.

\paragraph{Dimensional Analysis.}
The kinetic equations~\eqref{eq:eg2_nH_kinetic} involve three independent dimensional material parameters: the non-equilibrium shear modulus $\mu_i > 0$, the non-equilibrium fiber-reinforcement stiffness $\kappa_i \geq 0$, and the viscosity $\eta > 0$. These combine into the intrinsic time scale $\tau = {\eta}/{\mu_i}$, which governs the rate of viscous relaxation, and the dimensionless ratio $\beta_i = \kappa_i/\mu_i$, which reflects the effect of material anisotropy on the kinetic equations. Introducing the dimensionless time $\bar{t} = t/\tau$ recasts the kinetic system~\eqref{eq:eg2_nH_kinetic} into dimensionless form.
Stress~\eqref{eq:eg1nH_stress} is normalized by $\mu$, and the end force~\eqref{eq:eg2_AxialForce} by $\mu\pi R_0^2$. The resulting dimensionless equations involve the additional parameters $\beta=\kappa/\mu$ and $\nu_i=\mu_i/\mu$.
The complete set of dimensionless parameters together with the numerical values adopted for the simulations below are summarised in Table~\ref{tab:eg2_dimless}.

\vskip 0.2in
\begin{table}[h!]
\centering
\caption{Dimensionless parameters and numerical values for the disclination problem.}
\begin{tabular}{lllll}
\hline
Parameter & Definition & Physical significance & Value \\
\hline
$\tau$   & $\eta/\mu_i$ & Viscous relaxation time scale & ---    \\
$B$         & $2\pi/(2\pi+\Theta_0)$              & Disclination geometry                & $12/13$ \;($\Theta_0=\pi/6$) \\
$\beta$       & $\kappa/\mu$  & Equilibrium anisotropy       & $2$    \\
$\beta_i$     & $\kappa_i/\mu_i$                    & Non-equilibrium anisotropy & $2$    \\
$\nu_i$   & $\mu_i/\mu$   & Equilibrium/non-equilibrium moduli ratio & $1.11$ \\
$\bar{t}_l$ & $t_l/\tau$    & Load onset time                & $3$    \\
$\bar{t}_u$ & $t_u/\tau$    & Load removal time              & $6$    \\
$\bar{t}_0$ & $t_0/\tau$    & Ramp steepness parameter       & $0.1$  \\
$\lambda_0$ & $\lambda_0$ & Peak prescribed stretch        & $1.25$ \\
$\bar{F}_0$ & $F_0/(\mu\pi R_0^2)$                & Peak dimensionless force       & $5$    \\
\hline
\end{tabular}
\label{tab:eg2_dimless}
\end{table}

In what follows, we explore two loading scenarios: displacement control, in which the axial stretch $\lambda(t)$ is prescribed, and force control, in which the resultant axial force $F(t)$ is prescribed. Both share the same three-stage structure. First, the disclination is introduced and the bar is held at its initial state---unit stretch under displacement control, zero force under force control---allowing the disclination-induced residual stress to viscoelastically equilibrate. The controlled quantity is then smoothly ramped to its peak value and held until the transient non-equilibrium stress has fully relaxed. Finally, it is returned to its initial value via the reverse ramp. In both cases the ramps are prescribed as piecewise error-function profiles.

\paragraph{Displacement-controlled loading.}
The axial stretch $\lambda(\bar{t})$ is prescribed: the fixed-end configuration is first held at $\lambda=1$, then $\lambda(\bar{t})$ is ramped smoothly to its peak value $\lambda_0>1$, held at $\lambda_0$, and finally returned to $\lambda=1$ via the reverse ramp. This loading protocol is explicitly given by:
\begin{equation}\label{eq:lambda_loading_eg2}
\lambda(\bar{t}) =
\begin{dcases}
1\,, & \bar{t} < \bar{t}_l \,, \\[1ex]
1 + (\lambda_0 - 1)\,\erf\!\left(\dfrac{\bar{t} - \bar{t}_l}{\bar{t}_0}\right)\!, & \bar{t}_l \leq \bar{t} < \bar{t}_u \,, \\[1ex]
\lambda_0 - (\lambda_0 - 1)\,\erf\!\left(\dfrac{\bar{t} - \bar{t}_u}{\bar{t}_0}\right)\!, & \bar{t} \geq \bar{t}_u \,.
\end{dcases}
\end{equation}
The numerical values of $\bar{t}_l$, $\bar{t}_u$, $\bar{t}_0$, and $\lambda_0$ are given in Table~\ref{tab:eg2_dimless}.

\begin{figure}
\centering
\begin{subfigure}{0.49\textwidth}
    \centering
    \includegraphics[width=\linewidth]{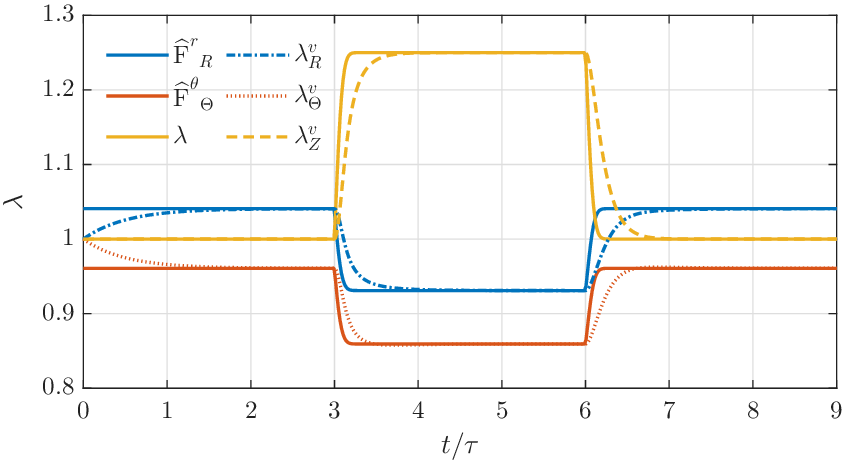}
    \caption{Prescribed stretch and viscous distortions}
    \label{fig:ex2-D-kin1}
\end{subfigure}
\begin{subfigure}{0.49\textwidth}
    \centering
    \includegraphics[width=\linewidth]{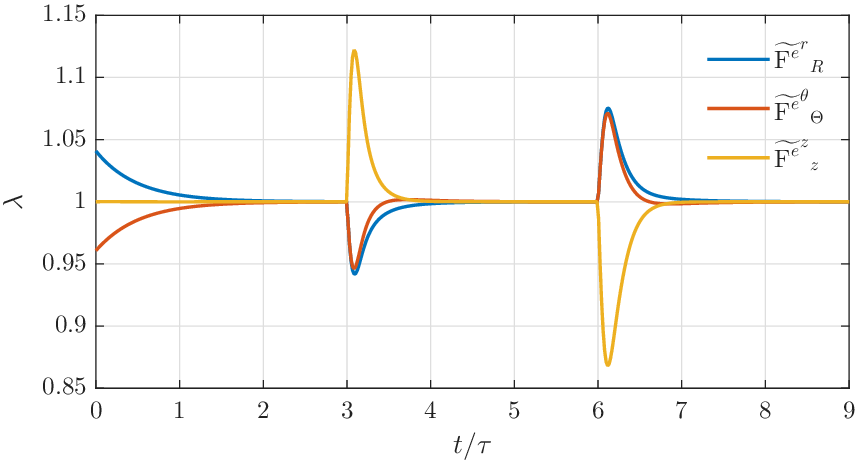}
    \caption{Elastic distortions}
    \label{fig:ex2-D-kin2}
\end{subfigure}
\caption{Displacement-controlled loading: evolution of the kinematics.}
\label{fig:ex2-D-kin}
\end{figure}

\begin{figure}
\centering
\begin{subfigure}{0.49\textwidth}
    \centering
    \includegraphics[width=\linewidth]{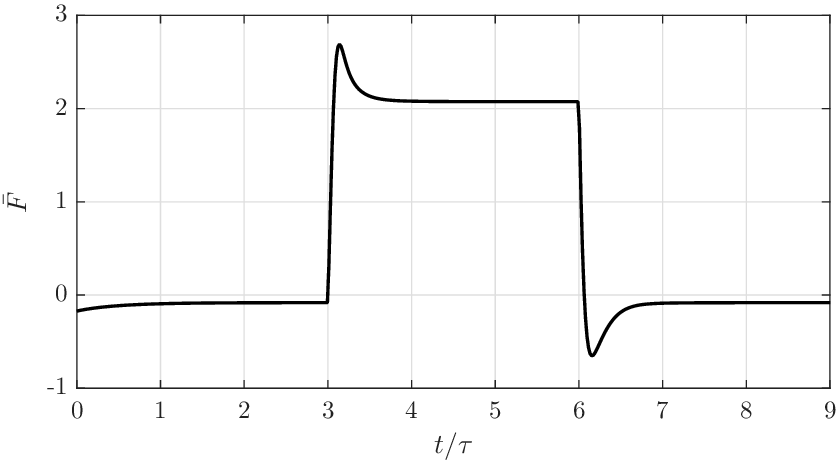}
    \caption{Axial force versus time}
    \label{fig:ex2-D-Ft}
\end{subfigure}
\begin{subfigure}{0.49\textwidth}
    \centering
    \includegraphics[width=\linewidth]{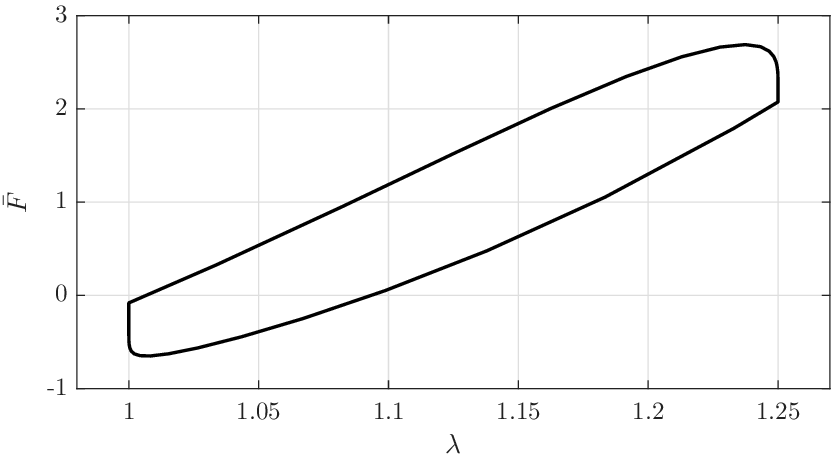}
    \caption{Force--stretch response}
    \label{fig:ex2-D-Flam}
\end{subfigure}
\caption{Displacement-controlled loading: global force response against time and axial stretch.}
\label{fig:ex2-D-force}
\end{figure}

\begin{figure}
\centering
\begin{subfigure}{0.32\textwidth}
    \centering
    \includegraphics[width=\linewidth]{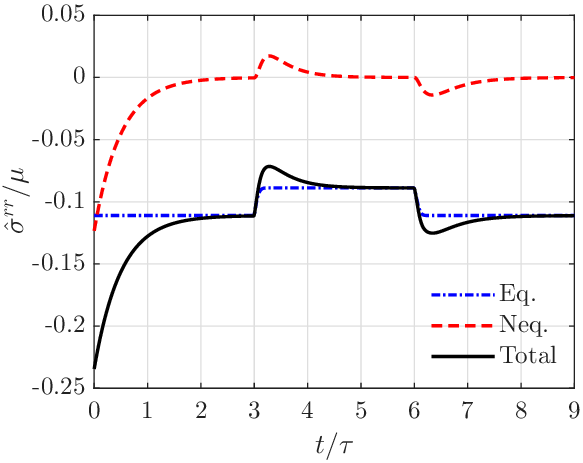}
    \caption{Radial stress at $R = R_0/2$}
    \label{fig:ex2-D-srr}
\end{subfigure}
\begin{subfigure}{0.32\textwidth}
    \centering
    \includegraphics[width=\linewidth]{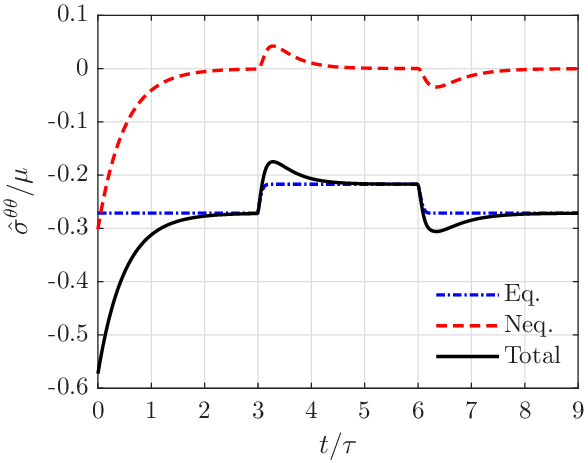}
    \caption{Azimuthal stress at $R = R_0/2$}
    \label{fig:ex2-D-stt}
\end{subfigure}
\begin{subfigure}{0.32\textwidth}
    \centering
    \includegraphics[width=\linewidth]{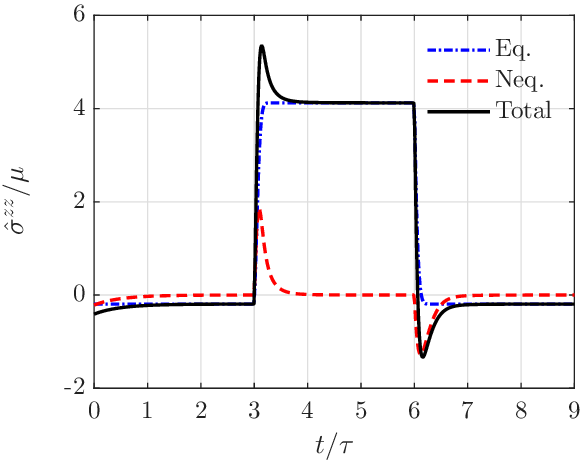}
    \caption{Axial stress at $R = R_0/2$}
    \label{fig:ex2-D-szz}
\end{subfigure}
    \\[1em]
\begin{subfigure}{0.32\textwidth}
    \centering
    \includegraphics[width=\linewidth]{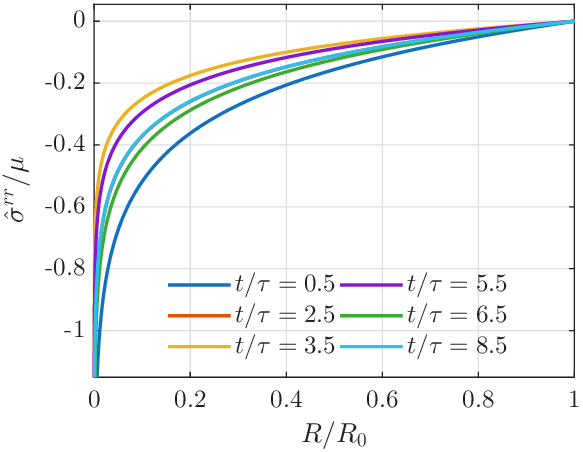}
    \caption{Radial stress}
    \label{fig:ex2-D-rad1}
\end{subfigure}
\begin{subfigure}{0.32\textwidth}
    \centering
    \includegraphics[width=\linewidth]{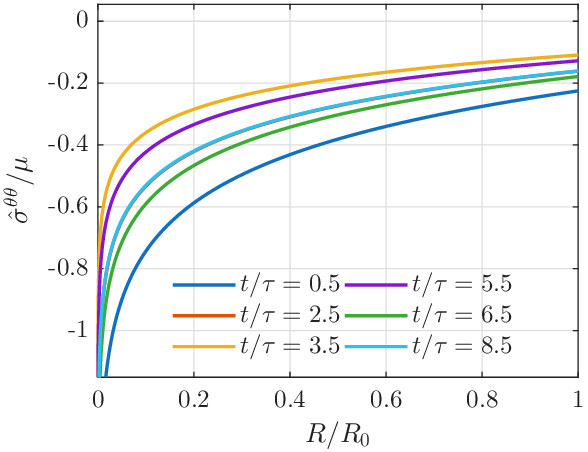}
    \caption{Azimuthal stress}
    \label{fig:ex2-D-rad2}
\end{subfigure}
\begin{subfigure}{0.32\textwidth}
    \centering
    \includegraphics[width=\linewidth]{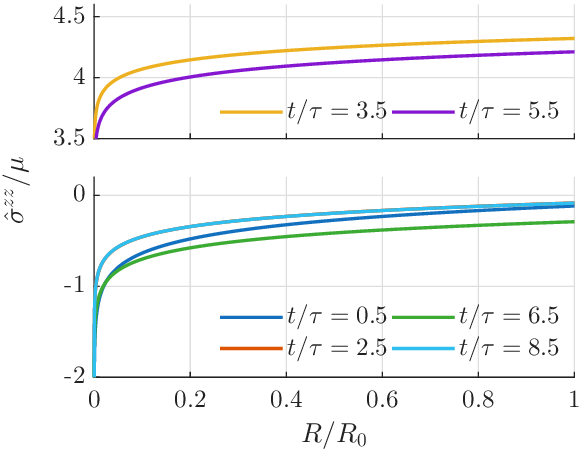}
    \caption{Axial stress}
    \label{fig:ex2-D-rad3}
\end{subfigure}
\caption{Displacement-controlled loading: stress evolution at fixed radius and stress radial profiles.}
\label{fig:ex2-D-Stress}
\end{figure}
 
The kinetic system~\eqref{eq:eg2_nH_kinetic} is solved for $\{\lvr,\lvz\}$ using a forward-Euler finite-difference scheme in dimensionless time $\bar{t}$. Figs.~\ref{fig:ex2-D-kin}--\ref{fig:ex2-D-Stress} report the resulting response, which may be naturally read in three stages, following the loading protocol~\eqref{eq:lambda_loading_eg2}.

\vskip 0.1 in \noindent\emph{i) Initial hold at fixed length ($\lambda=1$).}
Following the introduction of the disclination at $\bar t=0$, the bar is held at fixed length so that the defect can settle into the body at fixed total strain over the time scale $\tau=\eta/\mu_i$. The disclination instantaneously generates an incompatible, residually stressed configuration which then viscoelastically equilibrates toward its disclinated elastic equilibrium\textemdash the purely elastic response of the disclinated body~\citep{YavariGoriely2013a}.
This relaxation is mediated by the evolution of the viscous distortion $\Fv$: starting from unity ($\lvr=\lvt=\lvz=1$), the axial stretch $\lvz$ remains at unity while radial expansion and azimuthal contraction compensate one another to ensure incompressibility ($\lvr\lvt=1$ since $\lvz=1$), as the kinematics in Fig.~\ref{fig:ex2-D-kin1} display.
Crucially, the viscous distortion evolves to match the total deformation gradient (Fig.~\ref{fig:ex2-D-kin1}); equivalently, the elastic distortion relaxes to the identity, $\hat\Fe\to\mathbf{I}$, as confirmed by the elastic-distortion components in Fig.~\ref{fig:ex2-D-kin2}, in accordance with Remark~\ref{rmrk:longtime_distort}.
The axial force is nonzero from the outset (Fig.~\ref{fig:ex2-D-Ft}): the disclination instantaneously drives the cylinder into a state of axial tension while the fixed-end condition is maintained, and this reaction force mildly relaxes as the defect equilibrates and the cylinder settles onto its baseline residual state.
The stresses behave consistently: at $\bar t=0$ all three components attain their largest magnitudes of the hold, reflecting the combined contribution of the equilibrium and non-equilibrium branches, and are then drawn down onto their equilibrium values as the non-equilibrium contribution relaxes to zero (Figs.~\ref{fig:ex2-D-srr}--\ref{fig:ex2-D-szz}).
They remain compressive throughout this stage (Figs.~\ref{fig:ex2-D-rad1}--\ref{fig:ex2-D-rad3}), a direct consequence of $1-B^2>0$ for the negative disclination ($B<1$), which fixes the sign of the equilibrium stresses in~\eqref{eq:eg2nH_stress}: the inserted material wedge places the surrounding body in radial and azimuthal compression that intensifies logarithmically toward the axis.

\vskip 0.1 in \noindent\emph{ii) Loading--unloading ramps and final hold.}
As the bar is loaded toward the peak stretch $\lambda_0$, the kinematics mirror the initial hold: the viscous axial stretch tracks the imposed stretch with a rate-dependent lag set by $\tau$, $\lvz\to\lambda_0$; indeed the full viscous distortion tracks $\mathbf F$ and the elastic distortion returns to the identity, $\hat\Fe\to\mathbf{I}$ (Fig.~\ref{fig:ex2-D-kin}), in accordance with Remark~\ref{rmrk:longtime_distort}.
Viscous incompressibility is maintained throughout, with $\lvr$ and $\lvt$ contracting as $\lvz$ rises following the prescribed stretch $\lambda$ so that $\lvr\,\lvt\,\lvz=1$ (Fig.~\ref{fig:ex2-D-kin1}).
The axial stretching produces a pronounced overshoot in the end force and stress at ramp onset (Figs.~\ref{fig:ex2-D-Ft},\ref{fig:ex2-D-srr}--\ref{fig:ex2-D-szz}): on the loading time scale $\bar t_0$ the non-equilibrium branch responds strongly before it can relax, so the body is momentarily stiffer than its long-term equilibrium response.
During the hold at $\lambda_0$ this excess relaxes away and the stresses and end force settle onto their equilibrium plateaux\textemdash again a stress relaxation at constant stretch (Figs.~\ref{fig:ex2-D-Ft},\ref{fig:ex2-D-srr}--\ref{fig:ex2-D-szz}).
The reverse ramp produces a mirror-image undershoot (Figs.~\ref{fig:ex2-D-Ft},\ref{fig:ex2-D-szz}): the viscous distortion, now relaxed to the loaded configuration, momentarily over-accommodates the removed stretch and transiently reverses the sign of the non-equilibrium axial stress before relaxing.
During the final hold the cylinder returns to a clamped $\lambda=1$ configuration sustaining the same residual axial reaction as at the end of the initial hold\textemdash the purely elastic disclinated response~\citep{YavariGoriely2013a}\textemdash rather than the $\bar t=0$ instantaneous state.
The in-plane stresses respond more weakly to the axial load than $\hat\sigma^{zz}$ throughout (Figs.~\ref{fig:ex2-D-srr},\ref{fig:ex2-D-stt} vs.~\ref{fig:ex2-D-szz}), since their equilibrium parts depend on $\lambda$ only through the slowly varying prefactor in~\eqref{eq:eg2nH_stress}.

\vskip 0.1 in\noindent\emph{iii) Stress-relaxation-driven hysteresis.}
The distinct loading and unloading paths form a closed loop in the force--stretch plane (Fig.~\ref{fig:ex2-D-Flam}), whose enclosed area quantifies the energy dissipated over the cycle, with the transient (non-equilibrium) response scaling with $\tau$. The loop reflects the stress relaxation occurring during the holds at constant stretch, which separates the two branches; it would close onto a single curve in the quasi-static limit $\bar t_0\gg\tau$.

\paragraph{Force-controlled loading.}
The axial force $F(t)$ is prescribed: the unloaded state is first held at $F=0$, then $F(t)$ is ramped smoothly to its peak value $F_0>0$, held at $F_0$, and finally returned to $F=0$ via the reverse ramp. This loading protocol is explicitly given by:
\begin{equation}
\label{eq:force_loading_eg2}
F(t) =
	\begin{dcases}
	0\,, & \text{for } \bar{t} < \frac{t_l}{\tau}\,,\\[1ex]
	F_0 \, \erf\!\left( \dfrac{\bar{t} - \frac{t_l}{\tau}}{\frac{t_0}{\tau}} \right)\!\,, 
	& \text{for } \frac{t_l}{\tau} \leq \bar{t} < \frac{t_u}{\tau}\,,\\[1ex]
	F_0 \!\left( 1 - \erf\!\left( \dfrac{\bar{t} - \frac{t_u}{\tau}}{\frac{t_0}{\tau}} \right)\! \right)\!, & \text{for } \bar{t} \ge \frac{t_u}{\tau}\,.
	\end{dcases}
\end{equation}
The numerical values of $\bar{t}_l$, $\bar{t}_u$, $\bar{t}_0$, and $F_0$ are given in Table~\ref{tab:eg2_dimless}.

\begin{figure}[h!]
\centering
\begin{subfigure}{0.49\textwidth}
    \centering
    \includegraphics[width=\linewidth]{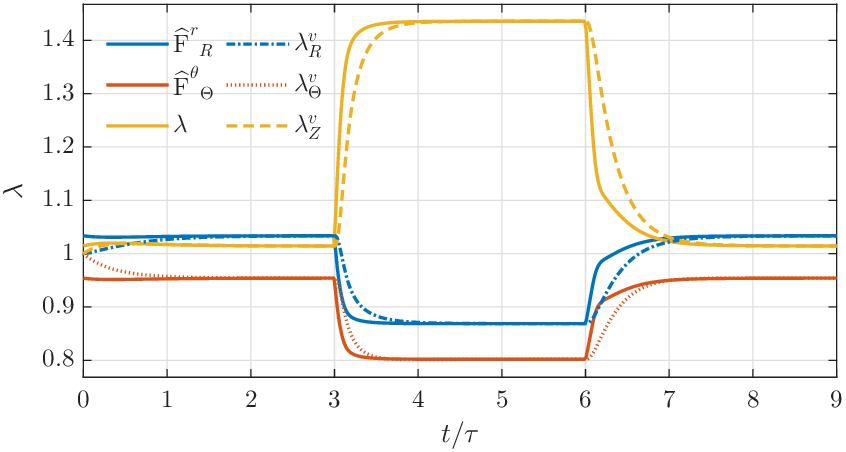}
    \caption{Stretch and viscous distortions}
    \label{fig:ex2-F-kin1}
\end{subfigure}
\begin{subfigure}{0.49\textwidth}
    \centering
    \includegraphics[width=\linewidth]{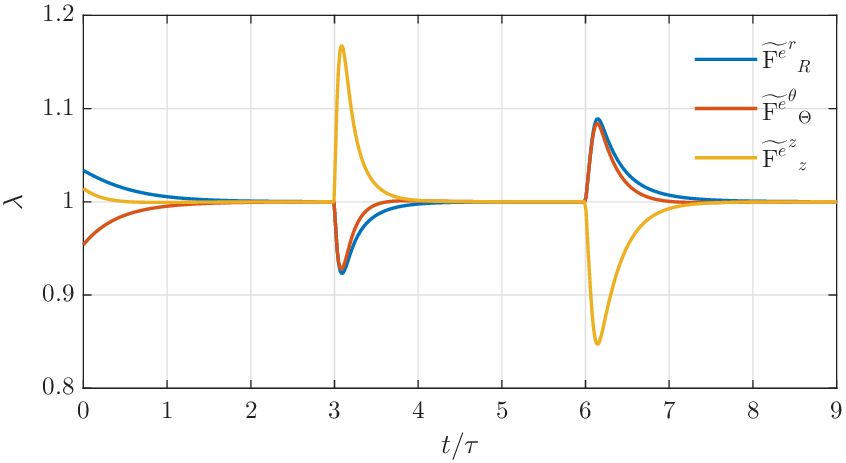}
    \caption{Elastic distortions}
    \label{fig:ex2-F-kin2}
\end{subfigure}
\caption{Force-controlled loading: evolution of the kinematics.}
\label{fig:ex2-F-kin}
\end{figure}

\begin{figure}[h!]
\centering
\begin{subfigure}{0.49\textwidth}
    \centering
    \includegraphics[width=\linewidth]{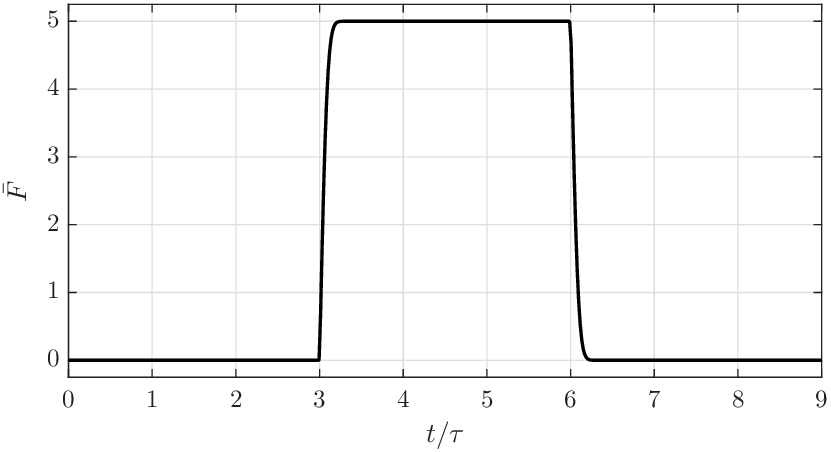}
    \caption{Prescribed axial force profile}
    \label{fig:ex2-F-Ft}
\end{subfigure}
\begin{subfigure}{0.49\textwidth}
    \centering
    \includegraphics[width=\linewidth]{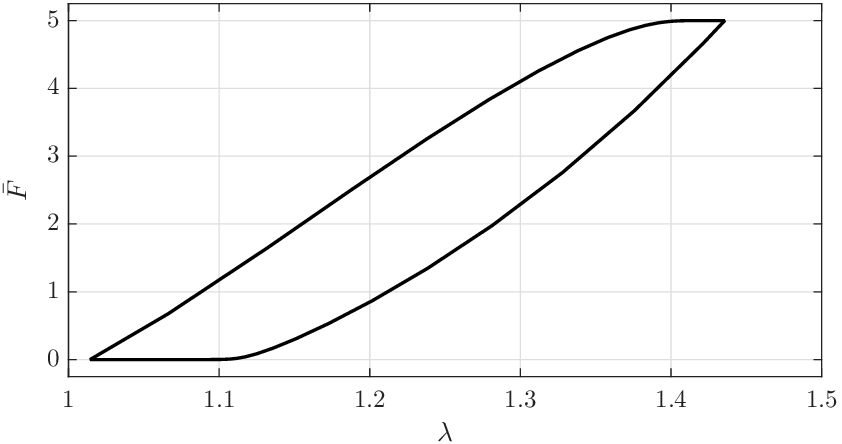}
    \caption{Force--stretch response}
    \label{fig:ex2-F-Flam}
\end{subfigure}
\caption{Force-controlled loading: global force response in time and in stretch space.}
\label{fig:ex2-F-force}
\end{figure}

\begin{figure}[h!]
\centering
\begin{subfigure}{0.32\textwidth}
    \centering
    \includegraphics[width=\linewidth]{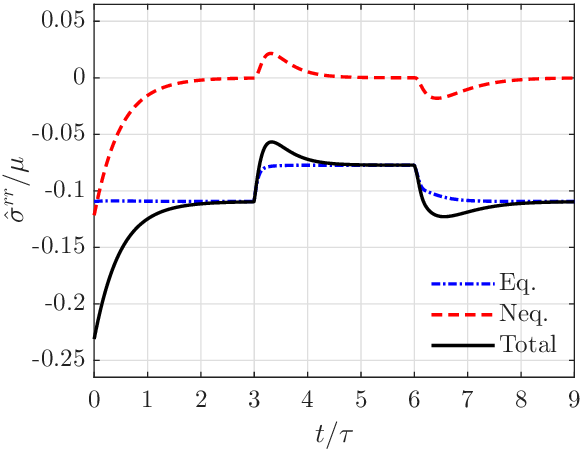}
    \caption{Radial stress at $R = R_0/2$}
    \label{fig:ex2-F-srr}
\end{subfigure}
\begin{subfigure}{0.32\textwidth}
    \centering
    \includegraphics[width=\linewidth]{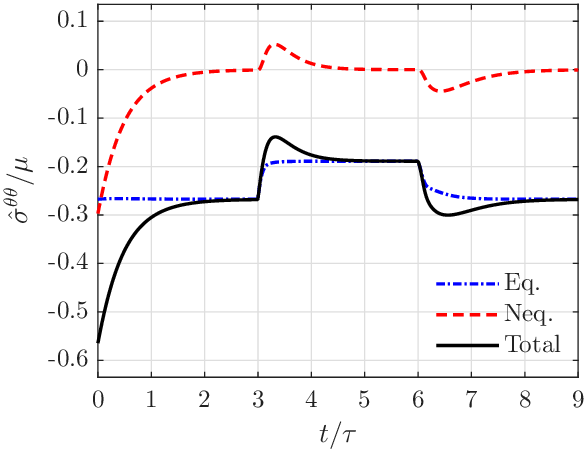}
    \caption{Azimuthal stress at $R = R_0/2$}
    \label{fig:ex2-F-stt}
\end{subfigure}
\begin{subfigure}{0.32\textwidth}
    \centering
    \includegraphics[width=\linewidth]{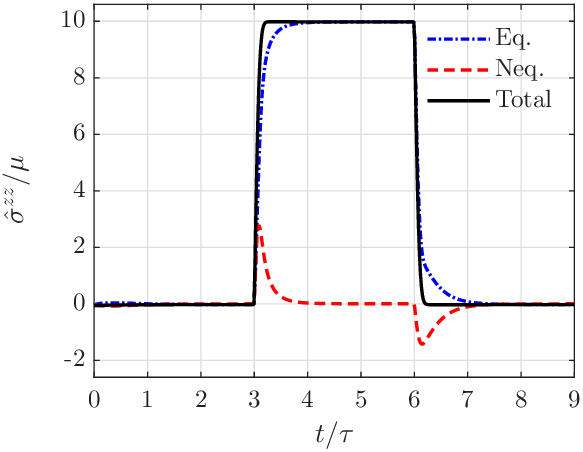}
    \caption{Axial stress at $R = R_0/2$}
    \label{fig:ex2-F-szz}
\end{subfigure}
    \\[1em]
\begin{subfigure}{0.32\textwidth}
    \centering
    \includegraphics[width=\linewidth]{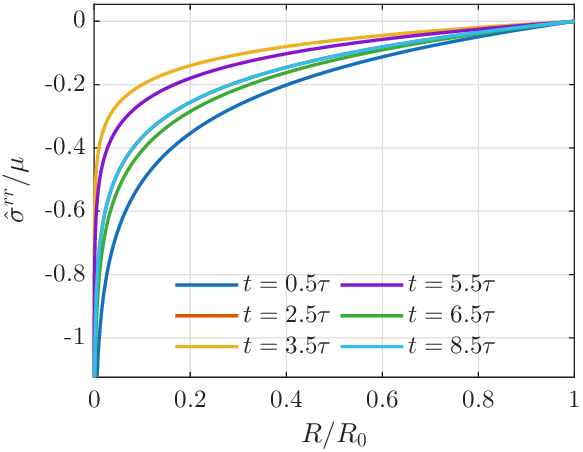}
    \caption{Radial stress}
    \label{fig:ex2-F-rad1}
\end{subfigure}
\begin{subfigure}{0.32\textwidth}
    \centering
    \includegraphics[width=\linewidth]{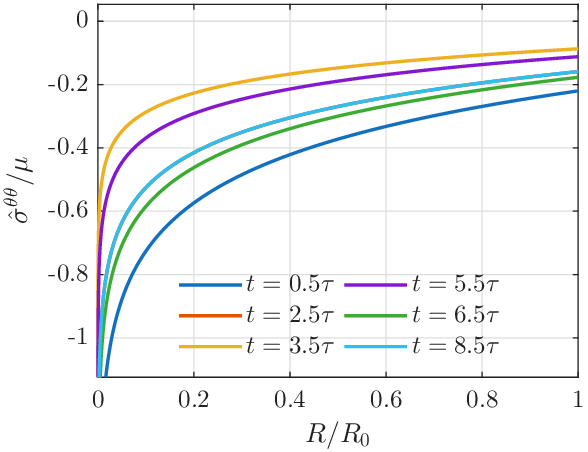}
    \caption{Azimuthal stress}
    \label{fig:ex2-F-rad2}
\end{subfigure}
\begin{subfigure}{0.32\textwidth}
    \centering
    \includegraphics[width=\linewidth]{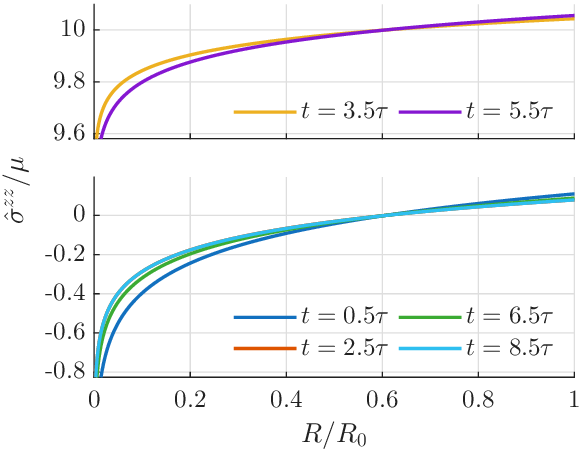}
    \caption{Axial stress }
    \label{fig:ex2-F-rad3}
\end{subfigure}
\caption{Force-controlled loading: stress evolution at fixed radius and stress radial profiles.}
\label{fig:ex2-F-Stress}
\end{figure}

The coupled problem, consisting of the kinetic equations~\eqref{eq:eg2_nH_kinetic} together with the global equilibrium condition~\eqref{eq:eg2_AxialForce}, is solved for $\{\lvr,\lvz,\lambda\}$ by advancing~\eqref{eq:eg2_kinetic} in dimensionless time $\bar{t}$ with a forward-Euler scheme and enforcing the prescribed force~\eqref{eq:force_loading_eg2} through~\eqref{eq:eg2_AxialForce} at each time step via a nonlinear solver in dimensionless space $\bar{R}$. Figs.~\ref{fig:ex2-F-kin}--\ref{fig:ex2-F-Stress} report the resulting response, which may again be read in three stages, following the loading protocol~\eqref{eq:force_loading_eg2}. The force-controlled case is the dual of the displacement-controlled one: the strain is now free and the force prescribed, so the transient at sustained load is creep rather than stress relaxation.

\vskip 0.1in\noindent \emph{i) Initial hold at zero force ($F=0$).}
With no axial force applied and the free ends, the disclination can settle into the body over the time scale $\tau=\eta/\mu_i$. The disclination instantaneously generates an incompatible, residually stressed configuration which then relaxes toward its disclinated elastic equilibrium\textemdash the purely elastic response of the disclinated body~\citep{YavariGoriely2013a}.
During the initial hold, the viscous distortion evolves close to unity with axial and radial expansion and azimuthal contraction compensating one another to maintain incompressibility ($\lvr\lvt\lvz=1$ Fig.~\ref{fig:ex2-F-kin1}).
Crucially, the viscous distortion evolves to match the total deformation gradient; equivalently, the elastic distortion relaxes to the identity, $\hat\Fe\to\mathbf{I}$ (Fig.~\ref{fig:ex2-F-kin2}), in accordance with Remark~\ref{rmrk:longtime_distort}.
The in-plane stresses behave as in displacement control: at $\bar t=0$ both $\hat\sigma^{rr}$ and $\hat\sigma^{\theta\theta}$ attain their largest magnitudes of this stage, reflecting the combined contribution of the equilibrium and non-equilibrium branches, and are then drawn down onto their equilibrium values as the non-equilibrium contribution decays to zero (Figs.~\ref{fig:ex2-F-srr}, \ref{fig:ex2-F-stt}). The axial stress $\hat\sigma^{zz}$, tied to the prescribed force through the equilibrium condition~\eqref{eq:eg2_AxialForce}, remains near zero throughout the hold (Fig.~\ref{fig:ex2-F-szz}). The in-plane stresses stay compressive over this stage (Figs.~\ref{fig:ex2-F-rad1}, \ref{fig:ex2-F-rad2}), a consequence of $1-B^2>0$ for the negative disclination ($B<1$), which fixes the sign of the equilibrium stresses in~\eqref{eq:eg2nH_stress}: the inserted material wedge places the surrounding body in radial and azimuthal compression that intensifies logarithmically toward the axis.

\vskip 0.1in\noindent \emph{ii) Loading--unloading ramps and final hold.}
When the force is ramped to its peak value $F_0$, the stretch responds in two steps (Fig.~\ref{fig:ex2-F-kin1}): an immediate elastic jump as the instantaneous stiffness carries the applied load, followed by a slower \emph{creep} at constant force as the non-equilibrium branch relaxes and the body softens toward its equilibrium response. This creep is the defining feature of the protocol\textemdash the stretch continues to grow while the load $F$ is sustained.
The kinematics otherwise mirror the holds: $\lvz$ tracks the rising $\lambda$ and the elastic distortion returns to the identity, $\hat\Fe\to\mathbf{I}$ (Fig.~\ref{fig:ex2-F-kin2}), with $\lvr$ and $\lvt$ contracting so that $\lvr\,\lvt\,\lvz=1$ is maintained (Fig.~\ref{fig:ex2-F-kin1}).
The stress components respond differently. The in-plane stresses $\hat\sigma^{rr}$ and $\hat\sigma^{\theta\theta}$ display a rise-and-decay transient over the hold that resembles stress relaxation (Figs.~\ref{fig:ex2-F-srr},\ref{fig:ex2-F-stt}); it is not, however, relaxation at fixed strain, but a byproduct of the viscous distortion evolving while the stretch creeps. The axial stress $\hat\sigma^{zz}$, by contrast, carries the applied load and is tied to it through the equilibrium condition~\eqref{eq:eg2_AxialForce}, so it simply tracks the prescribed force: it rises with $F$ and holds flat once $F$ reaches $F_0$, showing no transient of its own (Fig.~\ref{fig:ex2-F-szz}). The prescribed force shows no overshoot, being the controlled quantity (Fig.~\ref{fig:ex2-F-Ft}).
The reverse ramp produces the recovery: an immediate elastic contraction followed by a slower time-dependent return as $\Fv$ relaxes back toward its residual unitary value, and stresses drifting back to their residual baseline (Figs.~\ref{fig:ex2-F-srr}--\ref{fig:ex2-F-szz}).
Note that during the final hold the cylinder returns to the relaxed residual configuration of the initial stage\textemdash the purely elastic disclinated response~\citep{YavariGoriely2013a}\textemdash rather than the $\bar t=0$ instantaneous state.

\vskip 0.1in\noindent\emph{iii) Creep-driven hysteresis.}
The distinct loading and unloading paths form a closed loop in the force--stretch plane (Fig.~\ref{fig:ex2-F-Flam}), whose enclosed area quantifies the energy dissipated over the cycle, with the transient (non-equilibrium) response scaling with $\tau$. The loop reflects the creep occurring during the hold at constant force, which separates the two branches and widens the loop in stretch relative to the displacement-controlled case; it would close onto a single curve in the quasi-static limit $\bar t_0\gg\tau$.

\end{example}

\section{Conclusions} \label{Sec:Conclusions}

The motivation for this work stems from the long-standing challenge of formulating a general geometric and thermodynamically consistent theory that can describe the coupled viscous and anelastic behavior of solids undergoing large deformations. Classical finite viscoelastic theories, which model time-dependent response through rheological analogies or internal variables, and nonlinear anelastic theories, which describe the evolution of natural configurations through eigenstrains, have traditionally been treated as distinct frameworks. A major difficulty has been to reconcile these approaches within a single theory that consistently accounts for both dissipative and recoverable inelastic processes in a geometrically exact setting. In this paper, we propose a nonlinear geometric theory of visco-anelasticity that unifies these two descriptions. The theory is formulated using a multiplicative decomposition of the total deformation gradient into elastic, viscous, and anelastic parts, thereby generalizing the Bilby–Kr\"oner–Lee decomposition to visco-anelastic solids. The material metric is introduced as an evolving field that encodes the local stress-free configuration, enabling a natural representation of both viscous relaxation and permanent anelastic distortions. The governing equations are derived directly from the first and second laws of thermodynamics through a two-potential structure, ensuring full thermodynamic consistency without invoking frame invariance. The resulting framework provides a systematic approach to couple mechanical response, internal evolution, and dissipation in nonlinear solids. In particular, the proposed formulation clarifies the geometric roles of the elastic, viscous, and anelastic distortions and provides a consistent interpretation of the associated intermediate configurations.
A distinguishing feature of the theory is that the elastic, viscous, and anelastic distortions are assigned precise tensorial characters, which clarifies the geometric meaning of the corresponding intermediate configurations. The formulation also shows that in the presence of anelasticity the equilibrium response is governed by the viscoelastic distortion $\Fve=\Fe\Fv$ rather than the total deformation gradient $\mathbf{F}$. Together with the evolving material metric, these geometric ingredients provide a unified description of stress relaxation, residual stress formation, and anelastic distortion.

The numerical results illustrate the predictive capability of the proposed model for both isotropic and anisotropic visco-anelastic materials. They demonstrate the distinct yet coupled roles of viscous and anelastic distortions in stress relaxation, recovery, and the formation of residual stresses. In particular, the examples highlight how the evolution of the material metric captures irreversible effects while the viscous distortion governs time-dependent relaxation, together reproducing complex hysteresis and recovery behavior that cannot be represented by purely viscoelastic or purely anelastic models. These results also demonstrate that the framework is capable of capturing path-dependent responses arising from the interaction of dissipative and non-dissipative mechanisms within a unified setting.

It is worth noting that the practical application of the theory raises further questions concerning its numerical implementation and the calibration of its constitutive functions. The numerical results presented in this work are obtained using a semi-inverse method. The symmetry of the assumed universal-deformation ansatz reduces the governing equations to systems of ordinary and partial differential equations in the radial coordinate and time. These equations are integrated using a finite-difference discretization in space and an implicit time-stepping scheme. The solution of general initial-boundary-value problems, however, would require a finite element implementation of the theory. Such an implementation is expected to present several nontrivial challenges. In particular, it must address the time integration of the nonlinear kinetic equations governing the viscous and anelastic distortions, as well as the enforcement of any incompressibility constraints. Moreover, given the geometric structure of the theory, a standard displacement-based discretization may not be the most natural choice. Compatible-strain mixed finite element methods have been developed for compressible and incompressible nonlinear elasticity so as to respect the geometric structure of the governing equations \citep{angoshtari2017compatible,shojaei2018compatible,shojaei2019compatible}. Extending these methods to visco-anelasticity appears to us a natural point of departure. Such an extension must account for the time evolution and non-flatness of the material metric and must also discretize the viscous and anelastic distortions. These questions are best addressed in a dedicated computational study for which the present formulation provides a foundation. The semi-analytical solutions presented herein may then serve as benchmark problems for verifying candidate implementations.

Similar considerations apply to the constitutive functions adopted in the examples. The neo-Hookean energies and the quadratic isotropic dissipation potential are chosen solely for illustration, owing to their simplicity and compatibility with the thermodynamic admissibility conditions. They are not intended as calibrated models of any particular material. The admissibility conditions delimit \emph{a priori} the set of thermodynamically permissible material parameters and thereby constrain any identification procedure from the outset. Nevertheless, calibrating the free-energy and dissipation potentials for a given visco-anelastic solid remains a problem of experimental characterization. Such calibration would require experiments capable of separating the equilibrium, non-equilibrium, and anelastic contributions to the material response and therefore falls outside the scope of the present work.

In conclusion, the present formulation provides a general foundation for modeling a wide range of time-dependent and history-dependent material phenomena. Beyond the applications shown here, the theory can be extended to describe temperature-dependent visco-anelasticity, aging and damage in polymers, and active processes in biological or synthetic materials with evolving internal structures. In particular, the geometric structure of the theory makes it well suited for coupling with models of growth, remodeling, and defect evolution, where the material metric plays a central role.

\section*{Acknowledgement}

This work was partially supported by NSF -- Grant No. CMMI 1939901.

\bibliographystyle{abbrvnat}
\bibliography{ViscoAn_ref}

\section*{Appendices}

\appendix

\section{Derivatives of the principal invariants} \label{appendix:derivatives}

Given a scalar-valued function $f(\mathbf{C})$ differentiable at $\mathbf{C}$, an arbitrary material $\binom{2}{0}$-rank tensor $\mathbf{H}$,\footnote{In components, ${\mathbf{H}=\mathrm H_{AB}\,\mathrm{d}X^A \otimes \mathrm{d}X^B}$.} and a small non-negative scalar $\epsilon$, one may write
\begin{equation}
	f(\mathbf{C}+\epsilon\,\mathbf{H})=f(\mathbf{C})
	+\frac{\partial f}{\partial \mathbf{C}}\!:\!\mathbf{H}\,\epsilon+o(\epsilon)
	\,.
\end{equation}
For $I_1=\operatorname{tr}_{\Go}\Cve=\cCve_{AB}\mathring{\mathrm G}^{AB}$, one finds
\begin{equation}
	\frac{\partial I_1}{\partial\mathbf C}=\Go^{-1}\,.
\end{equation}
For $I_3=\det \Cve$, one may write
\begin{equation} \label{Determinant-Variation}
\begin{aligned}
	\det(\Cve+\epsilon\,\mathbf{H}) 
	&=\det\!\left[\Cve\!\left(\mathbf{I}+\Cve^{-1}\,\epsilon\,\mathbf{H}\right)\!
	\right]\! \\
	& =\det \Cve\,\det\!\left(\mathbf{I}+\epsilon\,\Cve^{-1}\mathbf{H}\right)\! \\
	& =\det \Cve\!\left[1+\epsilon\,\operatorname{tr}\!\left(\Cve^{-1}\mathbf{H}\right)\!+o(\epsilon)\right]\!\\
	& =\det \Cve + (\det \Cve)\,\Cve^{-1}\!:\mathbf{H}\,\epsilon+o(\epsilon)\,.
\end{aligned}
\end{equation}
Which then implies that
\begin{equation}
	\frac{\partial I_3}{\partial\cCve}=I_3\Cve^{-1}\,.
\end{equation}
In components, it reads ${\partial I_3}/{\partial \cCve_{AB}}=I_3(\cCve^{-1})^{AB}$.
For $I_2=(\det\Cve)\operatorname{tr}_{\Go}(\Cve^{-1})=I_3\,\operatorname{tr}_{\Go}(\Cve^{-1})$,\footnote{\label{Cayley-Hamilton}The characteristic polynomial of $\Cve$ is written as $\lambda^3-I_1\,\lambda^2+I_2\,\lambda -I_3=0$. The Cayley-Hamilton theorem tells us that $\Cve^3-I_1\,\Cve^2+I_2\,\Cve -I_3\,\Go=\mathbf{0}$. Multiplying both sides by $\Cve^{-1}$, one concludes that $I_3\,\Cve^{-1}=\Cve^2-I_1\,\Cve+I_2\,\Go $. This, in particular, implies that $I_2=\frac{1}{2}\!\left(I_1^2-\operatorname{tr}_{\Go}\Cve^2\right)\!=I_3\,\operatorname{tr}_{\Go}(\Cve^{-1})$.} one may write
\begin{equation}
\begin{split}
	\frac{\partial I_2}{\partial\Cve}
	&=\frac{\partial I_3}{\partial\Cve}\operatorname{tr}_{\Go}(\Cve^{-1})
	+I_3\,\frac{\partial}{\partial\Cve}[\operatorname{tr}_{\Go}(\Cve^{-1})]
	\\&
	=I_3\,\Cve^{-1}\operatorname{tr}_{\Go}(\Cve^{-1})
	+I_3\frac{\partial}{\partial\Cve}[\operatorname{tr}_{\Go}(\Cve^{-1})]
	\\&
	=I_2\,\Cve^{-1}+I_3\frac{\partial}{\partial\Cve}[\operatorname{tr}_{\Go}(\Cve^{-1})]\,.
\end{split}
\end{equation}
The second term on the right-hand side above may be calculated as follows:
\begin{equation}
\begin{aligned}
	\!\left(\Cve+\epsilon\,\mathbf{H}\right)^{-1}
	&=\!\left[\Cve\!\left(\mathbf{I}+\epsilon\,\Cve^{-1}\mathbf{H}\right)\!\right]^{-1}
	=\!\left(\mathbf{I}+\epsilon\,\Cve^{-1}\mathbf{H}\right)^{-1}\Cve^{-1}\\
	&=\!\left[\mathbf{I}-\!\left(-\epsilon\,\Cve^{-1}\mathbf{H}\right)\!\right]^{-1}\Cve^{-1}\\
	&=\!\left[\mathbf{I}+\!\left(-\epsilon\,\Cve^{-1}\mathbf{H}\right)\!+\!\left(-\epsilon\,\Cve^{-1}\mathbf{H}\right)^2+o(\epsilon^2)\right]\!\Cve^{-1} \\
	&=\Cve^{-1}-\epsilon\!\left(\Cve^{-1}\mathbf{H}\Cve^{-1}\right)\!
	+\epsilon^2\!\left(\Cve^{-1}\mathbf{H}\right)^2\Cve^{-1}+o(\epsilon^2)
	\,.
\end{aligned}
\end{equation}
Thus
\begin{equation}
	\operatorname{tr}_{\Go}\!\left(\Cve+\epsilon\,\mathbf{H}\right)^{-1}
	=\operatorname{tr}_{\Go}(\Cve^{-1})-\epsilon\operatorname{tr}_{\Go}\!\left(\Cve^{-1}\mathbf{H}\Cve^{-1}\right)\!
	+\epsilon^2\operatorname{tr}_{\Go}\!\left[\!\left(\Cve^{-1}\mathbf{H}\right)^2\Cve^{-1}\right]\!+o(\epsilon^2)\,.
\end{equation}
Note that since
\begin{equation}
	\operatorname{tr}_{\Go}\!\left(\Cve^{-1}\mathbf{H}\Cve^{-1}\right)\!
	=\operatorname{tr}_{\Go}
	\!\left(\Cve^{-1} \Go \Cve^{-1}\mathbf{H}\right)\!
	=\Cve^{-2}\!:\mathbf{H}\,,
\end{equation}
it follows that
\begin{equation}
	\frac{\partial}{\partial\Cve}[\operatorname{tr}_{\Go}(\Cve^{-1})] = -\Cve^{-2}\,.
\end{equation}
Therefore
\begin{equation}
	\frac{\partial I_2}{\partial\Cve}=I_2\,\Cve^{-1}-I_3\,\Cve^{-2}\,.
\end{equation}
Similarly,
\begin{equation}\label{eq:part_Ie}
	\frac{\partial \Ie_1}{\partial\Ce}=\Go^{-1}\,,\qquad 
	\frac{\partial \Ie_2}{\partial\Ce}=\Ie_2\Ce^{-\sharpo}-\Ie_3\Ce^{-2\sharpo}\,,\qquad
	\frac{\partial \Ie_3}{\partial\Ce}=\Ie_3\Ce^{-\sharpo}\,.
\end{equation}

\end{document}

%% file: Disclination.pdf_tex
\begingroup%
  \makeatletter%
  \providecommand\color[2][]{%
    \errmessage{(Inkscape) Color is used for the text in Inkscape, but the package 'color.sty' is not loaded}%
    \renewcommand\color[2][]{}%
  }%
  \providecommand\transparent[1]{%
    \errmessage{(Inkscape) Transparency is used (non-zero) for the text in Inkscape, but the package 'transparent.sty' is not loaded}%
    \renewcommand\transparent[1]{}%
  }%
  \providecommand\rotatebox[2]{#2}%
  \newcommand*\fsize{\dimexpr\f@size pt\relax}%
  \newcommand*\lineheight[1]{\fontsize{\fsize}{#1\fsize}\selectfont}%
  \ifx\svgwidth\undefined%
    \setlength{\unitlength}{559.30661179bp}%
    \ifx\svgscale\undefined%
      \relax%
    \else%
      \setlength{\unitlength}{\unitlength * \real{\svgscale}}%
    \fi%
  \else%
    \setlength{\unitlength}{\svgwidth}%
  \fi%
  \global\let\svgwidth\undefined%
  \global\let\svgscale\undefined%
  \makeatother%
  \begin{picture}(1,0.53564856)%
    \lineheight{1}%
    \setlength\tabcolsep{0pt}%
    \put(0,0){\includegraphics[width=\unitlength,page=1]{Disclination.pdf}}%
    \put(0.72887307,0.08175253){\color[rgb]{0,0,0}\makebox(0,0)[lt]{\lineheight{1.25}\smash{\begin{tabular}[t]{l}$\mathcal{B}$\end{tabular}}}}%
    \put(0.50093911,0.38322643){\color[rgb]{0,0,0}\makebox(0,0)[lt]{\lineheight{1.25}\smash{\begin{tabular}[t]{l}$\Theta_0$\end{tabular}}}}%
    \put(0.86096409,0.51887627){\color[rgb]{0,0,0}\makebox(0,0)[t]{\lineheight{1.25}\smash{\begin{tabular}[t]{c}$Z$\end{tabular}}}}%
    \put(0,0){\includegraphics[width=\unitlength,page=2]{Disclination.pdf}}%
    \put(0.00804574,0.08175253){\color[rgb]{0,0,0}\makebox(0,0)[lt]{\lineheight{1.25}\smash{\begin{tabular}[t]{l}$\mathcal{B}_0$\end{tabular}}}}%
    \put(0,0){\includegraphics[width=\unitlength,page=3]{Disclination.pdf}}%
    \put(0.13937197,0.51879411){\color[rgb]{0,0,0}\makebox(0,0)[t]{\lineheight{1.25}\smash{\begin{tabular}[t]{c}$Z$\end{tabular}}}}%
    \put(0,0){\includegraphics[width=\unitlength,page=4]{Disclination.pdf}}%
    \put(0.32045918,0.21503101){\color[rgb]{0,0,0}\transparent{0.80000001}\makebox(0,0)[t]{\lineheight{1.25}\smash{\begin{tabular}[t]{c}$\scalebox{2}{$\cup$}$\end{tabular}}}}%
    \put(0.85983238,0.00451525){\color[rgb]{0,0,0}\makebox(0,0)[t]{\lineheight{1.25}\smash{\begin{tabular}[t]{c}(b)\end{tabular}}}}%
    \put(0.26028105,0.00500152){\color[rgb]{0,0,0}\makebox(0,0)[t]{\lineheight{1.25}\smash{\begin{tabular}[t]{c}(a)\end{tabular}}}}%
  \end{picture}%
\endgroup%